\documentclass{JHEP3} 

\usepackage{graphicx}

\usepackage{amsmath}
\usepackage{amsfonts}   
\usepackage{amssymb}    

\def\e3{$\epsilon_3$}

\def\ch2{$\chi^2$}

\def\co#1{{\ifmmode{\cal O}_{#1}\else${\cal O}_{#1}$\fi}}

\newdimen\unit
\def\point#1 #2 #3{\vbox to0pt{\kern-#2\unit
 \hbox{\kern#1\unit#3}\vss}
\nointerlineskip}

\newcommand{\be}{\begin{equation}}
\newcommand{\ee}{\end{equation}}
\newcommand{\bea}{\begin{eqnarray}}
\newcommand{\eea}{\end{eqnarray}}

\newcommand\ps{\mbox{ ps}} 
\newcommand{\mev}{\mbox{ MeV}}
\newcommand{\gev}{\mbox{ GeV}}
\newcommand{\tev}{\mbox{ TeV}}

\newcommand{\cl}{\text{CL}}

\newcommand{\alphaemmz}{\alpha_{\text{em}}(M_Z)^{\overline{MS}}}
\newcommand{\alphas}{\alpha_s(M_Z)^{\overline{MS}}}
\newcommand{\like}{\mathcal{L}}

\newcount\hour
\newcount\minute
\newtoks\amorpm
\hour=\time\divide\hour by60 \minute=\time{\multiply\hour by60
\global\advance\minute by- \hour}
\edef\standardtime{{\ifnum\hour<12 \global\amorpm={am}%
   \else\global\amorpm={pm}\advance\hour by-12 \fi
   \ifnum\hour=0 \hour=12 \fi
   \number\hour:\ifnum\minute<100\fi\number\minute\the\amorpm}}
\edef\militarytime{\number\hour:\ifnum\minute<100\fi\number\minute}
\def\bold#1{\setbox0=\hbox{$#1$}%
    \kern-.025em\copy0\kern-\wd0
    \kern.05em\copy0\kern-\wd0
    \kern-.025em\raise.0433em\box0 }

\newcommand{\newc}{\newcommand}
\newc\eg{{\rm {e.g.}}}  \newc\etal{{\rm {et al.}}} \newc\ie{{\rm i.e.}}
\newc\etc{{\rm {etc}}}
\newcommand\lsim{\mathrel{\rlap{\lower4pt\hbox{\hskip1pt$\sim$}}
   \raise1pt\hbox{$<$}}}
\newcommand\gsim{\mathrel{\rlap{\lower4pt\hbox{\hskip1pt$\sim$}}
   \raise1pt\hbox{$>$}}}
\newc{\mhalf}{m_{1/2}}      \newc{\mzero}{m_0}
\newc{\tanb}{\tan\beta}
\newc{\azero}{A_0}
\newc{\at}{A_t} \newc{\ab}{A_b} \newc{\atau}{A_\tau}
\newc{\bmu}{B\mu}           \newc{\sgn}{{\rm sgn}}
\newc{\mone}{M_1}           \newc{\mtwo}{M_2}

\newc{\charone}{\chi_1^\pm} \newc{\mcharone}{m_{\chi_1^\pm}}

\newc{\hl}{h}               \newc{\mhl}{m_{\hl}}   \newc{\gammahl}{\Gamma_{\hl}}
\newc{\hh}{H}               \newc{\mhh}{m_{\hh}}   \newc{\gammahh}{\Gamma_{\hh}}
\newc{\ha}{A}               \newc{\mha}{m_{\ha}}   \newc{\gammaha}{\Gamma_{\ha}}
\newc{\hpm}{H^{\pm}}        \newc{\mhpm}{m_{\hpm}} \newc{\gammahpm}{\Gamma_{\hpm}}
\newc{\hp}{H^{+}} \newc{\mhp}{m_{\hp}} \newc{\hm}{H^{-}}
\newc{\mhm}{m_{\hm}}
\newc{\xt}{X_{t}}           \newc{\xb}{X_{b}}

\newc{\qzero}{Q_0}          \newc{\qstop}{Q_{\widetilde t}}
\newc{\amu}{a_{\mu}}        \newc{\amususy}{a_{\mu}^{\text{SUSY}}}
\newc{\amuexpt}{a_{\mu}^{\text{expt}}}        \newc{\amusm}{a_{\mu}^{\text{SM}}}
\newc{\deltaamususy}{\delta a_{\mu}^{\text{SUSY}}}
\newcommand{\gmt}{\deltaamususy}
\newcommand{\bsg}{\bsgamma}
\newc\gmtwo{(g-2)_{\mu}} \newc\deltaamu{\Delta a_{\mu}}
\newc{\msbar}{\overline{MS}} \newc{\drbar}{\overline{DR}}
\newc{\yt}{h_t} \newc{\yb}{h_b} \newc{\ytau}{h_{\tau}}

\newc{\mtop}{m_t}               \newc{\mtpole}{M_t}
\newc{\mtaupole}{m_{\tau}^{\text{pole}}}
\newc{\mtmtsmmsbar}{m_t(m_t)^{\msbar}_{{\text{SM}}}}
\newc{\mtmtsmdrbar}{m_t(m_t)^{\drbar}_{{\text{SM}}}}
\newc{\mtmtmssmdrbar}{m_t(m_t)^{\drbar}_{{\text{SUSY}}}}

\newc{\mbmbmsbar}{m_b(m_b)^{\msbar} }

\newc{\mbmbsmmsbar}{m_b(m_b)^{\msbar}_{{\text{SM}}}}
\newc{\mbmzsmmsbar}{m_b(\mz)^{\msbar}_{{\text{SM}}}}
\newc{\mbmzsmdrbar}{m_b(\mz)^{\drbar}_{{\text{SM}}}}
\newc{\mbmzmssmdrbar}{m_b(\mz)^{\drbar}_{{\text{SUSY}}}}

\newc{\mtaumzsmmsbar}{m_{\tau}(\mz)^{\msbar}_{{\text{SM}}}}
\newc{\mtaumzsmdrbar}{m_{\tau}(\mz)^{\drbar}_{{\text{SM}}}}
\newc{\mtaumzmssmdrbar}{m_{\tau}(\mz)^{\drbar}_{{\text{SUSY}}}}

\newc{\mgut}{M_{\rm GUT}}
\newc{\mplanck}{M_{\rm P}}      \newc{\mpl}{M_{\text{Pl}}}
\newc{\msusy}{M_{\rm SUSY}}      \newc{\ms}{M_{\text{S}}}
\newc{\jxf}{J({\xf})}
\newc{\jxfexact}{J_{\rm exact}({\xf})}  \newc{\jxfexp}{J_{\rm exp}({\xf})}
\newc{\VEV}[1]{\langle #1 \rangle}
\newc{\xf}{x_f}
\newc\vrel{v_{\rm rel}}
\newcommand\mchi{m_{\chi}}              
\newc\sell{{\widetilde e}_L}      \newc\msell{m_{\sell}}
\newc\selr{{\widetilde e}_R}      \newc\mselr{m_{\selr}}
\newc\snue{{\widetilde \nu}_e}      \newc\msnue{m_{\snue}}
\newc\snutau{{\widetilde \nu}_\tau}      \newc\msnutau{m_{\snutau}}
\newc\supl{{\widetilde u}_L}      \newc\msupl{m_{\supl}}
\newc\supr{{\widetilde u}_R}      \newc\msupr{m_{\supr}}
\newc\sdl{{\widetilde d}_L}      \newc\msdl{m_{\sdl}}
\newc\sdr{{\widetilde d}_R}      \newc\msdr{m_{\sdr}}

\newcommand\stopone{{\widetilde t}_1}   \newcommand\mstopone{m_{\stopone}}
\newcommand\stoptwo{{\widetilde t}_2}   \newcommand\mstoptwo{m_{\stoptwo}}

\newcommand\mgluino{m_{\widetilde g}}

\newc\sfermion{\tilde f}  \newc\msfermion{m_{\sfermion}}
\newc\cmeter{{\rm cm}} \newc\meter{{\rm m}} \newc\kmeter{{\rm km}}
\newc\second{{\rm sec}}
\newc{\gstar}{g_\ast}           \newc{\gsstar}{g_{s\ast}}
\newc{\geff}{g_{\rm eff}}
\newcommand\mz{m_{Z}}

\newc{\sthw}{\sin\theta_W}              \newc{\cthw}{\cos\theta_W}
\newc{\bino}{\widetilde B}              \newc{\wino}{\widetilde W_30}
\newc{\higgsinob}{{\widetilde H}^0_b}   \newc{\higgsinot}{{\widetilde H}^0_t}
\newc{\abund}{\Omega h^2}
\newc{\abundchi}{\Omega_\chi h^2}
\newc{\abundcdm}{\Omega_{\text{CDM}} h^2}
\newc{\omegam}{\Omega_{M}}       \newc{\abundm}{\Omega_{M} h^2}
\newc{\omegab}{\Omega_{b}}       \newc{\abundb}{\Omega_{b} h^2}
\newc{\omegacdm}{\Omega_{CDM}}
\newc{\omegatot}{\Omega_{TOT}}
\newc{\rhocrit}{\rho_{crit}}
\newc{\rhochi}{\rho_{\chi}}
\newcommand\pb{\,\mbox{pb}}

\newc\BR{BR}
\newc\bsgamma{b\rightarrow s \gamma }
\newc\bxsgamma{\overline{B}\rightarrow X_{s}\gamma}
\newc\brbsgamma{\BR(\overline{B}\rightarrow X_s\gamma)}

\newcommand\brbsmumu{\BR(\overline{B}_s\to\mu^+\mu^-)}

\newcommand\bbbarmix{\overline{B}_s\mbox{--}B_s}      
\newcommand\delmbs{\Delta M_{B_s}}
\newcommand\brbtaunu{\BR(\overline{B}_u\to \tau \nu)}

\newc{\beq}{\begin{equation}}
\newc{\eeq}{\end{equation}}

\newc\stoponetwo{{\widetilde t}_{1,2}}
\newc\sbotonetwo{{\widetilde b}_{1,2}}
\newc\stauonetwo{{\widetilde \tau}_{1,2}}

\newc{\sigsip}{\sigma^{SI}_{p}} \newc{\sigsin}{\sigma^{SI}_{n}}
\newc{\sigsiN}{\sigma^{SI}_{N}}
\newc{\sigsdp}{\sigma^{SD}_{p}} \newc{\sigsdn}{\sigma^{SD}_{n}}
\newc{\sigsiA}{\sigma^{SI}_{A}}

\newc\xilim{\xi_{\rm lim}} 
\newc\tlim{t_{\rm lim}} 
\newc\zetalim{\zeta_{\rm lim}} 

\newc\zetah{\zeta_h}
\newc{\relprobone}[1]{p({#1} \vert d)}
\newc{\relprobtwo}[2]{p({#1},{#2} \vert d)}

\long\def\begincomment#1\endcomment{%
       \begingroup\sf\baselineskip12pt#1\endgroup}

\newcommand{\squishlist}{
  \begin{list}{$\bullet$}
   { \setlength{\itemsep}{0pt}      \setlength{\parsep}{3pt}
     \setlength{\topsep}{3pt}       \setlength{\partopsep}{0pt}
     \setlength{\leftmargin}{1.em} \setlength{\labelwidth}{1em}
     \setlength{\labelsep}{0.5em} } }
\newcommand{\squishend}{
   \end{list}  }
        
\newcommand{\data}{d}
\newcommand{\nuis}{\psi}
\newcommand{\params}{\theta}
\newcommand{\basis}{m}
\newcommand{\derived}{\xi}

\newcommand{\sineff}{\sin^2 \theta_{\rm{eff}}}

\newcommand\rpp[3]  {
		{{\it Rept.\ Prog.\ Phys.\ }{\bf #1} (#2) #3}}

\newcommand{\dr}{{\rm d}}
\newcommand{\prof}{{\mathfrak L}}
\newcommand{\model}{{\scriptsize{ \rm model}}}

\newcommand{ \KL }{D_{\rm KL}}

\newcommand{\chisq}{\chi^2}
\newcommand{\chisqmin}{\chi^2_{\rm min}}
\newc{\ww}{0.49\linewidth} 
\newc{\ttr}{0.32\linewidth} 
\newc{\qq}{0.24\linewidth}

\newcommand{\fbasis}{\mathcal{F}}

\title{The impact of priors and observables on parameter inferences in the
 Constrained MSSM}

\author{Roberto Trotta\\
	  Astrophysics Group, Imperial College London \\
	  Blackett Laboratory, Prince Consort Road, London SW7 2AZ, UK \\ and \\
       Astrophysics Department, Oxford University \\
       Denys Wilkinson Building,  Keble Road, Oxford OX1 3RH, UK\\
        E-mail: \email{r.trotta@imperial.ac.uk}}
\author{Farhan Feroz\\
       Astrophysics Group, Cavendish Laboratory, University of Cambridge \\
       J.J. Thomson Avenue, 	Cambridge, CB3 0HE, UK \\
        E-mail: \email{ff235@mrao.cam.ac.uk}}

\author{Mike Hobson\\
       Astrophysics Group, Cavendish Laboratory, University of Cambridge \\
       J.J. Thomson Avenue, 	Cambridge, CB3 0HE, UK \\
        E-mail: \email{mph@mrao.cam.ac.uk}}

\author{Leszek Roszkowski\\
       Department of Physics and Astronomy, University of Sheffield,\\
       Sheffield S3 7RH, England
       E-mail: \email {L.Roszkowski@sheffield.ac.uk}}

\author{Roberto Ruiz de Austri\\
       Departamento de F\'{\i}sica Te\'{o}rica C-XI
       and Instituto de F\'{\i}sica Te\'{o}rica C-XVI,\\
       Universidad Aut\'{o}noma de Madrid, Cantoblanco,
       28049 Madrid, Spain\\
       E-mail: \email{rruiz@delta.ft.uam.es}}

\abstract{
We use a newly released version of the \texttt{SuperBayeS} code to analyze the impact of the
choice of priors and the influence of various constraints on the
statistical conclusions for the preferred values of the parameters of
the Constrained MSSM.
We assess the effect in a Bayesian framework and compare it with an
alternative likelihood-based measure of a profile likelihood. We
employ a new scanning algorithm (MultiNest) which increases
the computational efficiency by a factor $\sim 200$ with respect to
previously used techniques.  
We demonstrate that the currently
available data are not yet sufficiently constraining to allow one to
determine the preferred values of CMSSM parameters in a way that is completely independent of the choice of priors and
statistical measures. 
While $\brbsgamma$ generally favors large
$\mzero$, this is in some contrast with 
the preference for
low values of $\mzero$ and $\mhalf$ that is almost entirely a consequence of a
combination of prior effects and a single constraint coming from the
anomalous magnetic moment of the muon, which remains somewhat
controversial. 
Using an information-theoretical measure, we find
that the
cosmological dark matter abundance determination provides at least 80\% of the total constraining power of all available observables. Despite the
remaining uncertainties, prospects for direct detection in the CMSSM
remain excellent, with 
the spin-independent neutralino-proton cross section almost guaranteed
above $\sigsip\sim 10^{-10}\pb$, independently of the choice of priors
or statistics. Likewise, gluino and lightest Higgs discovery at the
LHC remain highly encouraging. While in this work we have used the
CMSSM as particle physics model, our formalism and scanning technique
can be readily applied to a wider class of models with several free
parameters.  }

\preprint{}
\begin{document}
\section{Introduction}\label{sec:intro}

Experiments at the Large Hadron Collider (LHC) will soon start testing
many frameworks of particle physics beyond the Standard Model
(SM). Particular attention will be given to the Minimal Supersymmetric
SM (MSSM) and other effective low-energy models involving
softly-broken supersymmetry (SUSY) which remain by far the most
theoretically developed and popular schemes. On another front, dark
matter (DM) experiments have by now reached the level of sensitivity that
would allow them to detect a signal from DM if it is
made up of the lightest neutralino, whose abundance as cold dark
matter (CDM) is now very well constrained thanks to WMAP and other
cosmic microwave background observations.  With enough effort, Tevatron experiments
may be able to improve the final LEP limit on the SM-like Higgs boson,
and perhaps even detect it. Heavy quark experiments continue improving
constraints on allowed contributions from ``new physics'' (be it SUSY
or some other framework) to several observables related to
flavor. Finally, an apparent discrepancy, at the level of about
$3\sigma$, between experiment and SM predictions (based on $e^+e^-$
data) for the anomalous magnetic moment of the muon, has now persisted
for several years.

In light of the expected vast improvement in the constraining power of
data from the LHC and DM searches, it is essential to develop a solid
formalism to allow one to fully explore properties of popular
low-energy SUSY and other models, and to reliably derive ensuing
experimental implications. Until a few years ago, a somewhat
oversimplified approach based on fixed-grid scans of subsets of
parameter space was sufficient. Such scans imposed observational
constraints on the grid in a rigid ``in-or-out'' fashion (e.g., points
outside some arbitrary 1 or $2\sigma$ experimental range of a given
observables were discarded), without paying attention to the varying
degree with which points could reproduce the data.  The points on the
grid surviving all the constraints were then used to qualitatively
evaluate the impact of thus applied data and ensuing predictions
for various observables. A major drawback of the approach was,
however, that it did not allow for a probabilistic interpretation of
results.  A step in the right direction was to employ a chi-square
analysis where, for example, the question of more properly weighting
experimental errors could be addressed~\cite{ellis,ehow06,
buchellislatest}. However, the approach remains of limited use as it
does not allow one to perform a full scan over all relevant
parameters. A major improvement in this direction has been provided by
employing a Markov Chain Monte Carlo (MCMC) algorithm~\cite{bg04}, linked with Bayesian
statistics~\cite{al05,rtr1}.

Bayesian methods coupled with MCMC technology are superior in many
respects to traditional, frequentist grid scans of the parameter
space.  (For an introduction, see,
e.g.,~\cite{Trotta:2008qt,BayesianRevs}.) For a start, they are much
more efficient, in that the computational effort required to explore a
parameter space of dimension $N$ scales roughly proportionally with
$N$. In contrast, on a grid scan with $k$ points per dimension, the
number of likelihood evaluations required goes as $k^N$, hence this
approach becomes computationally prohibitive even for parameter space
of moderate dimensionality. Secondly, the Bayesian approach allows one
to easily incorporate into the final inference all relevant sources of
uncertainty. For a given SUSY model one can include relevant SM
(nuisance) parameters and their associated experimental errors, with
the uncertainties automatically propagated to give the final
uncertainty on the SUSY parameters of interest.  In addition,
theoretical uncertainties can be easily included in the likelihood
(see~\cite{rtr1}). Thirdly, another key advantage is the possibility
to marginalize (i.e., integrate over) additional (``hidden'')
dimensions in the parameter space of interest with very little
computational effort. By ``hidden dimensions'' we mean here the
parameters others than the ones being plotted, for example in 1
dimensional or 2 dimensional plots. In this paper, we upgrade our
scanning technique to a much more efficient algorithm called
``MultiNest''~\cite{Feroz:2007kg}, which reduces very significantly
the computational burden of a full exploration of the parameter space.

These advantages are built into the Bayesian procedure. The latter
also requires the specification of a prior probability distribution function (or simply
prior), describing our state of knowledge about the problem before we
see the data. One of the main aims of this study is to assess the
influence of prior choice on the statistical conclusions on CMSSM
parameters.  A number of recent studies have
investigated the impact of several choices of priors on the parameter
inference ~\cite{bg04, al05,rtr1, alw06, rrt2, rrt3, Trotta:2006ew,
rrts, Allanach:2008tu, BenModelComp,allanach06,bclw07,Allanach:2008iq}
in the context of the Constrained Minimal Supersymmetric Standard
Model (CMSSM)~\cite{kkrw94}, and found it to be rather strong. The
CMSSM, because of its relative simplicity, is a model of much
interest.

The goal of the paper is twofold.  On one side we address the question
of the origin of the strong prior dependence. First, we point out and
examine the impact on SUSY parameter inference from the highly
non-linear nature of the mapping from the CMSSM parameters to the
observable quantities. Next, we adopt two different priors (flat on a
linear scale and flat on a log scale, see below). Within each we
explore in detail, and compare, the impact of several observables
which have been known to play a major role in constraining the CMSSM
parameter space, including LEP bounds on Higgs properties,
$\brbsgamma$, the relic abundance $\abundchi$ of the lightest
neutralino assumed to constitute most of CDM in the Universe, and the
anomalous magnetic moment of the muon $\gmtwo$. It is the last
observable that we find to play a singular role in favoring lower
values of superparners, in some tension with some other observables,
especially $\brbsgamma$ which favors larger scalar
masses~\cite{rrt3}.

The other major aim of our paper is to compare the Bayesian posterior
probability distribution with the statistical measure of a profile
likelihood in the context of prior dependence. We conclude that the
profile likelihood may provide a more robust assessment of the favored
regions of CMSSM parameters with respect to volume effects generated
by the prior choice. The coverage properties of this measure will be
studied elsewhere.  We focus here on the CMSSM which we treat as a
case study. The problem of prior dependence is likely to be even more
severe for more complicated SUSY models given present constraints,
although better data such as, e.g., sparticle and Higgs detection at LHC are
expected to cure it.

The paper is organized as follows. In section~\ref{sec:theory} we
review the statistical formalism used in this work. In
section~\ref{sec:cmssm} we focus on the CMSSM and introduce our
experimental constraints, before exploring in section~\ref{sec:priors}
the impact of priors and observables on inferences on the SUSY
parameter space. In section~\ref{sec:bsgvsgmt} we examine in more
detail the consistency of the various observational constraints and
focus in particular on the tension between $\gmtwo$ and
$\brbsgamma$. We also quantify the information content (i.e., the
constraining power) of each observable. Implications of parameter
inferences on gluino and light Higgs searches at the LHC and on direct
detection searches of DM are outlined in section~\ref{sec:det}, and
our conclusions are presented in section~\ref{sec:conclusions}.  In
Appendix~\ref{app:nest} we give a brief description of the MultiNest
algorithm.


\section{Statistical formalism}
\label{sec:theory}

\subsection{Statistical framework}
\label{sec:stats}

Let us denote a set of parameters of a model under
consideration by $\params$, and by $\nuis$  all other relevant (so-called
{\em ``nuisance parameters''}). Both sets form our {\em ``basis parameters''} 
\be
\basis = (\params,\nuis).
\label{basis:eq}
\ee
The cornerstone of Bayesian inference is provided by Bayes' theorem, which reads
\be \label{eq:bayes}
p(\basis | \data) = \frac{p(\data |
\derived) \pi(\basis)}{p(\data)}. \ee
The quantity $p(\basis | \data)$ on the l.h.s. of eq.~\eqref{eq:bayes}
is called a {\em posterior probability density function} (posterior
pdf, or simply a
{\em posterior}).  On the r.h.s., the quantity $p(\data | \derived)$,
taken as a function of $\derived$ for {\em fixed data} $\data$, is
called the {\em likelihood} (where the dependence $\derived(\basis)$
is understood).  The likelihood supplies the information provided by
the data. In the case of the CMSSM which we will consider below, it is
constructed in Sec.~3.1 of ref.~\cite{rtr1}.  The quantity
$\pi(\basis)$ denotes a {\em prior probability density function}
(prior pdf, or simply a {\em prior}) which encodes our state of knowledge about the
values of the parameters in $\basis$ before we see the data. The prior state
of knowledge is then updated to the posterior via the likelihood.  Much
care must be exercised in assessing the impact of priors on the final
inference on the model's properties. If the posterior strongly depends
on the choice of priors, then this is a signal that the available data
is not sufficiently constraining to override the prior, and hence the
information content of the posterior is strongly influenced by
the choice of the prior. Therefore judgement must be suspended until
more constraining data becomes available, unless there is a physically
strong motivation for a specific choice of priors. (For example, in some
simple situations the prior follows from considerations of the
invariance properties of the problem.)

Finally, the quantity in the denominator is called {\em evidence} or
{\em model likelihood}. If one is interested in constraining the
model's parameters, the evidence is merely a normalization constant,
independent of $\basis$, and can therefore be dropped. However, the
evidence is very useful in the context of Bayesian model
comparison (see e.g.~\cite{Trotta:2005ar}) but in this work we will use it instead to quantify the constraining power of each observable.  The evidence is a
multi-dimensional integral over the model's parameter space $\basis$
(including nuisance parameters),\footnote{More precisely, one should
write for the evidence $p(\data | \model)$, in order to show
explicitly that it is conditional on the assumption that the model is
the true theory. From there one can further employ Bayes' theorem to
obtain the posterior probability for the model's parameters given the
observed data, namely $p(\model | \data)$. This is the subject of
Bayesian model comparison (see e.g.~\cite{Trotta:2005ar} for an
illustration). Here we do not employ the evidence for this purpose
(see instead~\cite{alw06,BenModelComp} for applications to the CMSSM),
and therefore drop the explicit conditioning on the model under
study, although in the following one should always interpret $p(\data)
\equiv p(\data | \model)$.}  \be p(\data) = \int p(\data | \derived)
\pi(\basis) \dr \basis.
\label{evidence:eq}
\ee 

In our previous work~\cite{rtr1,rrt2,rrt3,Trotta:2006ew,rrts}, we
employed an MCMC algorithm to map out the posterior pdf via
eq.~\eqref{eq:bayes}. As extensively described in \cite{rtr1}, the
purpose of the MCMC algorithm is to construct a sequence of points in
parameter space (called {\em ``a chain''}), whose density is
proportional to the posterior pdf. The sequence of points thus
obtained gives a series of samples from the posterior, which are
weighted in such a way as to reflect the relative probability of the
various regions in parameter space.

In this work we upgrade our scanning technique to use a novel
algorithm, MultiNest~\cite{Feroz:2007kg}, which is based on the
framework of Nested Sampling, recently invented by
Skilling~\cite{SkillingNS}. MultiNest has been developed
in such a way as to be an extremely efficient sampler even for
likelihood functions defined over a parameter space of large
dimensionality with a very complex structure. This aspect is very
important for multi-parameter models. For example, previous MCMC scans
have revealed that the 8-dimensional likelihood surface of the CMSSM
can be very fragmented, and that it features many finely tuned regions
that are difficult to explore with conventional MCMC and grid
scans. Therefore we adopt MultiNest as an efficient sampler of the
posterior. We have compared the results with our MCMC algorithm and
found that they are identical (up to numerical noise). The main
motivation is the increased sampling efficiency (which improves
computational efficiency by a factor of $\sim 200$ with respect to our
previous MCMC algorithm) and the possibility of computing
automatically the Bayesian evidence, which we use in this work to
quantify the amount of information in the various
observables.\footnote{A new 
version of our code, including
MultiNest and a new interactive plotting routine (called
\texttt{SuperEGO}), is publicly available from
\texttt{www.superbayes.org}. The full lists of samples used in this
work are also available at the same location. An online plotting tool
is available at
\texttt{http://pisrv0.pit.physik.uni-tuebingen.de/darkmatter/superbayes/index.php}.}
We give a brief description of the MultiNest algorithm in
Appendix~\ref{app:nest}.

\subsection{Statistical measures }

Once a sequence of $M$ samples drawn from the posterior,
$\basis^{(t)}$ ($t=0, 1, \dots, M-1$), becomes available, it becomes a
trivial task to obtain Monte Carlo estimates of expectations for any
function of the parameters. For example, the posterior mean is given
by
\begin{equation} \label{eq:expectation}
\langle \basis \rangle =  \int  p(\basis|\data)\basis \dr\basis
  \approx \frac{1}{M} \sum_{t=0}^{M-1} \basis^{(t)},
\end{equation}
where $\langle \cdot \rangle$ denotes the expectation value with
respect to the posterior and the equality with the mean of the samples
follows because the samples $\basis^{(t)}$ are generated from the
posterior by construction. In general, one can easily obtain the
expectation value of any function of the parameters $f(\basis)$ as
\begin{equation} \label{eq:MC_estimate}
\langle f(\basis) \rangle \approx \frac{1}{M}\sum_{t=0}^{M-1}  f(\basis^{(t)}).
\end{equation}
It is usually interesting to summarize the results of the
inference by giving the 1--dimensional {\em marginal probability}
for $\basis_j$, the $j$--th element of $\basis$. Taking without
loss of generality $j=1$ and a parameter space of dimensionality
$N$, the {\em marginal posterior} for parameter $\basis_1$ is given by
\begin{equation} \label{eq:marginalisation_continuous}
p(\basis_1|\data) = \int  p(\basis|\data) \dr \basis_2 \dots \dr
\basis_N.
\end{equation} 
 From the samples it is trivial to obtain the marginal
posterior on the l.h.s. of eq.~\eqref{eq:marginalisation_continuous}:
since the samples are drawn from the full posterior,
$p(\basis|\data)$, their density reflects the value of the full
posterior pdf. It is then sufficient to divide the range of $\basis_1$
into a series of bins and {\em count the number of samples falling
within each bin}, simply ignoring the coordinates values $\basis_2,
\dots, \basis_N$.  A 2--dimensional posterior is defined in an
analogous fashion. A 1D 
2--tail
$\alpha\%$ credible region is given by the interval (for the
parameter of interest) within which fall $\alpha\%$ of the samples, obtained in such a way that a fraction $(1-\alpha)/2$ of the samples lie outside the interval on either side. In the
case of a 1--tail upper (lower) limit, we report the value of the
quantity below (above) which $\alpha\%$ of the sample are to be found.

An alternative statistical measure to the marginal posterior given by
\eqref{eq:marginalisation_continuous} is the {\em profile likelihood},
defined, say, for the parameter $\basis_1$ as 
\be \label{eq:profile}
\prof (\basis_1) \equiv \max_{\basis_2, \dots, \basis_N} \like(\data |
\basis), 
\ee 
where in our case $ \like(\data | \basis)$ is the full
likelihood function. Thus in the profile likelihood one maximises the
value of the likelihood along the hidden dimensions, rather than
integrating it out as in the marginal posterior. The profile
likelihood is obtained from the samples by maximising the value of the
likelihood in each bin, and it has been recently investigated in the context of MCMC scans of the CMSSM in~ \cite{bclw07}. 
The advantage is that the profile likelihood
is clearly independent of the prior. However, its numerical evaluation in a high--dimensional parameter space is in general very difficult, especially when finely tuned regions are present where the likelihood is large but whose volume is very small (for a given metric). For example, a log prior on the SUSY
masses will expand the volume of the low-mass parameter region and as
a consequence the algorithm will explore it in much finer detail than
it would be possible with a linear prior on the masses. This might
find points in parameter space that are good fits to the data and
that would have otherwise been missed by a scan performed using a
linear prior. This will be true of any scanning algorithm: scanning in one metric (in our language, for a given prior) might in general give a different value than the
numerical evaluation of the same quantity when scanning in another
metric. To the extent the different numerical evaluations of the same
quantity disagree, one must of course take with a grain of salt either value.\footnote{Notice that this is fundamentally different from the Bayesian perspective: a change of prior changes the posterior in Bayesian statistics, hence the mathematical function one wants to map out changes {\em independently on the numerical aspects of the scanning technique.}}  As we shall demonstrate below, the choice of priors
influences the numerical efficiency with which different regions of
parameter space are scanned. Therefore the numerical evaluation of the profile likelihood might in general be
different for different prior (i.e., metric) choices. In the following, when we refer to the profile likelihood in connection with the scanning results, we always mean ``our numerical evaluation of the profile likelihood''.

The profile likelihood can be directly interpreted as a likelihood
function, except of course that it does account for the effect of the
hidden parameters. Therefore one can think of plots of the profile
likelihood as analogous to what would be obtained by performing a more
traditional fixed-grid scan in 8--dimensions, computing the
chi--square at each point at then plotting the value maximised along
the hidden dimensions.  We report confidence intervals from the
profile likelihood obtained via the usual likelihood ratio test as
follows. Starting from the best-fit value in parameter space, an
$\alpha$\% confidence interval encloses all parameter values for which
the log--likelihood increases less than $\Delta \chisq(\alpha, n)$
from the best fit value. The threshold value depends on $\alpha$ and
on the number $n$ of parameters one is simultaneously considering
(usually $n=1$ or $n=2$), and it is obtained by solving
\be \alpha = \int_0^{\Delta \chisq} \chisq_n (x) \dr x, \ee 
where $\chisq_n(x)$ is the chi--square distribution for $n$ degrees of
freedom. The MultiNest algorithm we employ is much more efficient than
a standard grid scan in parameter space, and it allows one to explore the
full multi-dimensional parameter space at once. Therefore our scanning
algorithm when coupled with the profile likelihood can be understood
as an extremely efficient shortcut for the evaluation of the minimum
chi--square in a multi-dimensional parameter space. However, the MultiNest
technique (or indeed, any other Bayesian procedure) is not
particularly optimized to look for isolated points with large
likelihood in the parameter space. This means that the profile
likelihood is derived from a necessarily sparse sampling of our
8-dimensional parameter space, and it might well be that regions with
large likelihood that occupy a very small volume in parameter space
are missed altogether. This means that an analogous problem would
appear if the scan was done with a traditional grid technique, which
would find multiple maxima in the likelihood if executed in
8--dimensional parameter space (grid scans to date have never been
able to deal with sufficient resolution with such a high dimensional
parameter space). Nevertheless, Bayesian technology and the MultiNest
algorithm give several orders of magnitude improvement in the
efficiency of the scan, thereby allowing for the first time to
undertake a detailed analysis of the impact of the data when applied
one by one or simultaneously to the whole parameter space.

As an
alternative measure to the posterior, in our previous work we employed a quantity that we called the {\em mean quality of fit}  (see eq.~(3.1)
in~\cite{rrt3}), which is defined as the average (over the posterior) of the chi--square. Therefore the difference between the profile likelihood and the
mean quality of fit is that in the mean quality of fit the chi--square is averaged over the hidden
dimensions, while in the profile likelihood it is maximised. Numerical investigation
shows that the two quantities are very similar in the case of the
CMSSM. We have chosen to adopt in this work the profile likelihood because of its more
straightforward statistical interpretation, but we point out that our previous findings showing the mean quality of fit are very similar to what one would have obtained using the profile likelihood instead. 

In Bayesian statistics, the posterior pdf encodes the full information
coming from the data and the prior. Ideally, the information in the
data is much stronger than the information in the prior, so
effectively the posterior should be dominated by the likelihood
function and the prior choice ought to be irrelevant (see fig.~2
in~\cite{Trotta:2008qt} for an illustration). Furthermore, in this
case it is easy to show that the Bayesian posterior, the profile
likelihood and the mean quality of fit all become identical, and
therefore the conclusions from the different statistical measures
agree (and are uncontroversial). If the data are not strong enough,
the different statistical quantities encode different pieces of
information about the parameters and may in general disagree, and the
prior influence might come to dominate the result. This appears to be
the case with the CMSSM with currently available constraints.  One of
the main aims of this work is to clarify the reasons for this prior
and statistical measure dependence, and to assess how much one should
be worried about it.

\subsection{Information content and constraining power}

The Bayesian evidence returned by the MultiNest algorithm can be
employed in several ways, mainly as a tool for model comparison (see, 
e.g.~\cite{Trotta:2008qt}). Here we employ it to quantify the amount
of information (i.e., the constraining power) of the different
observables.  This is encoded in the {\em Kullback--Leibler (KL)
divergence} between the prior and the
posterior~\cite{Kullback:1951}. For ease of notation, let us denote
the posterior pdf by $p$ and the prior by $\pi$, as before. Then the KL
divergence is defined as
\be \label{eq:def_KL}
\KL (p,\pi) \equiv  \int p(\basis|\data)
\ln\frac{p(\basis|\data)}{\pi(\basis)} \dr\basis .
\ee
In virtue of Bayes' theorem
the KL divergence becomes the sum of the negative log evidence and the
expectation value of the log-likelihood under the posterior: \be
\label{eq:KL_derived} \KL (p,\pi) = - \ln p(\data) + \int
p(\basis|\data) \ln \like(\basis) \dr\basis = - \ln p(\data) - \langle
\chisq/2 \rangle .  \ee The first quantity on the r.h.s. is returned
by the MultiNest algorithm, while computing the expectation value of the
log-likelihood (i.e., the chi--square) is trivial from the samples. It
is sufficient to average the chi--square over the samples.

To gain a feeling for what the KL divergence expresses, let us
compute it for a 1--dimensional case, with a Gaussian prior around
0 of variance $\Sigma^2$ and a Gaussian likelihood centered on
$\basis_{\rm max}$ and variance $\sigma^2$. We obtain after a short
calculation
\begin{equation}
\KL (p,\pi) = - \frac{1}{2} - \ln \frac{\sigma}{\Sigma}
+ \frac{1}{2} \left[\left(\frac{\sigma}{\Sigma}\right)^2 \left(\frac{{\basis_{\rm max}}^2}{\sigma^2} -1  \right)
\right].
\end{equation}
The second term on the r.h.s. gives the reduction in
parameter space volume in going from the prior to the posterior.
For informative data, $\sigma/\Sigma \ll 1$, this terms is
positive and grows as the logarithm of the volume ratio. On the
other hand, in the same regime the third term is small unless the
maximum likelihood estimate is many standard deviations away from
what we expected under the prior, i.e. for $\basis_{\rm max}/\sigma \gg 1$.
This means that the maximum likelihood value is ``surprising'', in
that it is far from what our prior led us to expect. Therefore we
can see that the KL divergence is a summary of the amount of
information, or ``surprise'', contained in the data.

Other quantities can be used to assess the constraining power of the
data (see e.g.~\cite{Allanach:2008tu} for a recent application), but
the KL divergence has the advantage of being firmly grounded in
information theory and of having a clear interpretation.

\section{Implications for the Constrained MSSM}
\label{sec:cmssm}

As a theoretical particle physics framework to illustrate our
procedure we use the popular Constrained MSSM~\cite{kkrw94}. Some of
us have examined the model in the context of Bayesian statistics
before~\cite{rtr1,BenModelComp,rrt2,rrt3}. Here we summarize its relevant features here
for completeness. Below we also list, and update, where applicable, the
experimental constraints on the model.


\subsection{The Constrained MSSM}
\label{sec:cmssmdef}

In the CMSSM the parameters $\mhalf$, $\mzero$ and $\azero$, which are
specified at the GUT scale $\mgut\simeq 2\times 10^{16}\gev$, serve as
boundary conditions for evolving, for a fixed value of $\tanb$, the
MSSM Renormalization Group Equations (RGEs) down to a low energy scale
$\msusy\equiv \sqrt{\mstopone\mstoptwo}$ (where
$m_{\stopone,\stoptwo}$ denote the masses of the scalar partners of
the top quark), chosen so as to minimize higher order loop
corrections. At $\msusy$ the (1-loop corrected) conditions of
electroweak symmetry breaking (EWSB) are imposed and the SUSY spectrum
is computed. 

Our aim is to use experimental constraints on observational quantities
defined in terms of CMSSM parameters to infer the most probable values
of the CMSSM quantities themselves (and the associated errors). In
this paper with fix the sign of $\mu$ to be positive, in order for the
model to acommodate the apparent discrepancy of the anomalous magnetic
moment of the muon between experiment and SM predictions. We then
denote the remaining four free CMSSM parameters by the set
\be \label{indeppars:eq}
\params =  (\mzero, \mhalf, \azero, \tanb ).
\ee
As originally demonstrated in~\cite{al05,rtr1}, the values of the relevant SM
parameters can strongly influence some of the CMSSM predictions, and,
in contrast to common practice, should not be simply kept fixed at
their central values. We thus introduce a set $\nuis$ of so-called
{\em ``nuisance parameters''} of the SM parameters which are
relevant to our analysis,
\be \label{nuipars:eq} \nuis = ( \mtpole,
\mbmbmsbar, \alphaemmz, \alphas ), \ee
where $\mtpole$ is the pole top quark mass. The other three
parameters:  $\mbmbmsbar$ -- the bottom
quark mass evaluated at $m_b$, $\alphaemmz$ and $\alphas$ -- respectively the
electromagnetic and the strong coupling constants evaluated at the $Z$ pole mass
$M_Z$ --  are all computed in the $\msbar$ scheme.

The set of parameters $\params$ and $\nuis$ form an 8-dimensional set
$\basis$ of our {\em ``basis parameters''}~\eqref{basis:eq}. In terms
of the basis parameters we compute a number of collider and
cosmological observables, which we call {\em ``derived variables''}
and which we collectively denote by the set $\derived= (\derived_1,
\derived_2, \ldots)$.  The observables will be used to compare CMSSM
predictions with a set of experimental data $\data$, which is
available either in the form of positive measurements or as limits, as
discussed below.

\subsection{Priors, observables and data}\label{sec:constraints}

\begin{table}
\centering    .

\begin{tabular}{|l | l l | l|}
\hline
SM (nuisance)  &   Mean value  & \multicolumn{1}{c|}{Uncertainty} & Ref. \\
parameter &   $\mu$      & ${\sigma}$ (exper.)  &  \\ \hline
$\mtpole$           &  172.6 GeV    & 1.4 GeV&  \cite{topmass:mar08} \\
$m_b (m_b)^{\overline{MS}}$ &4.20 GeV  & 0.07 GeV &  \cite{pdg07} \\
$\alphas$       &   0.1176   & 0.002 &  \cite{pdg07}\\
$1/\alphaemmz$  & 127.955 & 0.03 &  \cite{Hagiwara:2006jt} \\ \hline
\end{tabular}
\caption{Experimental mean $\mu$ and standard deviation $\sigma$ 
adopted for the likelihood function for SM (nuisance) parameters,
assumed to be described by a Gaussian distribution.
\label{tab:meas}}
\end{table}

In order to estimate the impact of priors, we
adopt two different choices of priors:
\begin{itemize}
\item {\em\underline{flat priors}} in all the CMSSM parameters $\mhalf$,
 $\mzero$, $\azero$ and $\tanb$;

\item {\em \underline{log priors}},  that are flat in $\log\mhalf$ and
$\log\mzero$, while for the other two CMSSM parameters we keep flat priors.
\end{itemize}

As regards the ranges, in both cases we take $50\gev < \mhalf,\mzero <
4 \tev$, $|\azero| < 7\tev$ and $2 < \tanb < 62$, as
before~\cite{rtr1,rrt2,rrt3}.  Note that the above range of $\mzero$
includes the hyperbolic branch/focus point (FP)
region~\cite{Chan:1997bi,focuspoint-fmm} which will play an important
role in our discussion because it currently favored by the constraint
from $\brbsgamma$~\cite{rrt3}.

 The rationale for
our choice of priors is that they are distinctively different.  In
particular, the log prior gives equal {\em a priori} weights to all
decades for
the parameter. For example, with a log prior there is the same {\em a
priori} probability that $\mzero$ be in the range $10 \gev < \mzero <
100 \gev$ as in the range $100 \gev < \mzero < 1 \tev$. In contrast,
with a flat prior, the latter range of mass values has instead 10
times more {\em a priori} probability than the former. So the log
prior expands the low-mass region and allows a much more refined scan
in the parameter space region where finely tuned points can give a
good fit to the data (see below).  The reason why we apply different
priors to $\mhalf$ and $\mzero$ only is that both of them play a
dominant role in the determination of the masses of the superpartners
and Higgs bosons in the CMSSM.

Clearly a flat prior on a parameter set $\basis$ does not correspond
to a flat prior on some non-linear function of it,
${\fbasis}(\basis)$.The two priors are related by
\begin{equation}
\label{eq:prior_transformation}
\pi(\fbasis) = \pi(\basis) {\Big\arrowvert} \frac{\dr \basis}{\dr \fbasis} {\Big\arrowvert}.
\end{equation}
Thus, in the case of non-linear dependence of $\fbasis(\basis)$ the
term $|\dr \basis/\dr \fbasis|$ implies that an uninformative (flat)
prior on $\basis$ may be strongly informative about (\ie,
constraining) $\fbasis$. (In a multi-dimensional case, the derivative
term is replaced by the determinant of the Jacobian for the
transformation.)  It follows that a flat prior on $\log \basis$ (i.e.,
the log prior) corresponds to choosing a prior on $\basis$ of the form
$\pi(\basis) \propto \basis^{-1}$. Therefore we expect that the choice
of the log prior will give more statistical weight to lower values of
$\mhalf$ and $\mzero$ than in the case of flat priors.

Other choices of priors are possible, and indeed might be argued to be
more theoretically 
motivated from the point of view of penalizing finely tuned regions of
parameters space \cite{allanach06, bclw07, Allanach:2008iq}. However, one
would like the final inference to be as prior independent as possible,
and the constraints to be driven by the likelihood, rather than by
theoretical prejudices in the prior.

A related, although different issue is the choice of the parameters
with which to define the model. One particularly well-known
implementation of the CMSSM is one version of the so-called minimal
supergravity model~\cite{sugra-reviews} where the parameters
$\tanb$ and $\mz$ are replaced by $\mu$ and $B$.  This choice of
parameterization has been advocated in~\cite{bclw07,Allanach:2008iq}
as more ``fundamental''. This is questionable in the case of the CMSSM
which has originally been {\em defined} in ref.~\cite{kkrw94} in terms of the
parameters~\eqref{indeppars:eq} as an effective theory, without
necessarily any reference to any underlying supergravity theory. More
importantly, it is obvious that robust physical conclusions should not
strongly depend on one choice of parameters of the model or
another. If they do, this should serve as a warning bell that the
derived statistical implications for observable quantities, like
masses and cross sections, are not robust, in the same way as is the
case with the dependence on priors. (Note
that the impact of the same type of priors, e.g., flat, for different
choice of parameterization, may be very different, as implied by
eq.~\eqref{eq:prior_transformation}.)

\begin{table}
\centering
\begin{tabular}{|l | l l l | l|}
\hline
Observable &   Mean value & \multicolumn{2}{c|}{Uncertainties} & ref. \\
&   $\mu$      & ${\sigma}$ (exper.)  & $\tau$ (theor.) & \\\hline
$M_W$     &  $80.398\gev$   & $25\mev$ & $15\mev$ & \cite{lepwwg} \\
$\sineff{}$    &  $0.23153$      & $16\times10^{-5}$
               & $15\times10^{-5}$ &  \cite{lepwwg}  \\
$\deltaamususy \times 10^{10}$       &  29.5 & 8.8 &  1.0 & \cite{gm2}\\
$\brbsgamma \times 10^{4}$ &
3.55 & 0.26 & 0.21 & \cite{hfag} \\
$\delmbs$     &  $17.77\ps^{-1}$  & $0.12\ps^{-1}$  & $2.4\ps^{-1}$
& \cite{cdf-deltambs} \\
$\brbtaunu \times 10^{4}$ &  $1.32$  & $0.49$  & $0.38$
& \cite{hfag} \\
$\abundchi$ &  0.1099 & 0.0062 & $0.1\,\abundchi$& \cite{wmap5yr} \\\hline\hline
  &  Limit (95\%~\cl)  & \multicolumn{2}{r|}{$\tau$ (theor.)} & ref. \\ \hline
$\brbsmumu$ &  $ <5.8\times 10^{-8}$
& \multicolumn{2}{r|}{14\%}  & \cite{cdf-bsmumu}\\
$\mhl$  & $>114.4\gev$\ (SM-like Higgs)  & \multicolumn{2}{r|}{$3 \gev$}
& \cite{lhwg} \\
$\zetah^2$
& $f(m_h)$\ (see text)  & \multicolumn{2}{r|}{negligible}  & \cite{lhwg} \\
$m_{\tilde{q}}$ & $>375$ GeV  & & 5\% & \cite{pdg07}\\
$m_{\tilde{g}}$ & $>289$ GeV  & & 5\% & \cite{pdg07}\\
other sparticle masses  &  \multicolumn{3}{c|}{As in table~4 of
 ref.~\cite{rtr1}.}  & \\ \hline 
\end{tabular}
\caption{Summary of the observables used in the analysis. Upper part:
Observables for which a positive measurement has been
made. $\deltaamususy=\amuexpt-\amusm$ denotes the discrepancy between
the experimental value and the SM prediction of the anomalous magnetic
moment of the muon $\gmtwo$. 
As explained in the text, for each quantity we use a
likelihood function with mean $\mu$ and standard deviation $s =
\sqrt{\sigma^2+ \tau^2}$, where $\sigma$ is the experimental
uncertainty and $\tau$ represents our estimate of the theoretical
uncertainty. Lower part: Observables for which only limits currently
exist.  The likelihood function is given in
ref.~\cite{rtr1}, including in particular a smearing out of
experimental errors and limits to include an appropriate theoretical
uncertainty in the observables. $\mhl$ stands for the light Higgs mass
while $\zetah^2\equiv g^2(\hl ZZ)_{\text{MSSM}}/g^2(\hl ZZ)_{\text{SM}}$,
where $g$ stands for the Higgs coupling to the $Z$ and $W$ gauge boson
pairs.
\label{tab:measderived}}
\end{table}

For the SM parameters we assume flat priors over relatively
wide ranges: $167.0 \gev\leq \mtpole \leq 178.2\gev$, $3.92\gev \leq
\mbmbmsbar\leq 4.48\gev$, $127.835 \leq 1/\alphaemmz \leq 128.075$ and
$ 0.1096 \leq \alphas \leq 0.1256$.  This is expected to be irrelevant for the
outcome of the analysis since the nuisance parameters are
well-constrained by the data, as can be seen in table~\ref{tab:meas},
where for each of the SM parameters we adopt a Gaussian likelihood
with mean $\mu$ and experimental standard deviation $\sigma$.  Note
that, with respect to refs.~\cite{rrt2,rrt3}, we have updated the value of
$\mtpole$.

The experimental values of the collider and cosmological observables
that we apply (our derived variables) are listed in
table~\ref{tab:measderived}, with updates relative to~\cite{rrt3}
where applicable.  In our treatment of the radiative corrections to
the electroweak observables $M_W$ and $\sineff$, starting from
ref.~\cite{rrt2} we include full two-loop and known higher order SM
corrections as computed in ref.~\cite{awramik-acfw04}, as well as
gluonic two-loop MSSM corrections obtained in~\cite{dghhjw97}. We
further update an experimental constraint from the anomalous magnetic
moment of the muon $\gmtwo$ for which a discrepancy (denoted by
$\deltaamususy$) between measurement and SM predictions (based on
$e^+e^-$ data) persists at the level of
$3.2\sigma$~\cite{gm2}.\footnote{ Evaluations done by different groups
using $e^+e^-$ data give slighly different values but they all remain
close to the value given in
table~\ref{tab:measderived}~\cite{gm2_glasgow07}.  On the other hand,
using $\tau$ data leads to a much better agreement with experiment,
$\deltaamususy=(8.9\pm 0.95)\times 10^{-10}$.} We
will show that while this constraint on its own quite strongly prefers
lower values of $\mzero$ and $\mhalf$, this is in contradiction with
the impact of most other observables. Once they are also included,
this preference essentially disappears.

As regards $\brbsgamma$, with the central values of SM input
parameters as given in table~\ref{tab:meas}, for the new SM prediction
we obtain the value of $(3.12\pm 0.21)\times10^{-4}$.\footnote{The
value of $(3.15\pm 0.23)\times10^{-4}$ originally derived in
ref.~\cite{ms06-bsg,mm-prl06} was obtained for slightly different
values of $\mtpole$ and $\alphas$. Note that, in treating the error
bar we have explicitly taken into account the dependence on $\mtpole$
and $\alphas$, which in our approach are treated parametrically. This
has led to a slight reduction of its value.}  We compute SUSY contribution to
$\brbsgamma$ following the procedure outlined in
refs.~\cite{dgg00,gm01} which was extended in
refs.~\cite{or1+2,for1+2} to the case of general flavor mixing. In
addition to full leading order corrections, we include large
$\tanb$-enhanced terms arising from corrections coming from beyond the
leading order and further include (subdominant) electroweak
corrections.

The parametric uncertainty involved in the 
computation of $\brbtaunu$ comes from 
using $|V_{ub}| = (4.34 \pm 0.38) \times 10^{-3}$~\cite{hfag} obtained
from inclusive semileptonic B decays through the central value of 
$\mbmbmsbar$. For $\tau_B$ we use $1.643\pm 0.01\ps$~\cite{hfag} and 
$f_b = 0.216 \pm 0.022\gev$~\cite{lattice}, and obtain
$\brbtaunu^{\text{SM}} = 1.56 \pm 0.38 \times 10^{-4}$.
For the $\bbbarmix$ oscillations we use the SM parametric 
uncertainty given by the global fit from the UTfit collaboration~\cite{Utfit}.

Regarding cosmological constraints, we use the determination of the
relic abundance of cold DM based on the 5-year data from
WMAP~\cite{wmap5yr} to constrain the relic abundance $\abundchi$ of
the lightest neutralino. In order to be conservative, we employ the
constraint reported in table 1 of ref.~\cite{wmap5yr} (mean value),
obtained using WMAP data alone.  The relic abundance (assuming the
neutralino is the sole constituent of dark matter) is computed with
high precision, including all resonance and coannihilation effects,
through MicrOMEGAs~\cite{micromegas}, adding a 10\% theoretical error
in order to remain conservative. Note that our estimated theoretical
uncertainty is of the same order as the uncertainty from current
cosmological determinations of $\abundcdm$.

We further include in our likelihood function an improved 95\%~\cl\
limit on $\brbsmumu$ and a recent value of $\bbbarmix$ mixing,
$\delmbs$, which has recently been precisely measured at the Tevatron
by the CDF~Collaboration~\cite{cdf-deltambs}. In both cases we use
expressions from ref.~\cite{for1+2} which include dominant large
$\tanb$-enhanced beyond-LO SUSY contributions from Higgs penguin
diagrams.  Unfortunately, theoretical uncertainties, especially in
lattice evaluations of $f_{B_s}$ are still substantial (as reflected in
table~\ref{tab:measderived} in the estimated theoretical error for
$\delmbs$), which makes the impact of this precise measurement on
constraining the CMSSM parameter space rather limited.\footnote{On
the other hand, in the MSSM with general flavor mixing, even with the
current theoretical uncertainties, the bound from
$\delmbs$ is in many cases much more constraining than from other rare
processes~\cite{for3+4}.}

For the quantities for which positive measurements have been made (as
listed in the upper part of table~\ref{tab:measderived}), we assume a
Gaussian likelihood function with a variance given by the sum of the
theoretical and experimental variances, as motivated by eq.~(3.3) in
ref.~\cite{rtr1}. For the observables for which only lower or upper
limits are available (as listed in the bottom part of
table~\ref{tab:measderived}) we use a smoothed-out version of the
likelihood function that accounts for the theoretical error in the
computation of the observable, see eq.~(3.5) and fig.~1 in
ref.~\cite{rtr1}. In particular, in applying a lower mass bound from
LEP-II on the Higgs boson $\hl^0$ we take into account its dependence
on its coupling to the $Z$ boson pairs $\zetah^2$, as described in detail in
ref.~\cite{rrt2}. When $\zetah^2\simeq 1$, the LEP-II lower bound of
$114.4\gev$ (95\%~\cl)~\cite{lhwg} applies. For arbitrary values of
$\zetah$, we apply the LEP-II 95\%~\cl\ bounds on $\mhl$ and $\mha$,
which we translate into the corresponding 95\%~\cl\ bound in the
$(\mhl, \zetah^2)$ plane. We then add a conservative theoretical
uncertainty $\tau(\mhl) = 3\gev$, following eq.~(3.5) in
ref.~\cite{rtr1}. We will see that employing the full likelihood
function in the $(\mhl, \zetah^2)$ plane will allow us to discover
some regions that evade the $114.4\gev$ lower bound, and which would
not have been seen in a scan that would have simply cut off all the
points below the limit.

Finally, points that do not fulfil the conditions of radiative
EWSB and/or give non-physical (tachyonic) solutions are discarded.

\section{Effect of priors and of different observables }
\label{sec:priors}

We now turn to the discussion of the effects of priors and
experimental observables on the CMSSM parameter inference using
Bayesian statistics and profile likelihood.  We begin with some
general remarks.

The choice of a prior pdf implies a certain measure on the parameter
space defined by $\basis$. For
example, the log prior will give less {\em a priori} weight to larger
values of $\mhalf$ and $\mzero$, thus reducing the preference for the
FP region. What is most important is that the flat parameter space measure
imposed on the basis parameter space via the choice of priors {\em
does not correspond to a flat measure over the space of the
observables quantities} $\derived$, since these are in general a strongly
non-linear function of the chosen set of model's
parameters. Conversely, comparing observables quantities with
experimental data leads to rather complicated implications for the
basis parameters. 

If the data are constraining enough, the effect of the likelihood
dominates over that of the prior and one expects the prior dependence
to be negligible in the final inference (based on the posterior
pdf). Below we examine to what extent this is the case in the
CMSSM. We note that the CMSSM is one of the most economical
phenomenological models on the table -- more complex models (with more
free parameters) are qualitatively expected to compound the problem,
given that, as we will show below, current constraints are not
sufficiently strong to allow drawing prior-independent conclusions.

As regards experimental observables, since we will be interested in comparing
the constraining power of different combinations of data, it is
convenient to use shortcuts to designate them in shorthand. Those are
given in table~\ref{tab:datacombinations}.

\begin{table}
\centering  
\begin{tabular}{| l | l | }
\hline
Shortcut & Observables included in data set \\ \hline
PHYS	 & Physicality constraints (no tachyons, EWSB, neutralino LSP) \\
NUIS	    &  $\mtpole,    m_b (m_b)^{\overline{MS}}, \alphas, 1/\alphaemmz$ \\
COLL	    &  $\mhl$ and sparticle masses (limits) \\
CDM 	    & $\abundchi$ \\
BSG	    & $\brbsgamma$\\
GM2  	    &  $\deltaamususy=\amuexpt-\amusm$ \\
EWO       &  $\sineff$, $M_W$ \\
BPHYS 	     &  $\delmbs, \brbsmumu$, $\brbtaunu$\\
ALL			& All of the above  \\\hline
\end{tabular}
\caption[aa]{Shortcuts for different data combinations applied in the analysis. The
 actual data employed in the numerical analysis are given in tables \ref{tab:meas} and
 \ref{tab:measderived}.}  
\label{tab:datacombinations}
\end{table}

\subsection{Impact of priors}

In this subsection we explore the impact of the flat and the
log priors on the CMSSM parameters and on the predictions for the
observable quantities.  To set the stage, we perform a scan of the
basis parameter space without imposing any experimental constraints at all,
i.e., we take a
constant likelihood function. We only discard points suffering from
unphysicalities: no self-consistent solutions to the RGEs, no EWSB and 
tachyonic states. Furthermore, we require the  neutralino to be the
LSP in order to be the dark matter.  Therefore the
final list of samples only contains physical points in parameter
space.  Without the physicality constraint, we would have expected
that such a scan would return a posterior identical to the prior,
i.e., flat in the variables over which a flat prior has been imposed.

\begin{figure}[tbh!]
\begin{center}
\includegraphics[width=\ww]{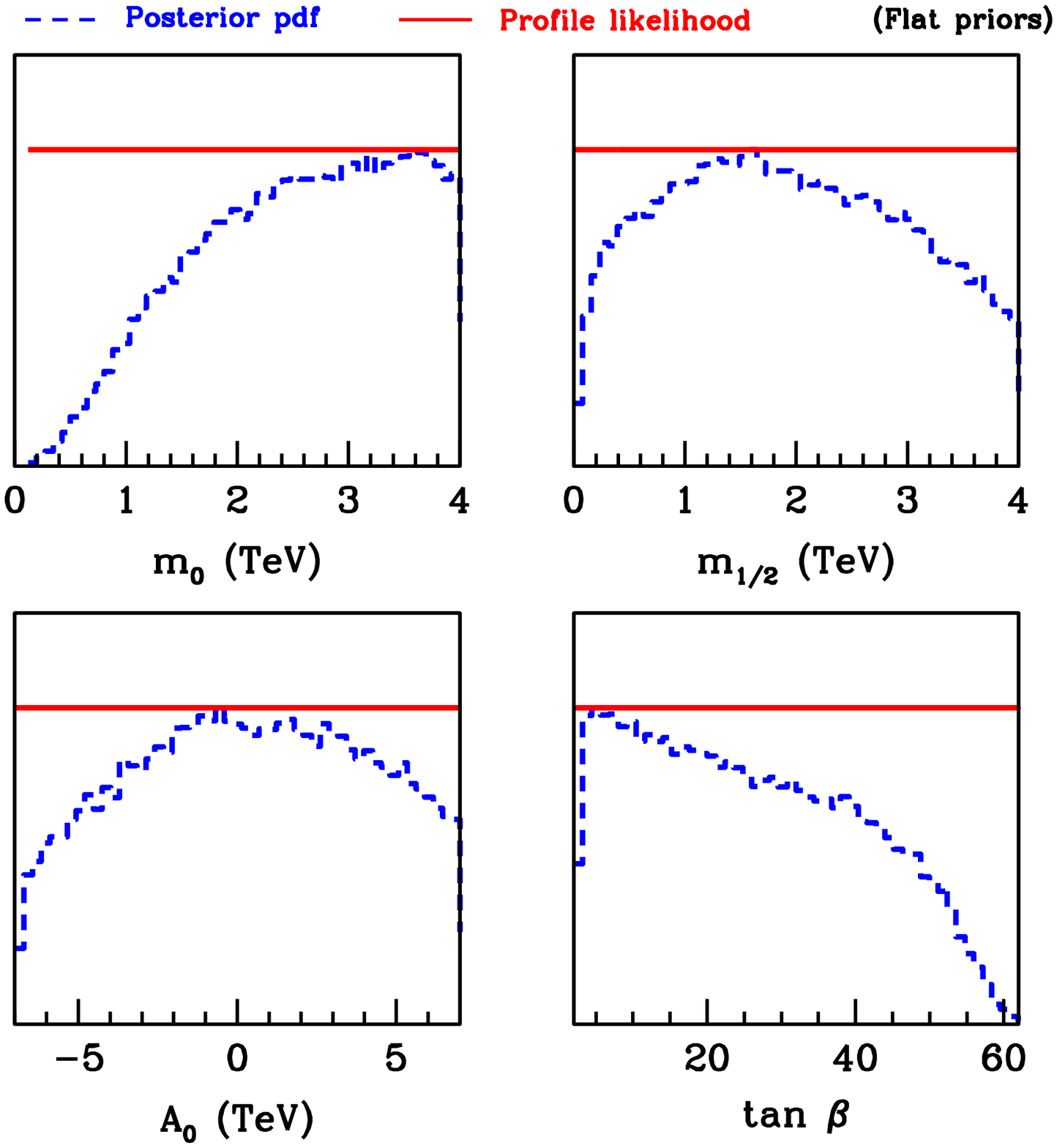}
\includegraphics[width=\ww]{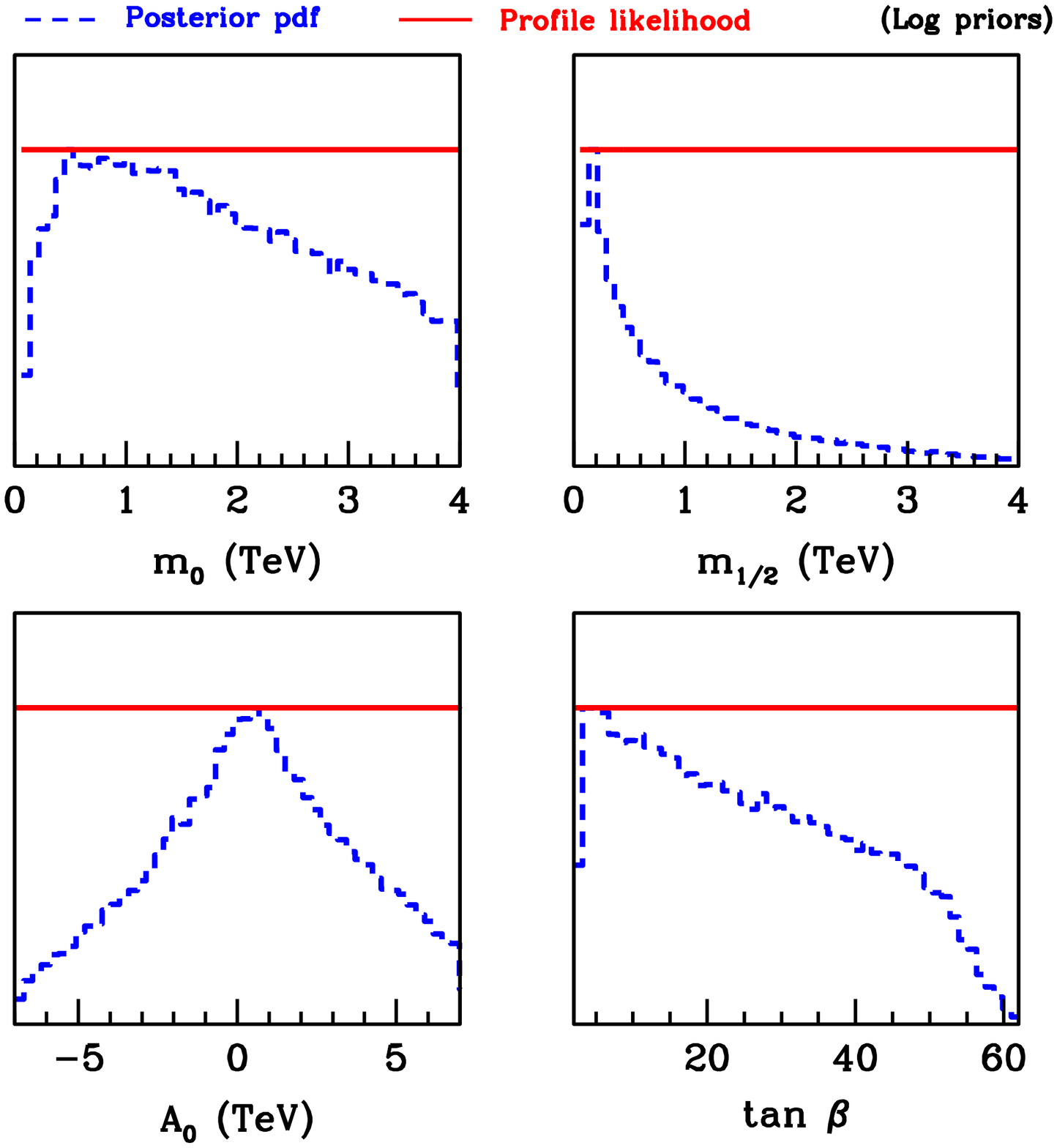}
\caption[test]{A scan including no experimental data, but only the
requirement of physicality (\texttt{PHYS}). Two columns of panels on
the left: 1D posterior distribution (dashed blue) and 1D profile
likelihood (solid red) for the CMSSM parameters for the flat priors
case. Two columns of panels on the right: the same quantities but for
the log priors case. The plots reflect the prior distributions alone
of the CMSSM parameters and the physicality constraints.}
\label{fig:1Dpriors_base}
\end{center}
\end{figure}

In fig.~\ref{fig:1Dpriors_base} we present the implication for 1D
distributions of the posterior (dashed blue) and the profile
likelihood (solid red) for the CMSSM parameters with only the
physicality constraint imposed (\texttt{PHYS}). In the four leftmost
panels we assume flat priors  while in the four rightmost panels we
assume log priors. (For all the SM
nuisance parameters both distributions are basically flat over the prior range of the SM parameters, and we do
not show them here.)  Notice that the lack of samples in certain
regions of parameter space, as induced by the physicality constraints,
shows up in the posterior pdf as a reduction of the marginalised
probability for that region. Thus for the flat priors case, the drop
at low $\mzero$ and large $\mhalf$ is primarily caused by the fact
that in that region the LSP is the stau and hence our assumed
requirements for physical points are not met. On the other hand, a
gradual decrease in the posterior of $\tanb$ is a reflection of
increasing difficulty for the RGEs to find self-consistent
solutions. Eventually, at large $\tanb$ over about 62, the Yukawa coupling
of the top quark grows to non-perturbative values before the GUT scale
is reached and no solutions are found anymore, as was explained
in~\cite{rtr1}. For the log priors case, the increased {\em a priori}
probability for small values of $\mzero$ compensates the above
effects, while the large $\mzero$ region is now suppressed. The same
trend is even more evident for $\mhalf$, where the marginal posterior pdf
follows closely the expected dependence $\propto 1/\mhalf$
characteristic of a log prior.  In contrast, the profile likelihood
remains flat across all the CMSSM parameters. This is precisely what
one would have expected since no data have been employed.

\begin{figure}[tbh!]
\begin{center}
\includegraphics[width=\qq]{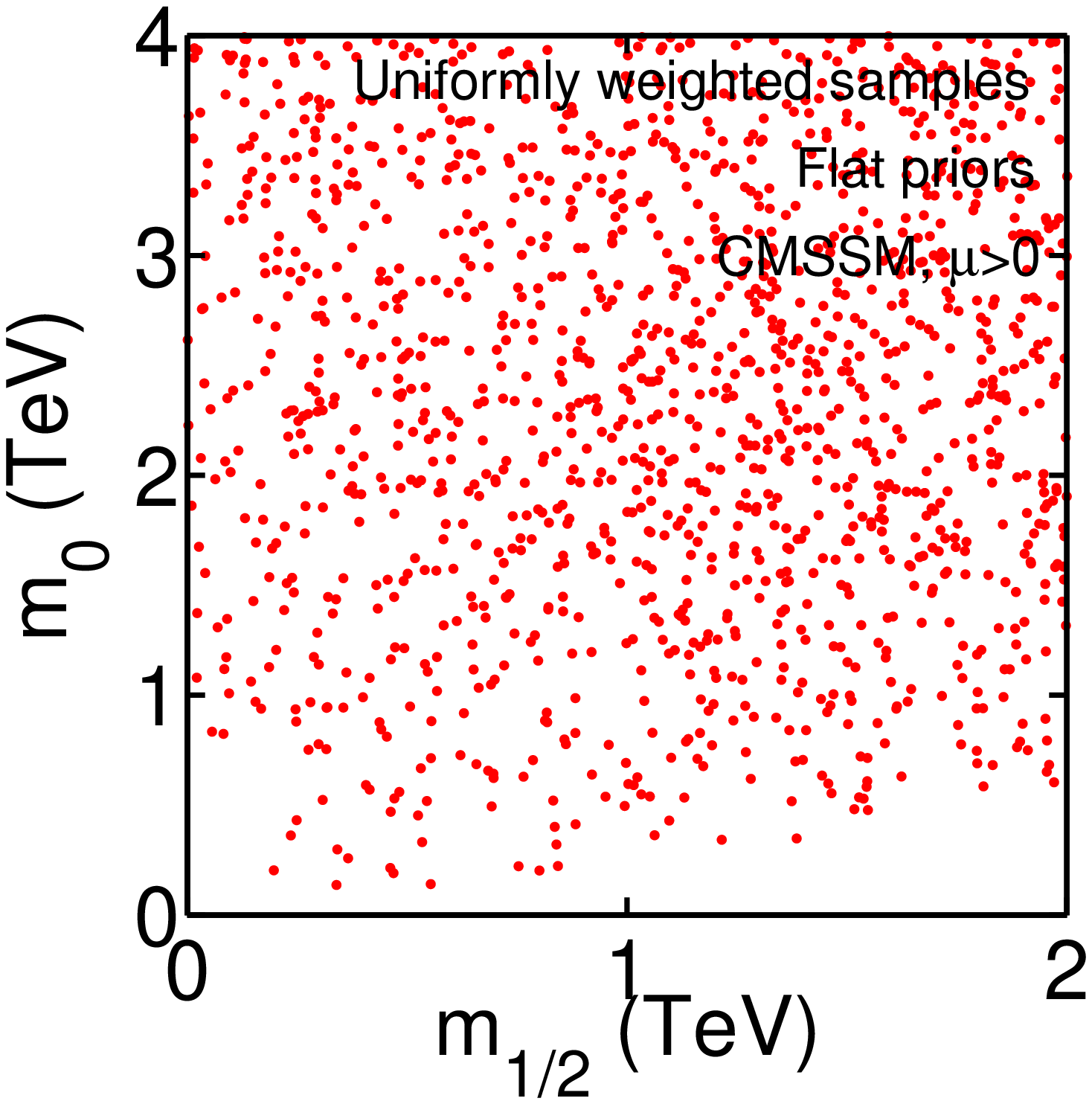}
\includegraphics[width=\qq]{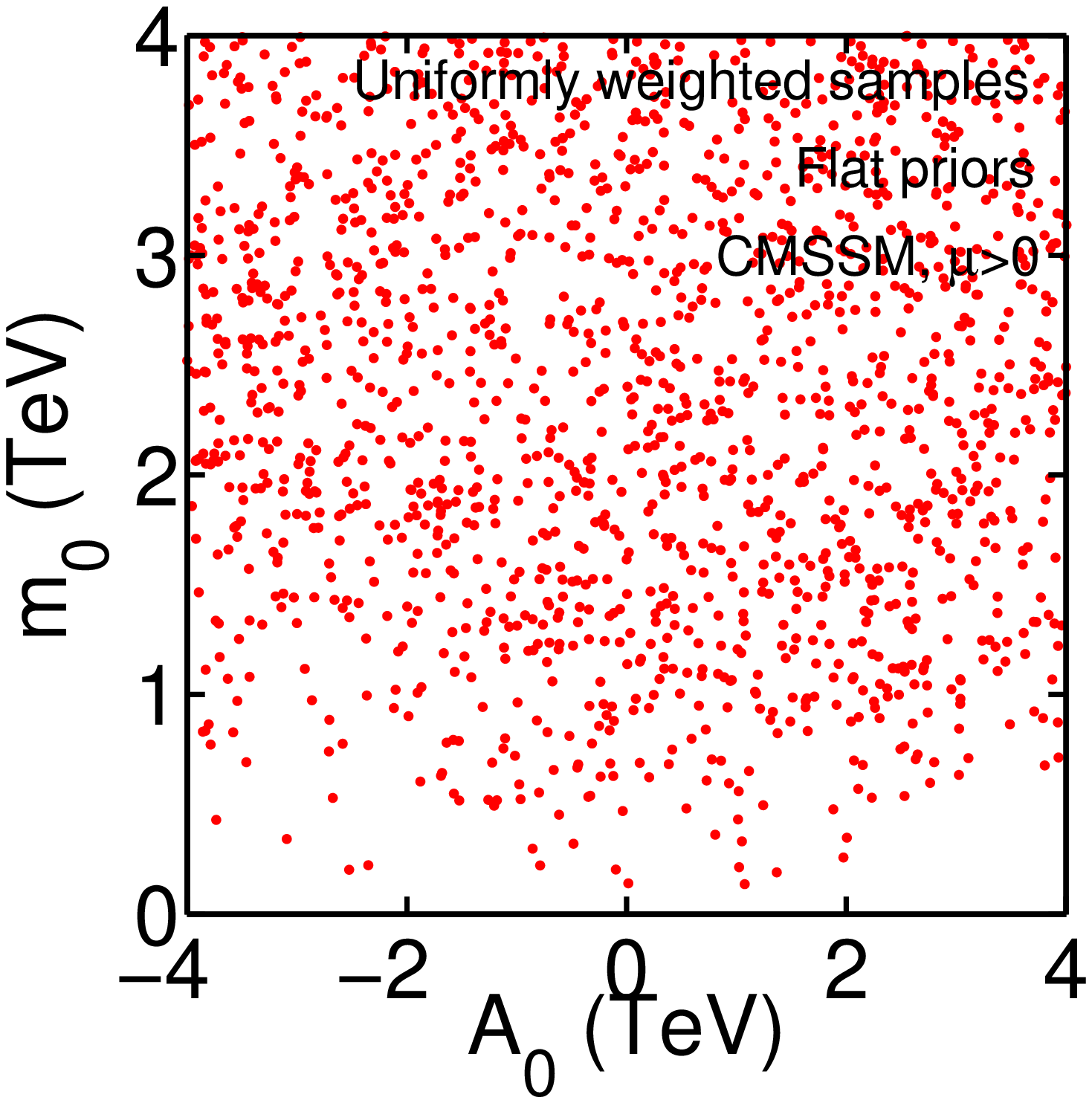}
\includegraphics[width=\qq]{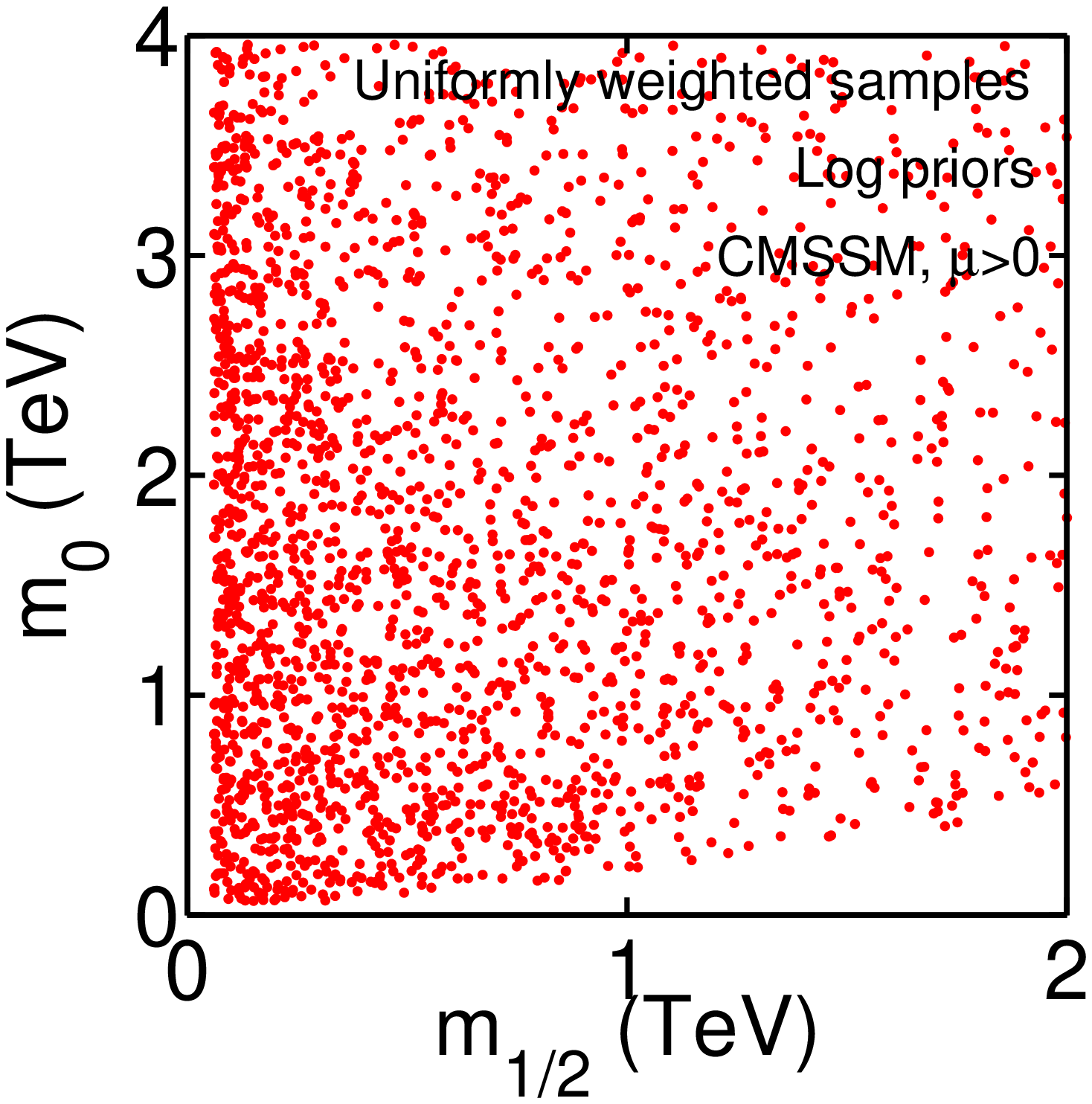}
\includegraphics[width=\qq]{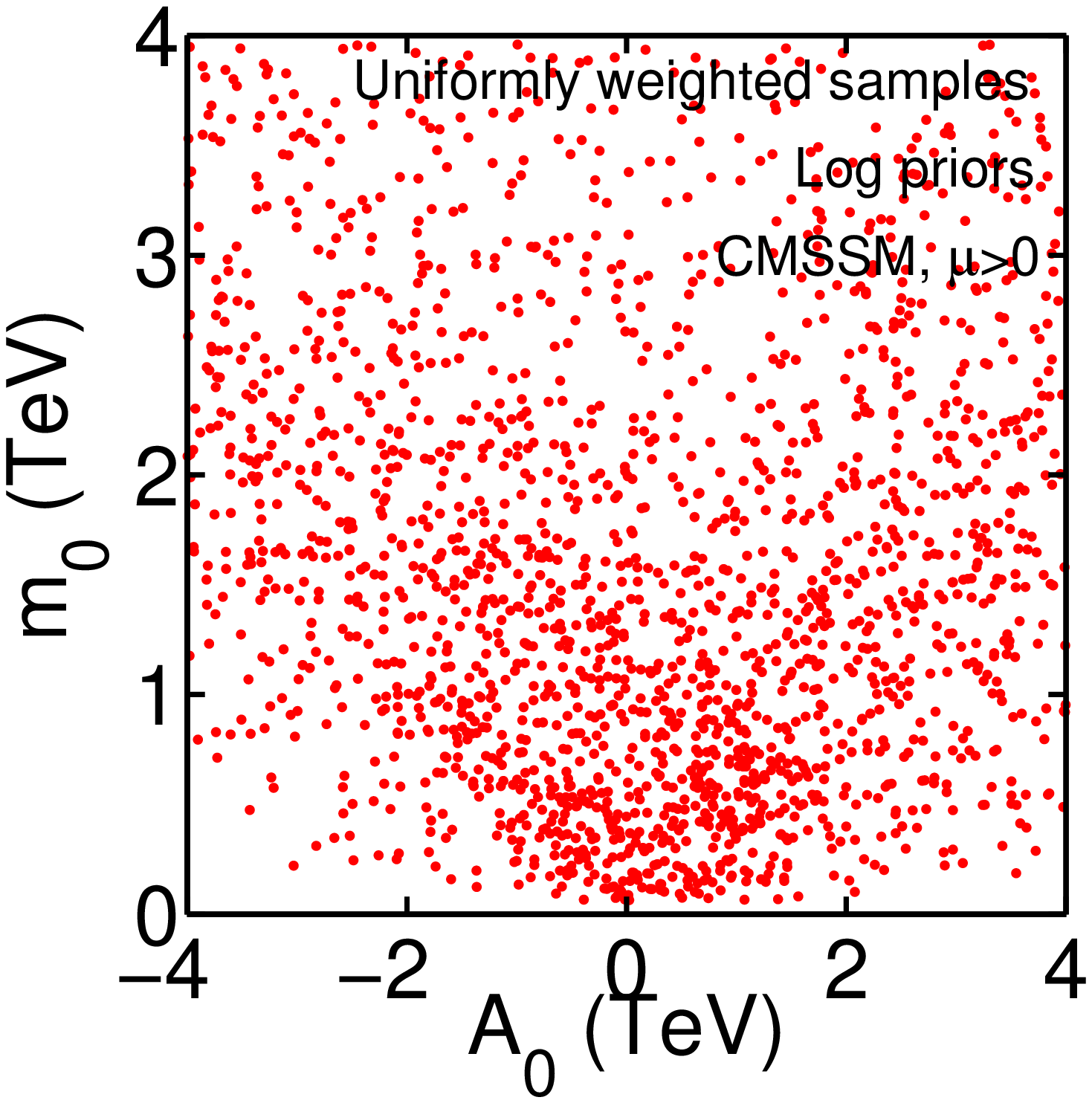} \\
\includegraphics[width=\qq]{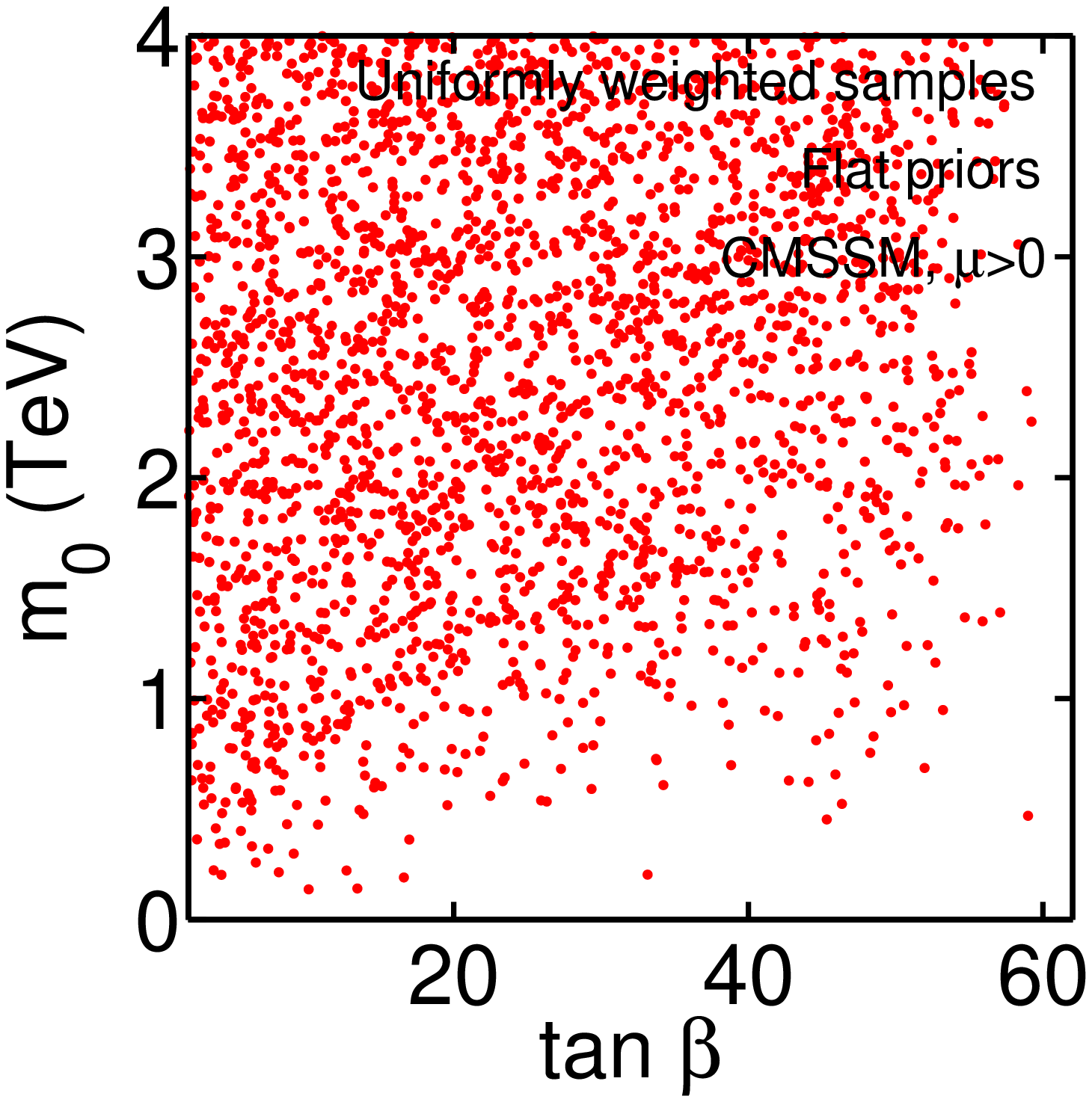}
\includegraphics[width=\qq]{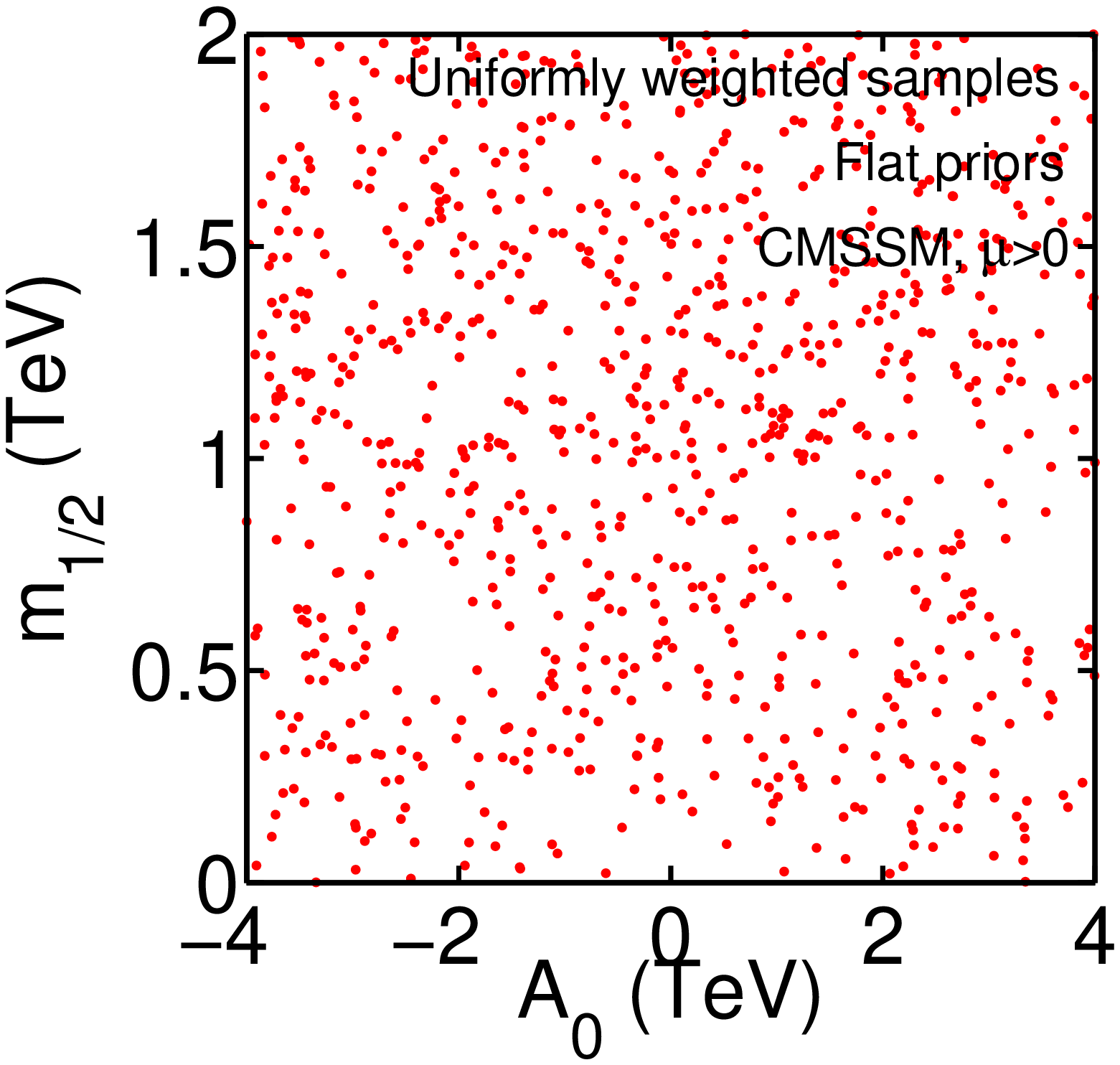}
\includegraphics[width=\qq]{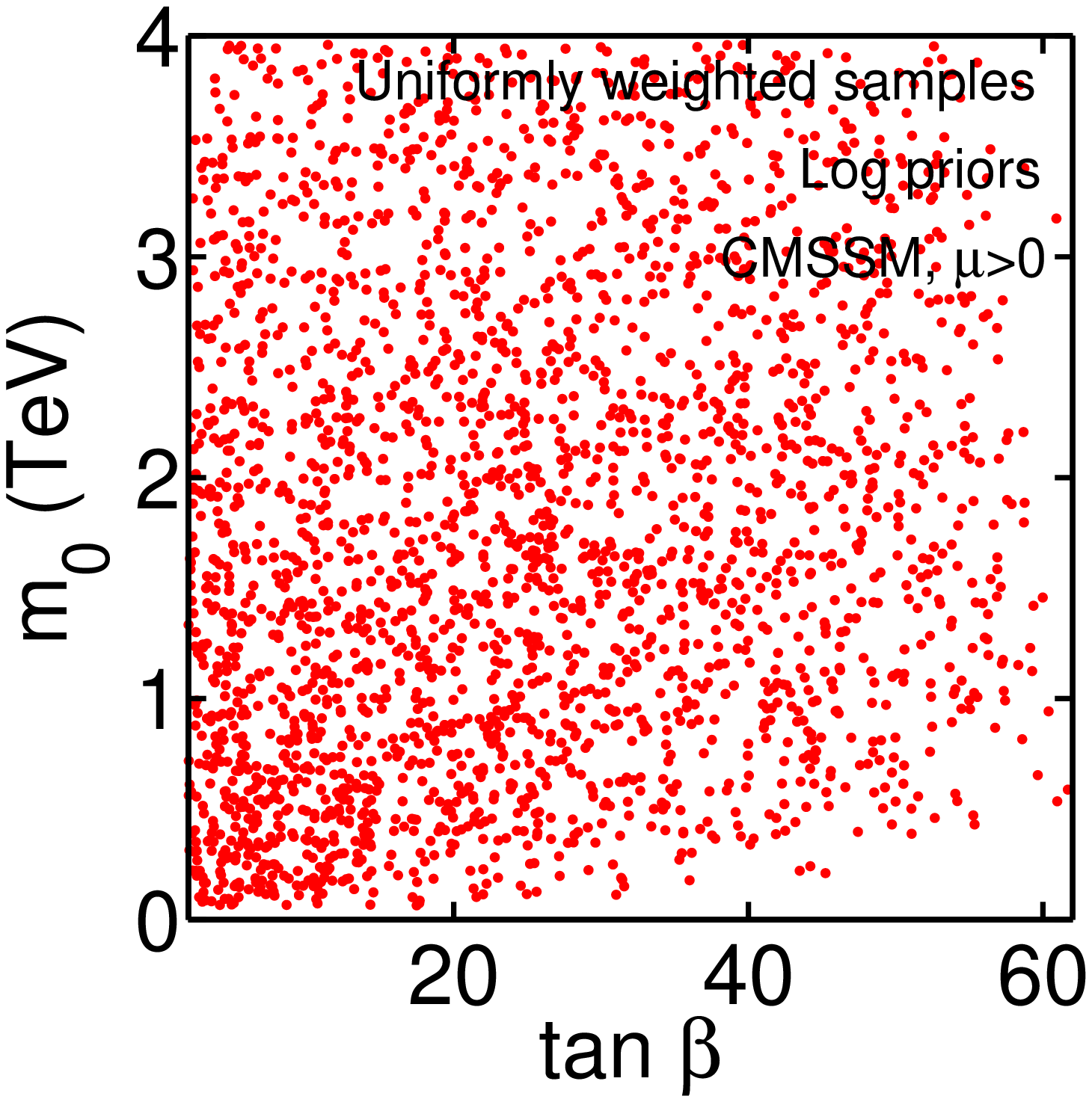}
\includegraphics[width=\qq]{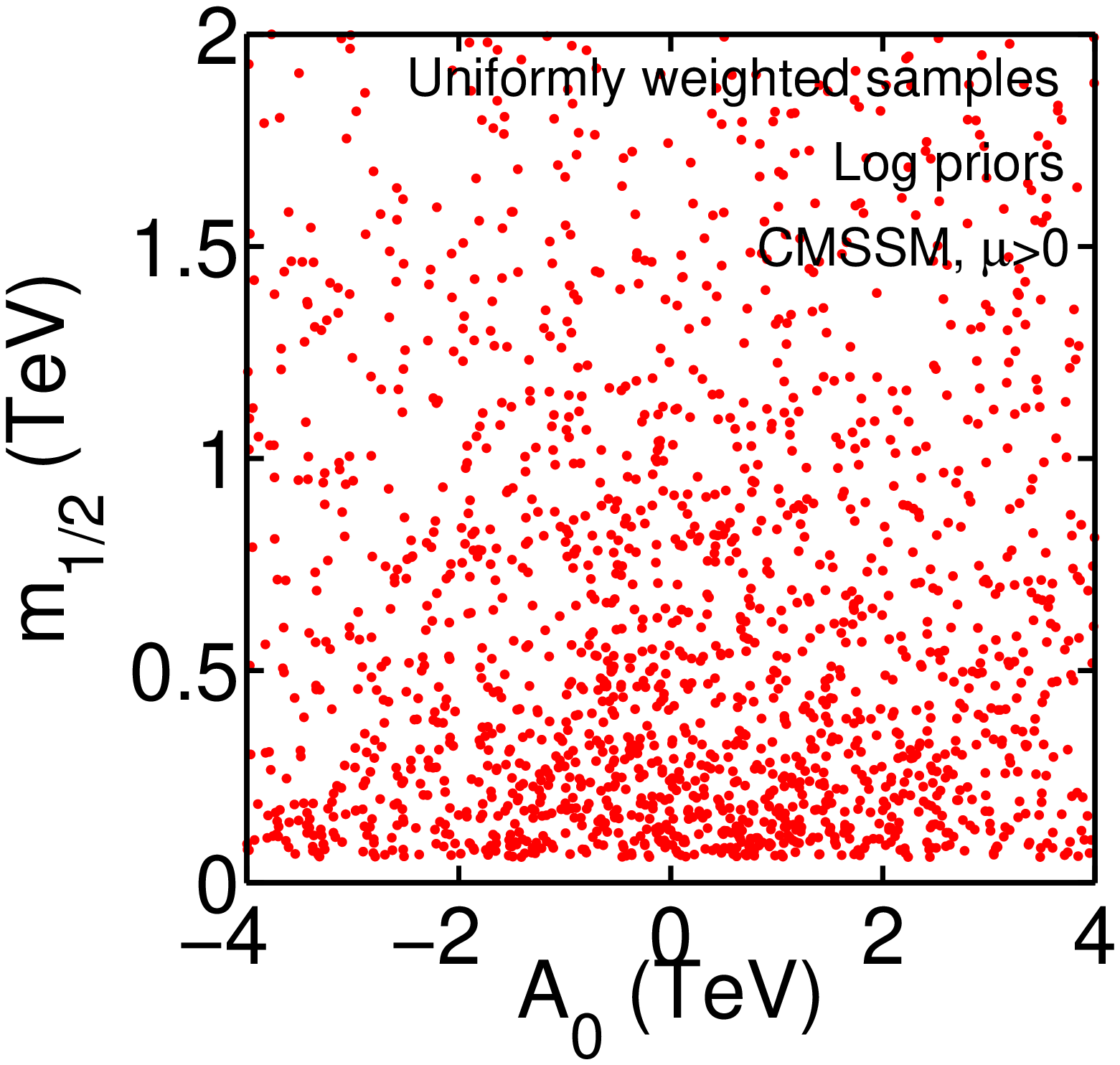} \\
\includegraphics[width=\qq]{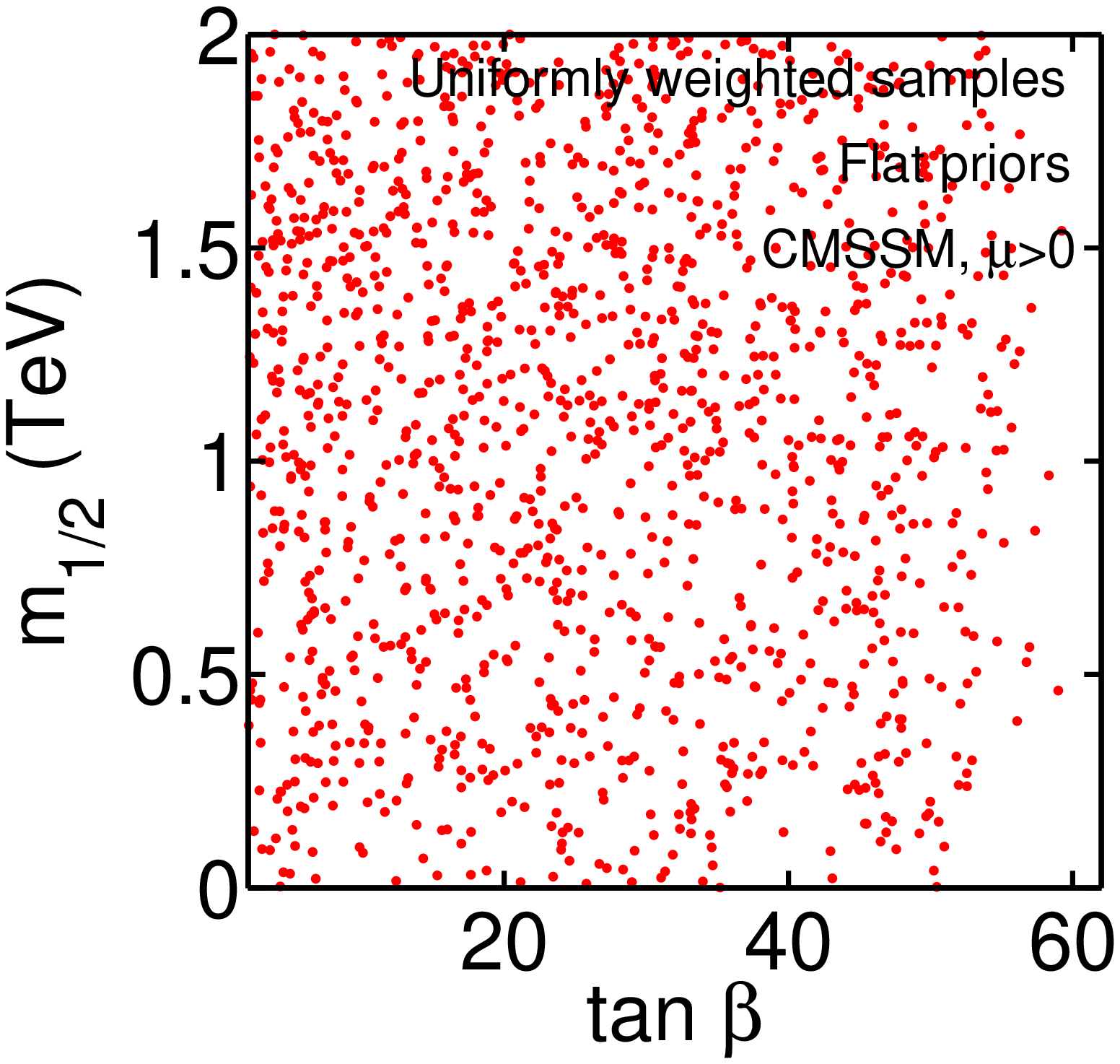}
\includegraphics[width=\qq]{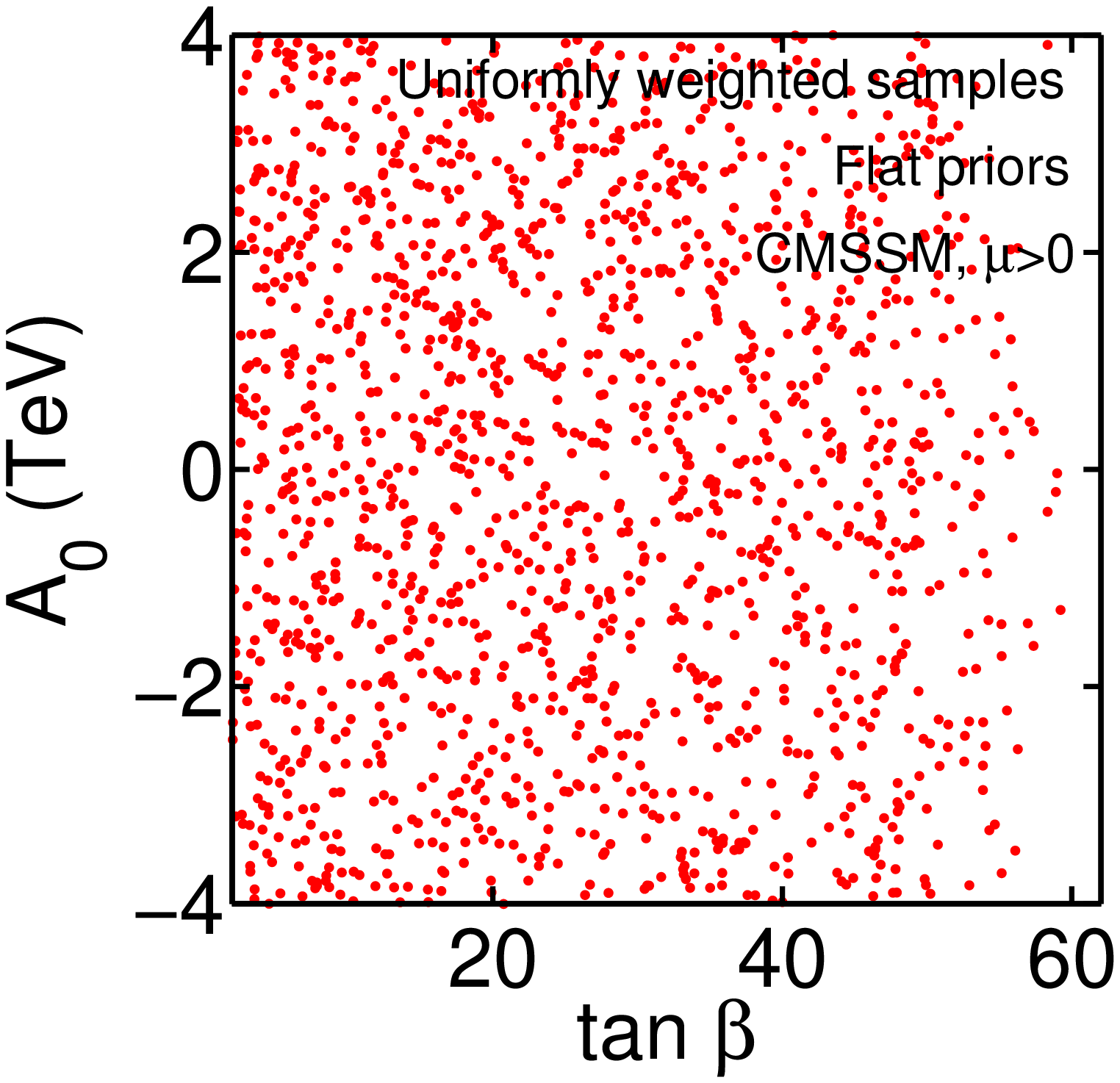}
\includegraphics[width=\qq]{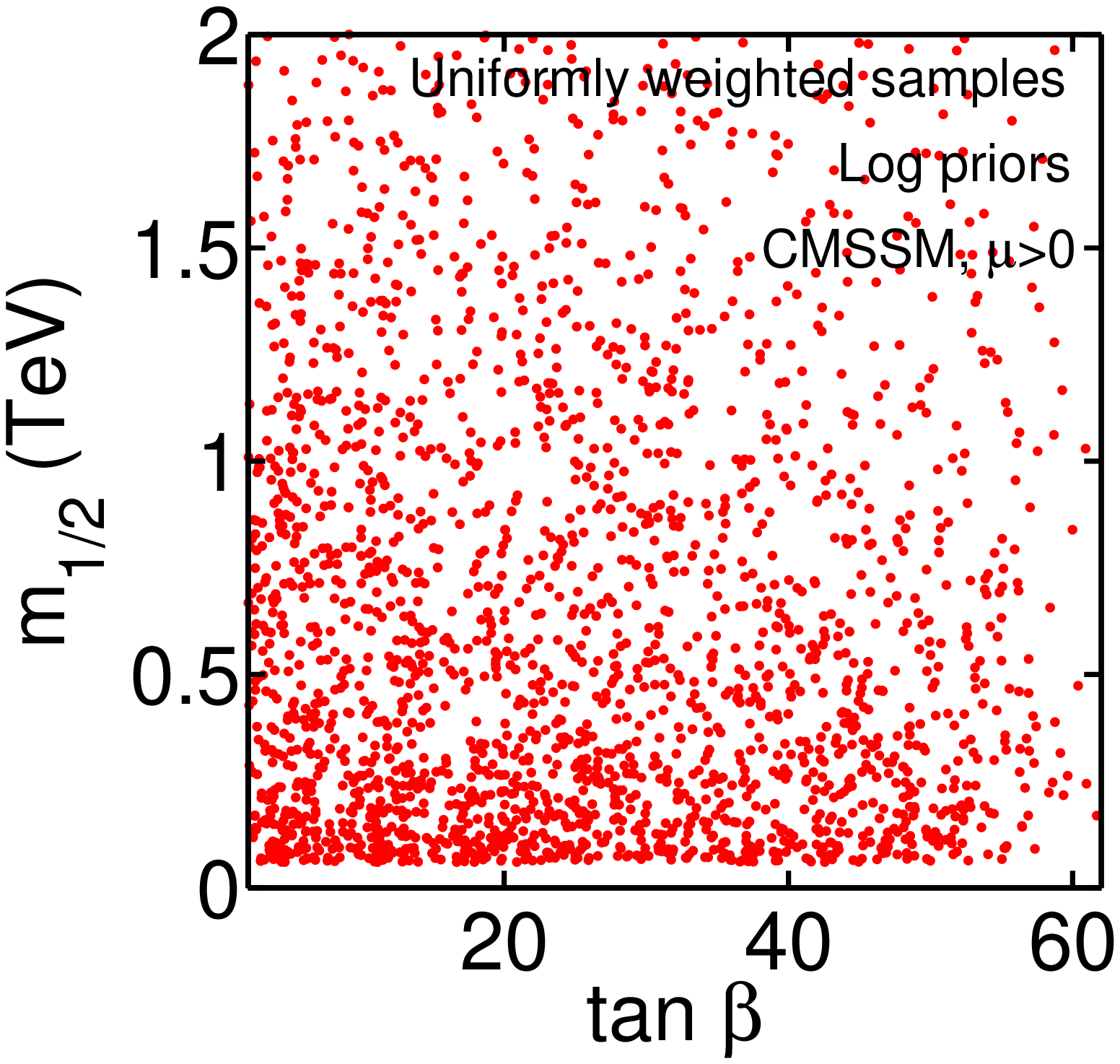}
\includegraphics[width=\qq]{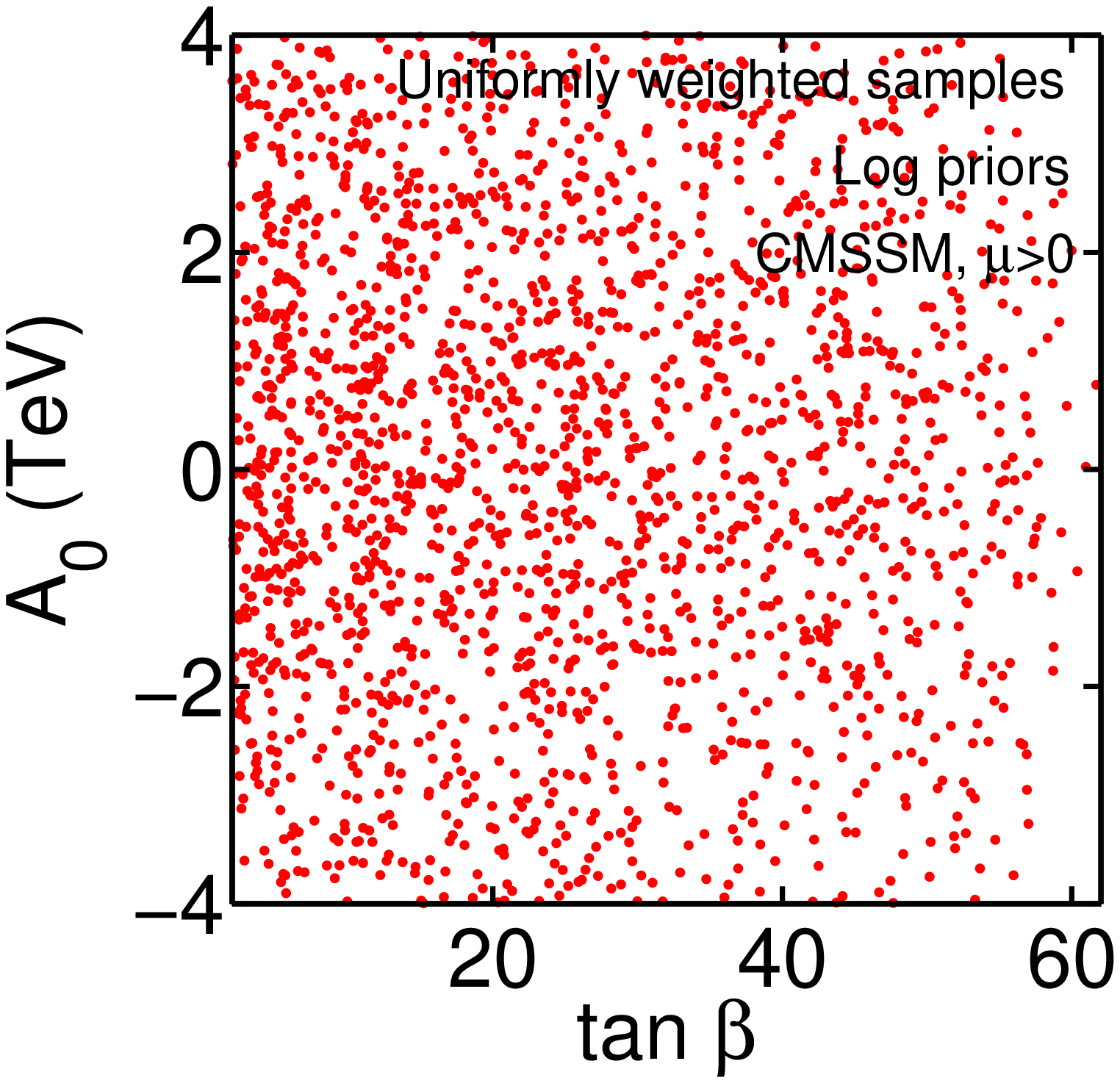} 
\caption[short]{A scan including no experimental data, but only the
requirement of physicality (\texttt{PHYS}), for flat priors (panels in
the left two columns) and log priors (panels in the right two
columns). Samples are drawn with equal weight from the prior, hence
their density reflects 2D probability for different projections on the
CMSSM parameters.}
\label{fig:2D_priors}
\end{center}
\end{figure} 

The above points can be confirmed by looking at the corresponding 2D
distributions, which are shown in fig.~\ref{fig:2D_priors}. There we
plot samples drawn with uniform weight from the prior (once the
physicality constraints have been imposed), hence the density of
samples reflect the prior pdf.

\begin{figure}[tbh!]
\begin{center}
\includegraphics[width=\ww]{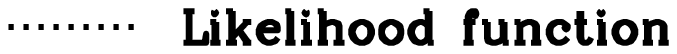} \includegraphics[width=\ww]{figs/BlackLike.ps}\\
\includegraphics[width=\ww]{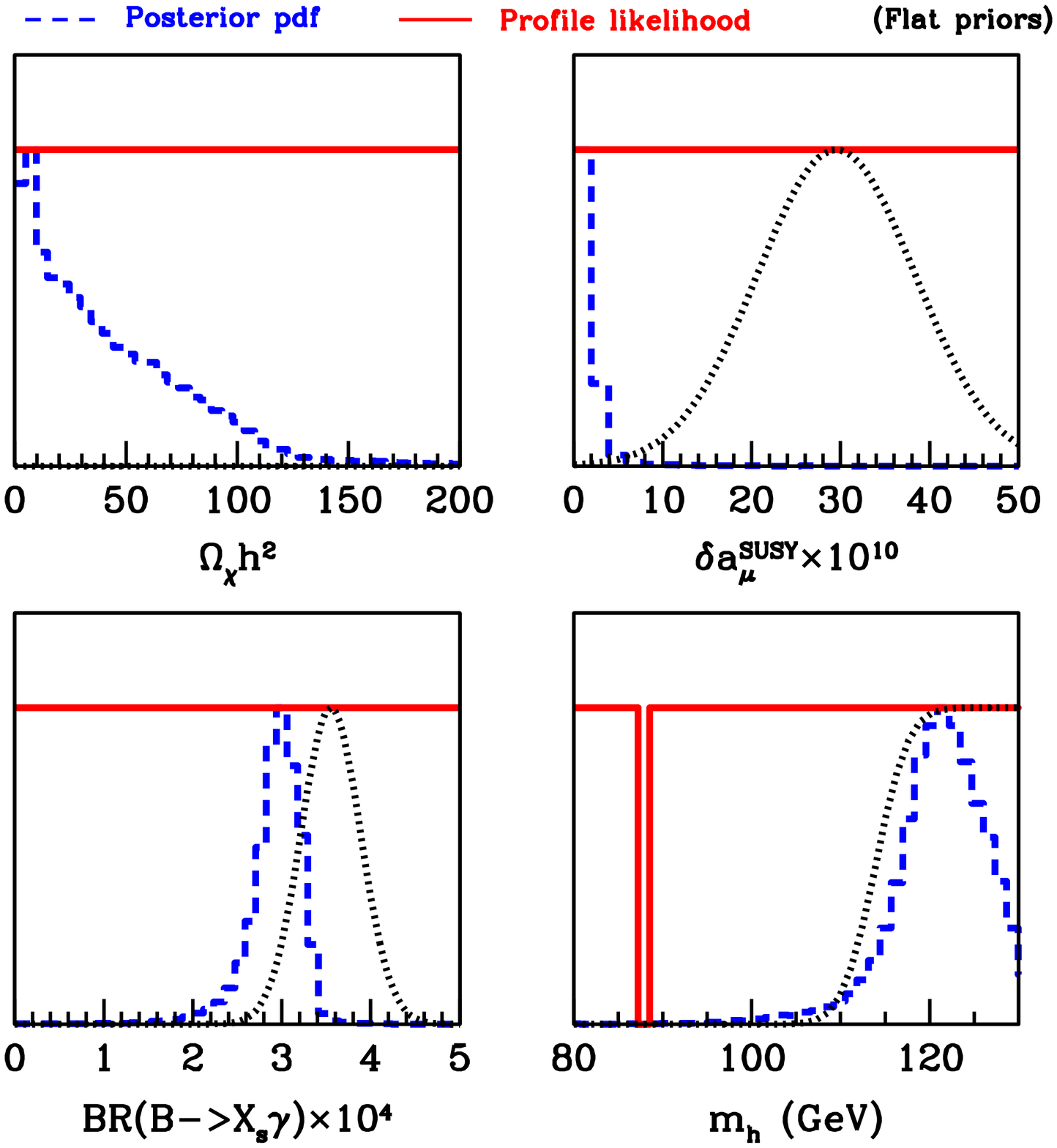}
\includegraphics[width=\ww]{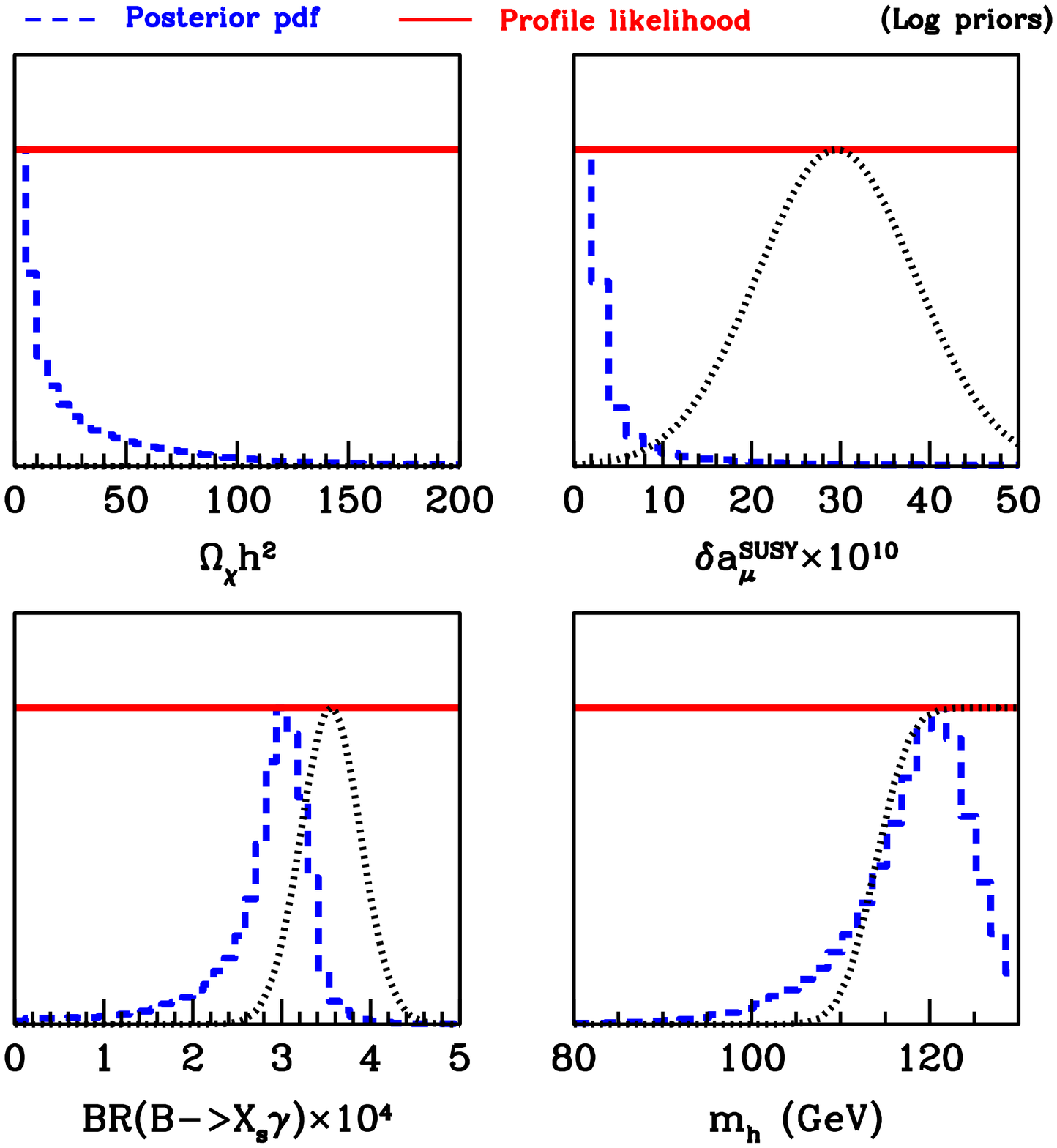} 
\caption[test]{A scan including no experimental data, but only the
requirement of physicality (\texttt{PHYS}). The posterior probability
distribution (dashed blue) and the profile likelihood (solid red) for
the most constraining observables (with flat priors on the left, and
log priors on the right): the DM relic abundance $\abundchi$ of the
neutralino, the excess in the anomalous magnetic moment of the
muon $\deltaamususy$, the $\brbsgamma$ and the lightest Higgs mass
$\mhl$. For comparison, the dotted black, smooth curves give the
likelihood function for the plotted observable (not imposed in this
scan). For the DM abundance, the likelihood function plotted shows
only the experimental error (i.e., it does not include the theoretical
error employed in the scan).
\label{fig:1Dpriors_obs}
}
\end{center}
\end{figure}

\begin{figure}[tbh!]
\begin{center}
\includegraphics[width=\ww]{figs/BlackLike.ps} \includegraphics[width=\ww]{figs/BlackLike.ps}\\
\includegraphics[width=\ww]{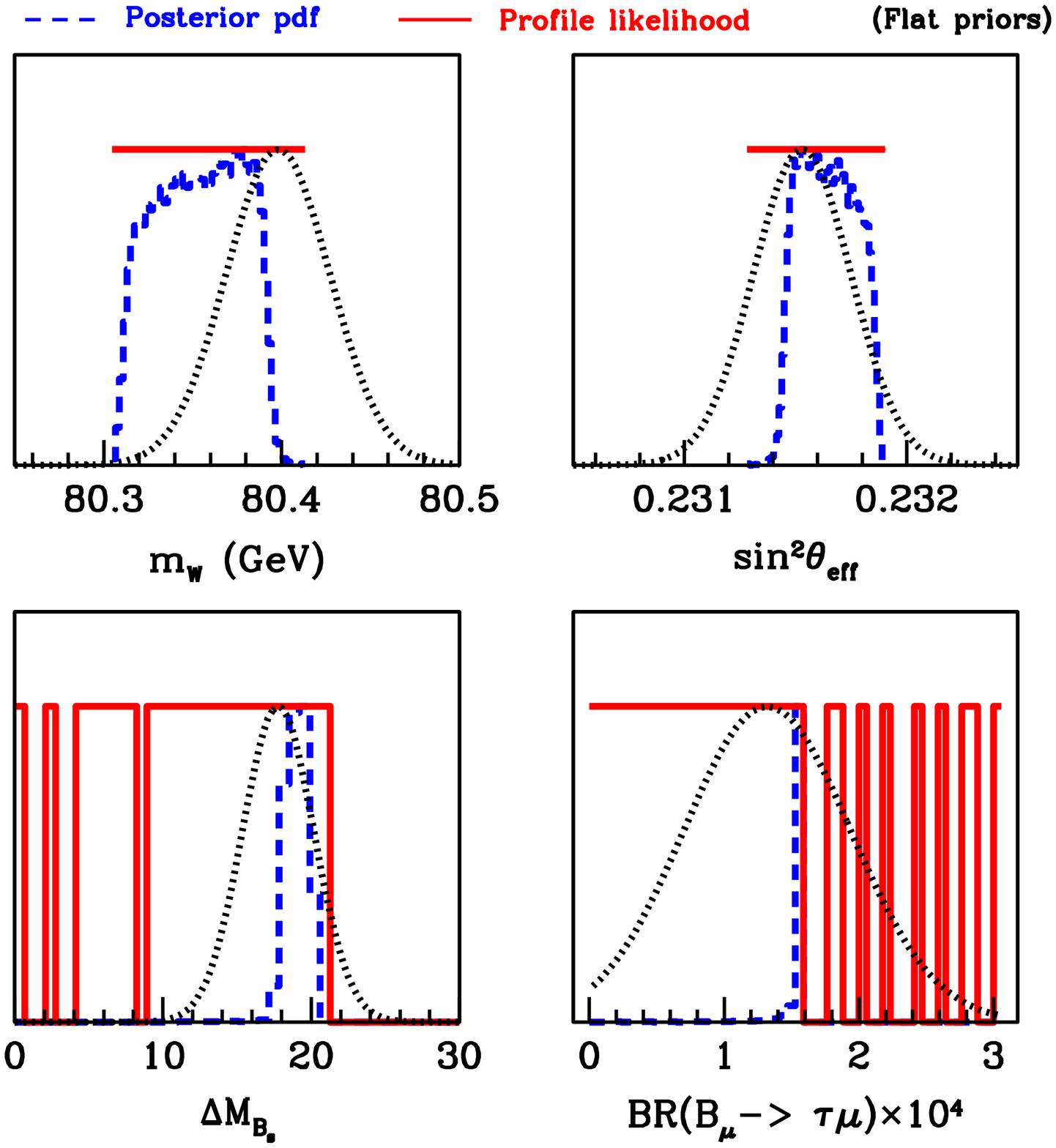}
\includegraphics[width=\ww]{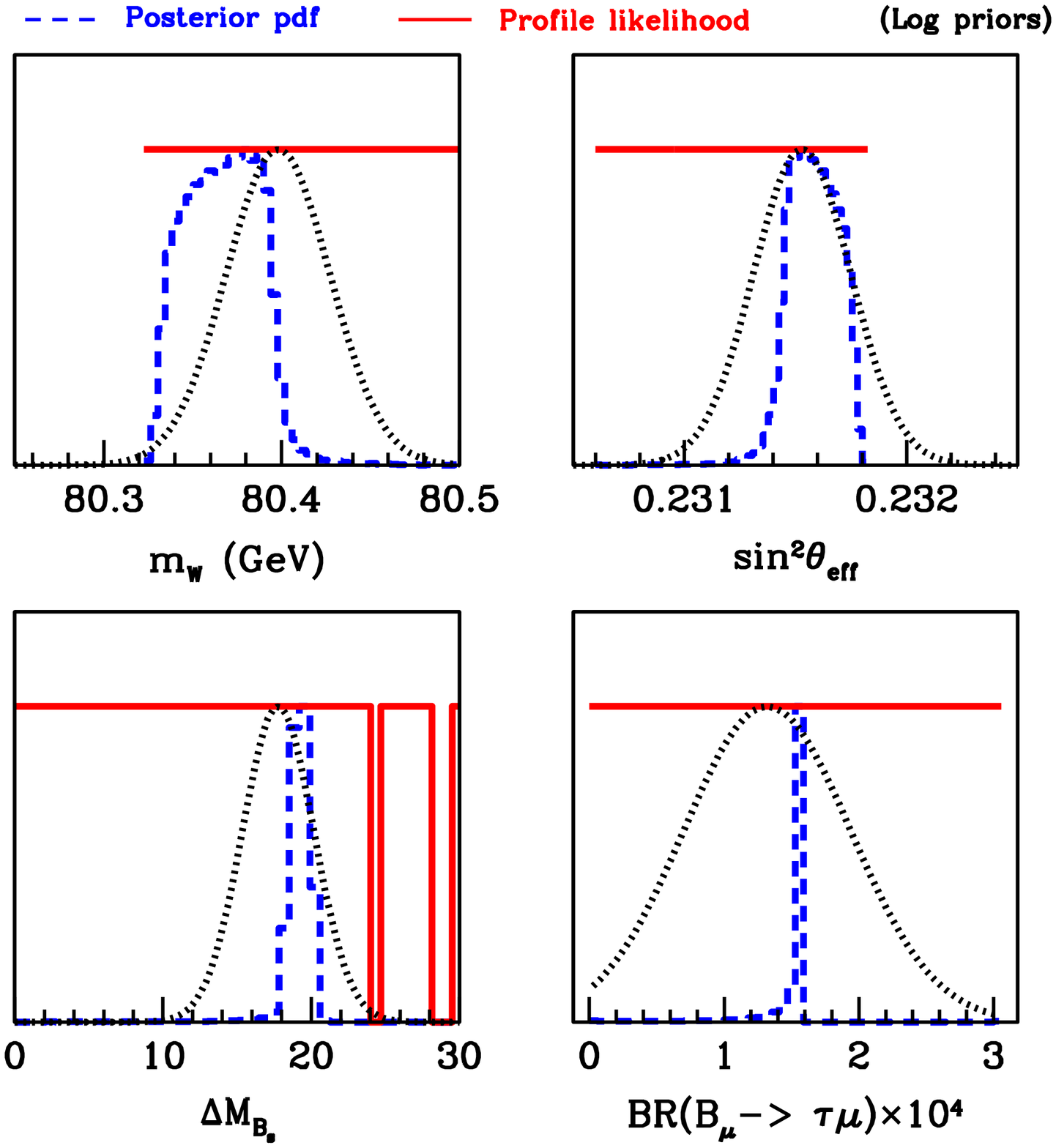}
\caption[test]{As in fig.~\ref{fig:1Dpriors_obs}, but for some other
 observables. No experimental constraints have been imposed but only the
 requirement of physicality (\texttt{PHYS}) for both flat priors
 (left panels) and log priors (right panels). We plot the posterior
 probability distribution (dashed blue) and the profile likelihood
 (solid red). For comparison, the dotted black, smooth curves give
 the likelihood function for the plotted observable (not imposed in
 this scan). The range of the profile likelihood (solid red line) gives the range of values for the quantities covered by the scan, as a consequence of the priors presented in section~\ref{sec:priors}.
\label{fig:1Dpriors_obs_2}
}
\end{center}
\end{figure}

It is interesting to consider the implied distribution for the
observable quantities. This can be understood as a predictive
distribution from the priors and the physicality constraints for the
observables.  In fig.~\ref{fig:1Dpriors_obs} we present the 1D
distributions of the posterior (dashed blue) and the profile
likelihood (solid red) for the quantities which will play the most
important role in constraining base parameters. For comparison, for
each observable we also display the likelihood function (dotted
black), which however has {\em not} been imposed in this scan.  The
two left (right) columns are for the flat (log) prior.

Starting from the CDM abundance, we note that, in the absence of
constraints from the data, for both choices of priors, the neutralino
relic density is typically much larger than unity, as is well
known. When we later impose the WMAP constraint (see below), we will
therefore expect that the posterior will be dominated by the
likelihood, since the prior is much wider (by orders of magnitude)
than the likelihood. We also note that, in contrast, the profile
likelihood remains flat out to much larger values --- a reflection of
the fact that the Bayesian posterior is suppressed because only a small number
of samples is found with an extremely large relic abundance ($\abundchi \gg
100$). 

On the other hand, the posterior for $\deltaamususy$ is very strongly
peaked around zero. This is a consequence of the overwhelming number
of samples in the FP region, where the large superpartner masses lead
to a strong suppression in the SUSY contribution to
$\deltaamususy$. Even the log prior can only give a
slight extra weight to the pdf for larger values of
$\deltaamususy$. Again, the profile likelihood is unaffected by the
choice of priors.

Similar reasoning can also explain the fairly strong peak in the
posterior for $\brbsgamma$ at $\sim 3 \times 10^{-4}$, below the SM
central value. This is the result of the negative (for $\mu>0$)
chargino/stop contribution often overriding the always positive
charged Higgs/top contribution.  Finally, a large concentration of
samples at large $\mhalf$ and $\mzero$ also accounts for the fairly
strongly peaked distribution in the pdf of the lightest Higgs mass
$\mhl$. In contrast, the profile likelihood is not affected by such
volume effects, and remains flat, except for small dip at
$\mhl\sim88\gev$, well below the LEP limit (where the scan has not
found any point satisfying the physicality constraints). This is
likely to be the consequence of the finite number of samples we could
gather.

In fig.~\ref{fig:1Dpriors_obs_2} we plot the predictive distribution
from the prior for the EW precision observables and $b$--physics
quantities. Notice how for both choices of priors the marginal pdf
implied by the prior (dashed blue) is typically much more strongly
peaked than the likelihood function (dotted black). This means that
the constraining power of the data for these quantities is expected to
be smaller than the information already implied by the prior (see
section~\ref{sec:infocontent} for more details). Therefore, as we
shall explicitely show below, the impact of including them in the
likelihood will be fairly limited.

To summarize, the key point is that, as we have emphasized at the beginning of this
section, in the CMSSM (and, more generally, in a class of effective
SUSY models where input parameters are defined at some high scale),
the connection between the basis parameters and the observable
quantities (other than the nuisance parameters, which obviously are
directly constrained) is {\em highly non-linear}. Therefore the data,
although constraining fairly strongly some of the observables, can only
give indirect constraints on the parameters of the
model. This is because one can move them around in order to satisfy a
given constraint. Therefore plotting the posterior for the obervables
in the absence of data gives the amount by which the prior measure
impacts on the observable quantities. Another way of
interpreting the above behavior is as the prior-predictive 
 distribution for the
observable quantities, i.e., the probability distribution for the
observables implied by the choice of priors.

\subsection{Impact of collider data, CDM abundance,
 \boldmath$\bsg$ and $\gmt$}\label{sec:colcdmbsggmt}

We now move on to adding the other constraint sets from
table~\ref{tab:datacombinations} and investigate how they influence
the conclusions obtained above for the two statistical measures and
for our choices of priors.

\begin{figure}[tbh!]
\begin{center}
\includegraphics[width=\ww]{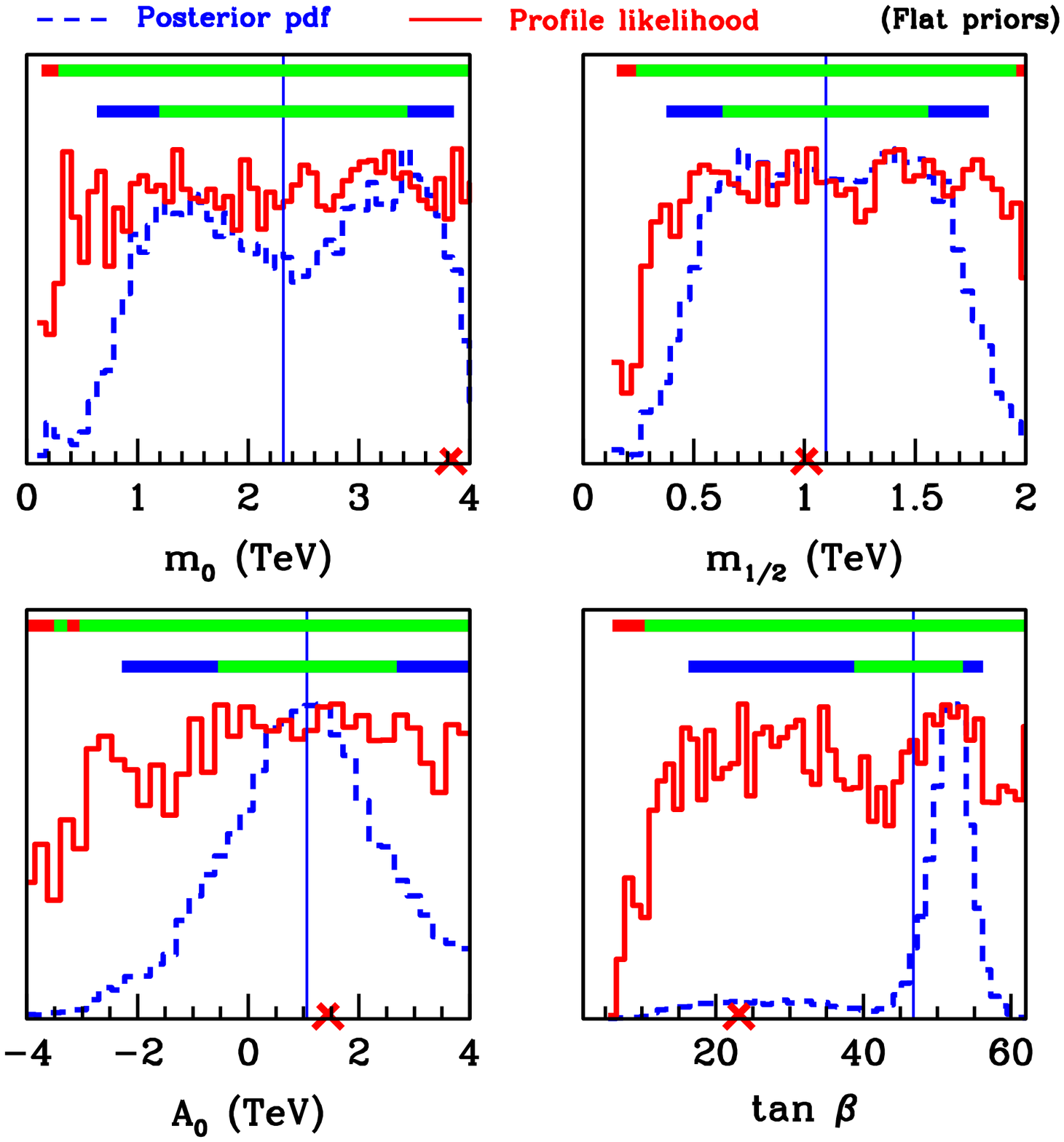}
\includegraphics[width=\ww]{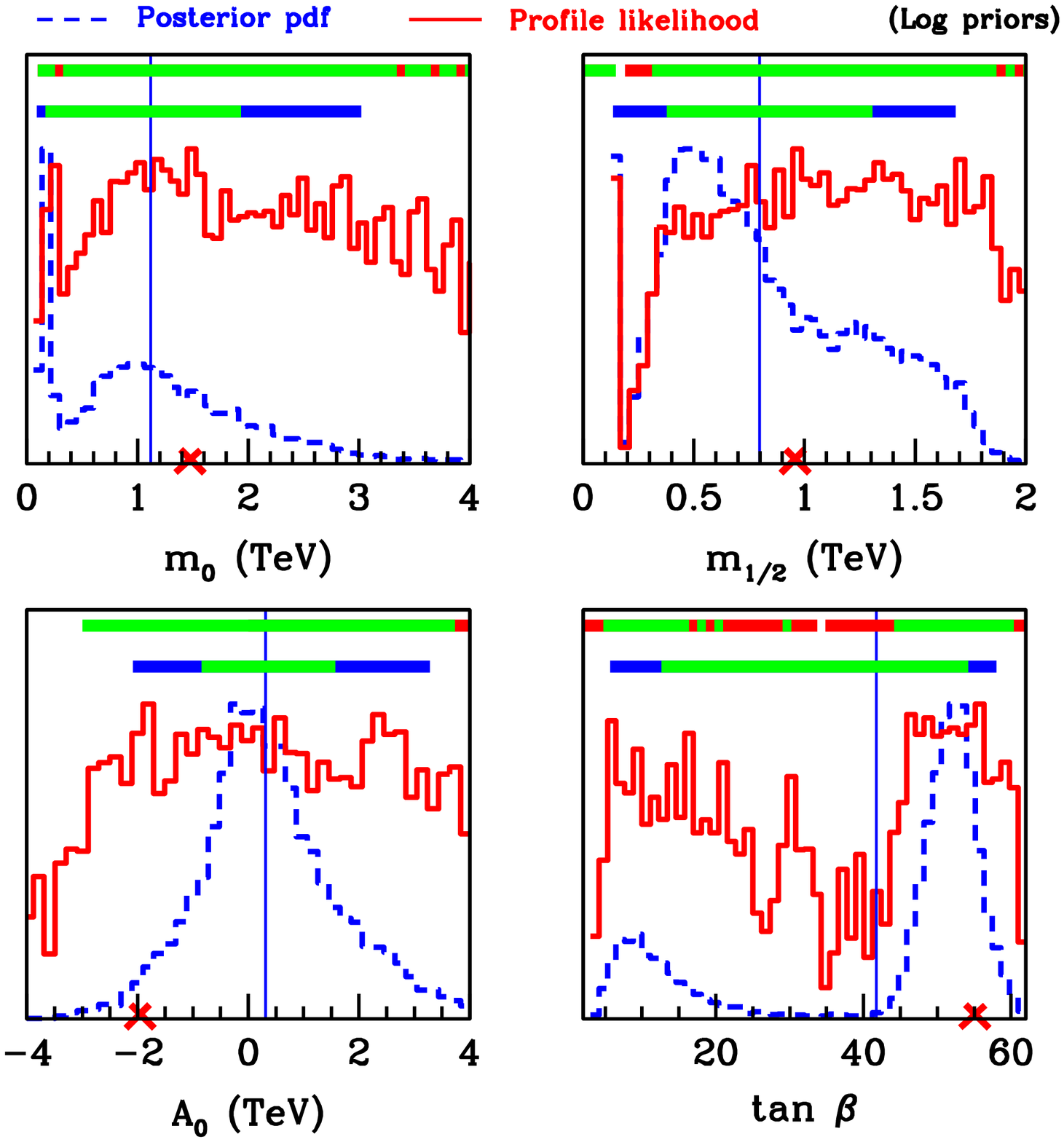}\\
\caption[test]{As in fig.~\ref{fig:1Dpriors_base}, but now adding the
 constraint on SM nuisance parameters, collider limits on Higgs and
 superpartner masses and the WMAP5 CDM abundance determination
 (\texttt{PHYS+NUIS+COLL+CDM}), for flat/log priors (panels in the
 two left/right columns).  The vertical, thin line is the posterior
 mean, the red cross the best-fit point. The horizontal bars on the
 top express in a graphical way the constraints on the parameters:
 the top bar gives 68\% (green) and 95\% (red) limits from the
 profile likelihood, while the bar below it gives 68\% (green) and
 95\% (blue) intervals from the marginal pdf.}
\label{fig:1Dpriors+nuis+lep+cdm}
\end{center}
\end{figure}

\begin{figure}[tbh!]
\includegraphics[width=\qq]{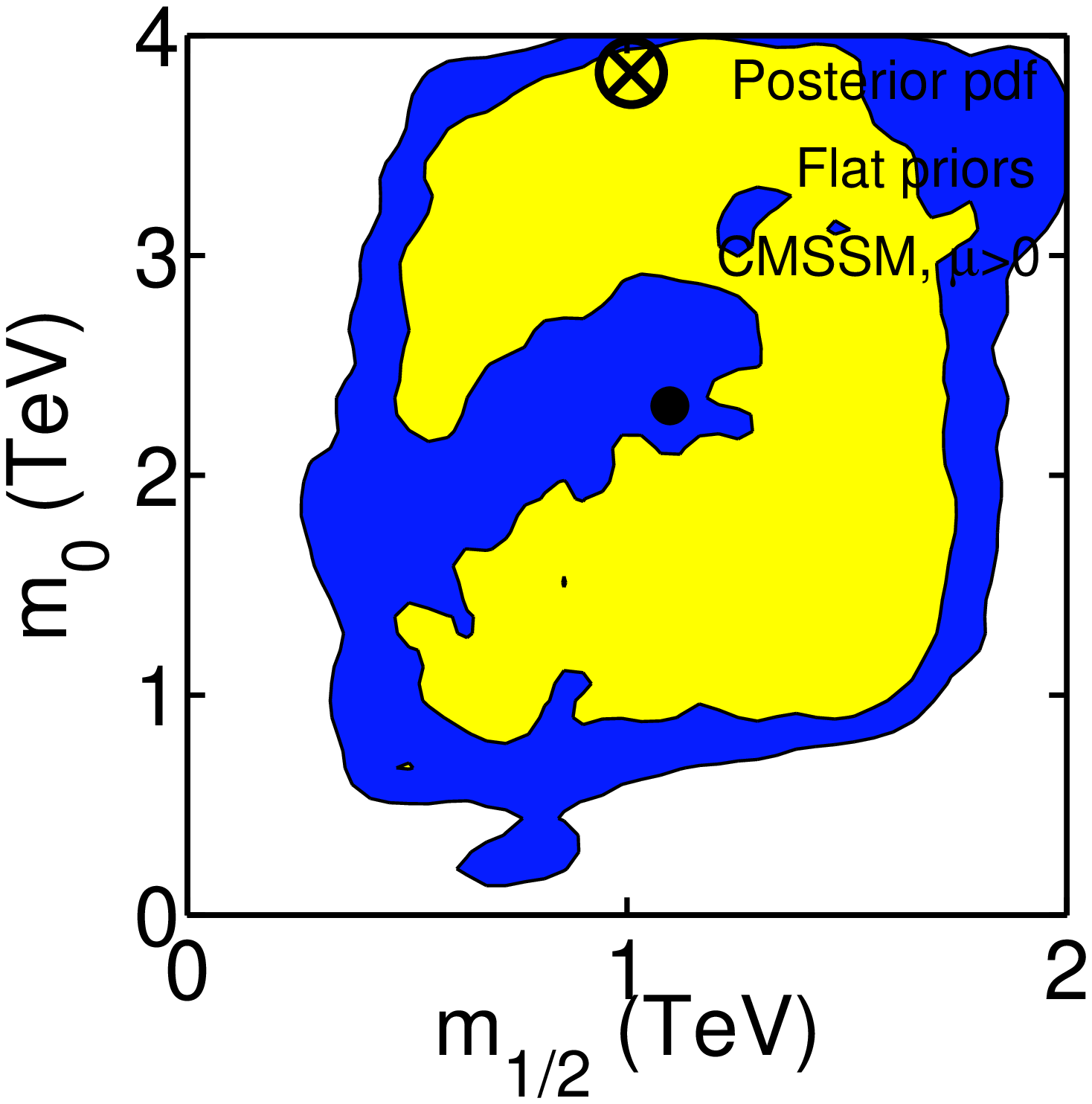}
\includegraphics[width=\qq]{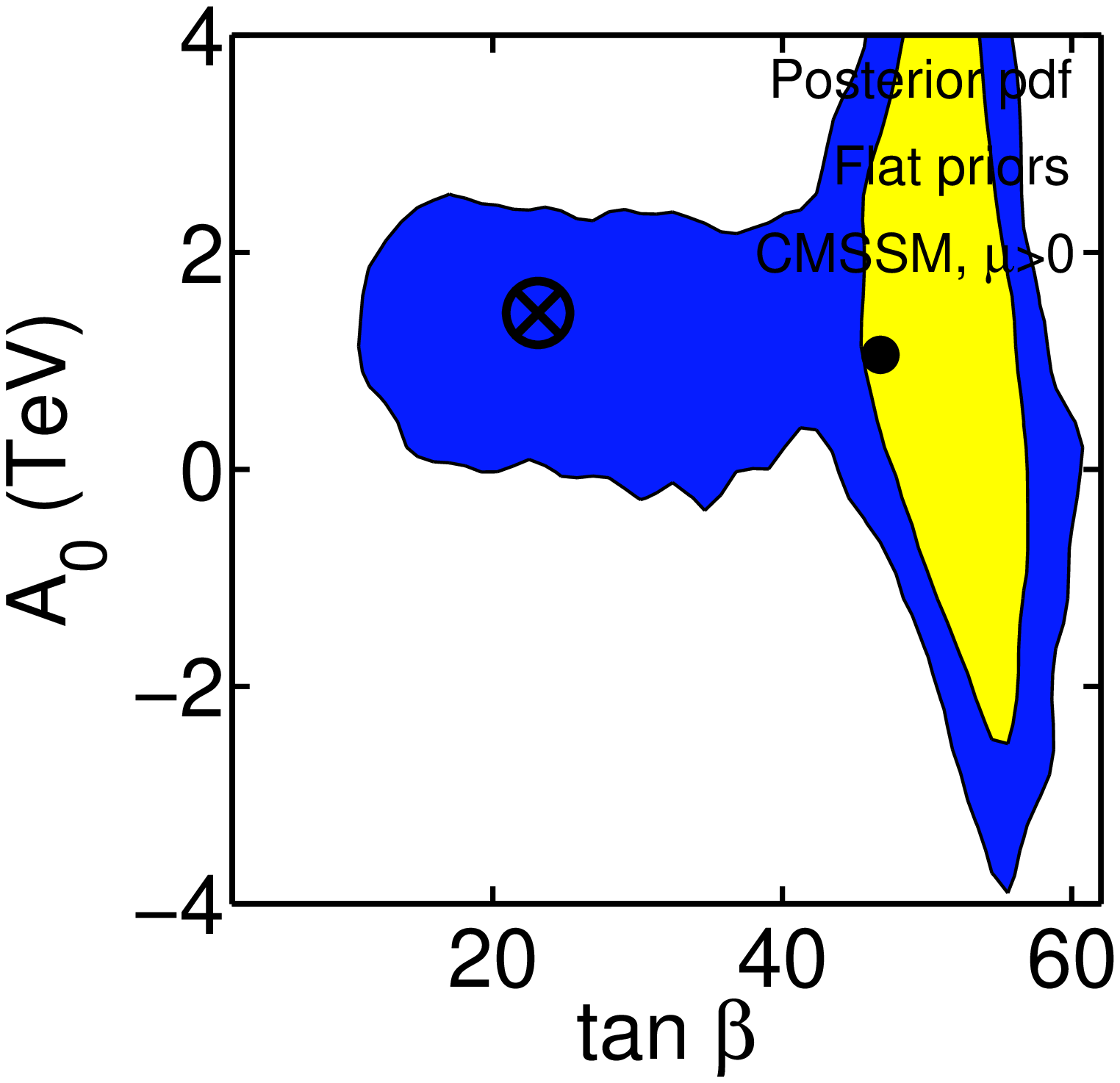}
\includegraphics[width=\qq]{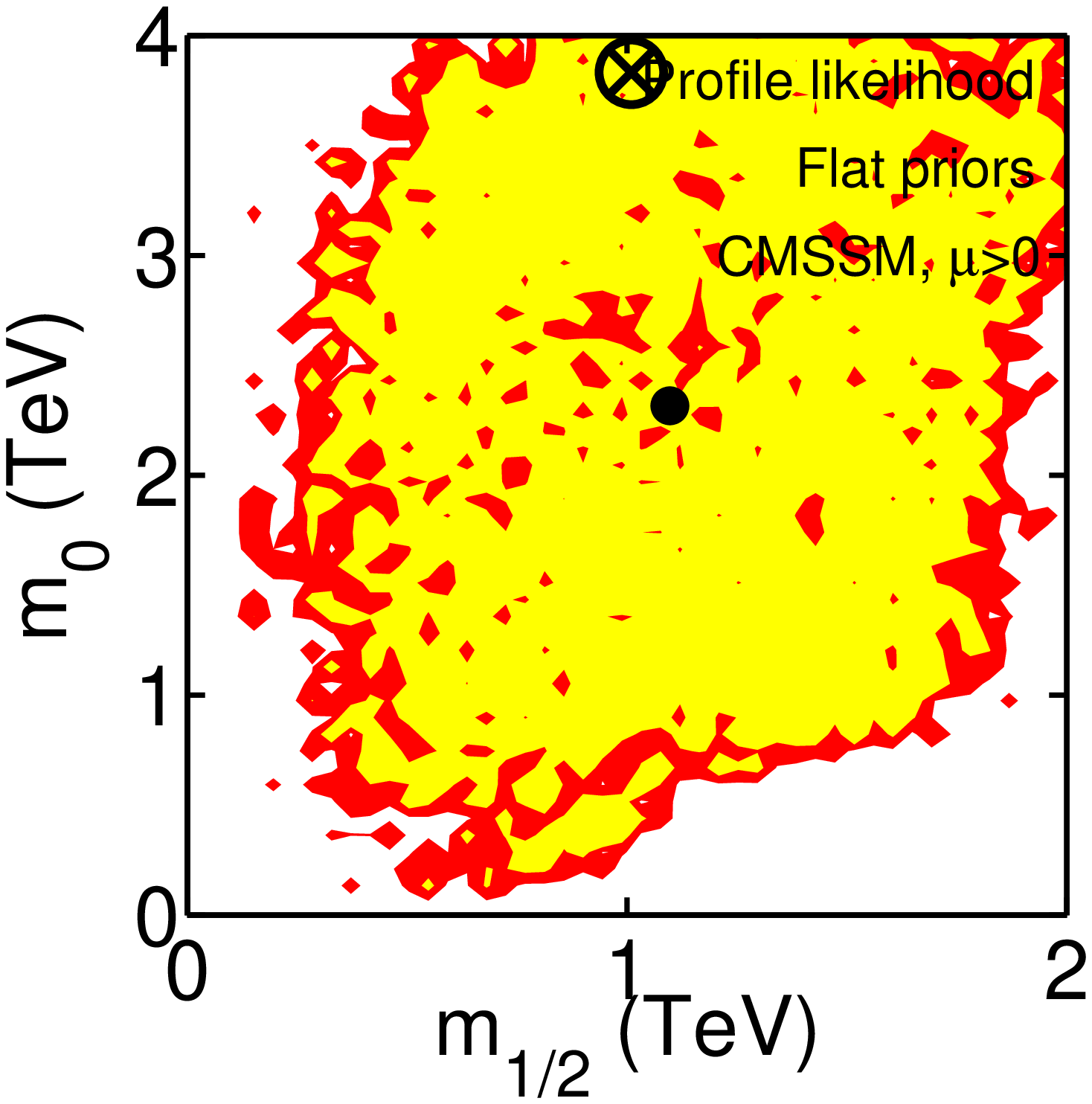}
\includegraphics[width=\qq]{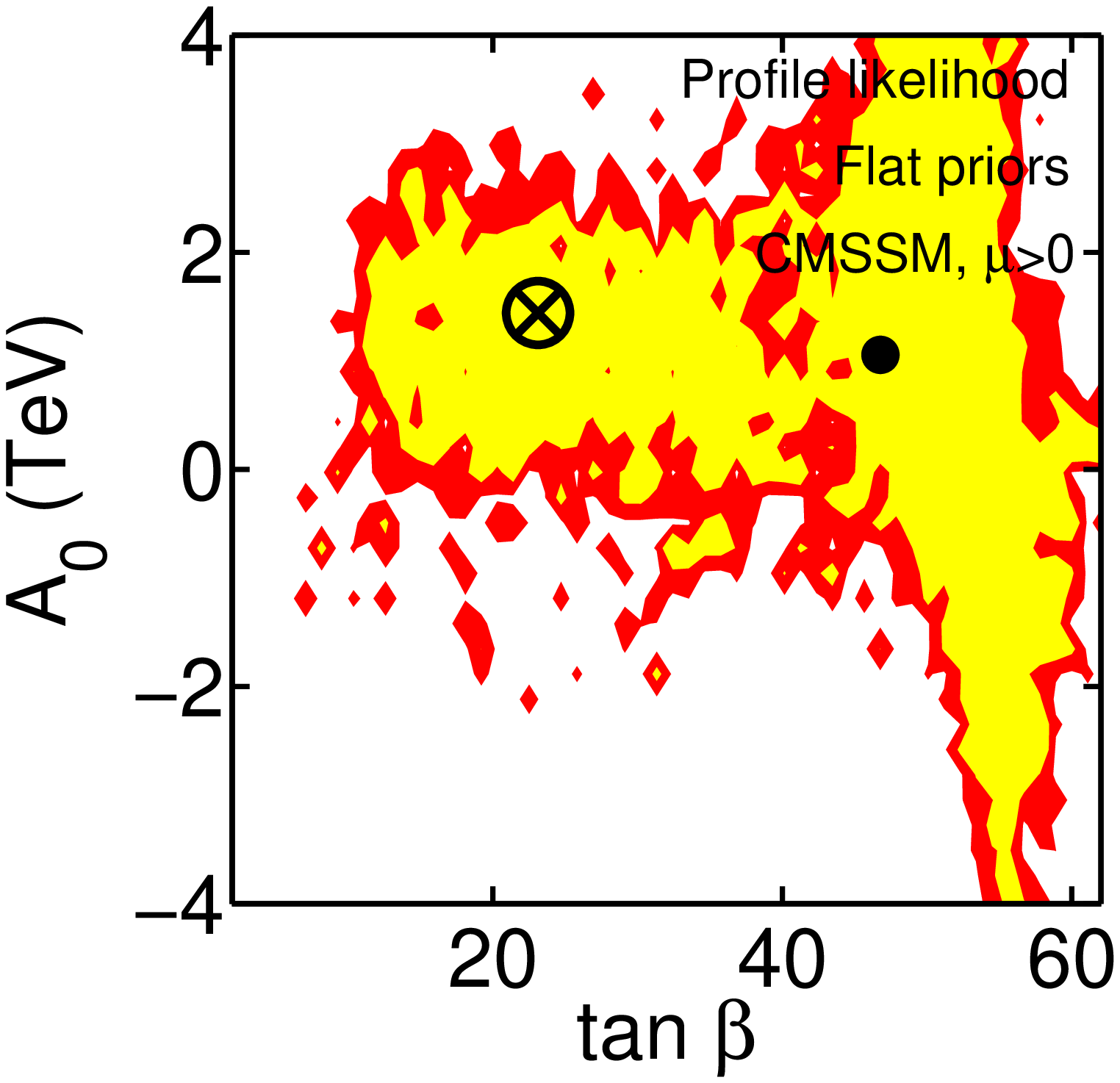}
\\
\includegraphics[width=\qq]{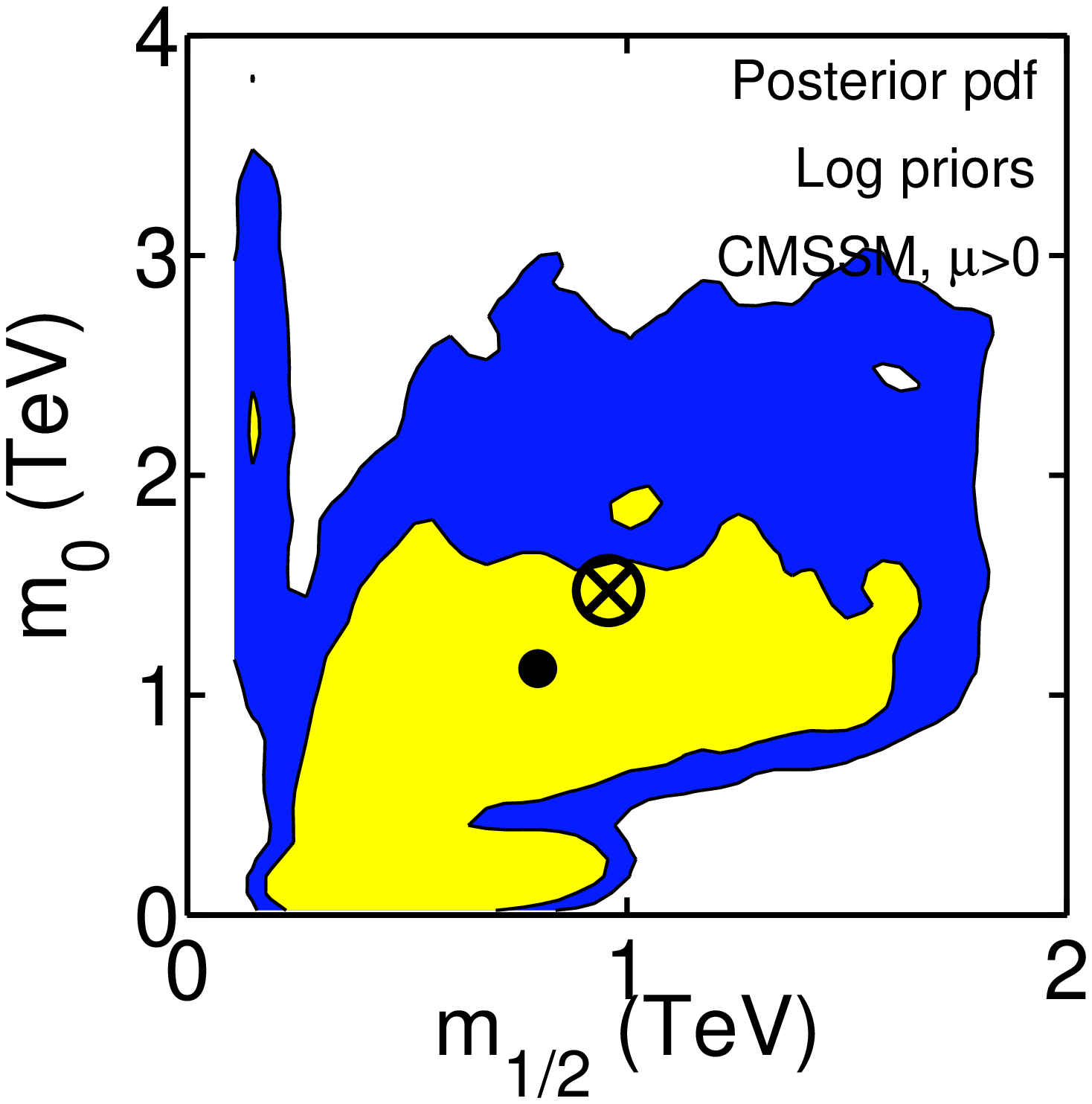}
\includegraphics[width=\qq]{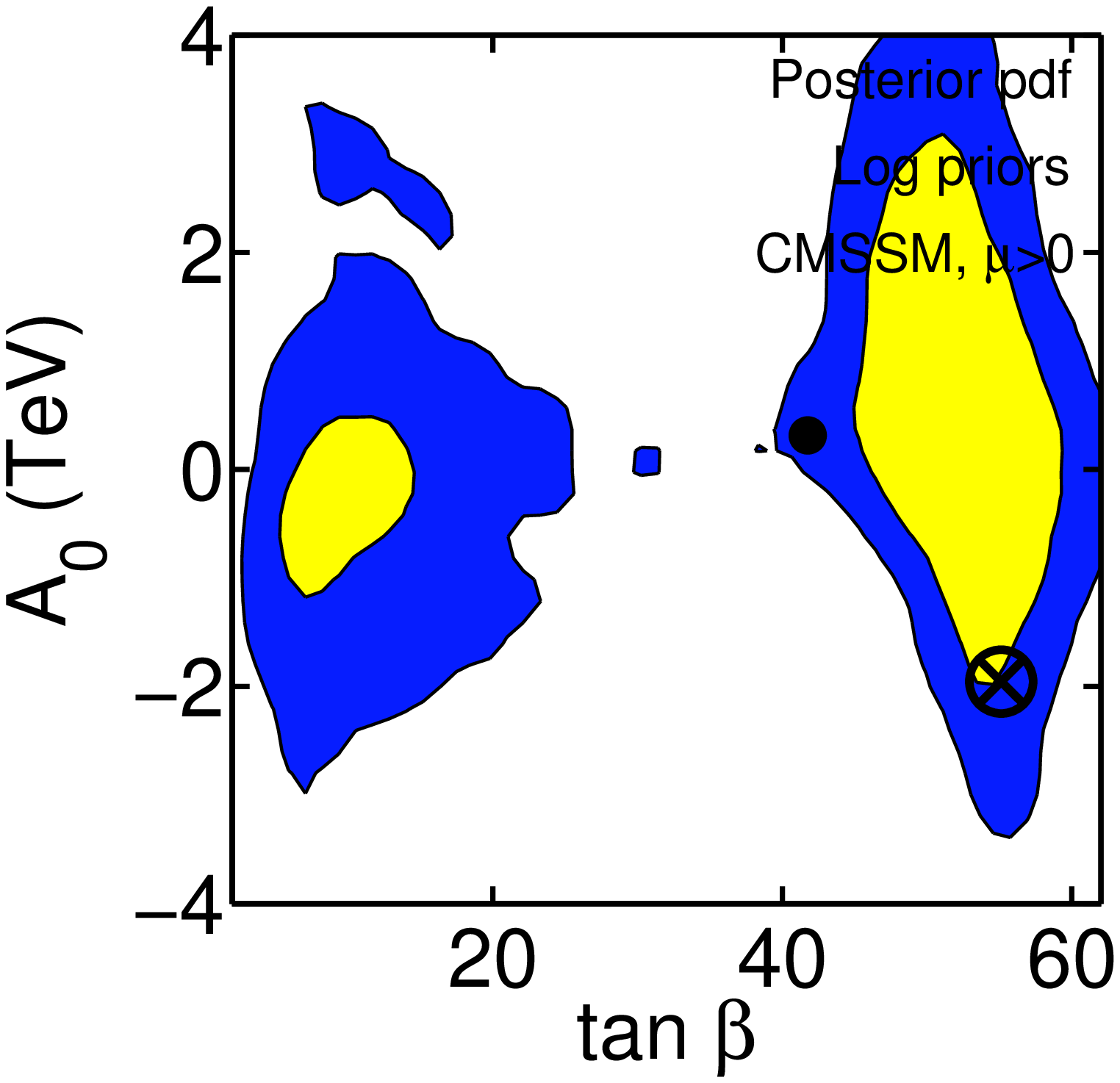}
\includegraphics[width=\qq]{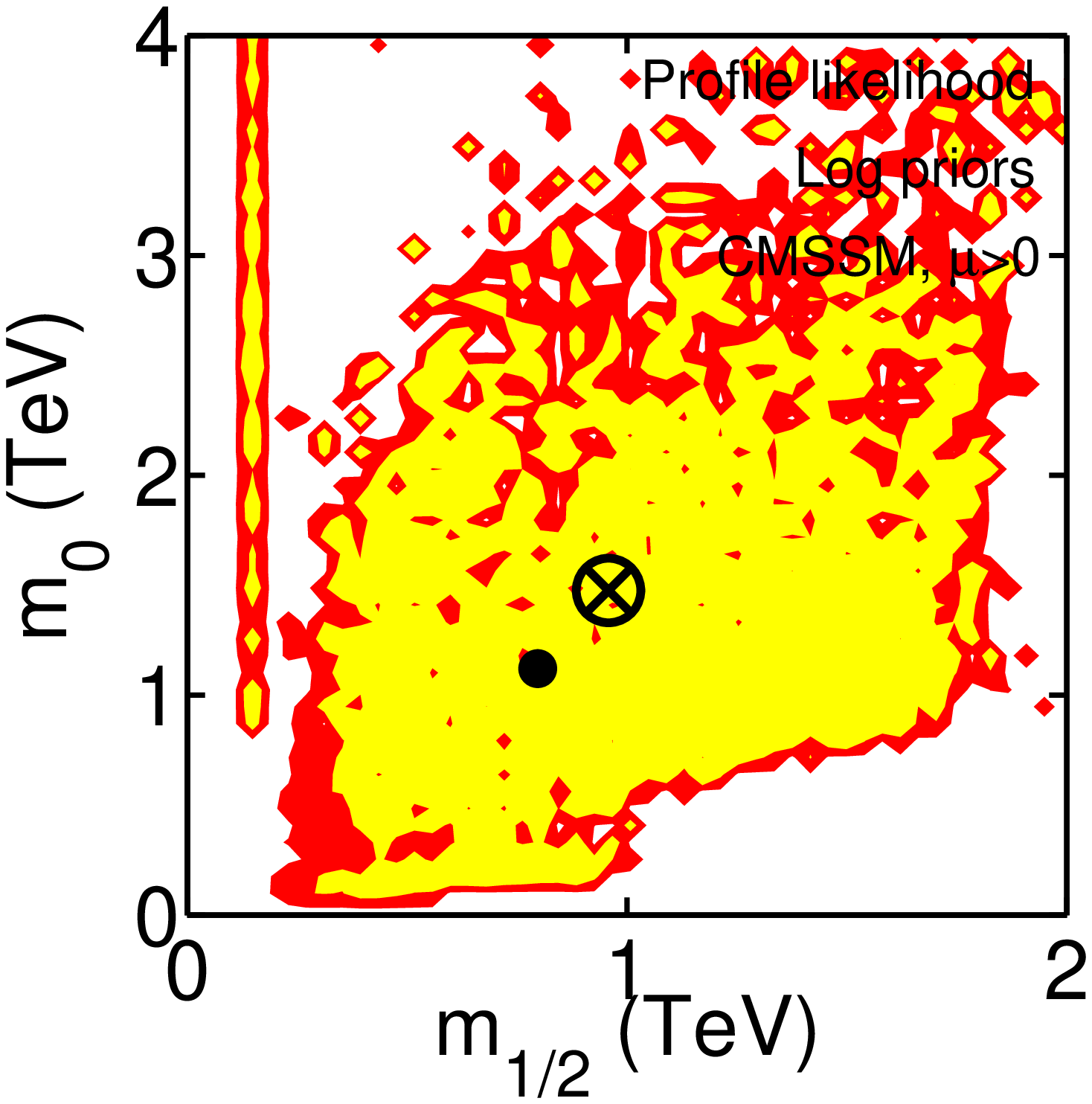}
\includegraphics[width=\qq]{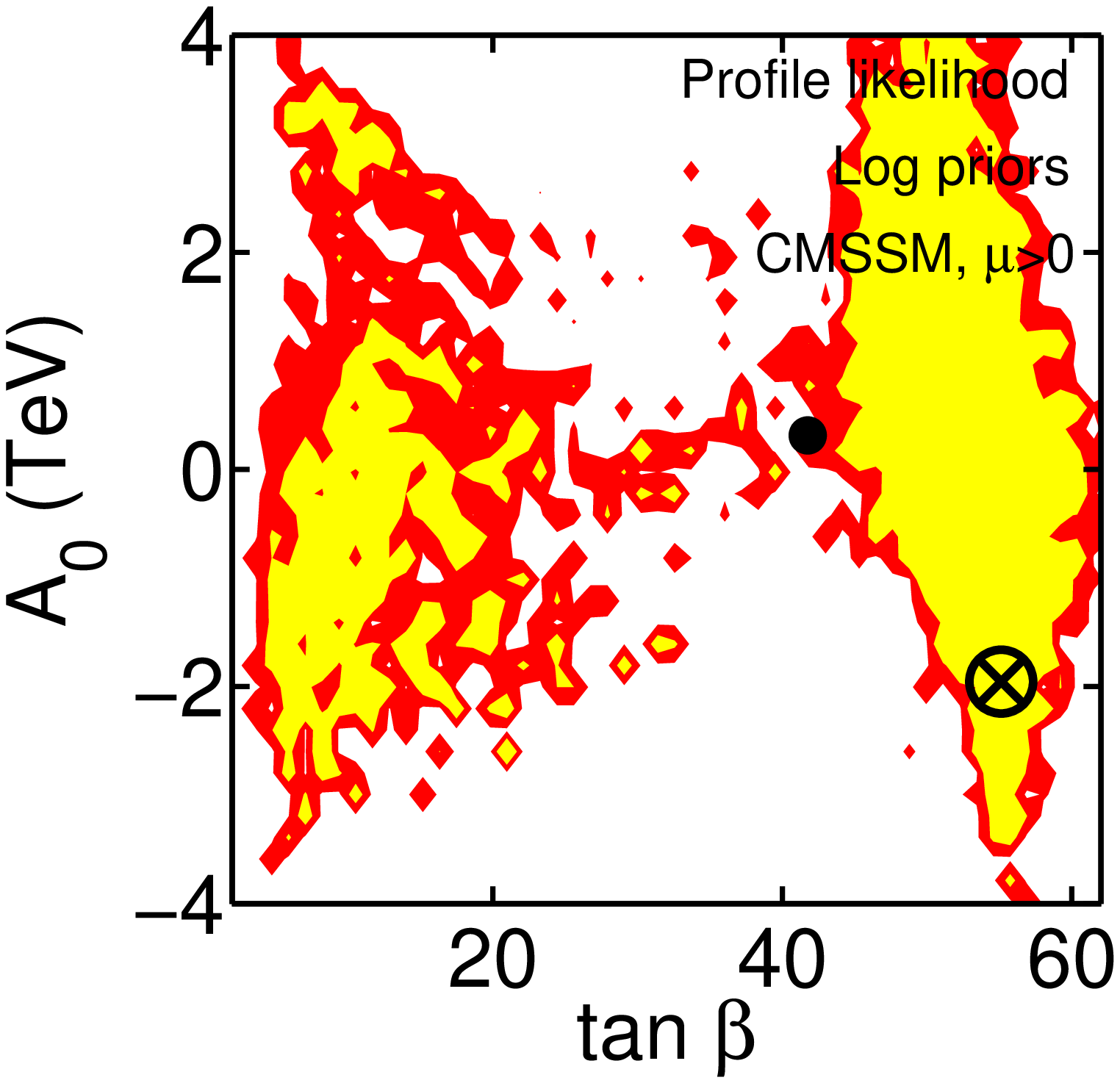}
\caption[test]{Posterior pdf (left two columns) and profile likelihood
(right two columns) for flat priors (top row) and log priors (bottom
row) for a scan including SM nuisance parameters constraints, collider
limits on Higgs and superpartner masses and the WMAP5 CDM abundance
determination (\texttt{PHYS+NUIS+COLL+CDM}). The inner and outer
contours enclose respective 68\% and 95\% joint regions for both
statistics. The posterior pdf has been smoothed with a Gaussian kernel
of 1 bin width for display purposes. The cross gives the best-fit
point, the filled circle is the posterior mean. \label{fig:2D_priors+nuis+lep+cdm}}
\end{figure}

First, in fig.~\ref{fig:1Dpriors+nuis+lep+cdm} we show the CMSSM
parameters (as in fig.~\ref{fig:1Dpriors_base}) but now with data on SM
nuisance parameters, collider limits on Higgs and superpartner masses
and the WMAP5 CDM abundance determination added to the likelihood
(\texttt{PHYS+NUIS+COLL+CDM}). Corresponding 2D posterior pdf and
profile likelihood for some of the CMSSM variable combinations
are shown in fig.~\ref{fig:2D_priors+nuis+lep+cdm}.

By examining both figures, it is clear that the resulting constraints
on the CMSSM parameters depend very much on the chosen statistical
measure. For example, while in the log prior case
the posterior pdf shows a stronger preference smaller $\mzero$ than
with the flat prior (and a strong peak at small $\mzero$), the profile
likelihood remains essentially flat across all CMSSM parameters for
both choices of priors.
This is an indication that the data employed are not providing
sufficient constraints on the parameters. More generally, we can see
that the profile likelihood gives more conservative limits than the
posterior pdf. These features can also be seen in
fig.~\ref{fig:2D_priors+nuis+lep+cdm} (2D distributions).  The
95\% contours are broadly similar for both statistics for a given
choice of prior, but are quite different for the two different
priors. In general, the log prior favors more strongly the low energy
region. We have also found that the chi-square of the best fit point
(indicated by a cross) is lower for the log prior scan than the flat
prior scan. There are also evident differences between the location of
the best fit point and the posterior mean (indicated by a filled
dot). This results from the fact the the posterior mean is influenced
by the posterior distribution and its associated volume whose distribution depends fairly strongly on
the chosen prior.

On the other hand, the nuisance parameters are already at this
point extremely well
constrained by the Gaussian likelihood, for both the Bayesian pdf and
the profile likelihood statistics. The two statistics are almost
identical for those variables and equal to the experimental
likelihood, hence we do not show them here.

\begin{figure}[tbh!]
\begin{center}
\includegraphics[width=\ww]{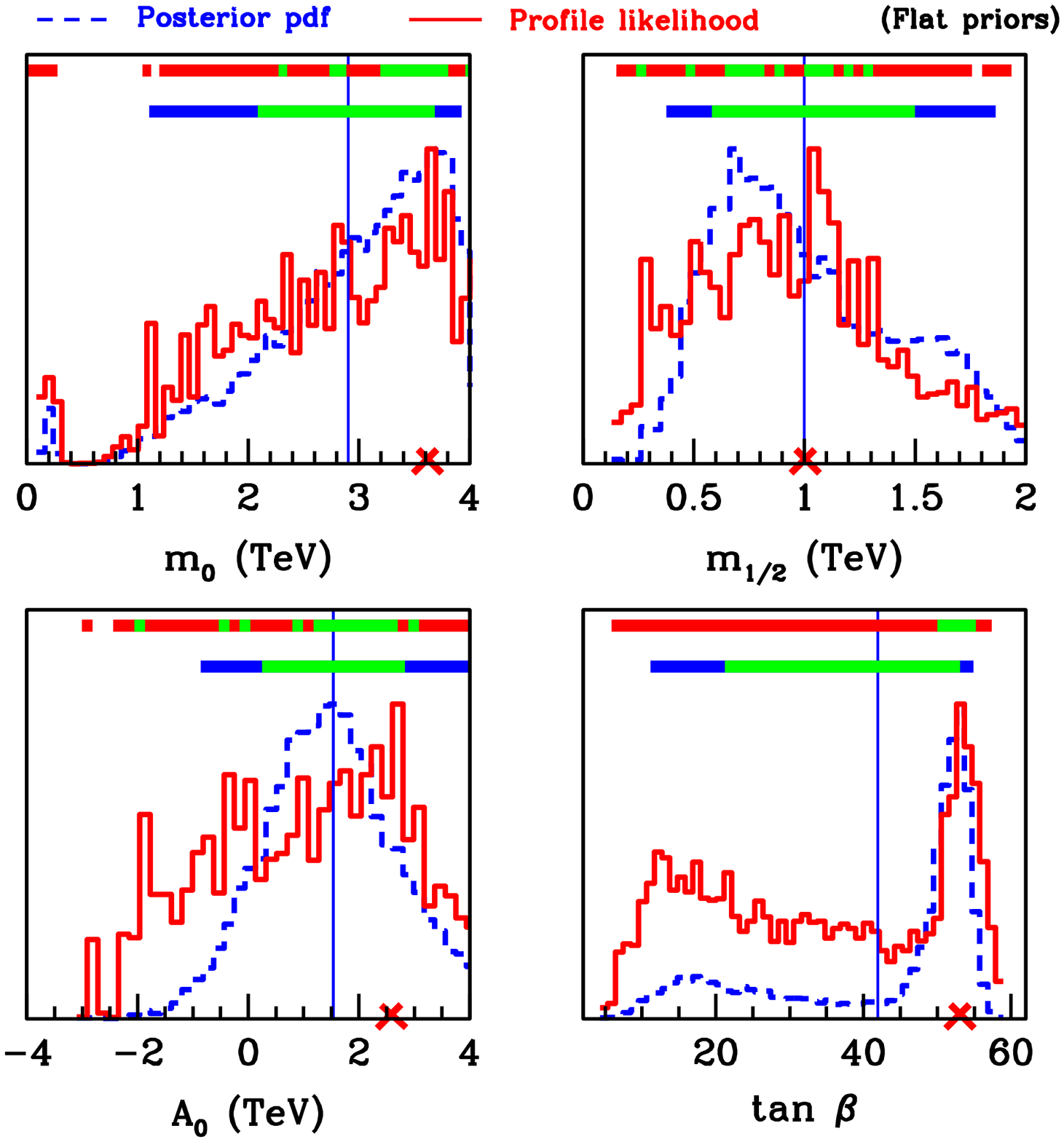} \includegraphics[width=\ww]{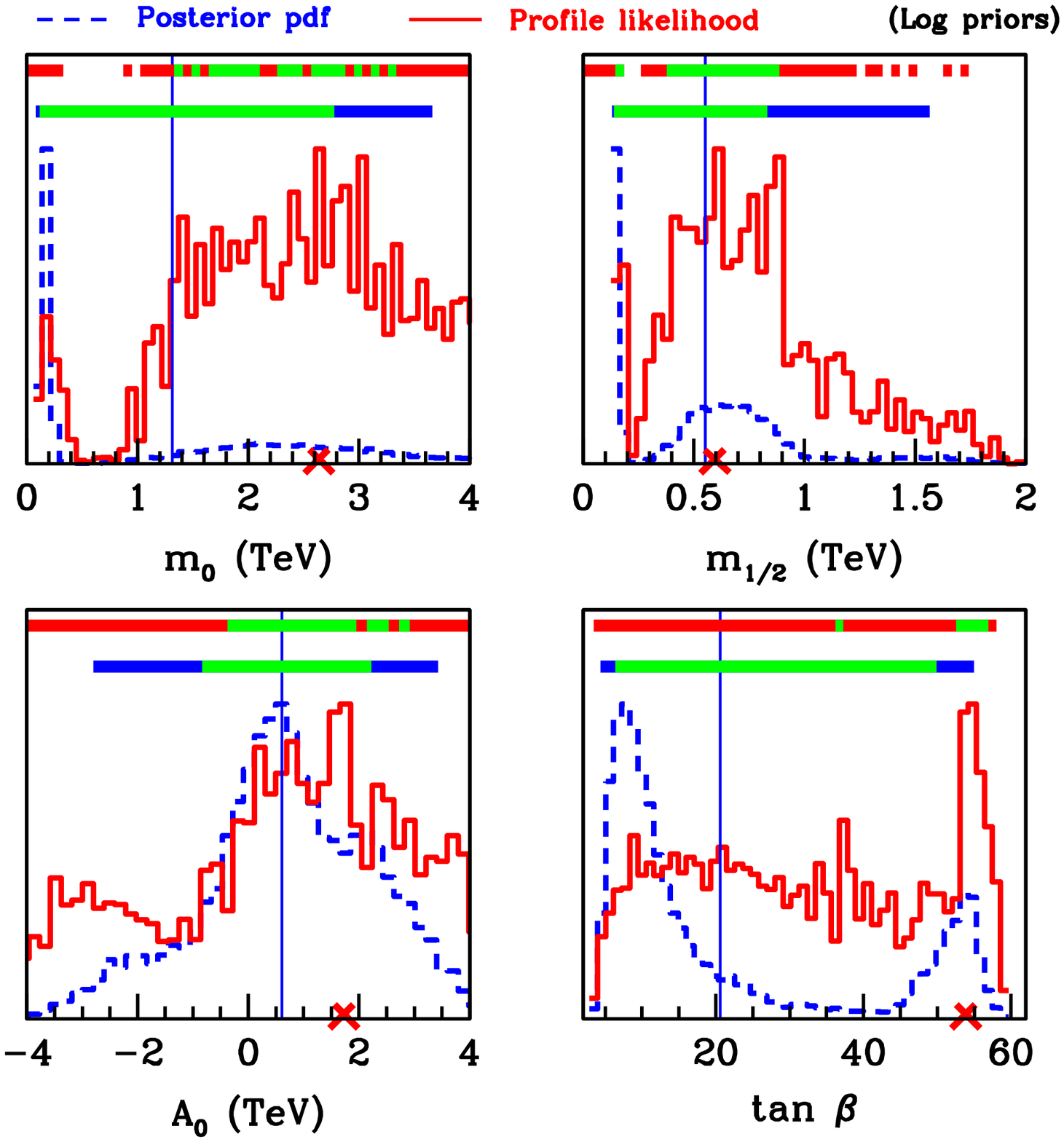}\\
\caption[test]{As in fig.~\ref{fig:1Dpriors+nuis+lep+cdm}, but with an
 additional constraint from $\brbsgamma$
 (\texttt{PHYS+NUIS+COLL+CDM+BSG}).}
\label{fig:1Dpriors+nuis+lep+cdm+bsg}
\end{center}
\end{figure}

\begin{figure}[tbh!]
\includegraphics[width=\qq]{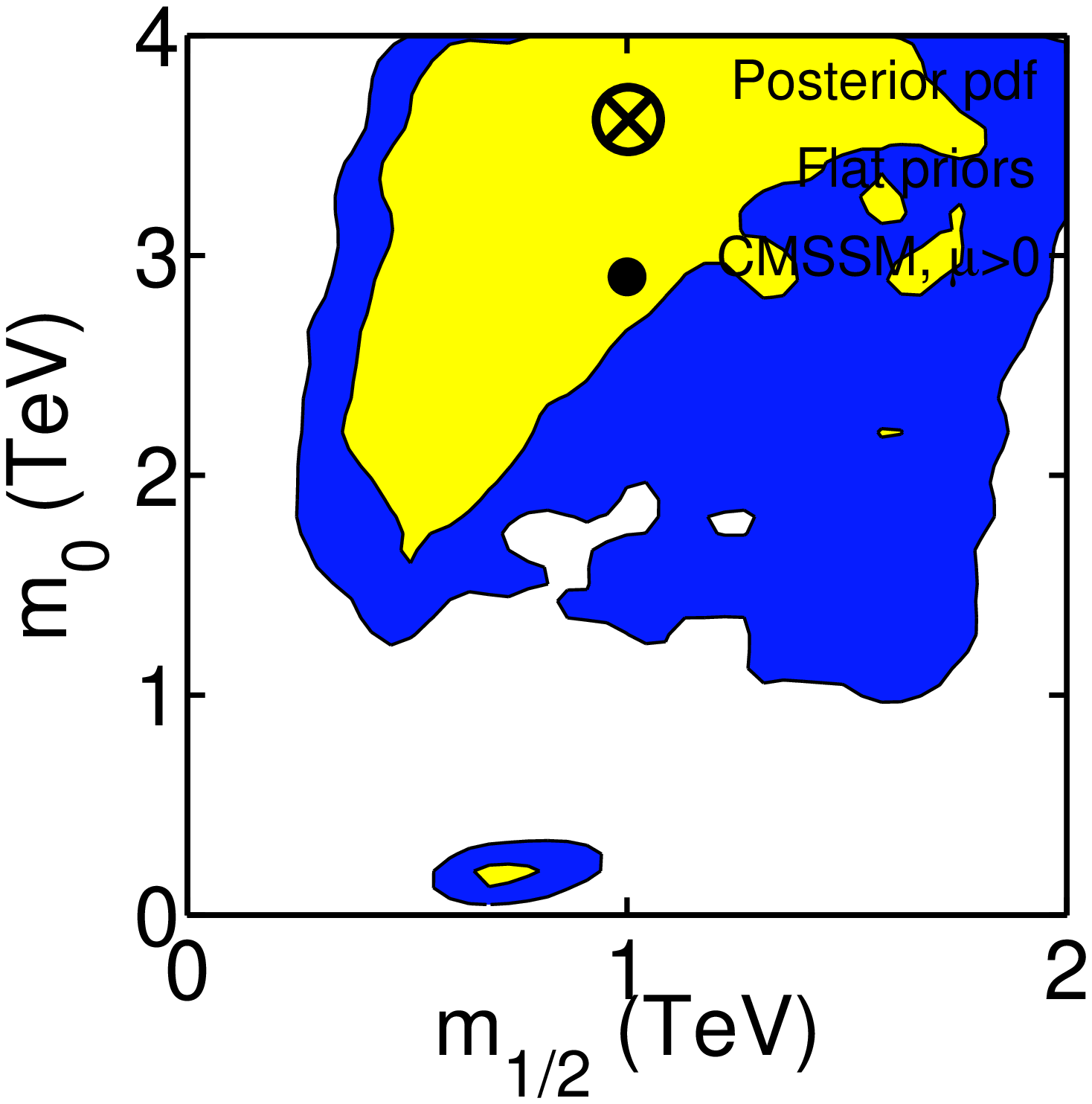} \includegraphics[width=\qq]{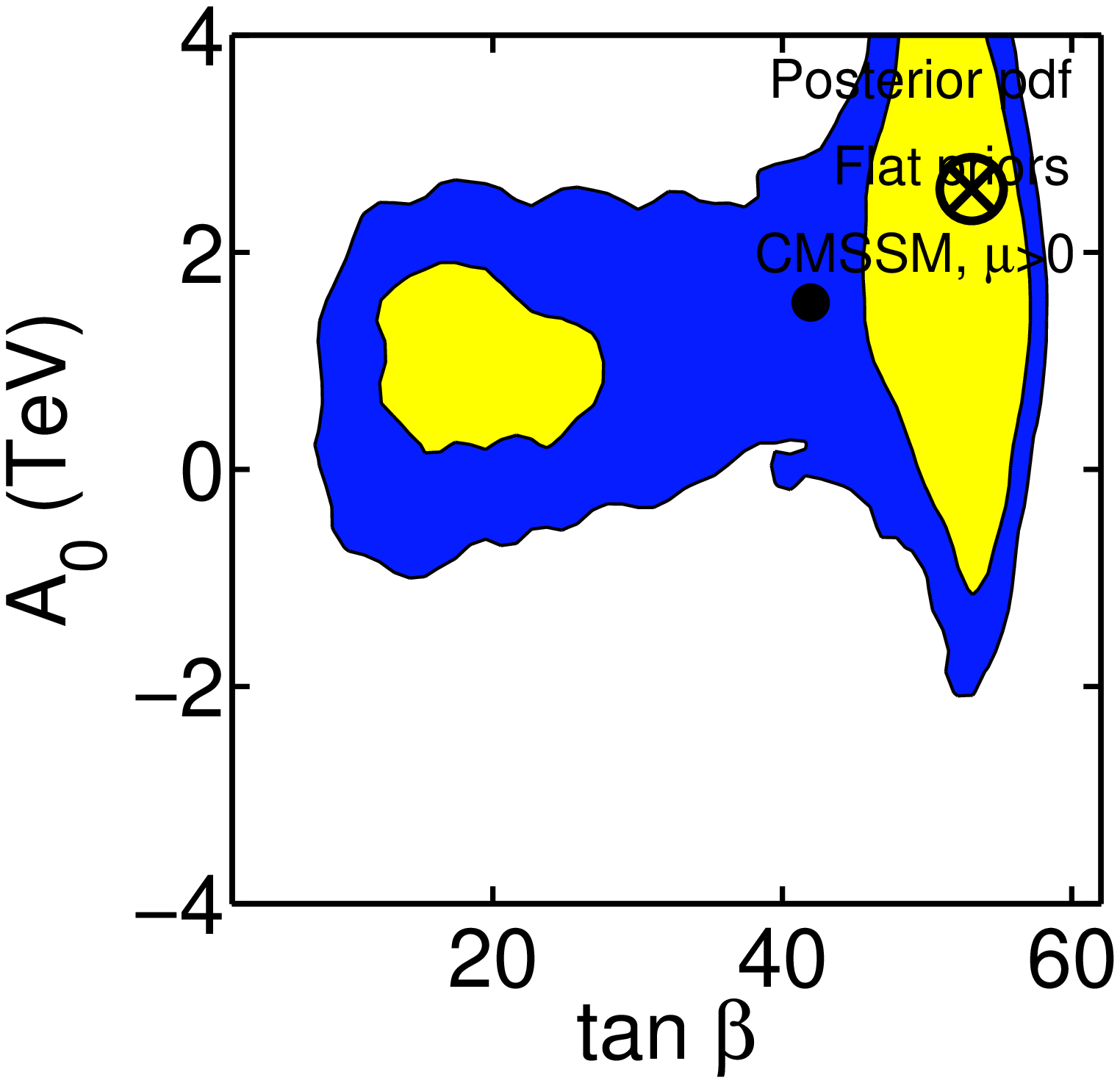}
\includegraphics[width=\qq]{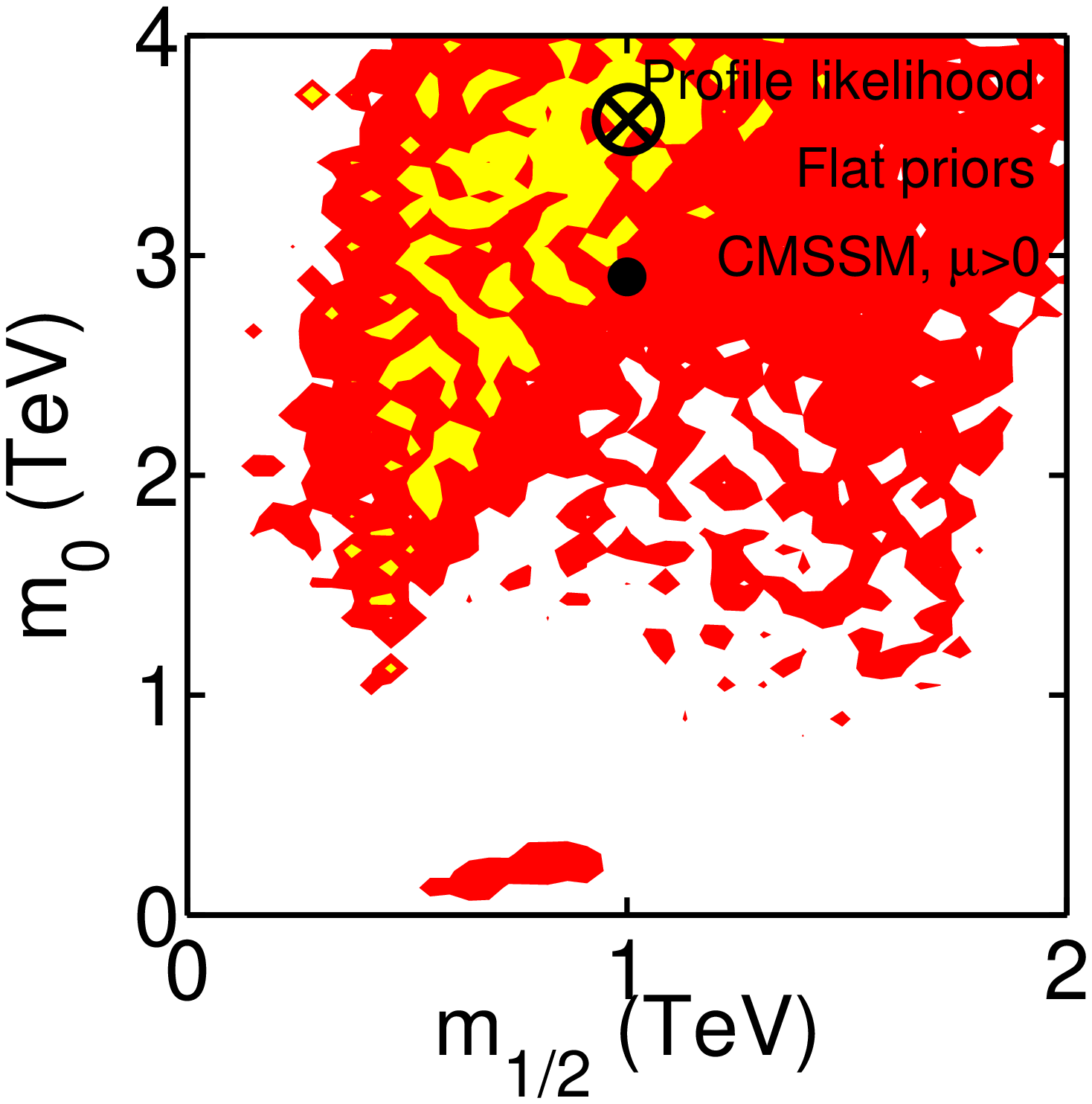} \includegraphics[width=\qq]{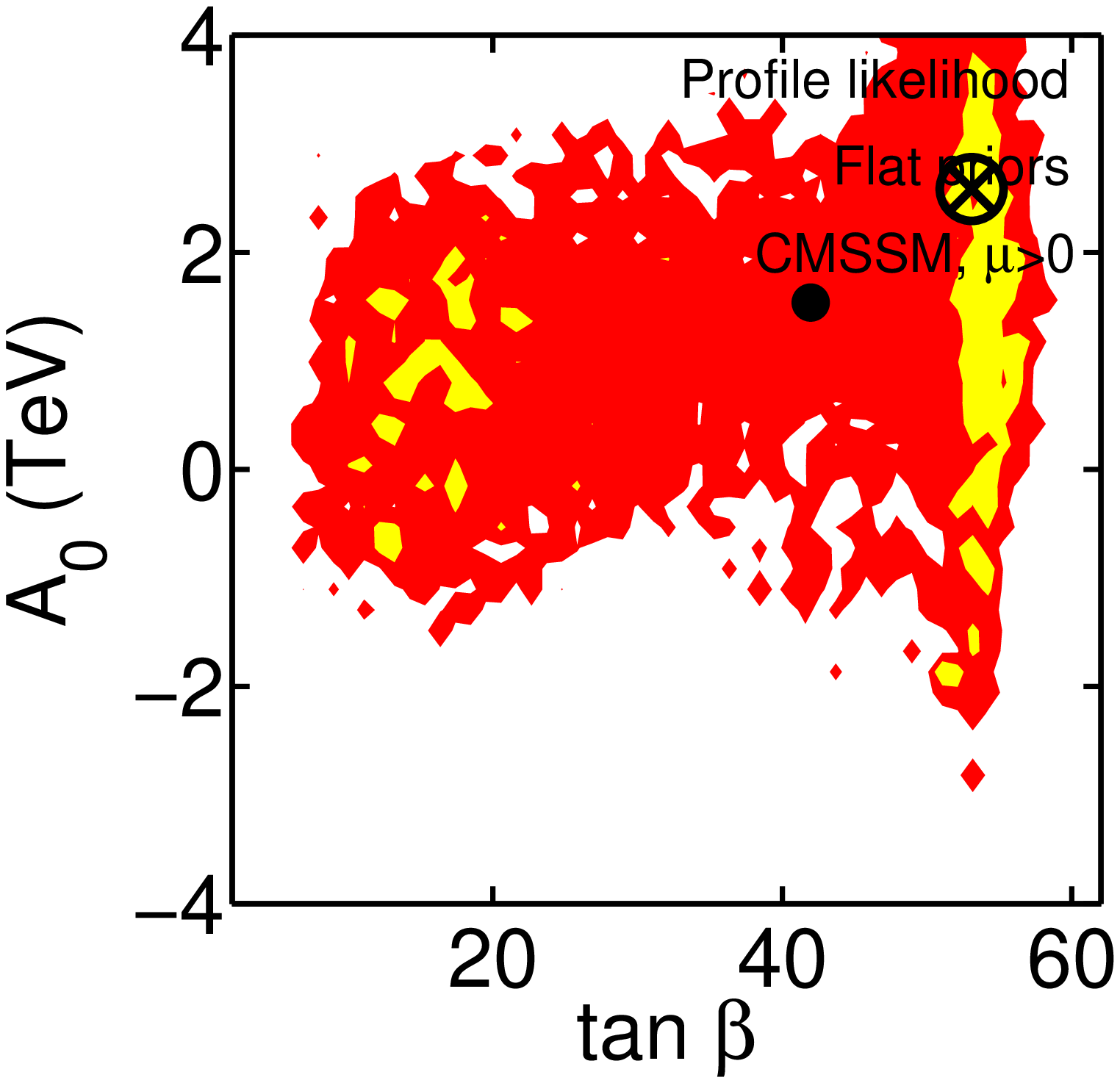} \\
\includegraphics[width=\qq]{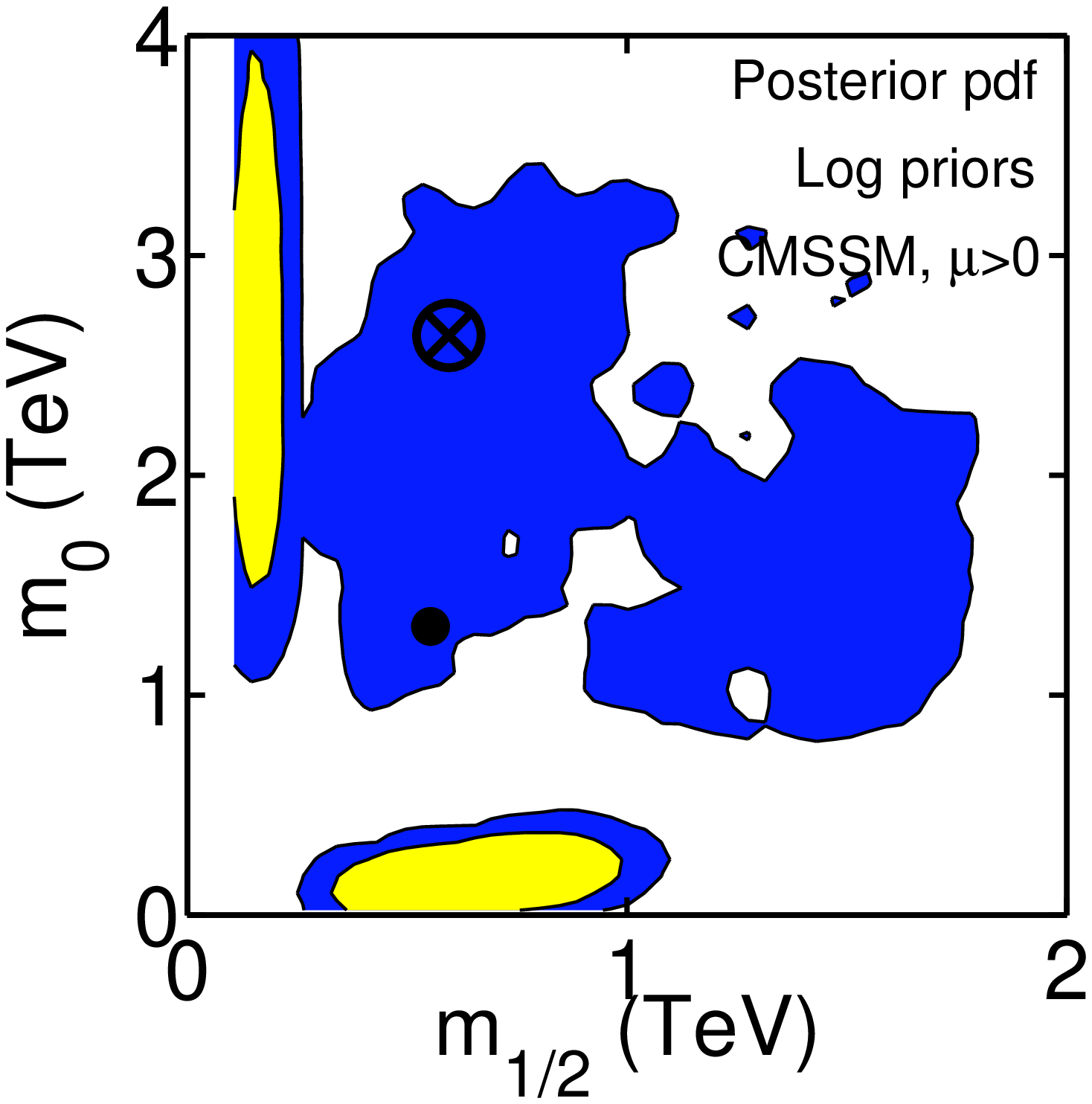} \includegraphics[width=\qq]{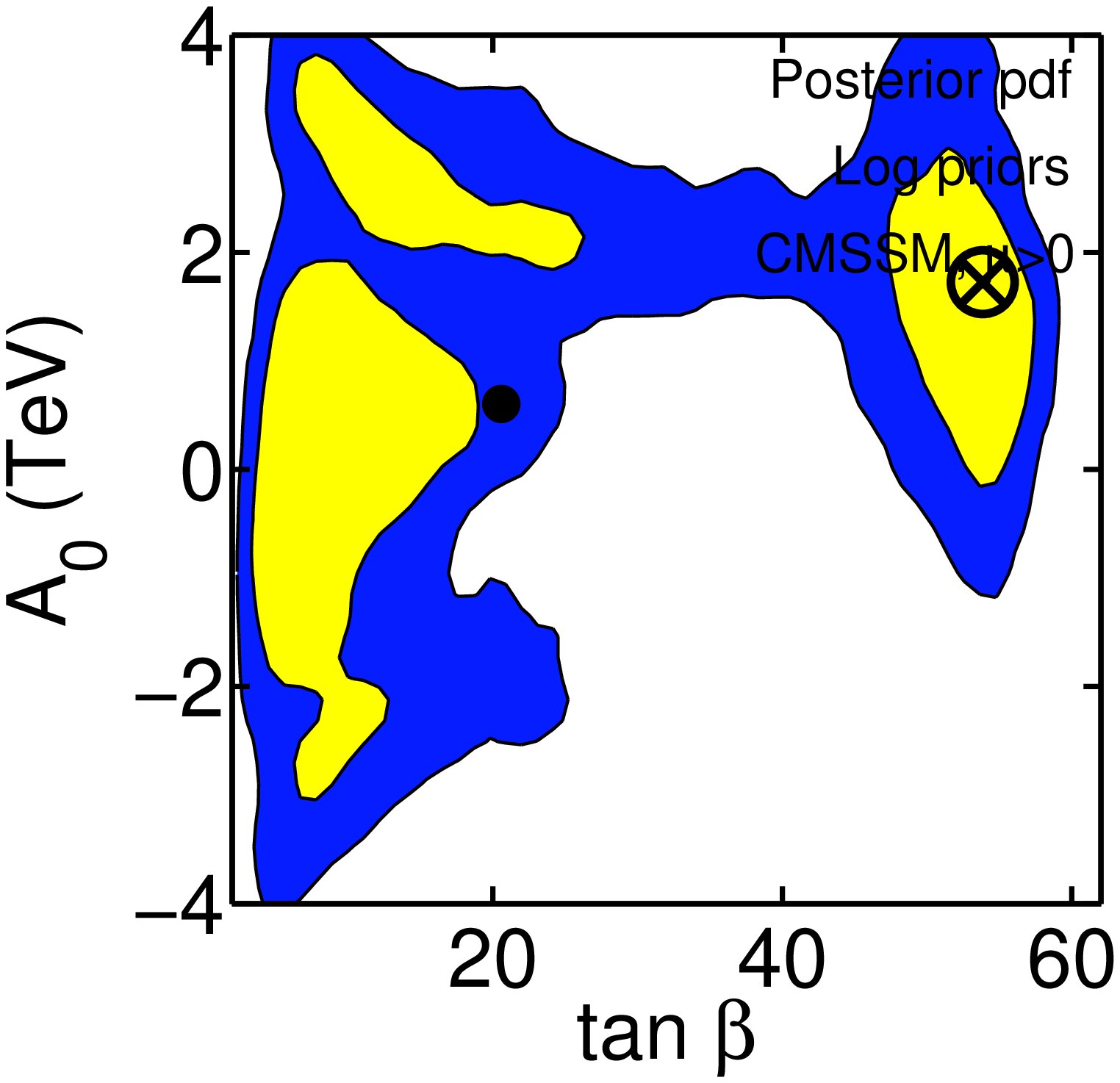}
\includegraphics[width=\qq]{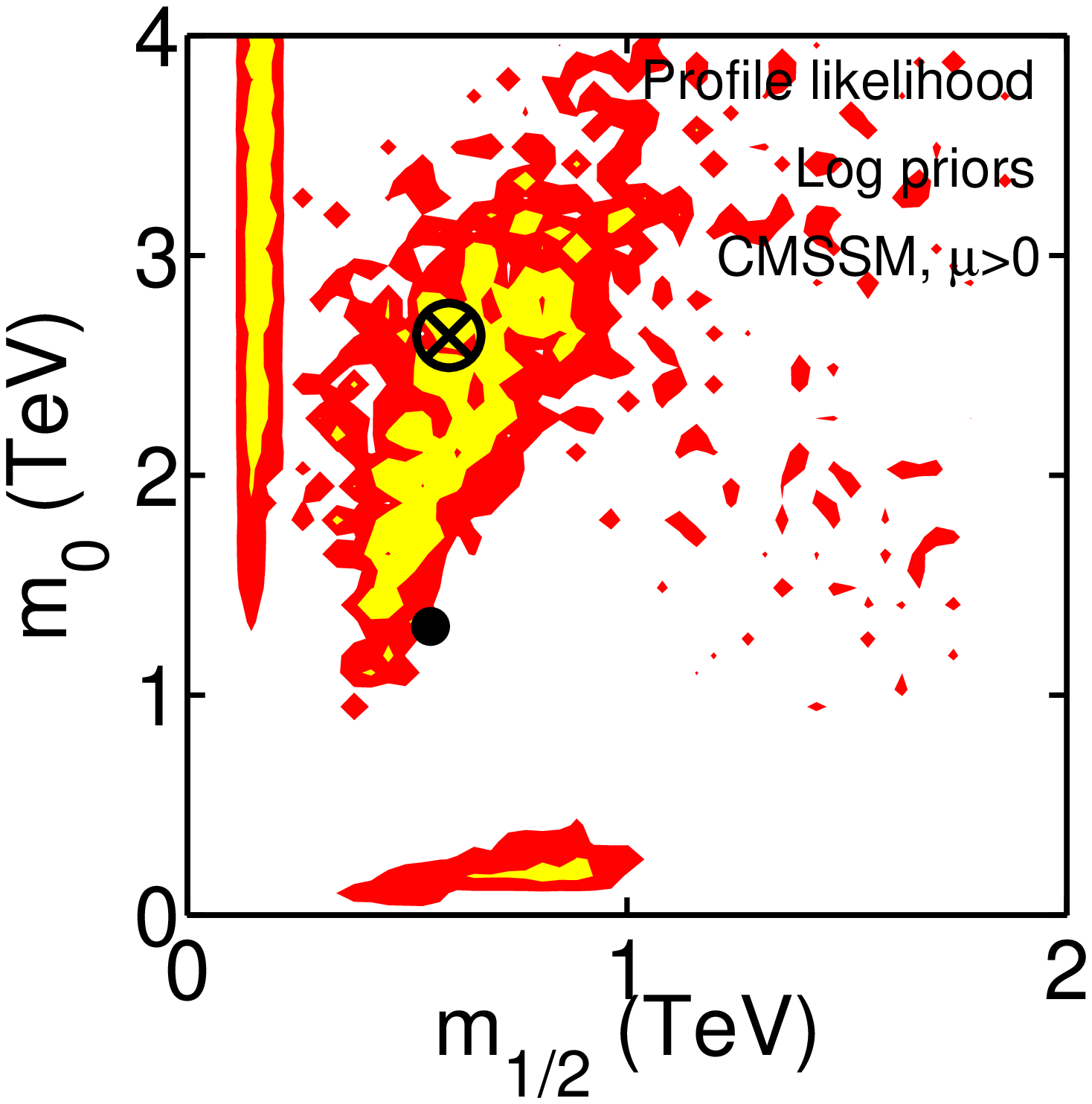} \includegraphics[width=\qq]{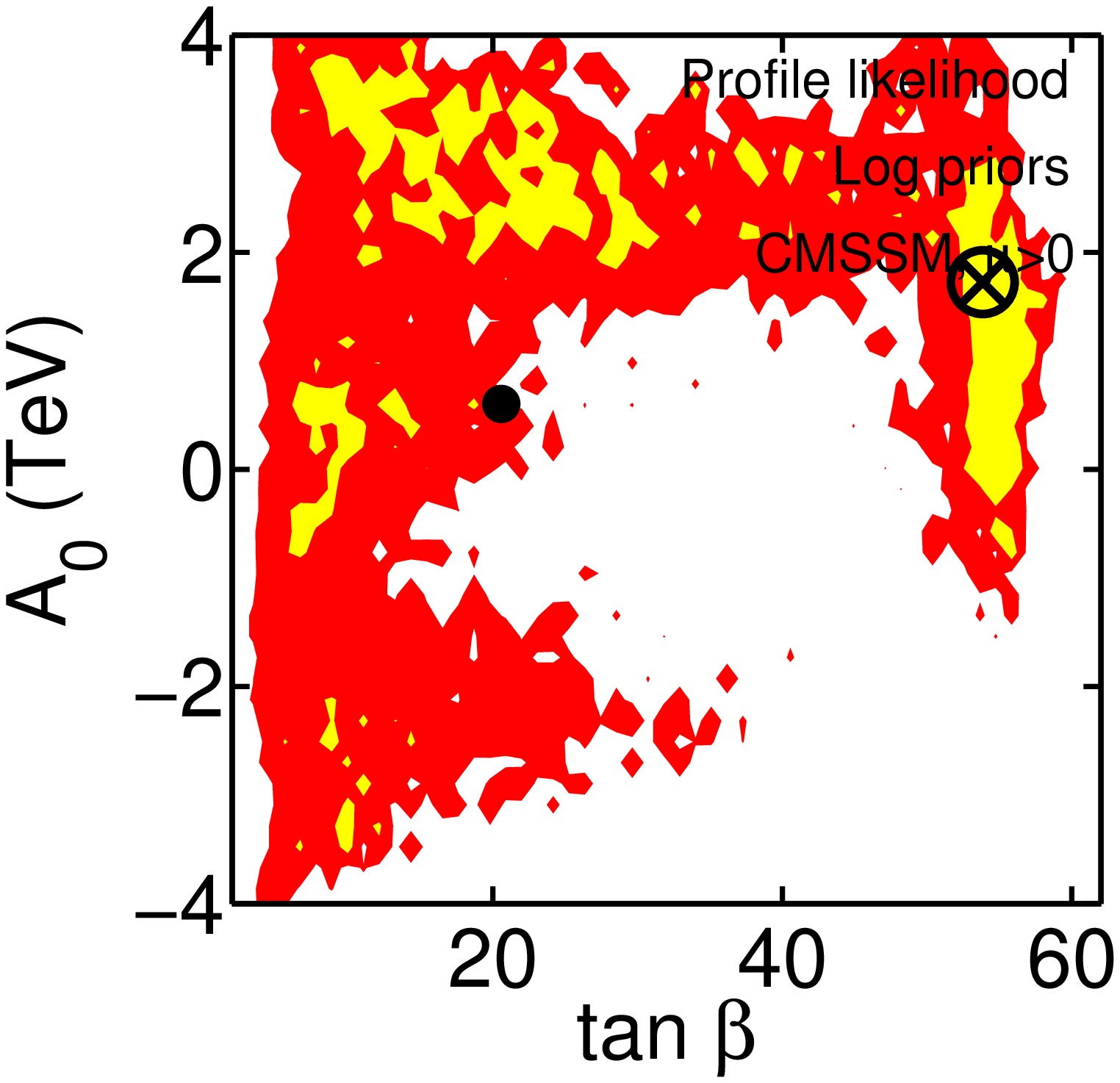}
\caption[test]{As in fig.~ \ref{fig:2D_priors+nuis+lep+cdm}, but with
 an additional constraint from $\brbsgamma$ (\texttt{PHYS+NUIS+COLL+CDM+BSG}).
\label{fig:2D_priors+nuis+lep+cdm+bsg}}
\end{figure}

Next we add the $\brbsgamma$ constraint
(\texttt{PHYS+NUIS+COLL+CDM+BSG}) in
figs.~\ref{fig:1Dpriors+nuis+lep+cdm+bsg} (1D distribution)
and~\ref{fig:2D_priors+nuis+lep+cdm+bsg} (2D distribution). This has
the effect of moving the region preferred by the profile likelihood
towards large $\mzero$ (the FP region), for both the flat and, to a
lesser extent, log prior.\footnote{The reason why the $\brbsgamma$
constraint favors the FP can be seen as follows.
Starting from the SM central value of $3.12\times10^{-4}$, the always positive
charged Higgs/top contribution has to be large enough so that, when
combined with 
the negative (for $\mu>0$) chargino/stop contribution the total ends up
around the experimental central value of $3.55\times10^{-4}$. This requires
the charged Higgs to be light enough and also the stop (or chargino) to be
heavy enough. Both conditions are satisfied in the FP region. Of
course the above argument is somewhat oversimplified, as it does not
take into account the associated error bars on the above values but it
does explain the basic mechanism, which remains dominant in a full
numerical analysis~\cite{rrt3}.}
However, the posterior pdf still suffers
from a strong prior dependence, with the flat prior clearly giving
more weight to larger $\mzero$, while the log prior case strongly
preferring lower $\mzero$ and, to a lesser extent, $\mhalf$, a
reflection of the larger {\em a priori} probability given to lower
ranges of both parameters. Constraints on $\tanb$ are also dependent
on the prior and the choice of the statistical measure.

\begin{figure}[tbh!]
\begin{center}
\includegraphics[width=\ww]{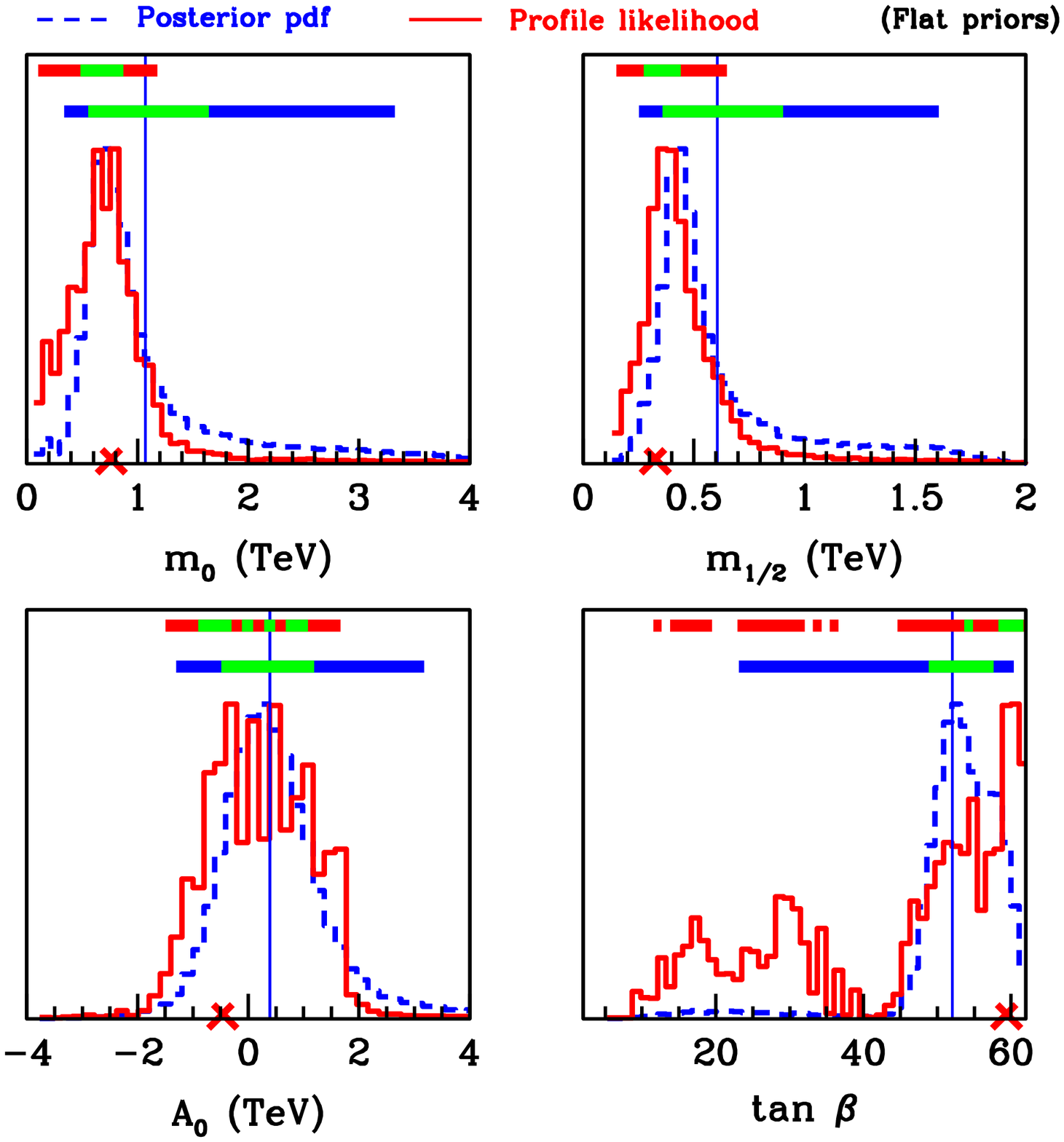} \includegraphics[width=\ww]{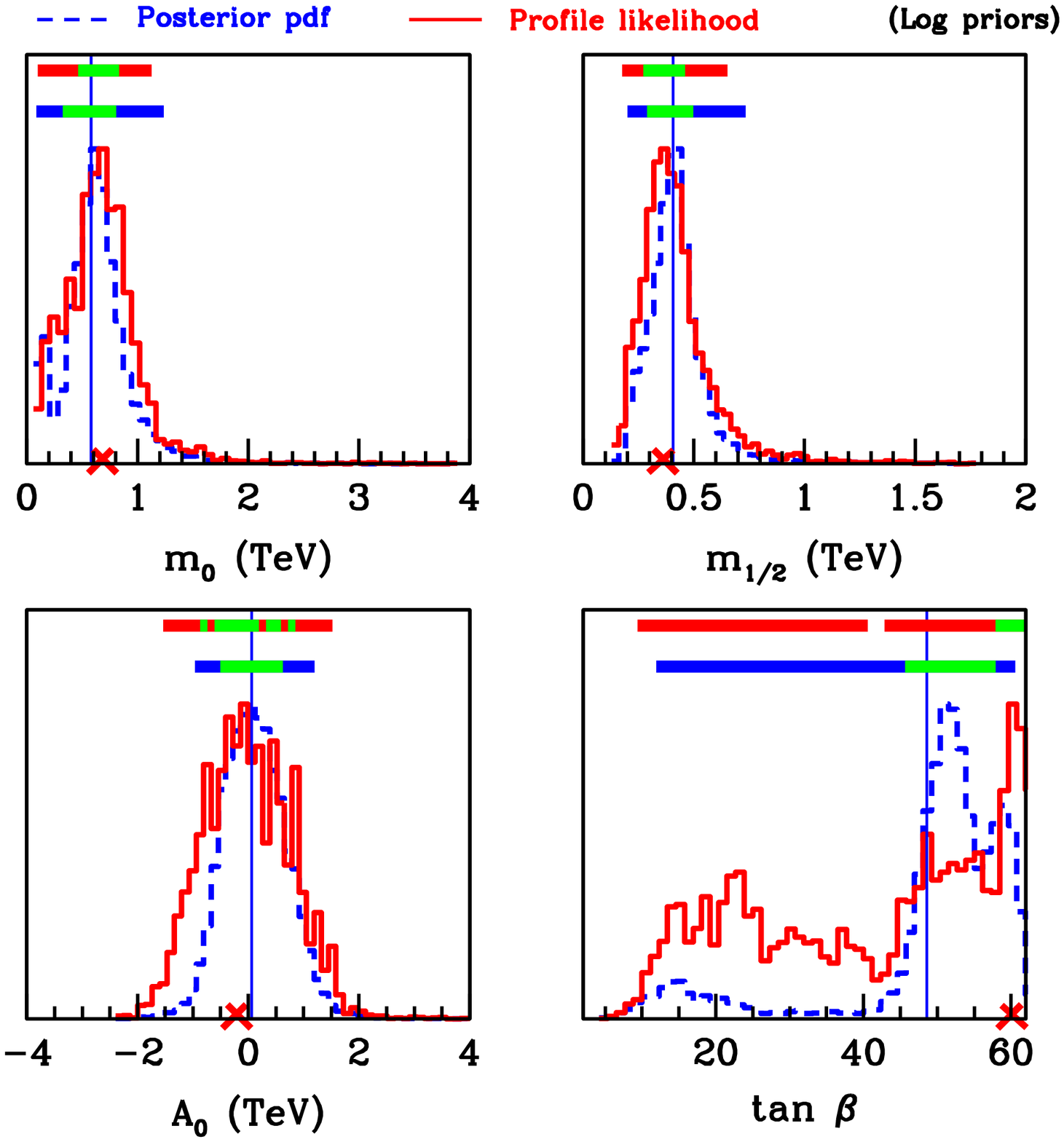}\\
\caption[test]{As in fig.~\ref{fig:1Dpriors+nuis+lep+cdm}, but with an
 additional constraint from $\deltaamususy$, instead of $\brbsgamma$
 (\texttt{PHYS+NUIS+COLL+CDM+GM2}).
\label{fig:1Dpriors+nuis+lep+cdm+gm2}}
\end{center}
\end{figure}

\begin{figure}[tbh!]
\includegraphics[width=\qq]{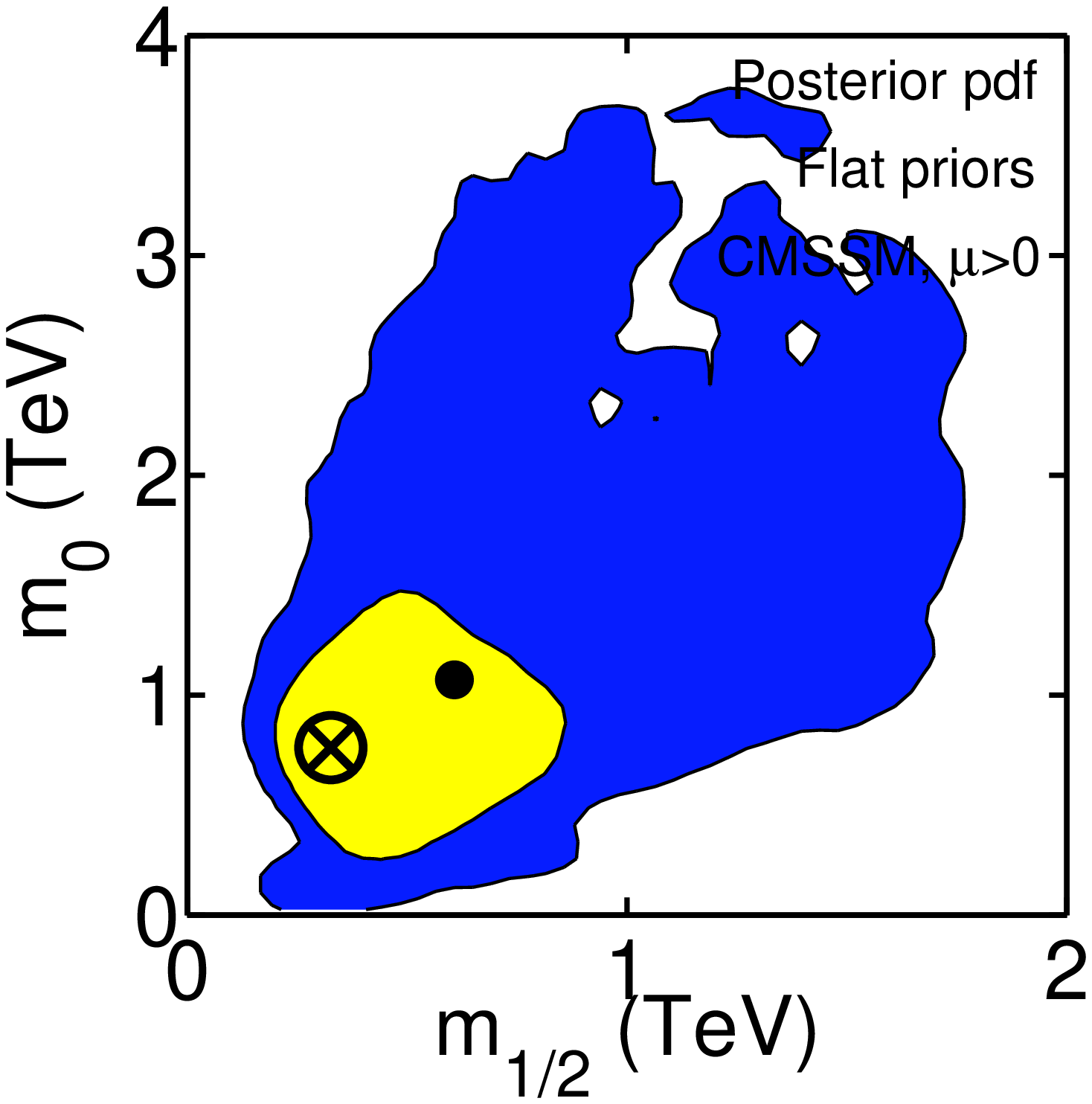} \includegraphics[width=\qq]{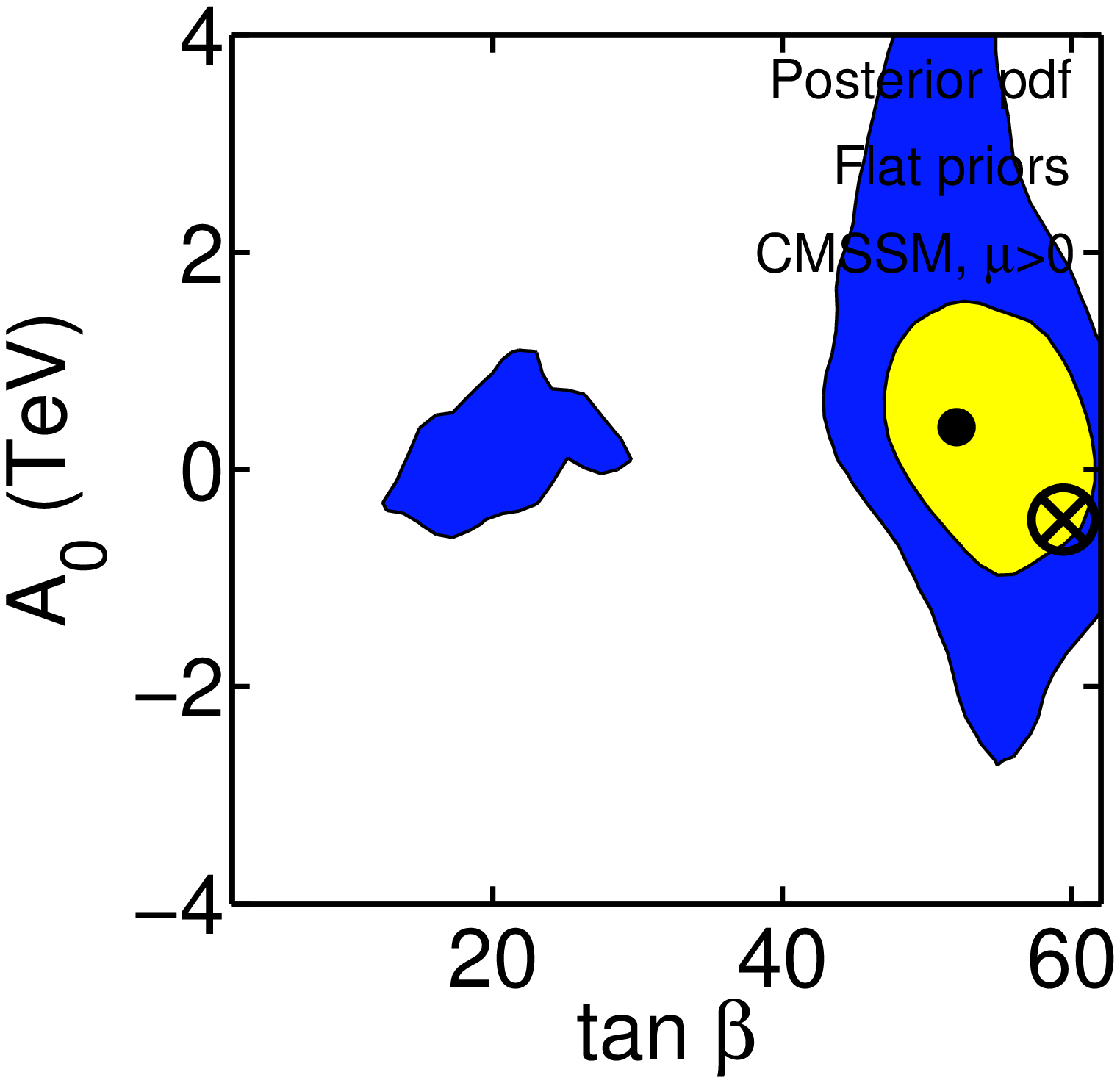}
\includegraphics[width=\qq]{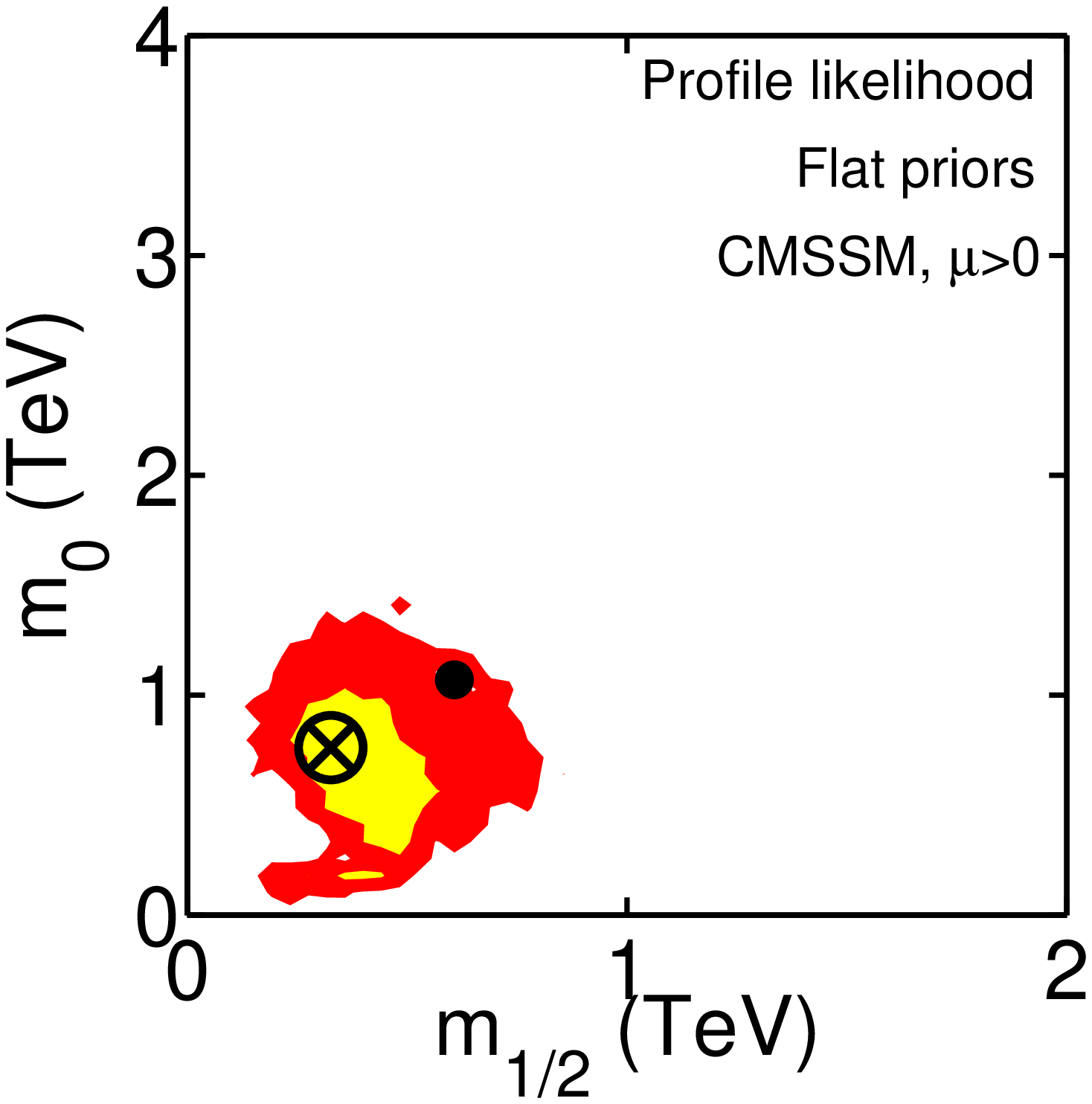} \includegraphics[width=\qq]{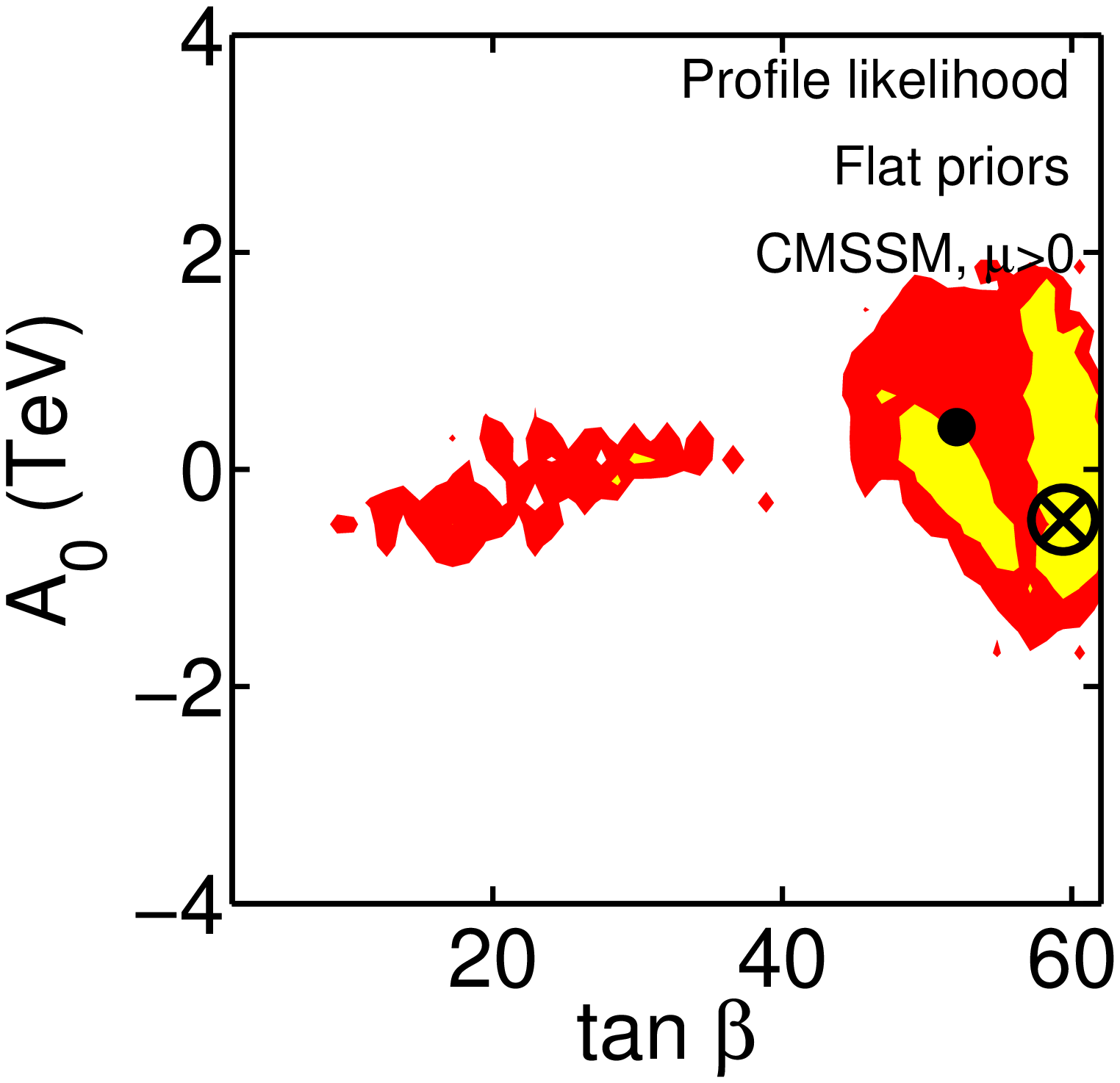} \\
\includegraphics[width=\qq]{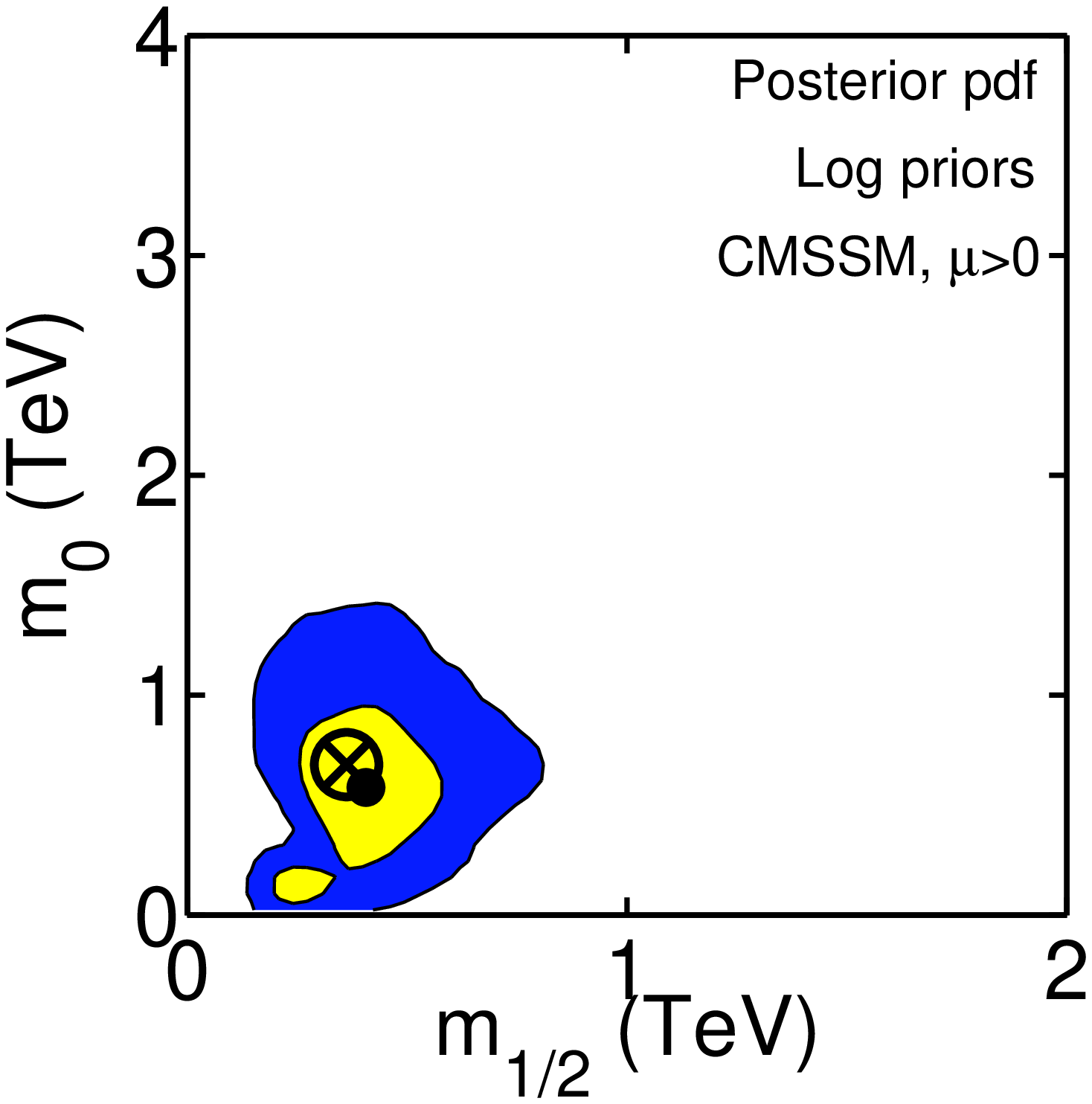} \includegraphics[width=\qq]{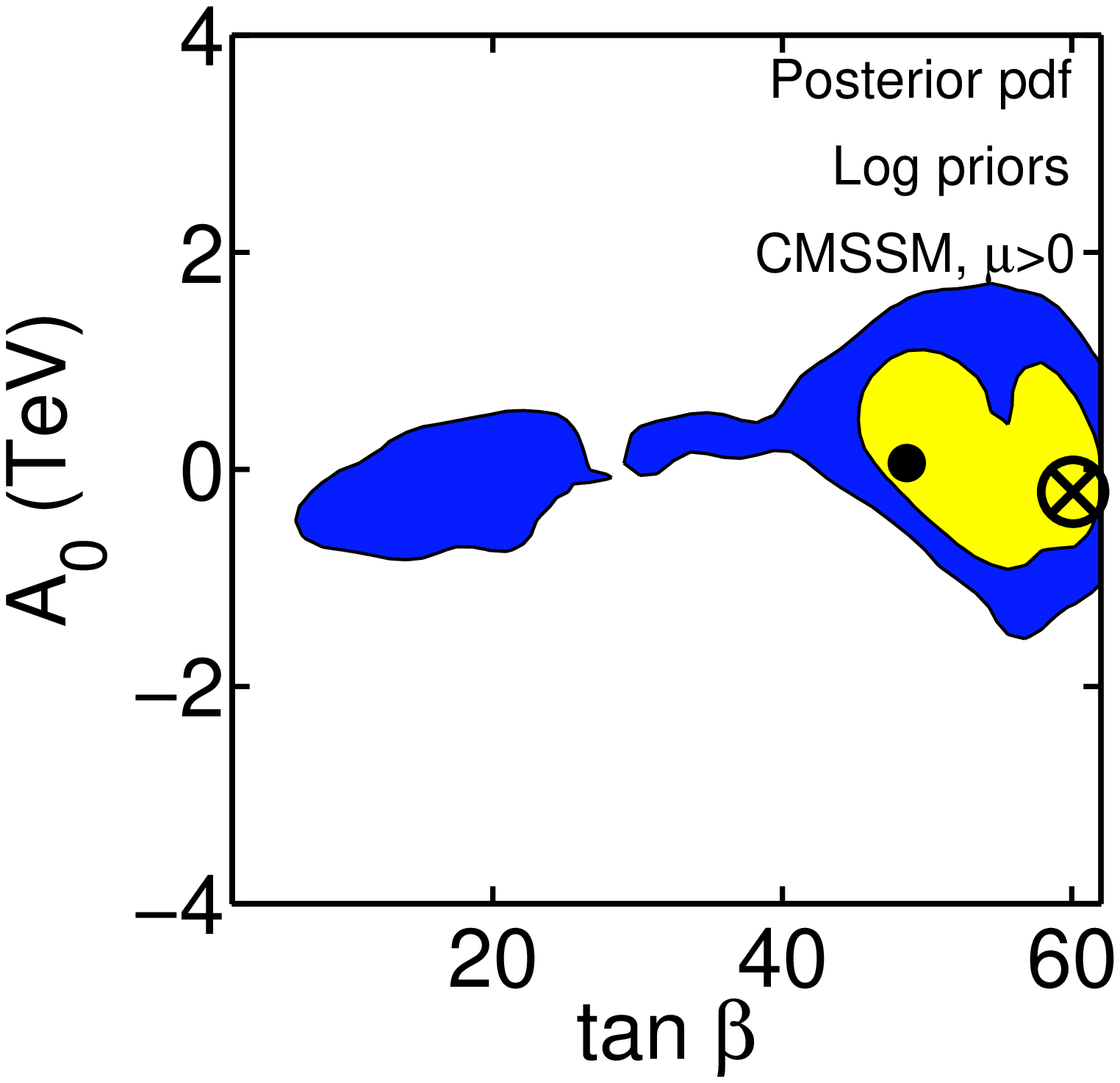}
\includegraphics[width=\qq]{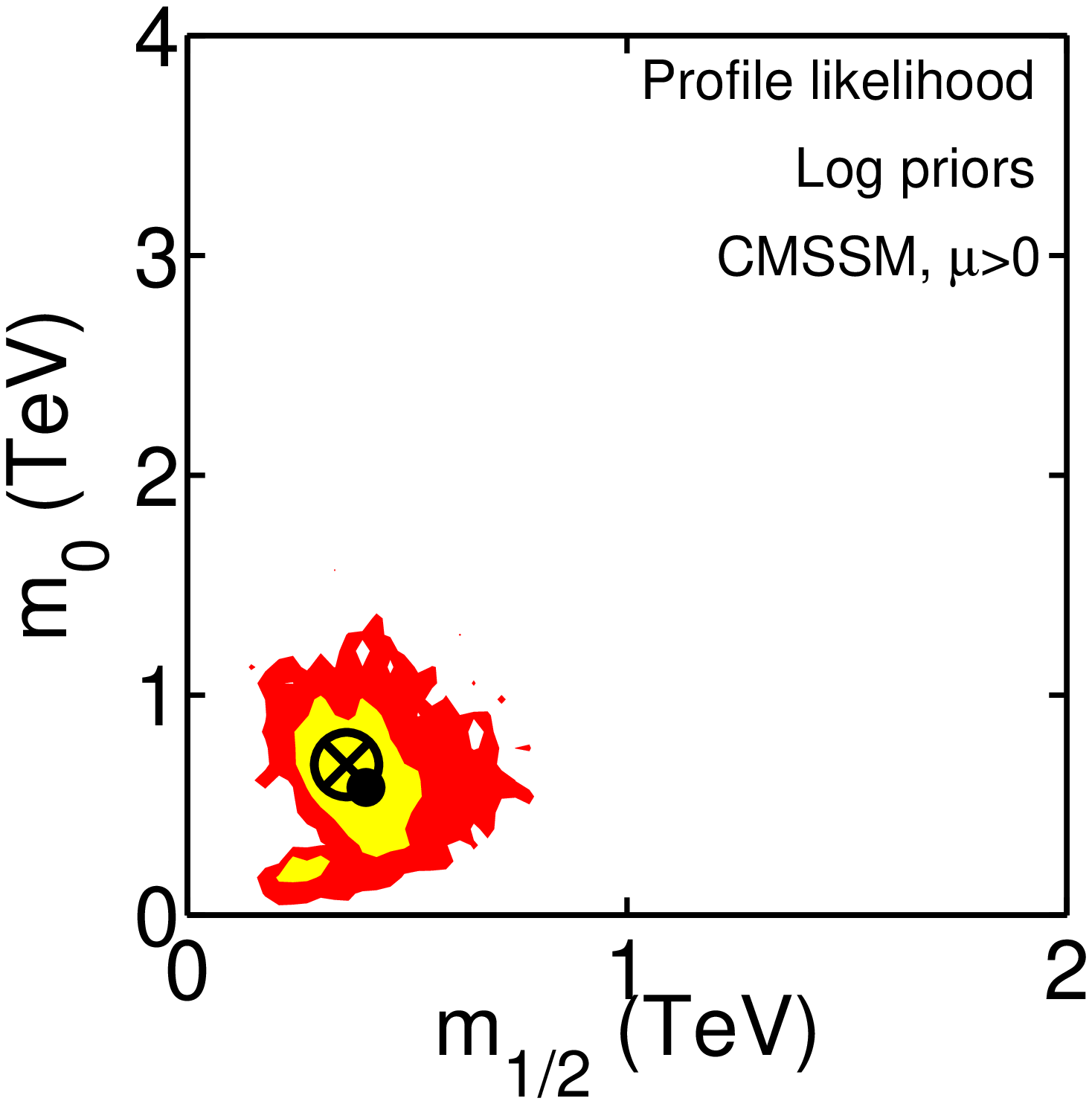} \includegraphics[width=\qq]{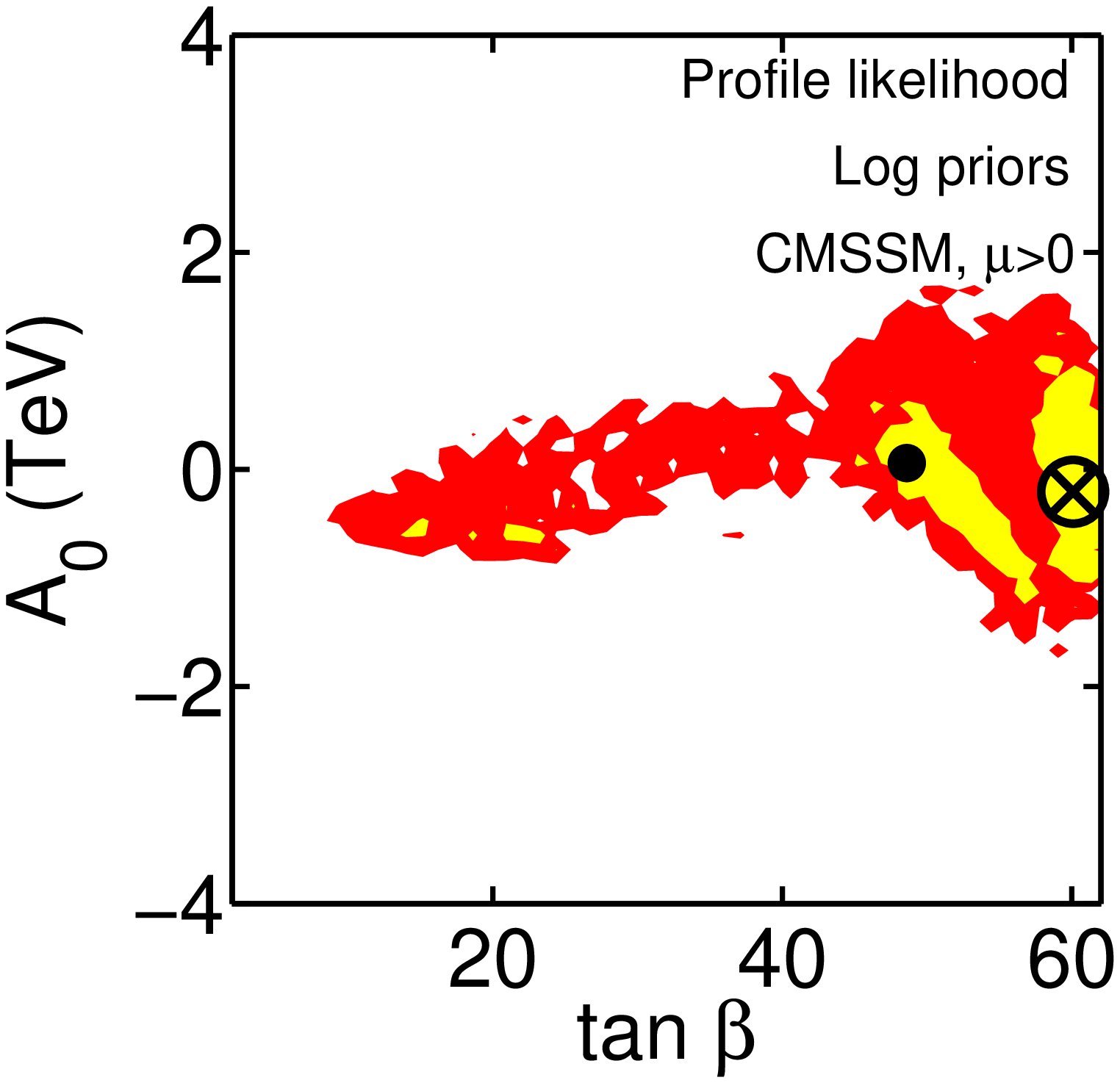}
\caption[test]{As in fig.~ \ref{fig:2D_priors+nuis+lep+cdm}, but with
 an additional constraint from $\deltaamususy$, instead of
 $\brbsgamma$ (\texttt{PHYS+NUIS+COLL+CDM+GM2}).
\label{fig:2D_priors+nuis+lep+cdm+gm2}}
\end{figure}

In order to examine the impact of the anomalous magnetic moment of the
muon, in figs.~\ref{fig:1Dpriors+nuis+lep+cdm+gm2} (1D distribution)
and~\ref{fig:2D_priors+nuis+lep+cdm+gm2} (2D distribution) we replace
the constraint from $\brbsgamma$ with $\deltaamususy$
(\texttt{PHYS+NUIS+COLL+CDM+GM2}). This has the effect of moving, for
both statistical measures, the prefered regions to lower masses,
$\mzero, \mhalf \lsim 1\tev$. While there is some residual prior
dependence in the posterior pdf, the profile likelihood is now almost
independent of the prior and the constraints on all parameters are
largely reconciled for both statistics and prior measures. This means
that, in the absence of the constraint from $\brbsgamma$, the
constraining power of the $\gmt$ observable is rather strong.

\begin{figure}[tbh!]
\begin{center}
\includegraphics[width=\ww]{figs/BlackLike.ps} \includegraphics[width=\ww]{figs/BlackLike.ps}\\
\includegraphics[width=\ww]{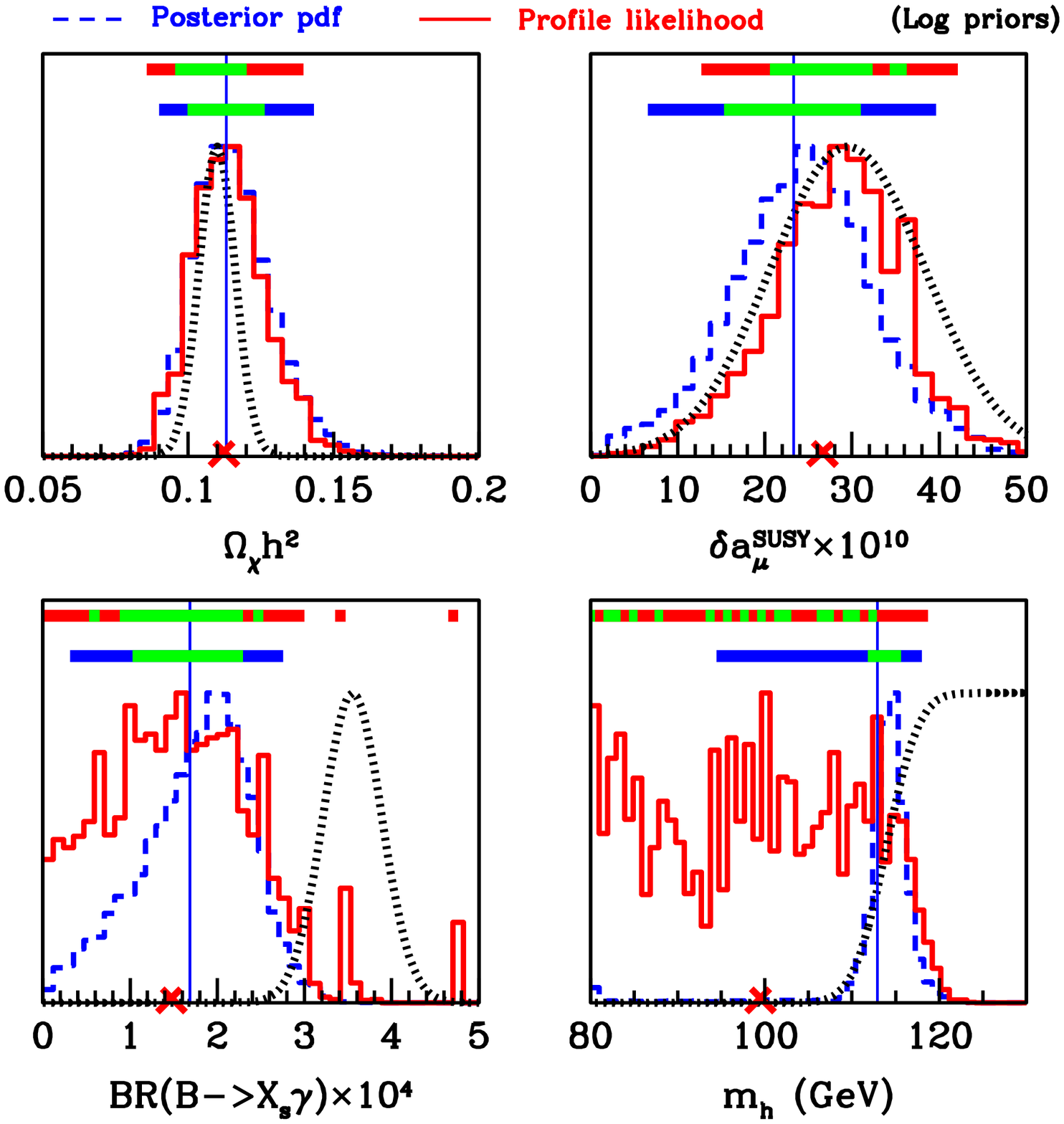}\includegraphics[width=\ww]{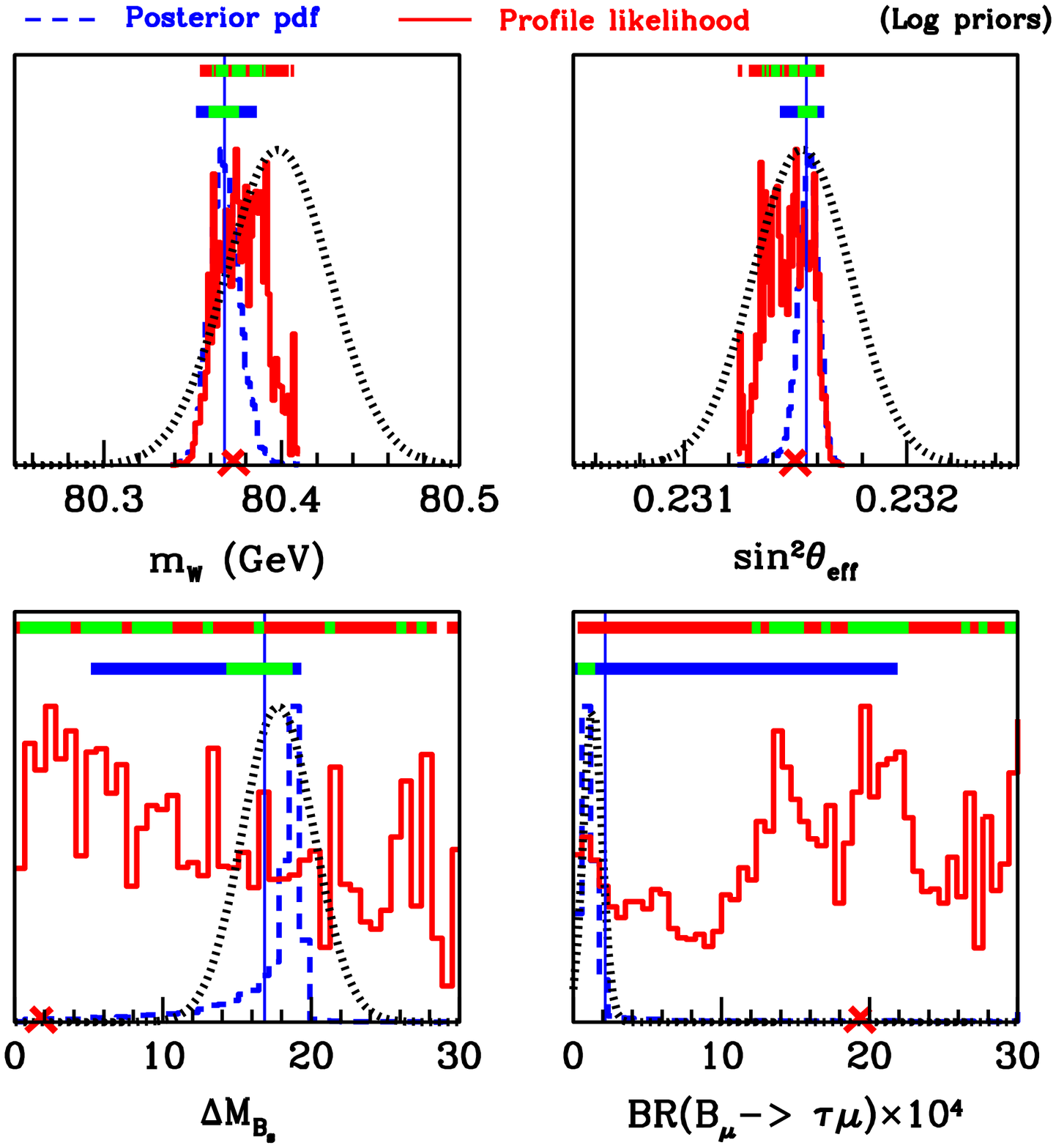}
\caption[test]{As in fig.~\ref{fig:1Dpriors+nuis+lep+cdm+gm2}
 (\texttt{PHYS+NUIS+COLL+CDM+GM2}), but for several obervable
 quantities. Only the log priors case is shown here, the flat prior case is qualitatively similar.
\label{fig:1Dpriors+nuis+lep+cdm+gm2_obs_2}}
\end{center}
\end{figure}

However, such a strong constraint comes at the price of a tension with
other observables which have {\em not} been included in this scan,
especially $\brbsgamma$.  This is shown in
fig.~\ref{fig:1Dpriors+nuis+lep+cdm+gm2_obs_2} for the log prior (the
case of the flat prior is qualitatively similar). As before, the
posterior pdf is shown in dashed blue, the profile likelihood in solid
red and the likelihood (data) in dotted black.  The DM abundance and
the $\gmt$ are well constrained and both statistics are in agreement
with the likelihood. But both the posterior and the profile likelihood
for $\brbsgamma$ peak at a very low value, well below the SM value,
reflecting a sizeable negative contribution of SUSY corrections. This
is in strong diagreement with the observed likelihood. The other two
$b$--physics observables exhibit a similar tension, as well. Hence we
expect that, once $\brbsgamma$ and the other constraints are applied
both the pdf and the profile likelihood will shift considerably and
the $\gmt$ constraint will produce a tension with the other
data.\footnote{An interesting oddity is the long tail of the profile
likelihood for values $\mhl\lsim114\gev$. This is caused by the fact
that, in that case the light Higgs coupling $\zetah^2$ becomes
suppressed, thus evading LEP limits on the SM-like Higgs mass (and
also corresponding to large values of $\brbtaunu$, well above the
observed value, which however has not been imposed in this scan). Note
that this does not show up in the Bayesian pdf, because there is only
a small number of samples with non-SM-like coupling.}  We will
discuss the tension between $\gmt$ and the other observables in more
detail in the next section.

\subsection{Combined impact of all observables}

\begin{figure}[tbh!]
\begin{center}
\includegraphics[width=\ww]{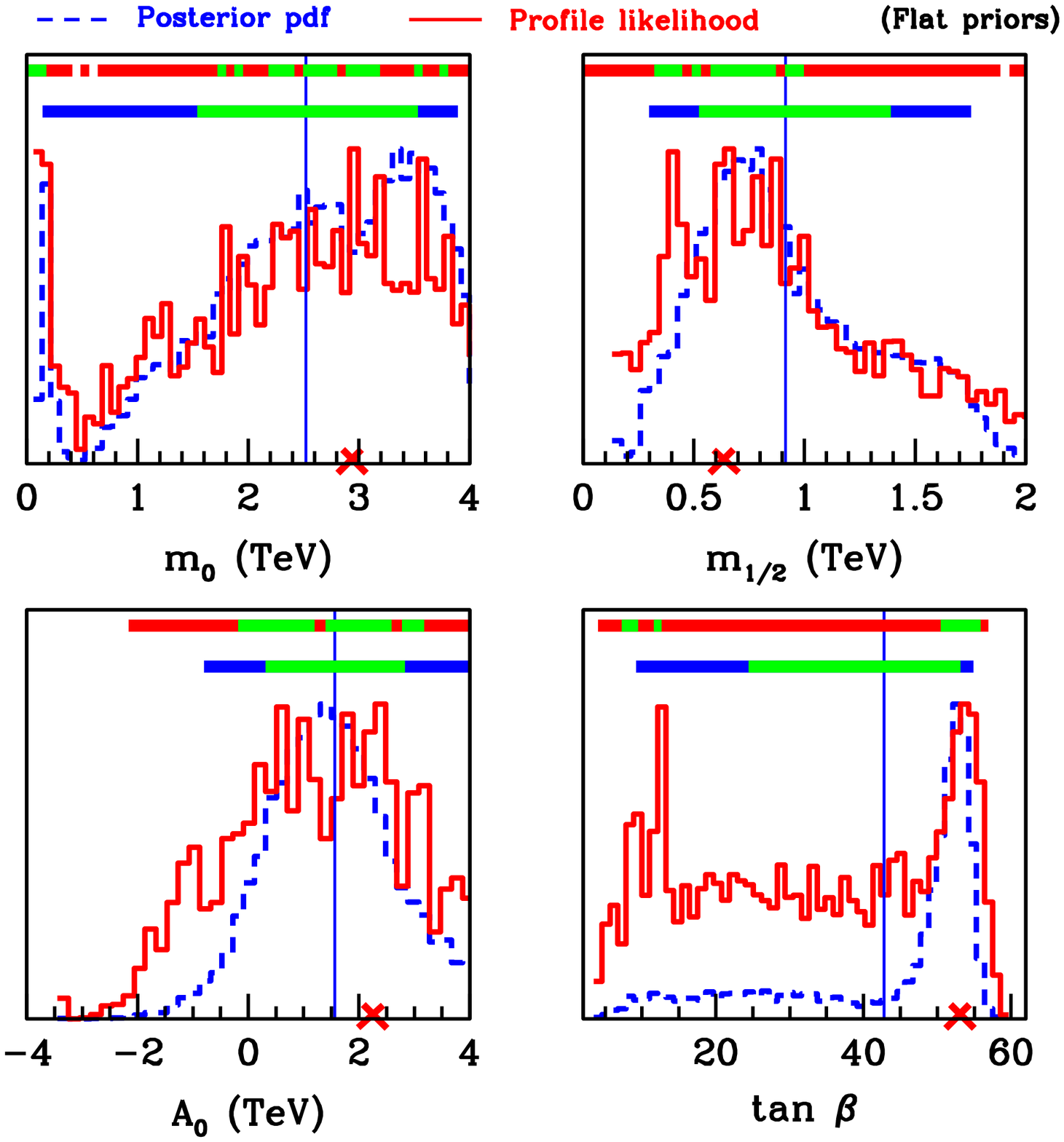}
\includegraphics[width=\ww]{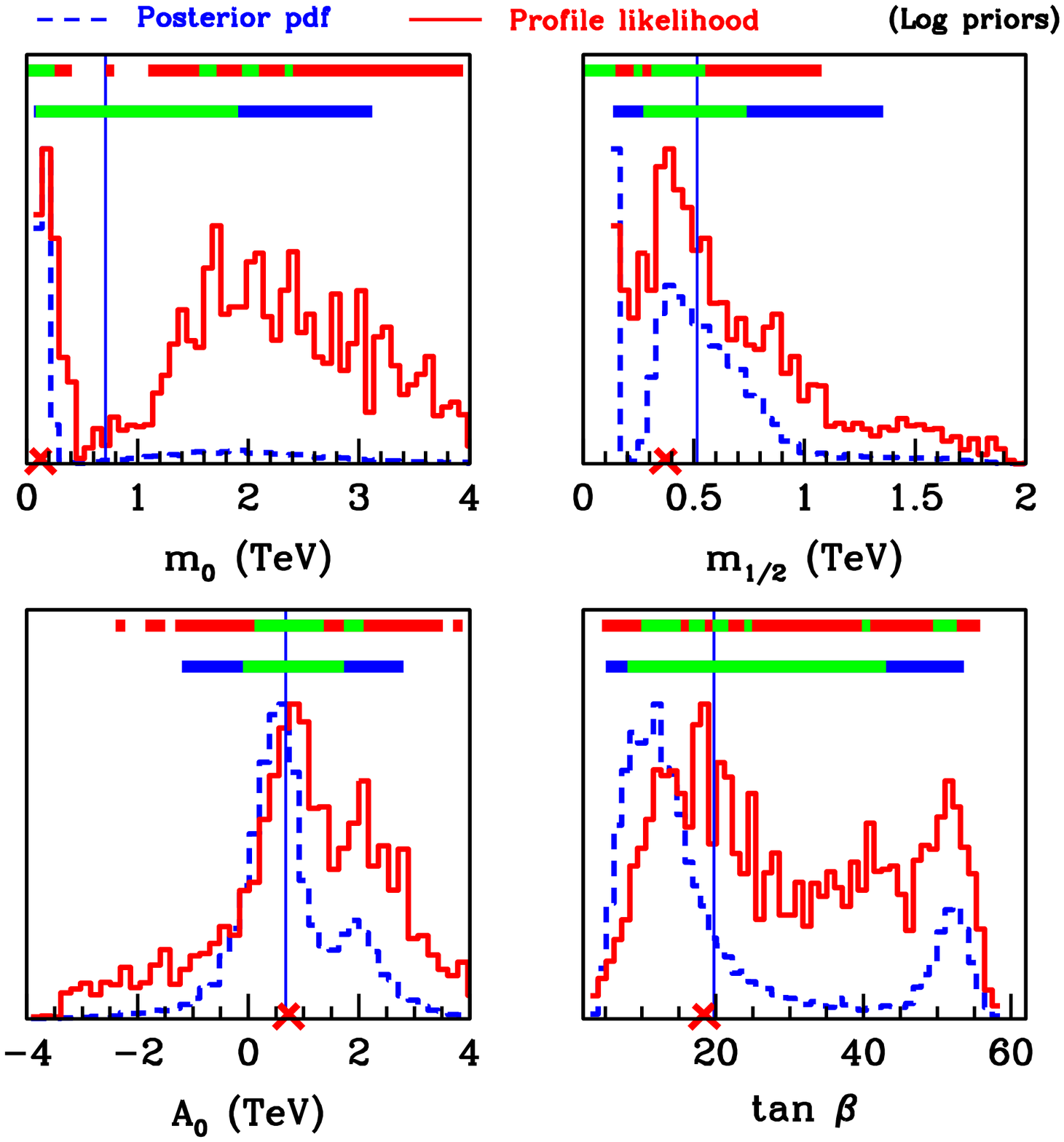}\\
\caption[test]{As in fig.~\ref{fig:1Dpriors+nuis+lep+cdm}, but for a scan
 including all the constraints listed in table~\ref{tab:datacombinations}
 (\texttt{ALL}).}
\label{fig:1Dpriors+nuis+lep+cdm+bsg+gm2+ewo+bphys_base}
\end{center}
\end{figure}

\begin{figure}[tbh!]
\begin{center}
\includegraphics[width=\qq]{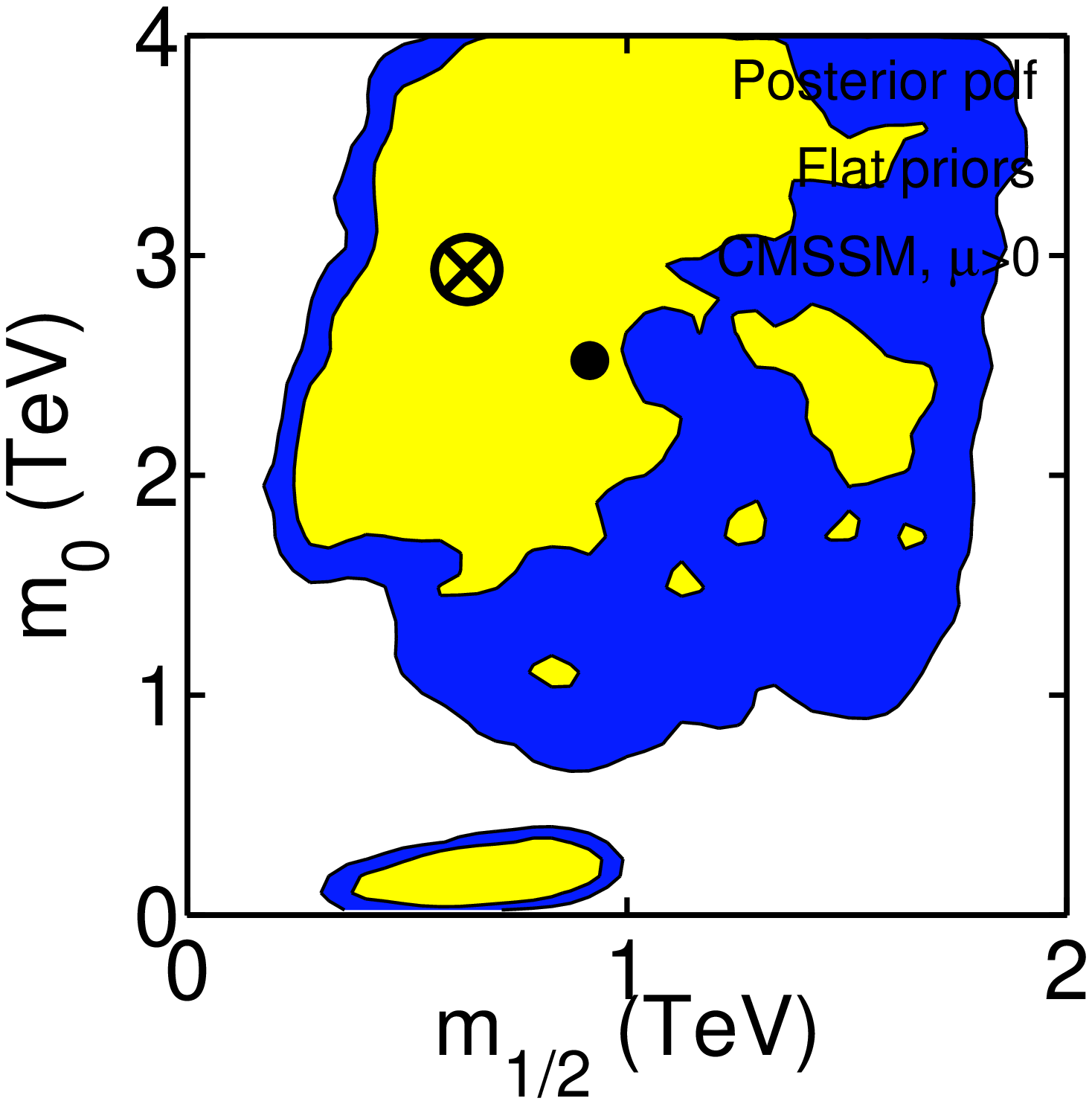} \includegraphics[width=\qq]{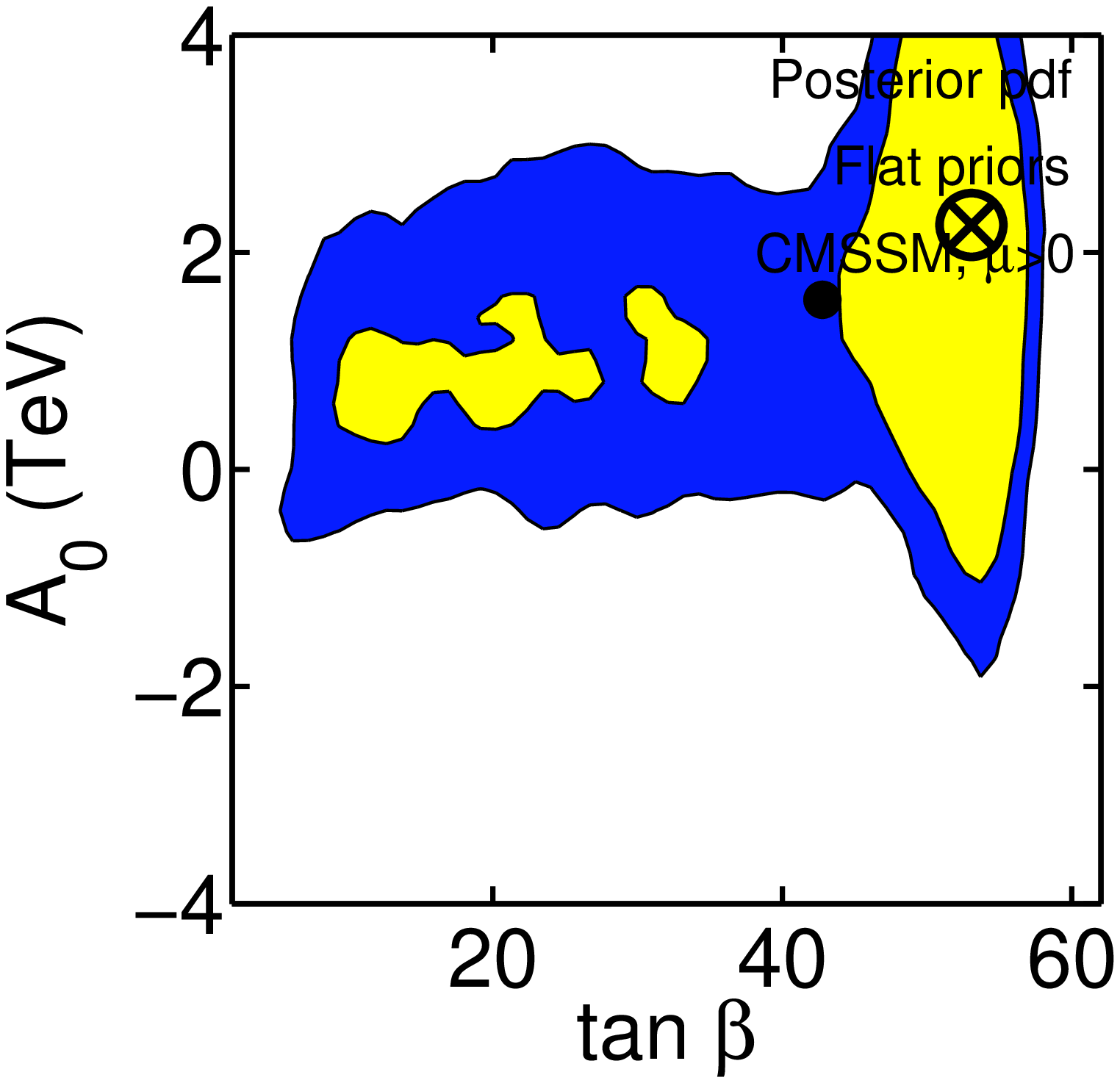}
\includegraphics[width=\qq]{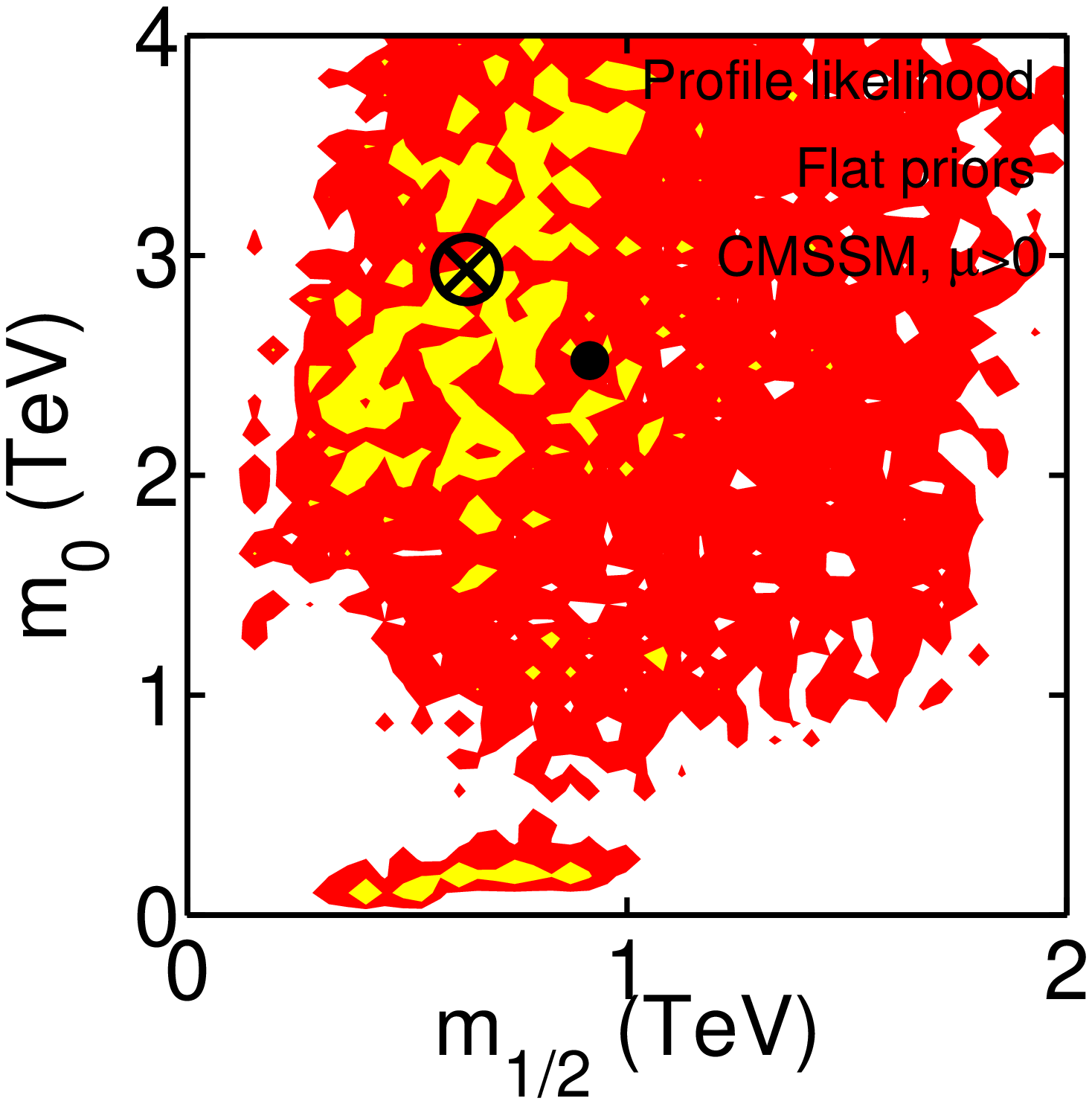} \includegraphics[width=\qq]{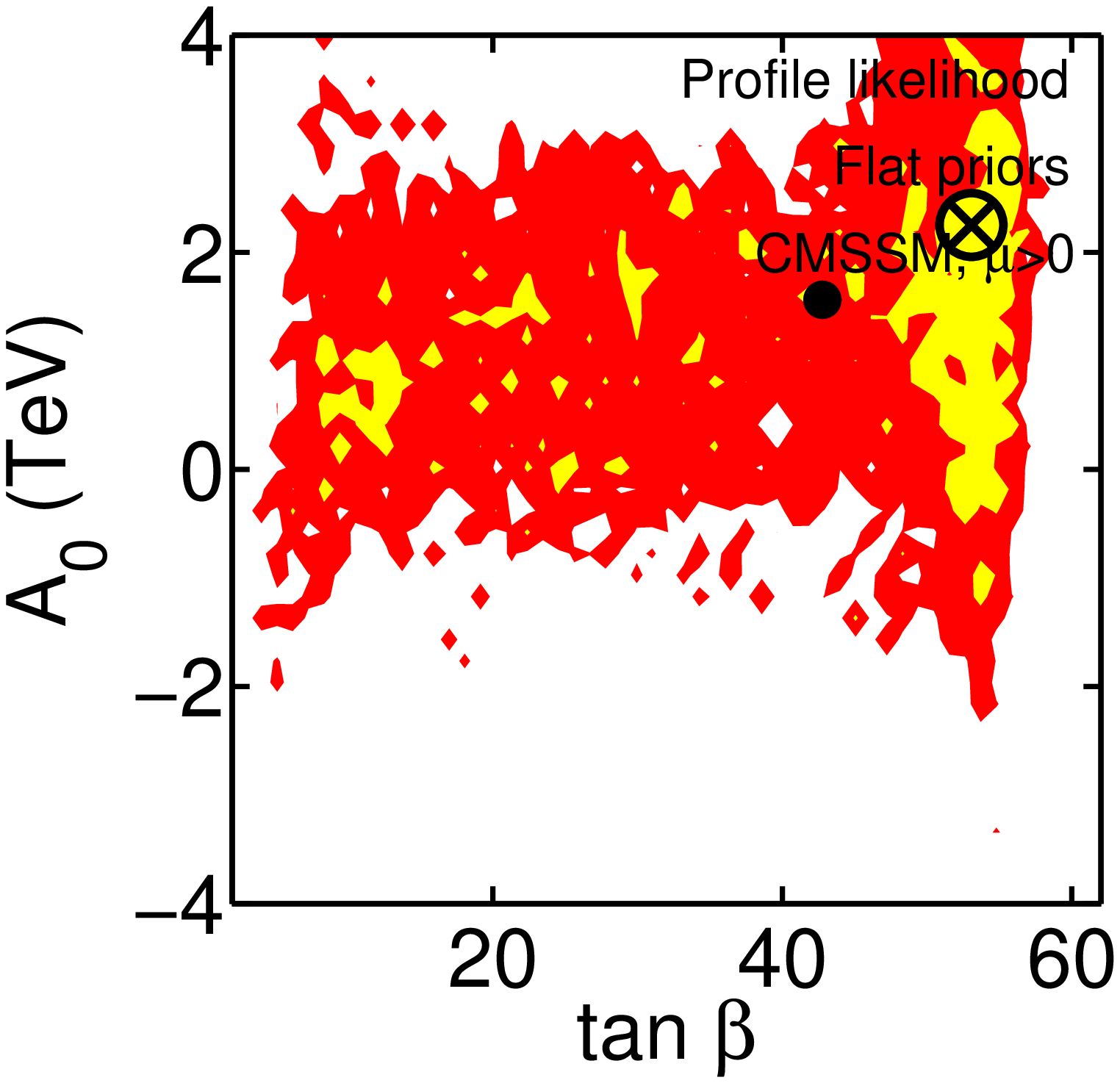} \\
\includegraphics[width=\qq]{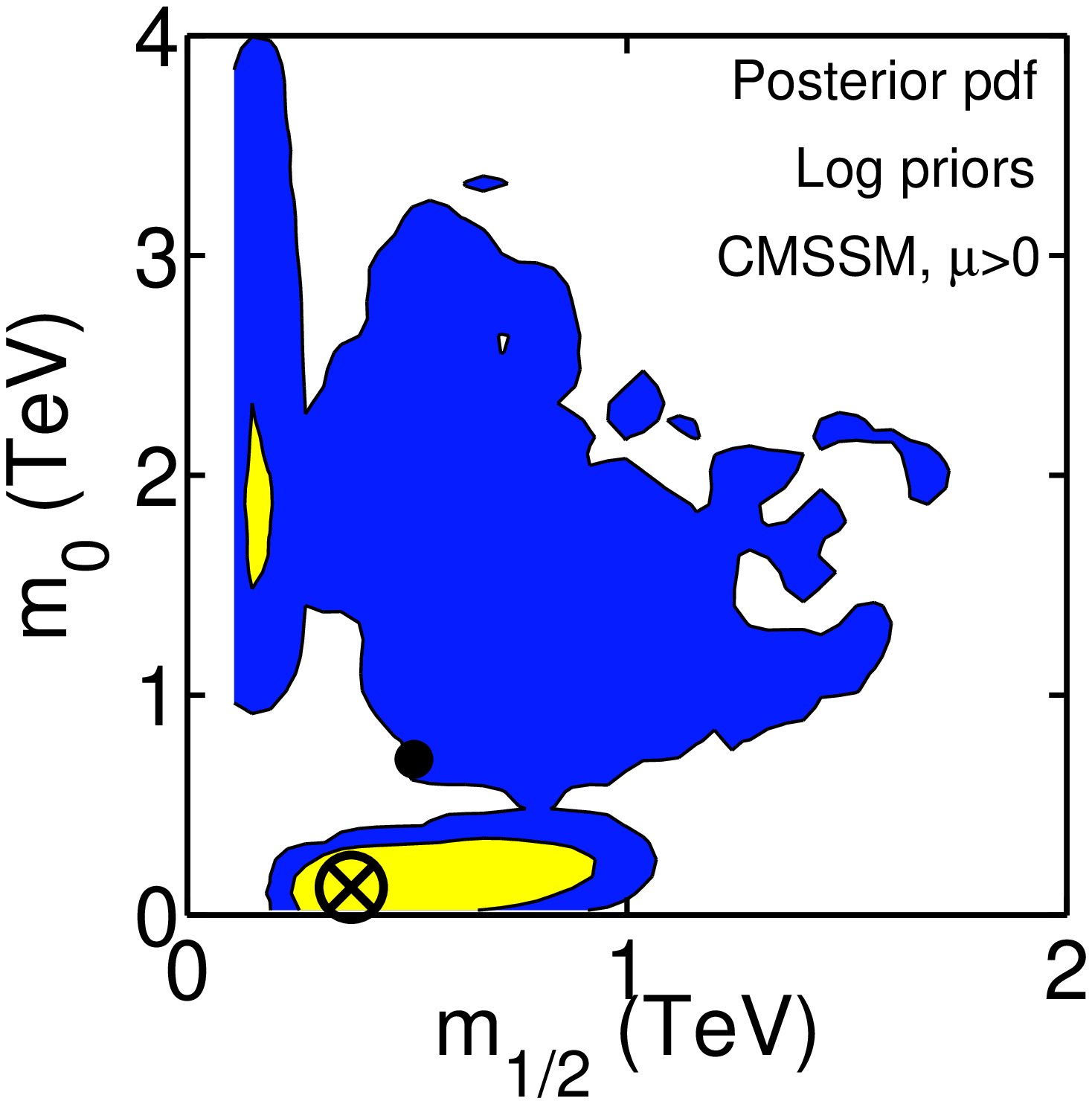} \includegraphics[width=\qq]{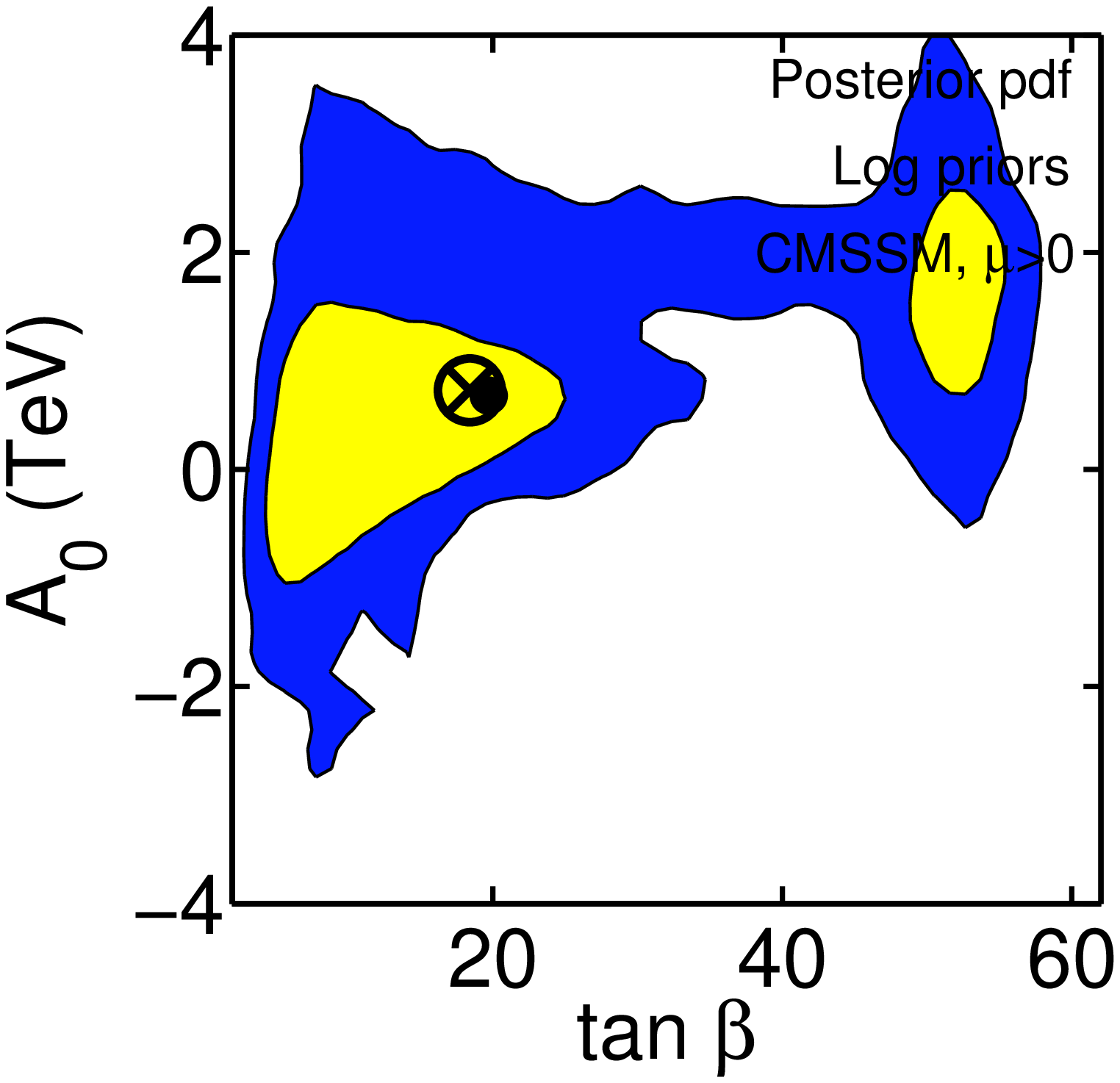}
\includegraphics[width=\qq]{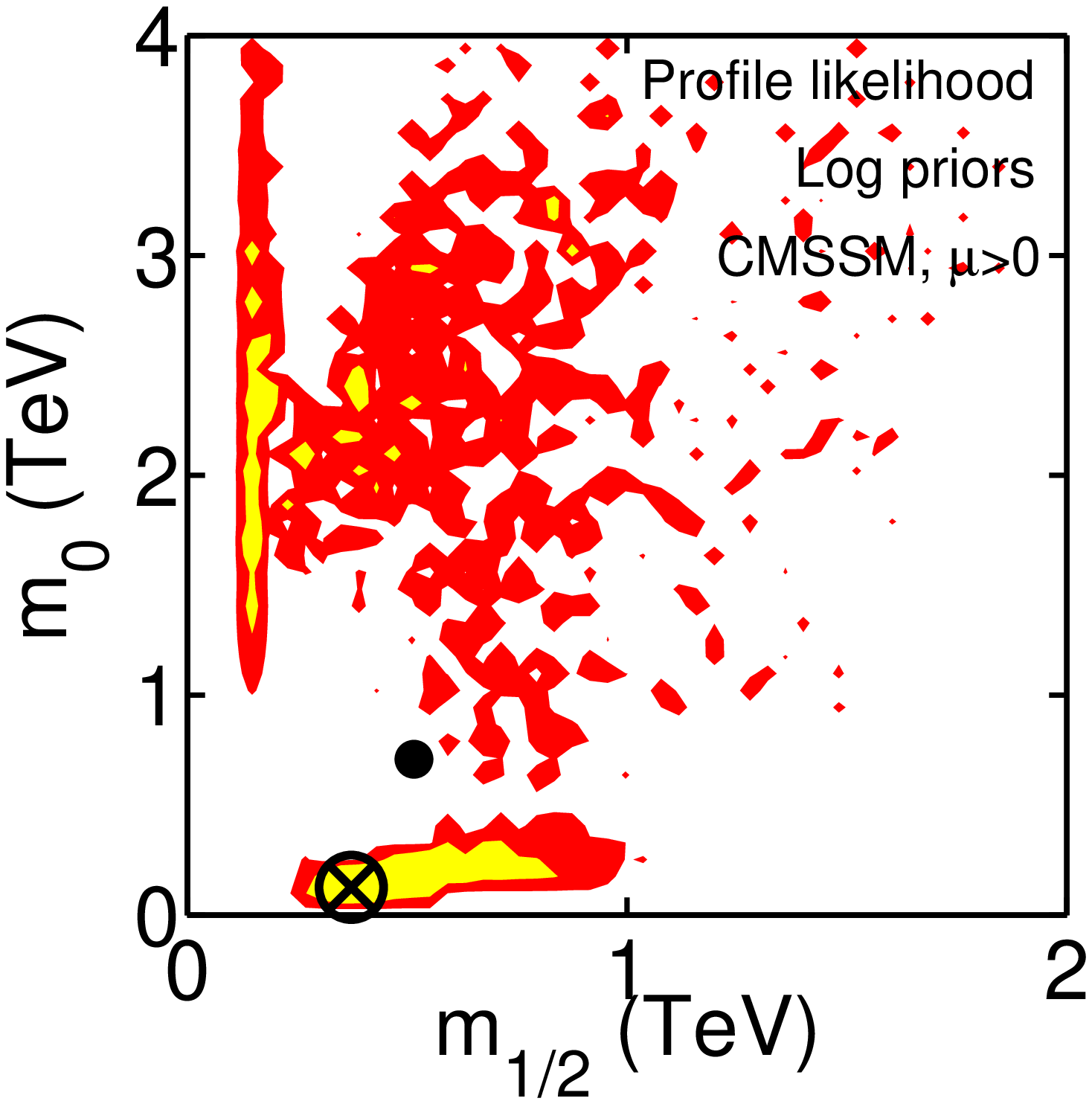} \includegraphics[width=\qq]{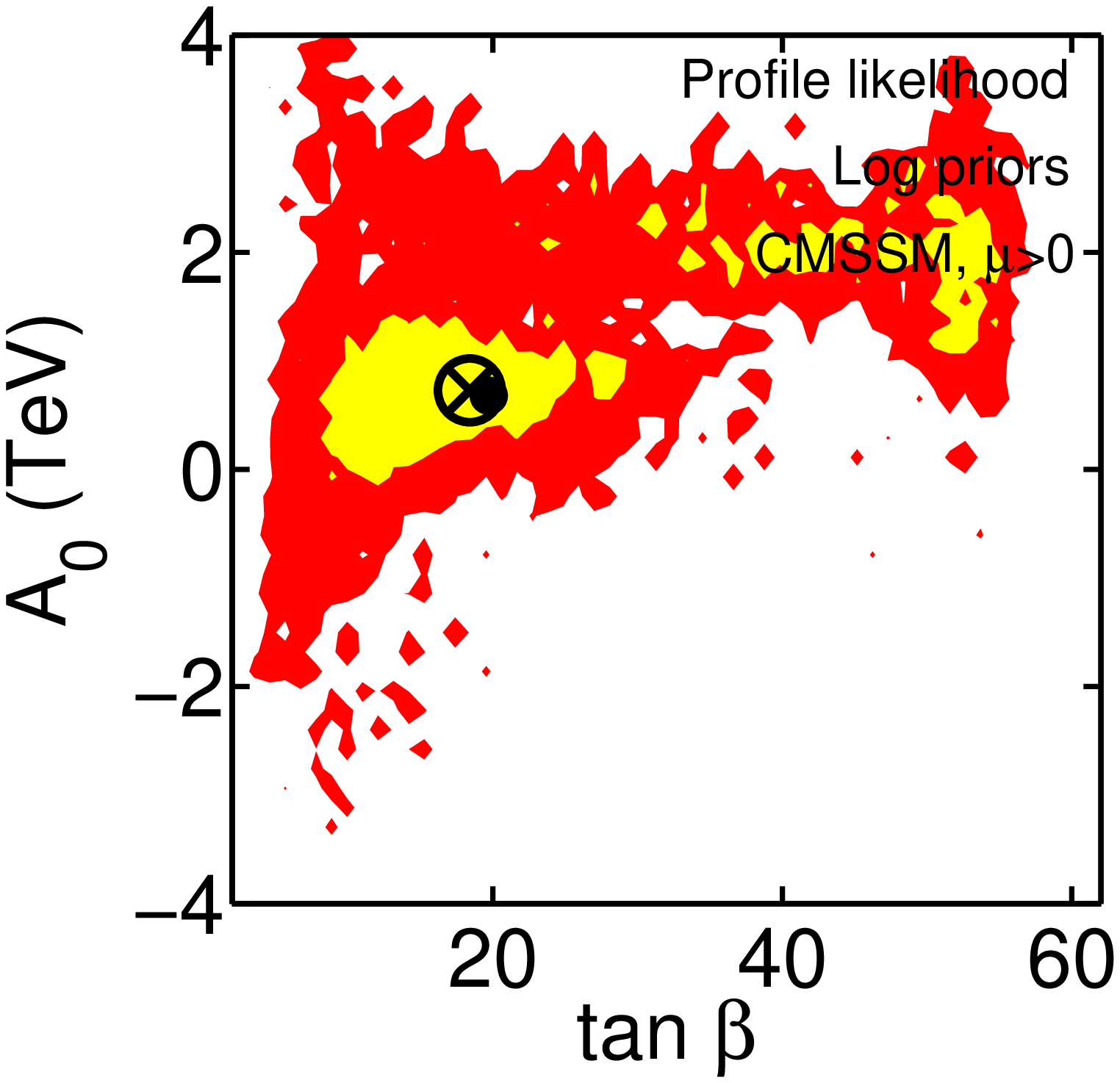}
\caption[short]{As in
 fig.~\ref{fig:2D_priors+nuis+lep+cdm}, but but for a scan
 including all the constraints listed in table~\ref{tab:datacombinations}
 (\texttt{ALL}). The change in the numerical evaluation of the profile likelihood for scans with different priors is due to 
  the change in the efficiency with which the algorithm finds good--fitting points for the two different choices of metric, especially for small SUSY masses.
\label{fig:2D_priors+nuis+lep+cdm+bsg+gm2+ewo+bphys}}
\end{center}
\end{figure}

Finally, we examine the combined effect of all the constraints listed
in table~\ref{tab:datacombinations} 
(\texttt{ALL}).  The corresponding plots for the CMSSM parameters are
shown in figs.~\ref{fig:1Dpriors+nuis+lep+cdm+bsg+gm2+ewo+bphys_base}
(1D distributions)
and~\ref{fig:2D_priors+nuis+lep+cdm+bsg+gm2+ewo+bphys} (2D
distributions). In the case of the flat prior (two leftmost columns),
both posterior pdf and profile likelihood show a clear preference for
large $\mzero$ and large, but not as much, $\mhalf$ (the FP region), 
as well as a fairly narrow peak at small $\mzero$ (the stau coannihilation
region).  Both statistical measures also appear to favor non-zero,
positive $\azero$.  On the other hand, the posterior shows a peak at
large $\tanb\sim55$, although at 95\% confidence both the posterior
and especially the profile likelihood allow a wide spread of values,
down to small values of about 10 (where the profile likelihood shows
another peak), and even less. Turning next to the log prior (two rightmost
columns), the posterior for $\mzero$ is now more strongly peaked at
small values while the probability for larger values is suppressed
(again as expected from a log prior). In contrast, the profile
likelihood continues to indicate a preference for large
$\mzero\gsim1\tev$, in the FP region. On the other hand, the prefered
ranges of $\mhalf$ have for both statistical measured moved towards
smaller values, as expected from the log prior, although the profile
likelihood is qualitatively similar to the flat prior case. In
contrast, the distributions for $\azero$ have not changed
dramatically, while the bi-modality in the ones for $\tanb$ is
somewhat stronger and showed more preference for lower values. We
remind the reader that, for both choices of priors, we have used flat
distributions in both $\azero$ and $\tanb$.

It is clear that
figs.~\ref{fig:1Dpriors+nuis+lep+cdm+bsg+gm2+ewo+bphys_base}
and~\ref{fig:2D_priors+nuis+lep+cdm+bsg+gm2+ewo+bphys} are
qualitatively similar to figs.~\ref{fig:1Dpriors+nuis+lep+cdm+bsg}
and~\ref{fig:2D_priors+nuis+lep+cdm+bsg} (which show the impact of
including $\brbsgamma$ but not $\gmt$), and significantly different
from figs.~\ref{fig:1Dpriors+nuis+lep+cdm+gm2}
and~\ref{fig:2D_priors+nuis+lep+cdm+gm2} (which show impact of
including $\gmt$ but not $\brbsgamma$). This is yet another reflection
of the strong tension between $\gmt$ and the other constraints, mostly
$\brbsgamma$, which at the end override to a large extent the impact of
$\gmt$.

\begin{figure}[tbh!]
\begin{center}
\includegraphics[width=\ww]{figs/BlackLike.ps} \includegraphics[width=\ww]{figs/BlackLike.ps}\\
\includegraphics[width=\ww]{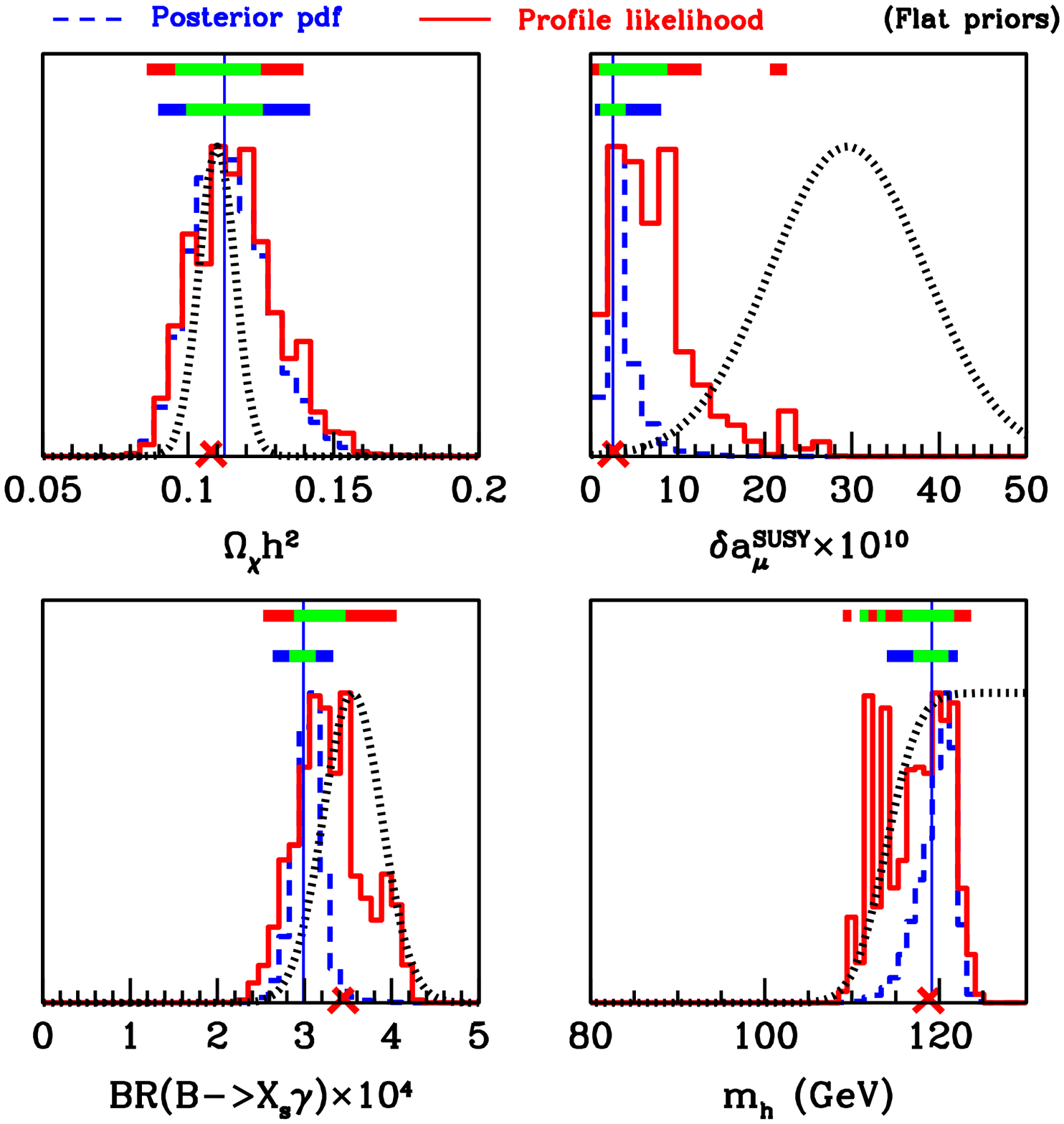} \includegraphics[width=\ww]{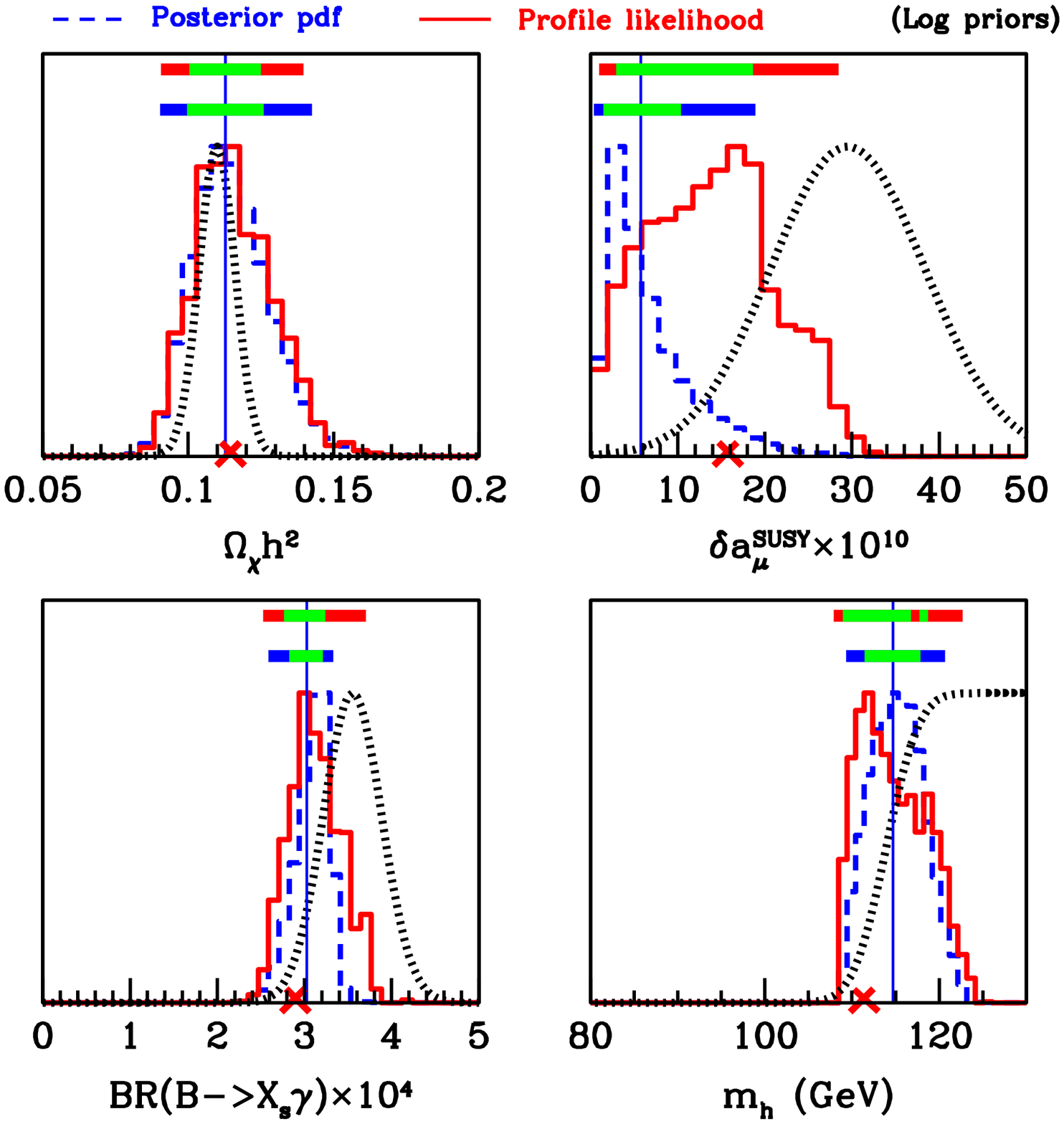}\\
\includegraphics[width=\ww]{figs/BlackLike.ps} \includegraphics[width=\ww]{figs/BlackLike.ps}\\
\includegraphics[width=\ww]{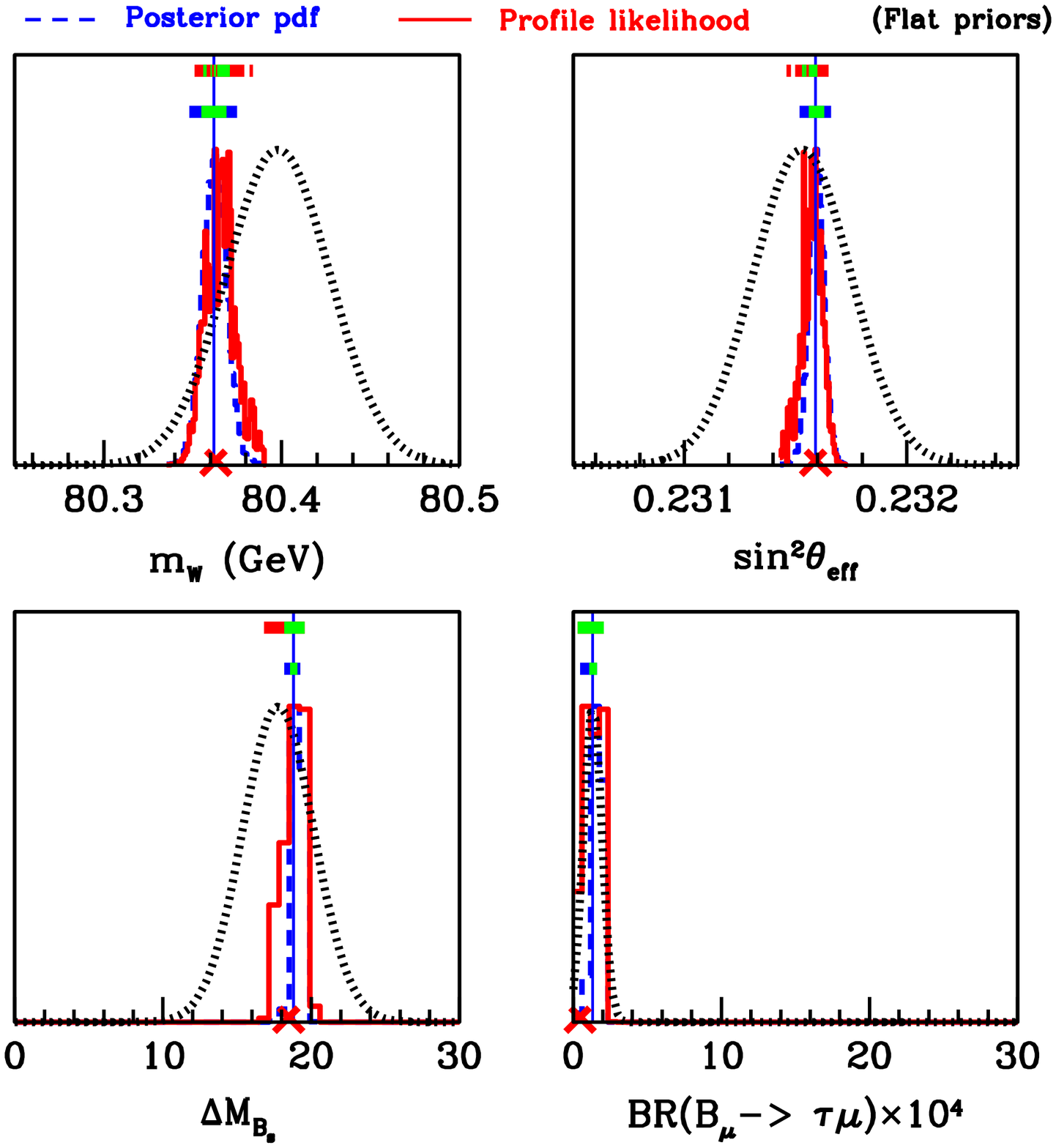} \includegraphics[width=\ww]{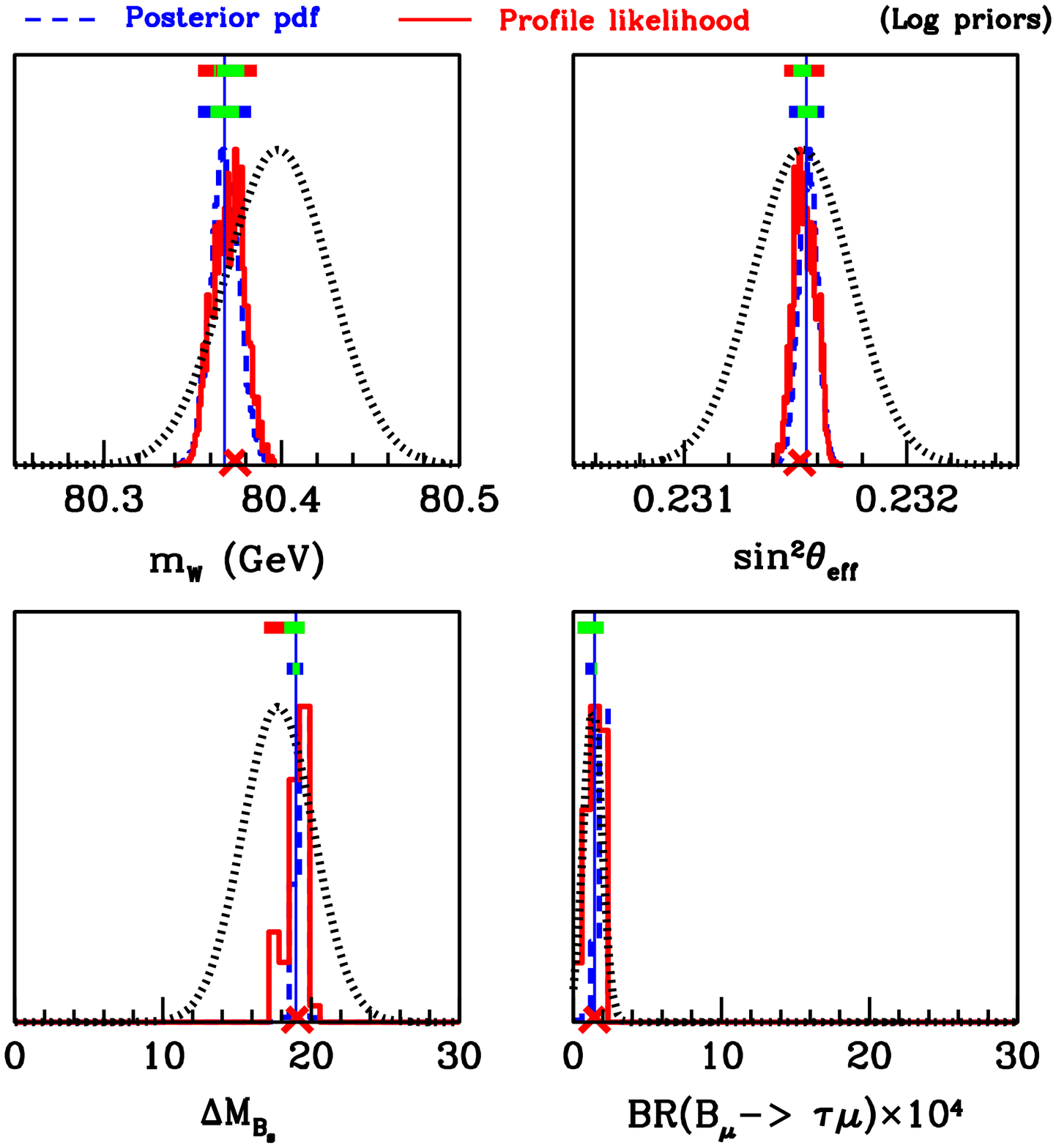}
\caption[test]{As in
 fig.~\ref{fig:1Dpriors+nuis+lep+cdm+bsg+gm2+ewo+bphys_base}  
 (\texttt{ALL}), but for the main observables.}
\label{fig:1Dpriors+nuis+lep+cdm+bsg+gm2+ewo+bphys_obs}
\end{center}
\end{figure}

The corresponding plots for several observables are shown in
fig.~\ref{fig:1Dpriors+nuis+lep+cdm+bsg+gm2+ewo+bphys_obs}. It is
instructive to compare them with the corresponding panels in
fig.~\ref{fig:1Dpriors+nuis+lep+cdm+gm2_obs_2} (where $\gmt$ was
included but not $\brbsgamma$) for the log prior. Again, we see a
large shift in the distributions of $\gmt$ (which now shows a strong
peak in the posterior pdf near zero and a more spread-out distribution
for the profile likelihood). On the other hand, the distributions for
$\brbsgamma$ and $\mhl$ now agree much better with the experimental
data (for both statistical measures). The same remains broadly true
also for the other obervables shown in
fig.~\ref{fig:1Dpriors+nuis+lep+cdm+bsg+gm2+ewo+bphys_obs}.

By examining the combined effect of all the constraints on both the
CMSSM parameters on the observables themselves
(figs.~\ref{fig:1Dpriors+nuis+lep+cdm+bsg+gm2+ewo+bphys_base},
~\ref{fig:2D_priors+nuis+lep+cdm+bsg+gm2+ewo+bphys} and
\ref{fig:1Dpriors+nuis+lep+cdm+bsg+gm2+ewo+bphys_obs}), we conclude
that the precise constraints are dependent on both the statistics and
on the prior choice, although broad trends are apparent. This means
that the combined data are not yet sufficiently strong to completely
override the prior dependence. By comparing the profile likelihood for
the two priors, we see that it suffers much less from prior
dependence. From
fig.~\ref{fig:1Dpriors+nuis+lep+cdm+bsg+gm2+ewo+bphys_obs} we notice
that both the posterior and the profile likelihood for all of the EW
and $b$-physics observables are much narrower than the likelihood, a
clear sign that they are dominated by the prior distribution and that
the effect of the data is solely to cut away the points preferred by
$\gmt$ (compare with
fig.~\ref{fig:1Dpriors+nuis+lep+cdm+gm2_obs_2}). On the other hand,
the CDM abundance, $\brbsgamma$ and the Higgs mass limit are all in
good agreement with both statistics. In contrast, the $\gmt$
constraint cannot be easily fullfilled simultaneously, as shown by the fact
that the posterior and the profile likelihood do not match with the
likelihood function.

Given the tension between $\gmt$ and the other observables
we have also carried out a scan applying all observables but omitting
the $\gmt$ constraint. The results are qualitatively similar to the
ones presented here, with the difference that the preference for low
masses is further reduced. This further implies that indeed the $\gmt$
constraint is to a large extent overridden by all other data
preferring a different region in parameter space.

\section{Consistency and constraining power of the
 observables}\label{sec:bsgvsgmt} 

We now come back to examining in more detail the tension between the
constraints from $\gmt$ and $\brbsgamma$ which we have already
emphasized above. (Compare figs.~\ref{fig:1Dpriors+nuis+lep+cdm+bsg}
and~\ref{fig:2D_priors+nuis+lep+cdm+bsg} with
figs.~\ref{fig:1Dpriors+nuis+lep+cdm+gm2}
and~\ref{fig:2D_priors+nuis+lep+cdm+gm2}, respectively.)

\subsection{Priors and a tension between \boldmath$\gmt$ and  $\brbsgamma$}

\begin{figure}[tbh!]
\begin{center}
\includegraphics[width=\ww]{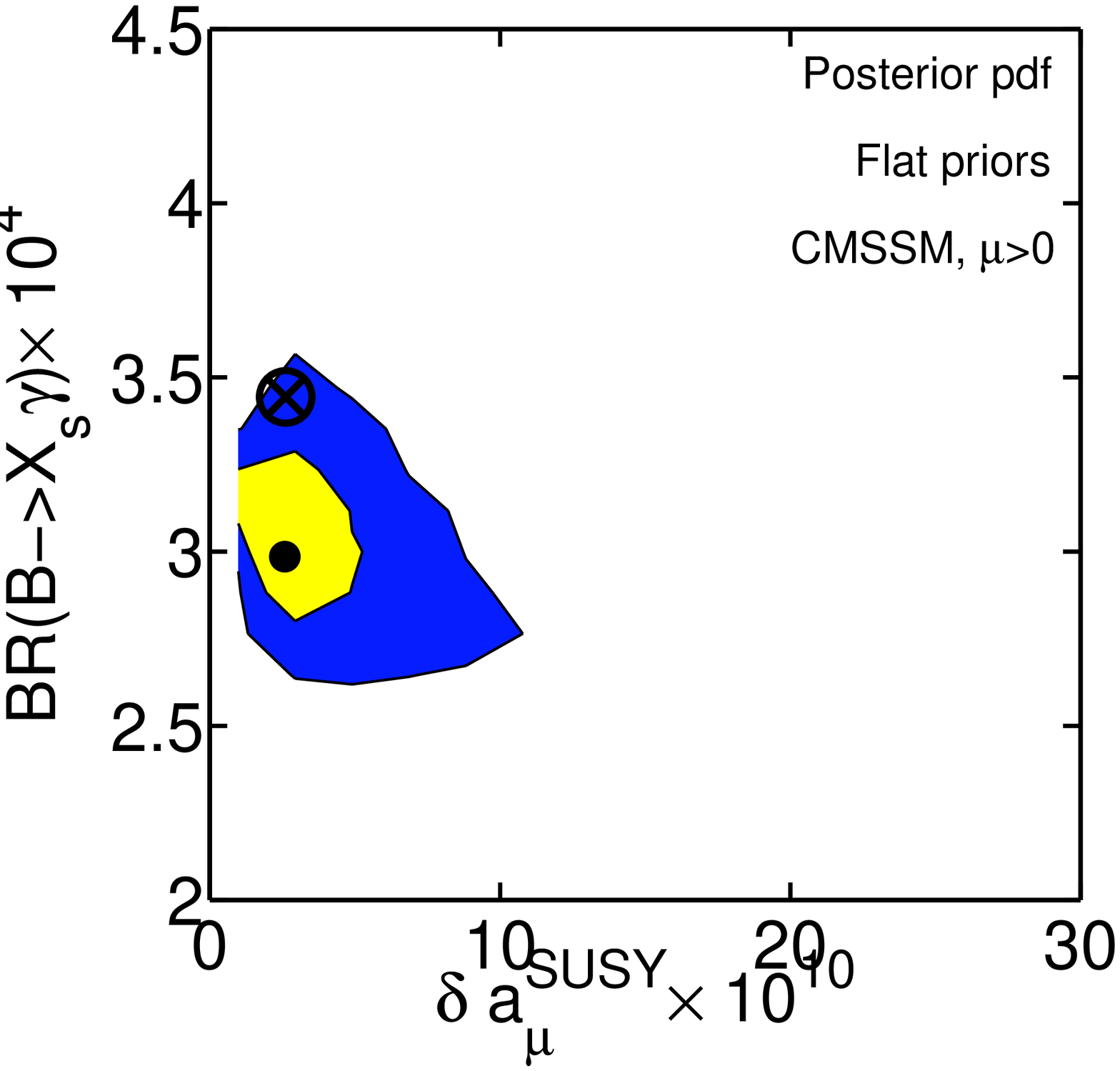} 
\includegraphics[width=\ww]{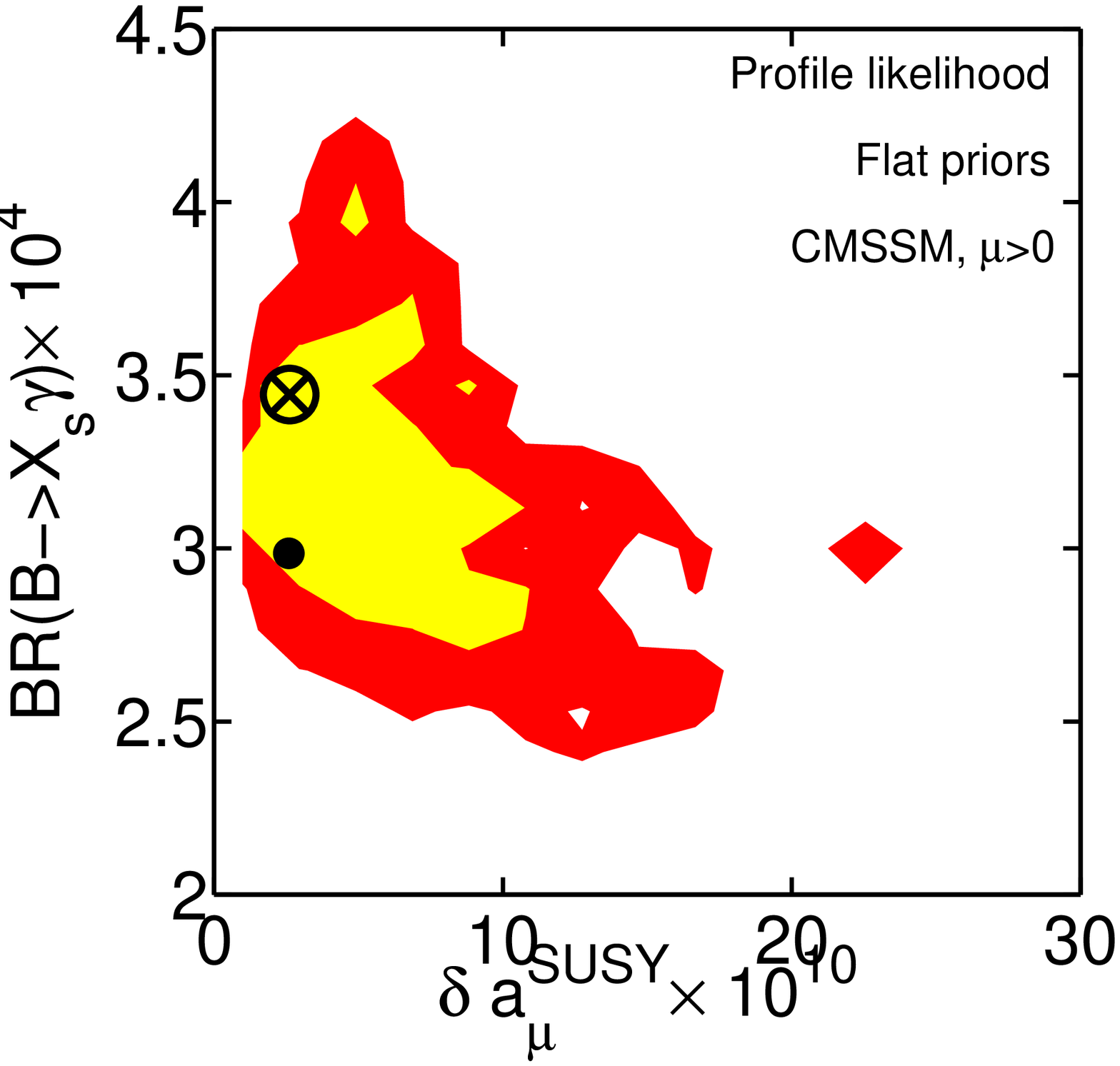}\\
\includegraphics[width=\ww]{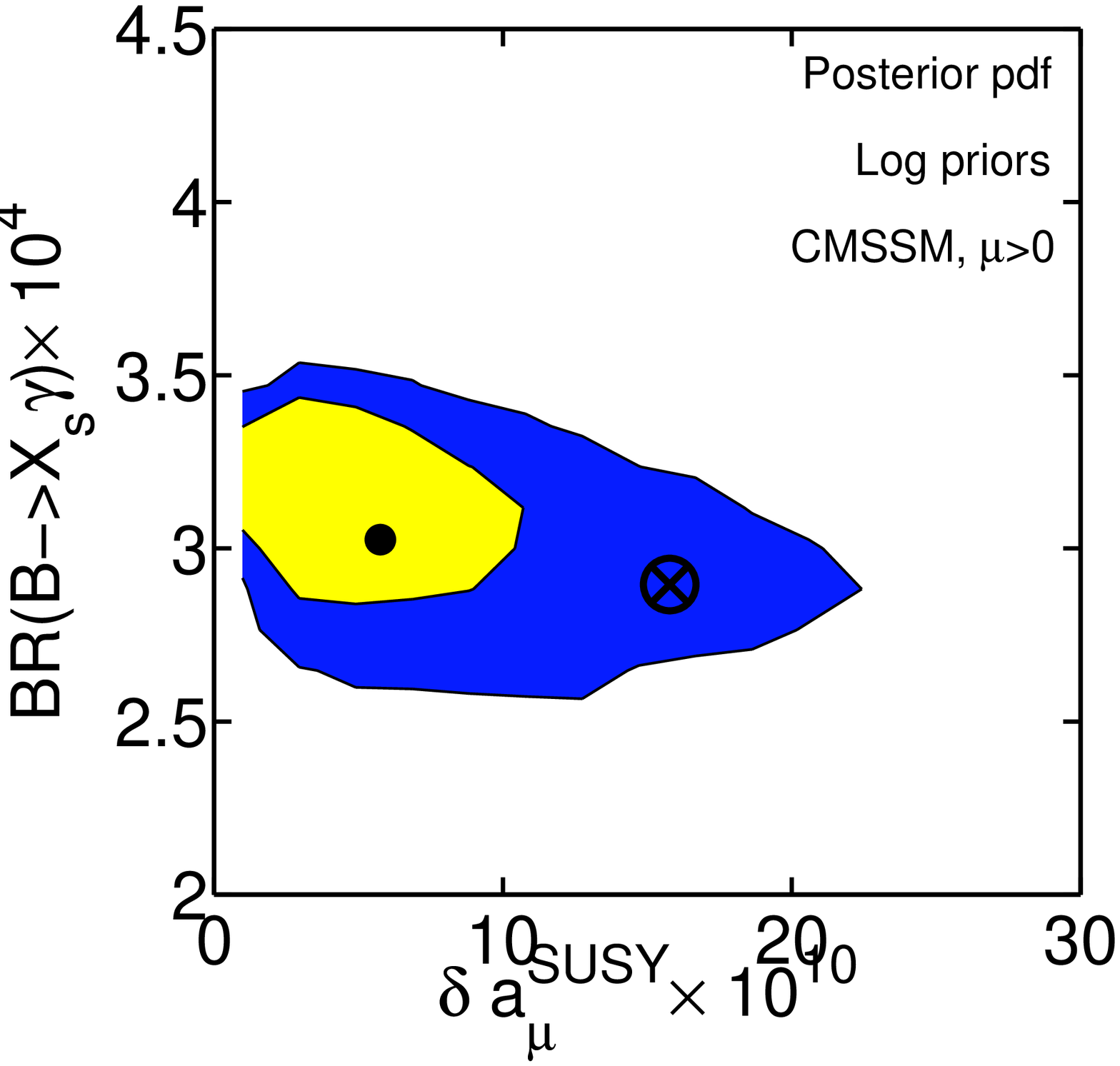} 
\includegraphics[width=\ww]{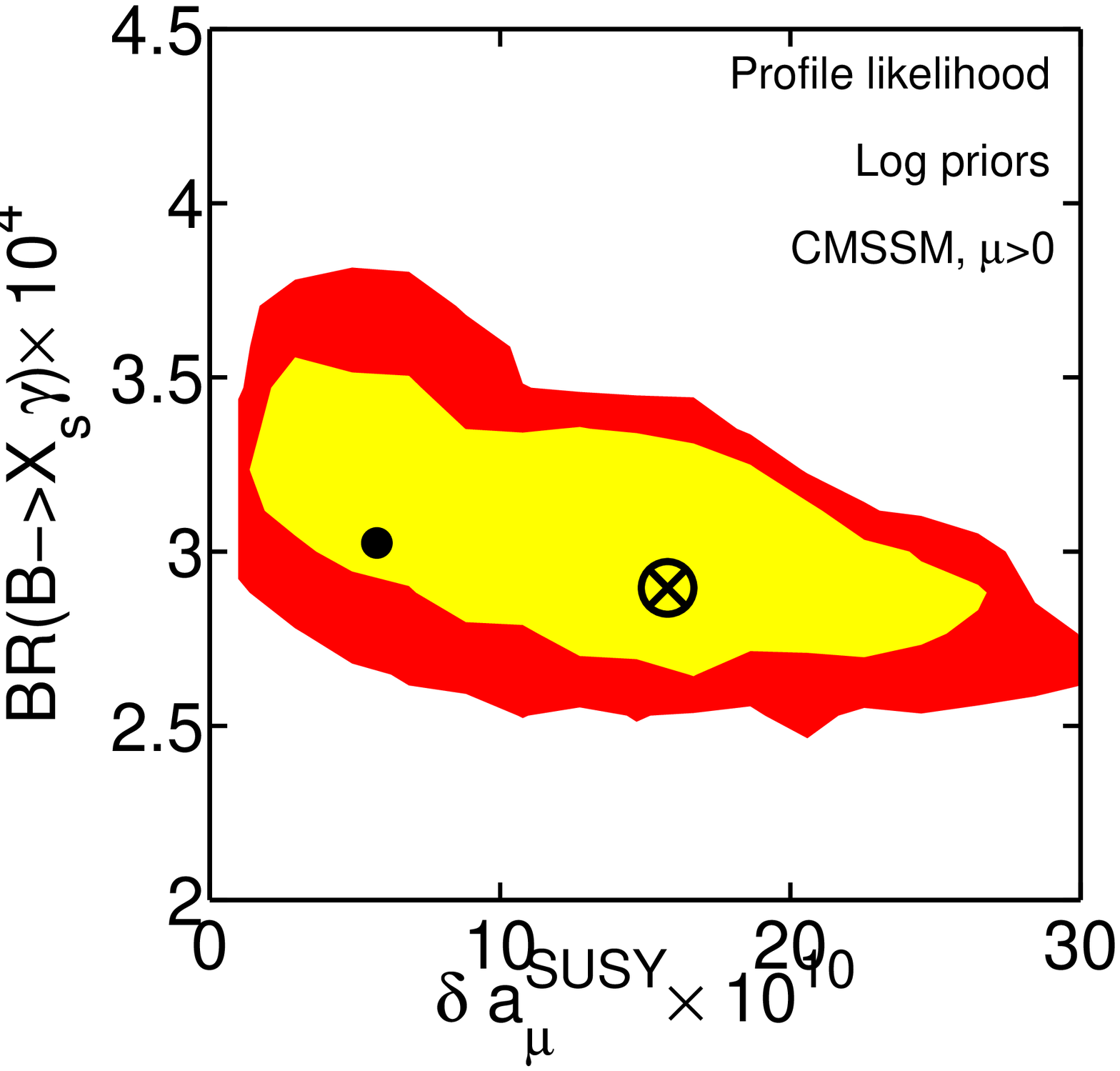}
\caption[short]{2D posterior pdf (left column) and profile likelihood
 (right column) for $\gmt$ and $\brbsgamma$ for the flat (upper row) and
 log priors (lower row) from a scan including constraints from all
 available observations
 (\texttt{ALL}). Notice that the change in the numerical evaluation of the profile likelihood for different priors is a consequence of the implicit change of metric in which the scan is executed. E.g., in the region of small SUSY masses (i.e., large $\gmt$ values) the log prior scan is much more detailed and can find better fitting points in that region that might have been missed by the linear prior scan.}
\label{fig:2D_gm2_vs_bsg_priors+nuis+lep+cdm+bsg+gm2+ewo+bphys}
\end{center}
\end{figure} 

The tension is clearly exposed in
fig.~\ref{fig:2D_gm2_vs_bsg_priors+nuis+lep+cdm+bsg+gm2+ewo+bphys}
where we include all the constraints
(\texttt{ALL}). It is stronger with
flat priors but remains substantial also in the case of log priors,
and therefore stronger for the posterior pdf than for the profile
likelihood since the former is more strongly prior dependent. 
We notice that the best fit point (cross) depends on the choice of prior quite strongly, with the log prior case able to find a point that has lower value of the masses and hence larger SUSY contributions to $\gmt$. On the contrary, the posterior mean (circle) is very similar in both cases. This is because the posterior distribution tends to favor regions with low $\gmt$ once all constraints are taken into account, and even the change of priors can extend the 95\% contour only mildly towards larger $\gmt$ values.

\begin{figure}[tbh!]
\begin{center}
\includegraphics[width=\qq]{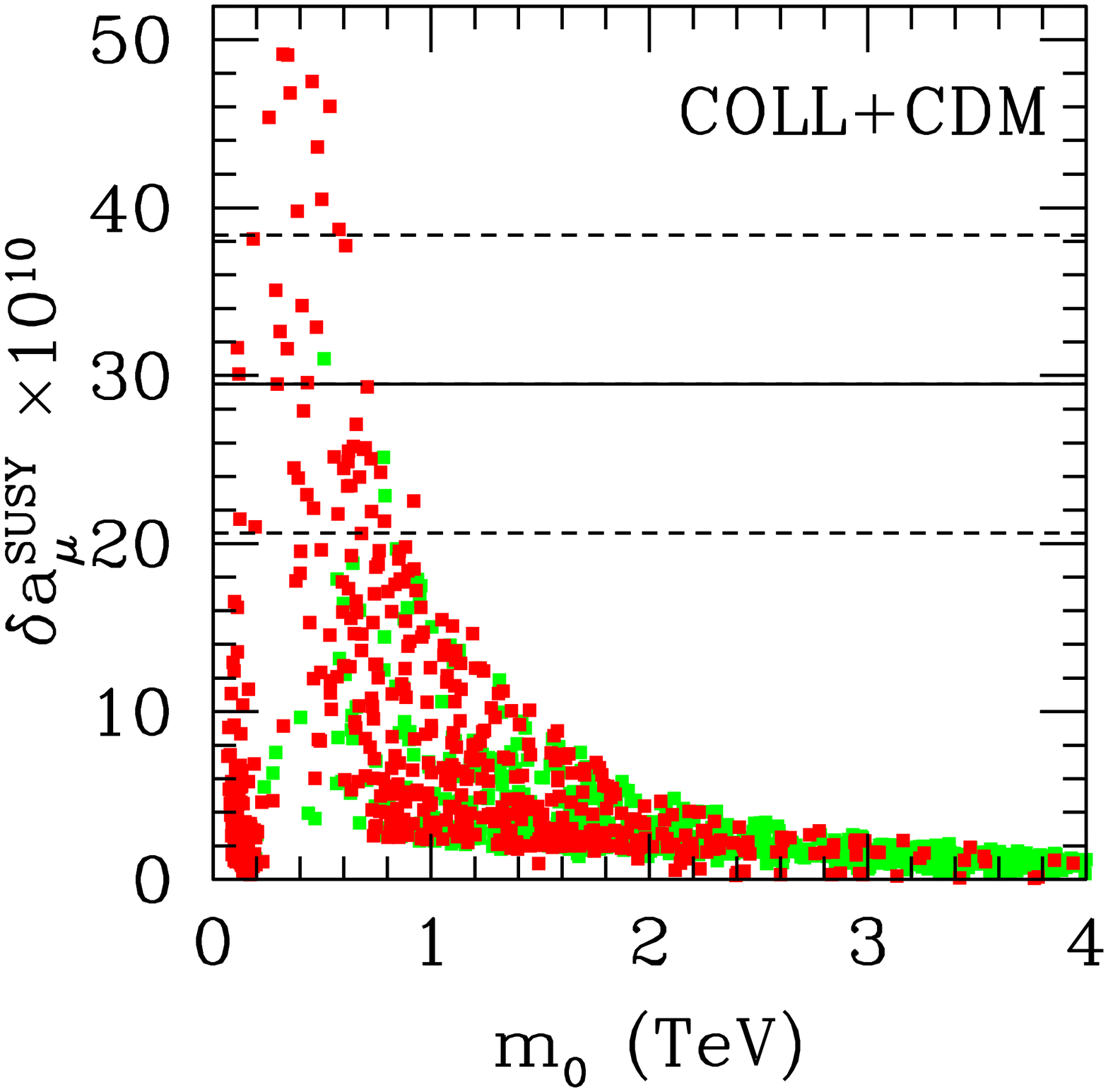} 
\includegraphics[width=\qq]{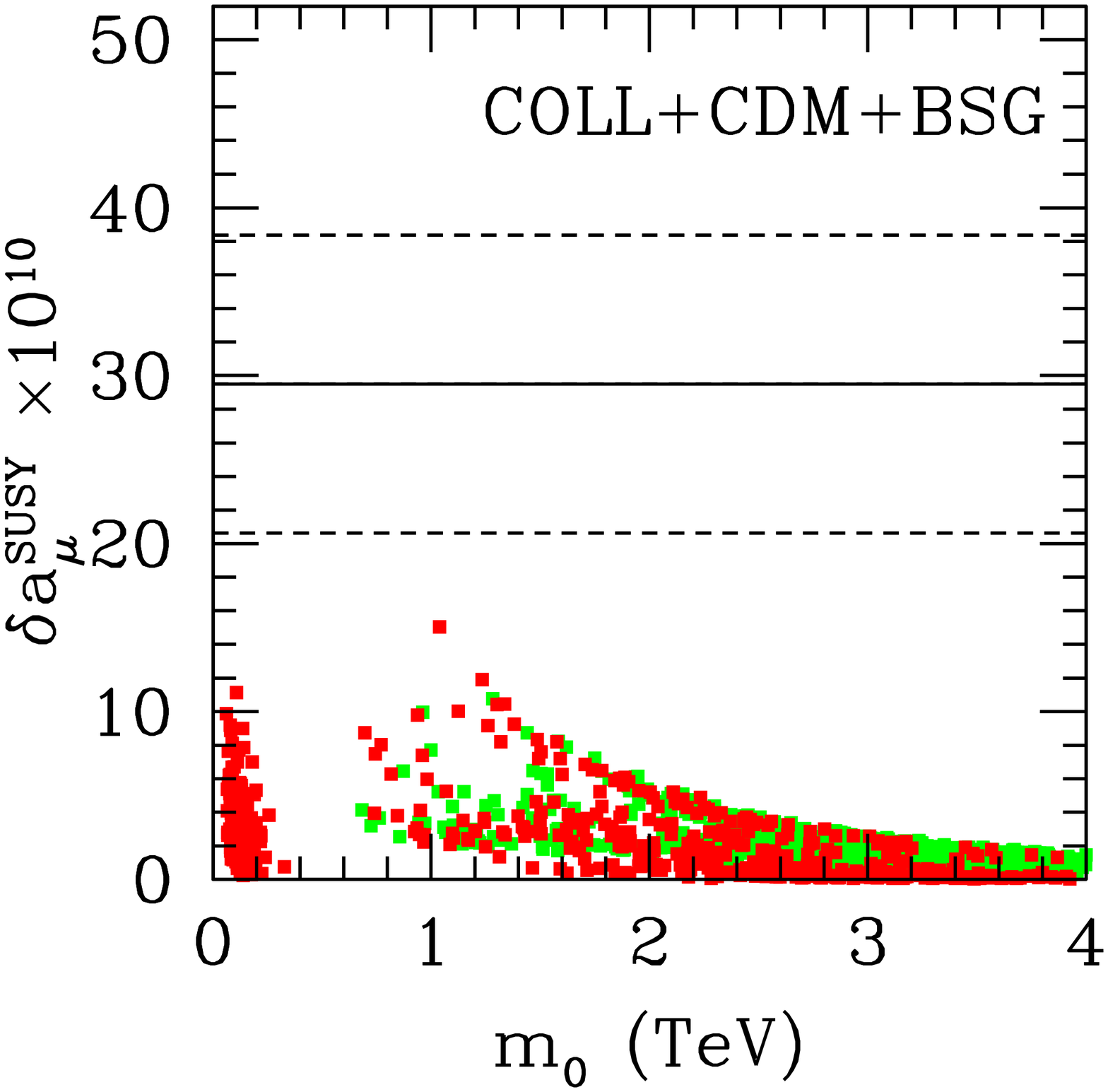} 
\includegraphics[width=\qq]{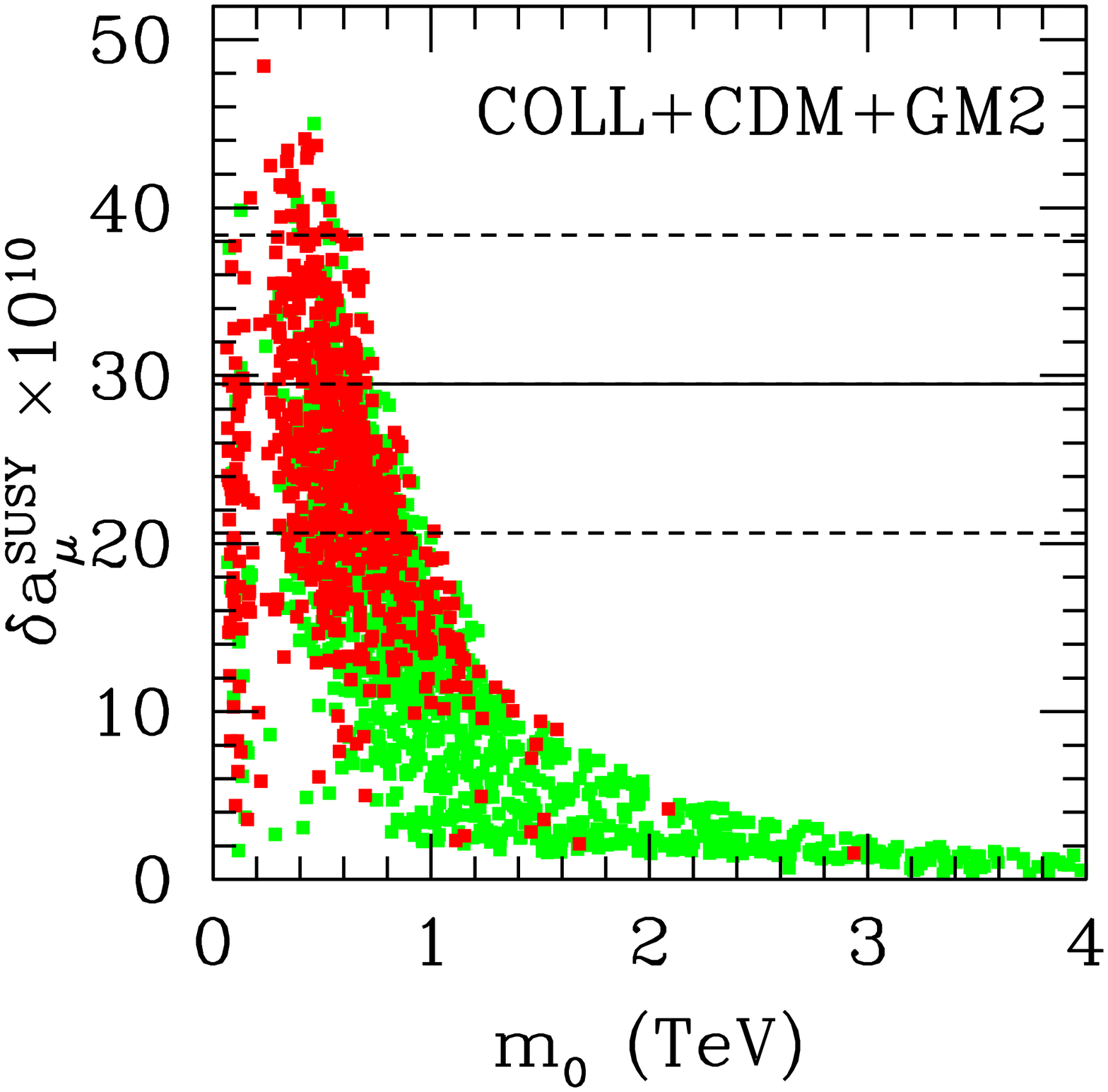} 
\includegraphics[width=\qq]{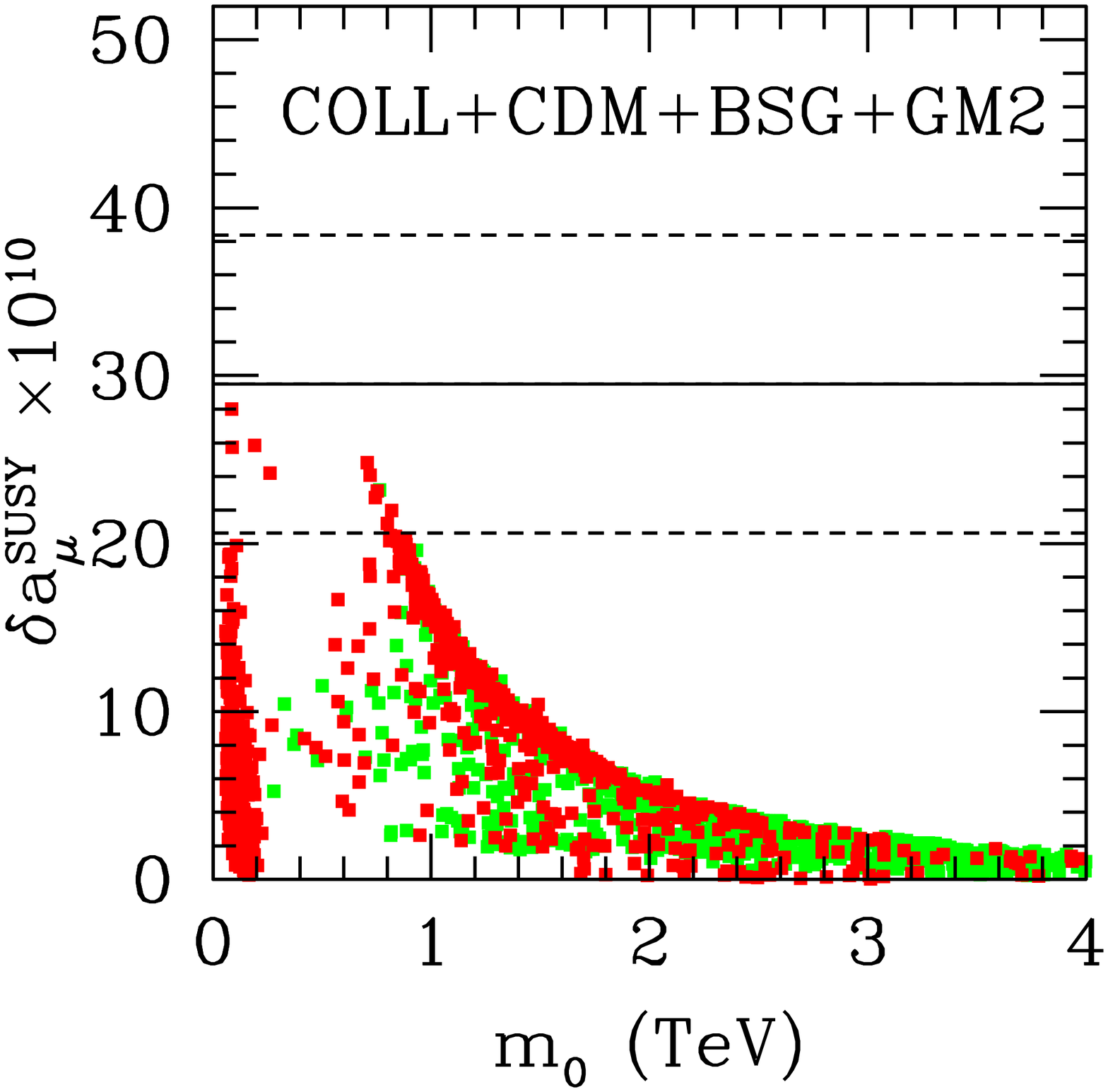} \\
\includegraphics[width=\qq]{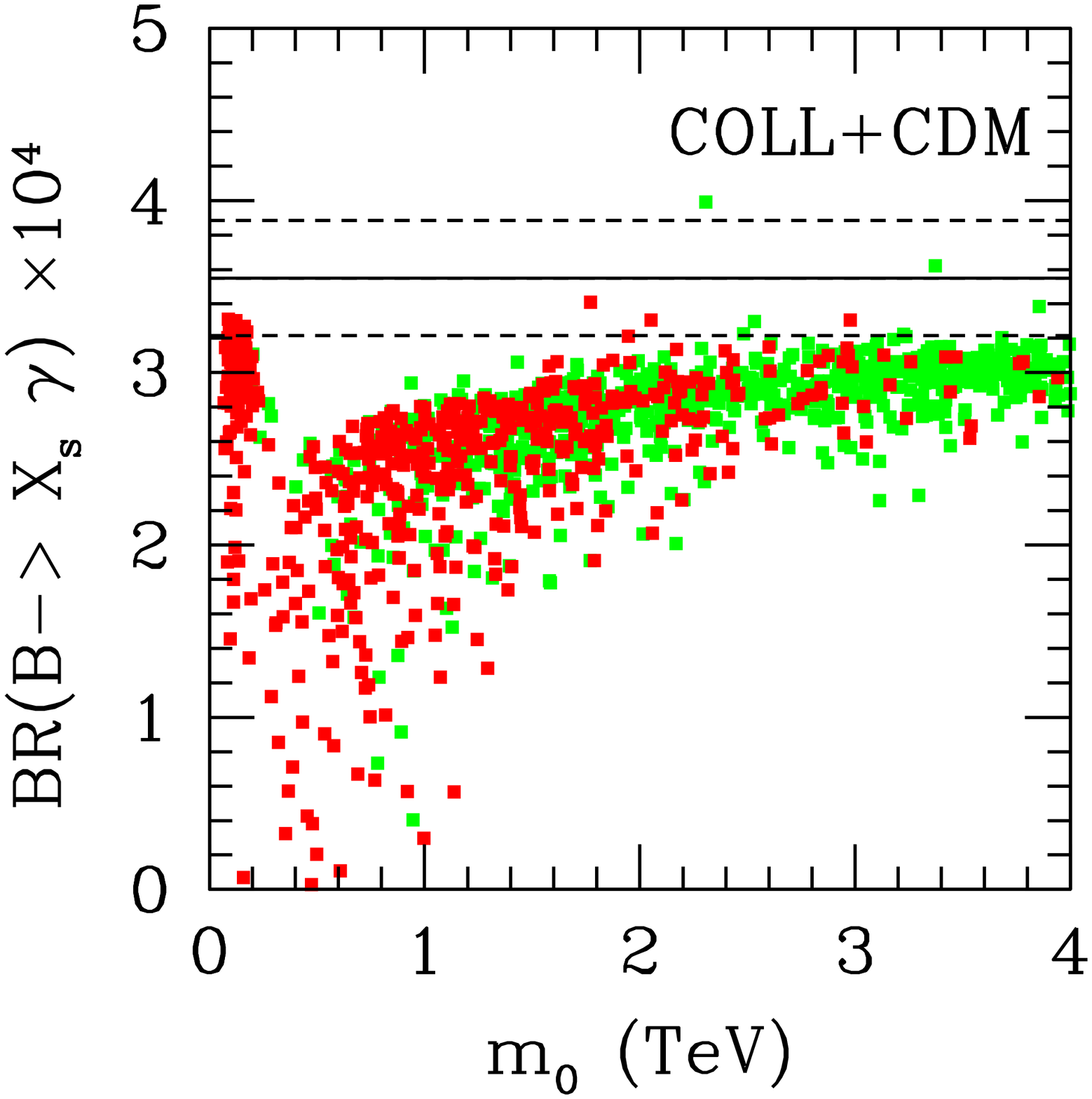} 
\includegraphics[width=\qq]{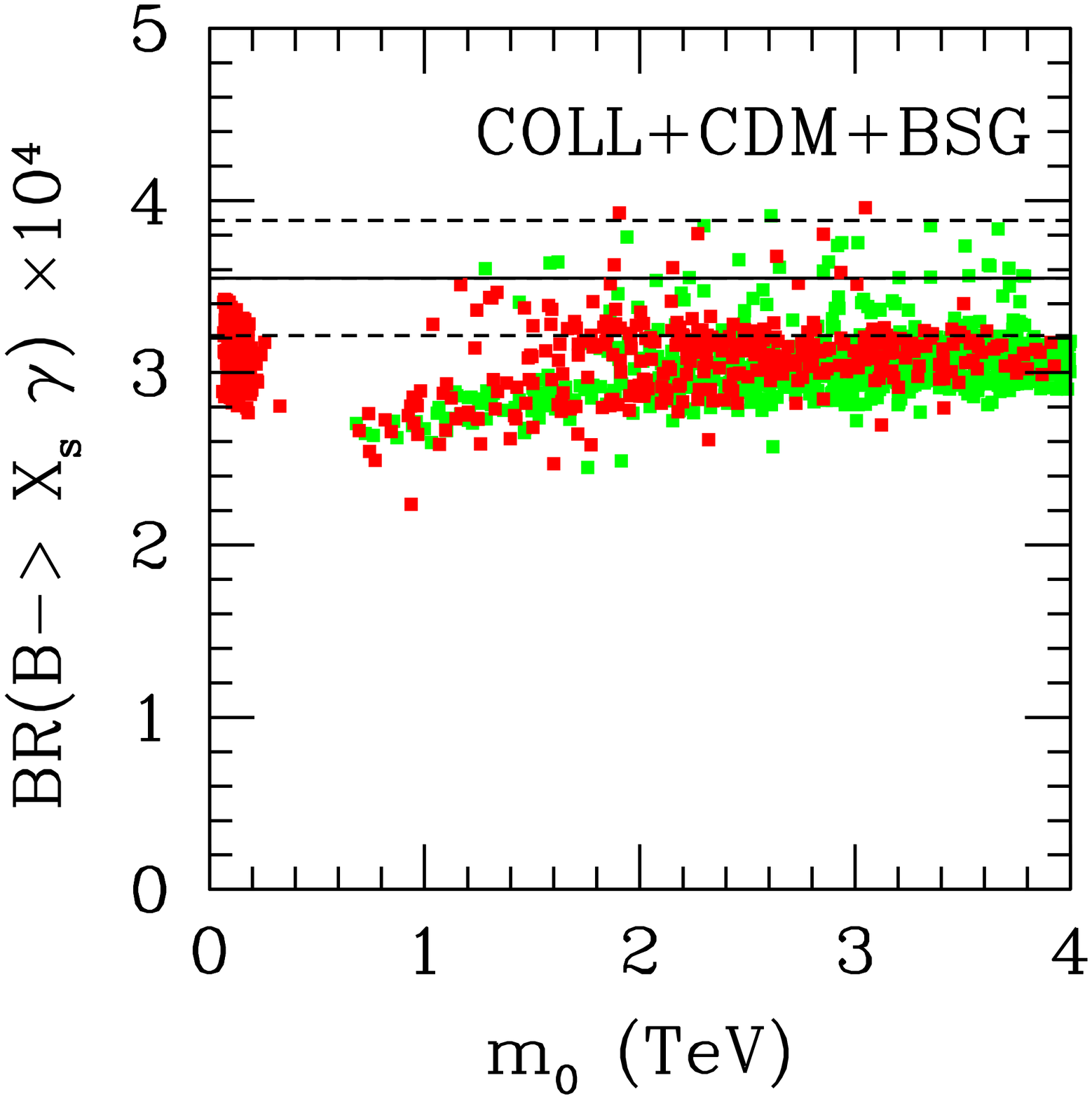} 
\includegraphics[width=\qq]{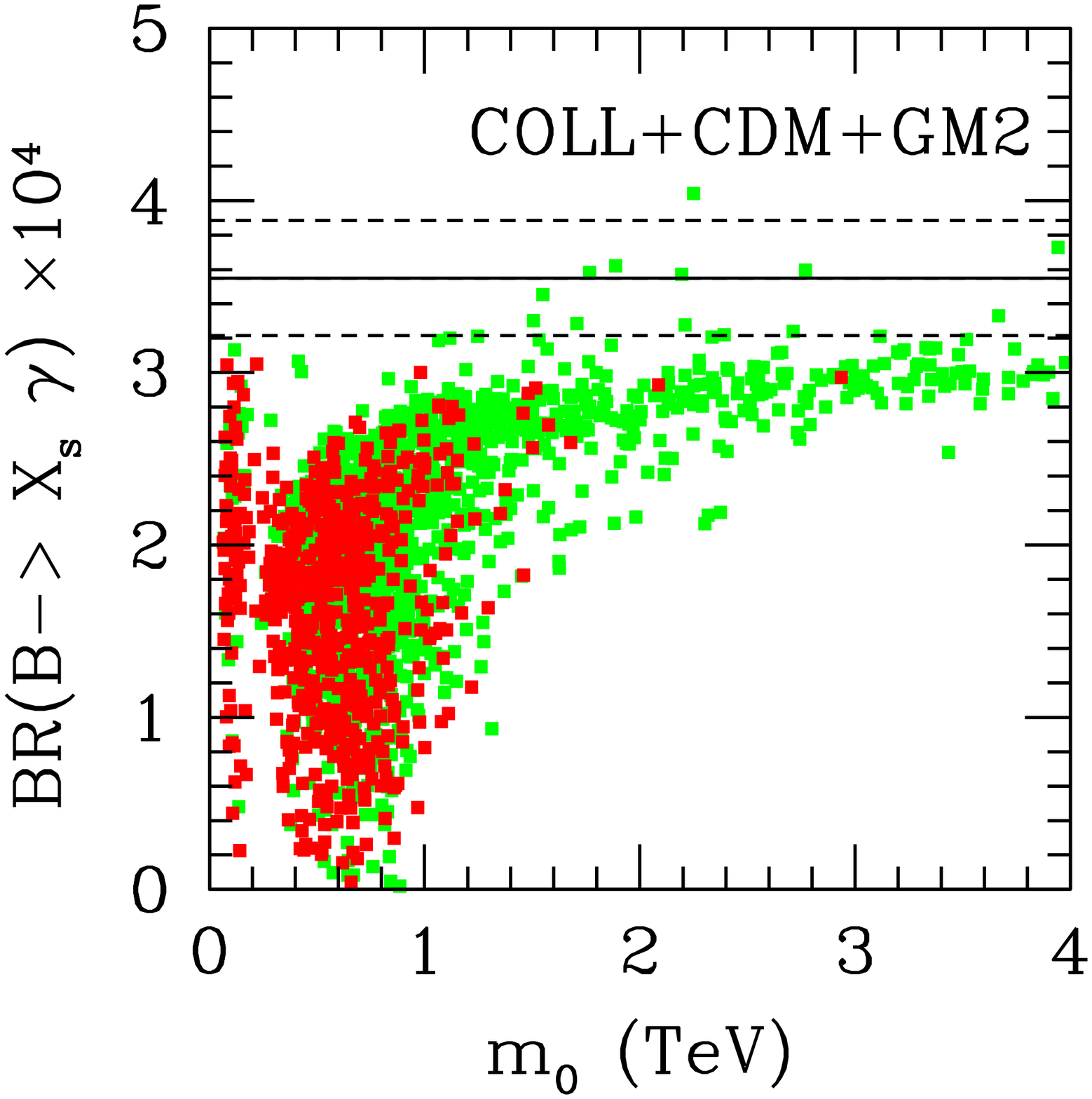} 
\includegraphics[width=\qq]{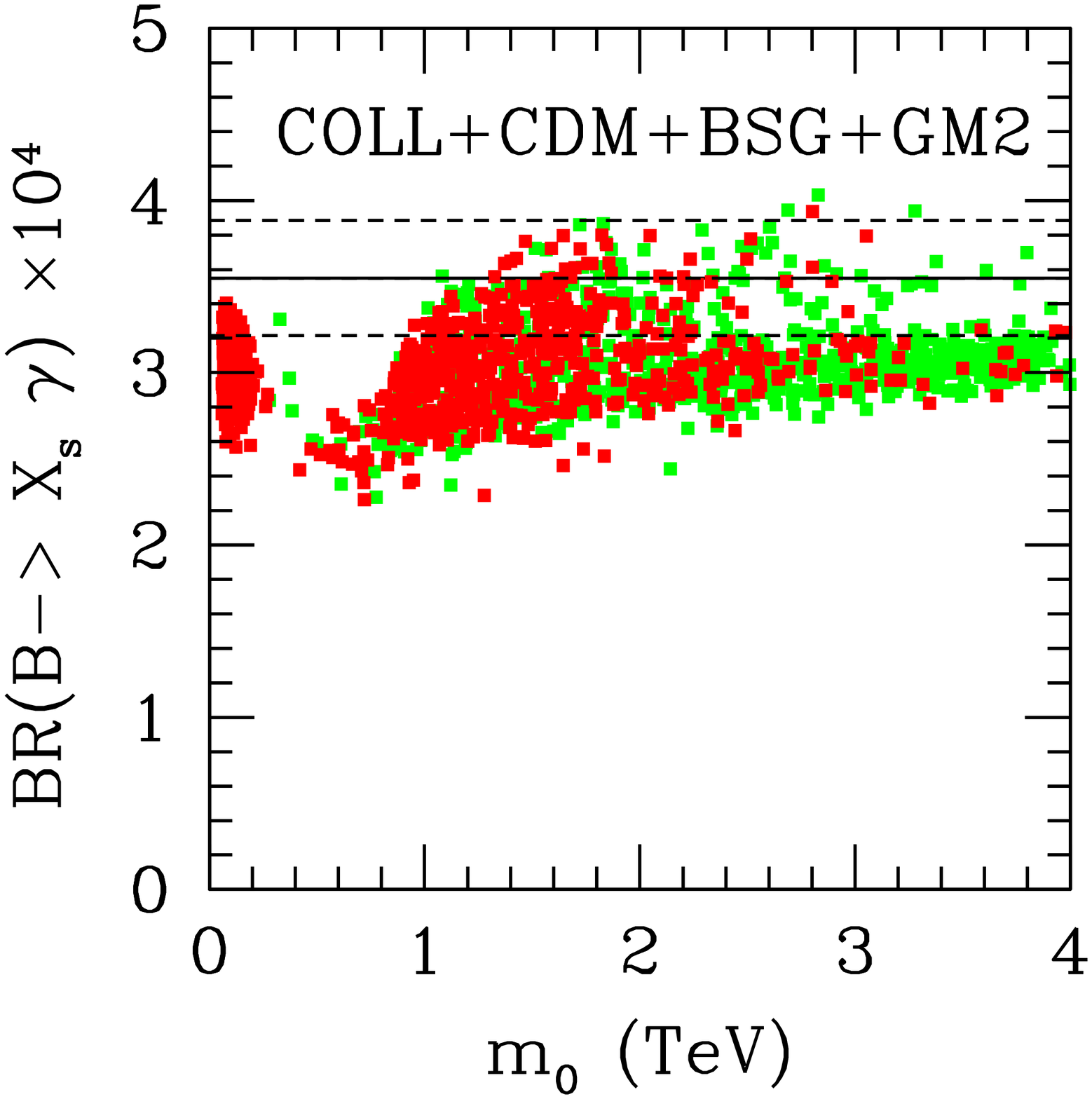} 
\caption[aa]{Distribution of samples from the posterior pdf, showing
 the preferred values for $\deltaamususy$ (top panels) and
 $\brbsgamma$ (bottom panels) for different combinations of
 constraints. Since the samples are drawn from the pdf, their density
 reflects the region's probability. Green points are for flat priors,
 red for log priors. The horizontal dashed lines give the $1\sigma$
 interval preferred by observations, the solid line is the central
 value. The samples have been thinned by a factor of 20 for
 visualisation purposes. }
\label{fig:2D_FP}
\end{center}
\end{figure}

The influence of priors and their interaction with the $\gmt$ and
$\brbsgamma$ constraints is further investigated in
fig.~\ref{fig:2D_FP}, where we plot equally weighted samples from
the posterior pdf, hence the density of points represents probability
density. The top panels show the probability density for
$\deltaamususy$ vs $\mzero$, while the bottom row shows $\brbsgamma$ vs
$\mzero$. Red points are for the log prior case, green for the flat
prior. From left to right, we change the sets of constraints being
imposed. The panels in the first column on the left have only physicality
constraints, nuisance parameters constraints, Higgs and superpartner
masses limits and the CDM abundance constraint imposed. The flat
priors give a fairly large mass to the FP region, hence the
predictions are dominated by the asymptotic SM value, $\gmt \sim 0$
while $\brbsgamma \sim 3.11 \times 10^{-4}$. Both observational
constraints (the horizontal dashed lines give $1\sigma$ regions from
the likelihood) prefer different values --- hence the tension between
the prior structure (and the CDM constrain) and both $\deltaamususy$
and $\brbsgamma$.

Once the $\brbsgamma$ constrain is further imposed (second column from the left), this
has the effect of strongly shifting the preference towards the FP
region, as pointed out in~\cite{rrt3} and explained above.  
Notice how, as a consequence, the favored range of $\deltaamususy$
collapses even further towards zero, hence making the observed anomalous
magnetic moment even more discrepant with the CMSSM favored range. 

In contrast, imposing the
$\deltaamususy$ constraint instead of $\brbsgamma$ (third column from
the left) has the effect of shifting the bulk of the probability to
smaller values of $\mzero$, as low enough smuon and/or sneutrino
masses are needed to produce a sufficiently large SUSY contribution to
$\gmtwo$. This, on the other hand, has the effect of selecting values
of $\brbsgamma$ (which has not been imposed in this case) below the SM
prediction, in strong disagreement with the experimental
determination.
 
Finally, once both the $\deltaamususy$ and
the $\brbsgamma$ observations are imposed (rightmost column), the
posterior settles in a compromise region, which is in fair agreement
with the $\brbsgamma$ observation but still quite discrepant with
$\gmt$. This comes about because the likelihood for $\deltaamususy$ is
large in the region where the other constraints, and in particular
$\brbsgamma$ (combined with the flat prior) give a very low
probability.

Hence we conclude that the only observable favoring smaller values of $\mhalf$ and
$\mzero$ is $\gmt$, while all the ones are either neutral or, as is
the case with especially $\brbsgamma$, favor the FP
region~\cite{rrt3}.

\subsection{Quality of fit and information content}
\label{sec:infocontent}

\begin{table}
\centering 
\begin{tabular}{| l l |  l l l | l l l | }
\hline
Constraints & Data &\multicolumn{3}{|c|}{Flat priors} & \multicolumn{3}{|c|}{Log priors} \\ 
		     &      points &  $\chisqmin$ & $\langle \chisq \rangle$ & $\KL$ &   $\chisqmin$ & $\langle \chisq \rangle$ & $\KL$ \\\hline
		     
PHYS+NUIS 			 &  4   & $0.06$    & $3.89$  & 1.00 &  0.02 & 3.88 & 1.00  \\\hline
+CDM 		&  5   & $0.05$   & $4.36$  & 3.22 & 0.10 & 4.32 & 2.59 \\
+BSG 		&  5   & $0.31$   & $6.48$  & 1.11 & 0.10 & 5.48 & 1.21 \\
+GM2			&  5   & $0.27 $ & $11.55$  & 1.35 & 0.13 & 6.38 & 1.20 \\ \hline
+COLL+CDM	&  5+ & $0.28 $ & $4.60$   & 3.20 & 0.15 & 5.04 & 2.98 \\
+COLL+BSG	&  5+ & $0.99$ & $6.82$ & 1.11& 0.45 & 6.54 & 1.24 \\
+COLL+GM2	&  5+ & $1.79$ & $13.43$ & 1.10& 0.17 & 9.92 & 1.49 \\\hline
+COLL+CDM+BSG	&  6+ & $0.75$ & $7.15$ & 3.36 & 0.68 & 7.72 & 3.29\\
+COLL+CDM+GM2	&  6+ & $0.62$ & $9.24$ &2.90 & 0.43 & 7.49 & 3.23\\
+COLL+CDM+BSG+GM2	&  7+ & $6.27$ & $15.83$ & 3.48 & 4.67 & 14.89 & 3.39\\ \hline
ALL but GM2				&10+ & $3.51$ & $9.45$ & 3.42 & 3.22 & 9.51 & 3.28\\
ALL but CDM				&10+ & $12.17$ & $18.86$ & 1.10 & 4.14 & 18.30 & 1.24\\
ALL 						&  11+ & $13.51$ & $19.29$ & 3.38 & 11.90 & 18.41 & 3.26\\ \hline

\end{tabular}
\caption[aa]{Best-fit chi--square, $\chisqmin$,  average chi-square over the posterior, $\langle\chisq\rangle$, and amount of information contained in the data, quantified using the KL divergence criterion ($\KL$ column, given by eq.~\eqref{eq:KL_derived}). The information content has been normalized to the information from priors alone with physicality and nuisance constraints imposed (\texttt{PHYS+NUIS}). The column ``Data points'' gives the number of constraints applied, where a $+$ indicates that collider limits on the Higgs and superpartner masses have been applied.}
\label{tab:KL}
\end{table}

In the light of the different constraining power of the observables,
it is interesting to investigate summary statistics for the
information content and the quality of fit including different
combinations of data and for the two choices of priors. This is given
in table~\ref{tab:KL}. The information content is quantified using the
KL divergence, which gives the information increase in going from the
prior to the posterior, and for each prior is normalized to the
information from priors alone with physicality and nuisance
constraints imposed.

First, looking at the quality of fit statistics (both the minimum
$\chisq$ and the average of the $\chisq$ over the posterior), we
notice that when the $\gmt$ constraint is added on top of 
$\brbsgamma$, the quality of fit worsens dramatically, for both choices
of priors. This reflects the tension between the two observables. Even
when the $\gmt$ constraint is applied on its own (cases
\texttt{+GM2} and \texttt{+COLL+GM2}), the fit can only
achieve a fairly poor average $\chisq$, with the situation being worse
for the linear prior scan which gives more weight to the FP region,
which is at odds with the $\gmt$ experimental value. Also, the best-fit $\chisq$ is
around 3 for both priors when we include all observables but $\gmt$
(case \texttt{ALL but GM2}). Such a fit has nominally 2 degrees of
freedom (dof), if we neglect the effect of imposing the collider
limits. So a classical quality of fit test would give a 
$\chisq/{\rm dof}$ of 1.5
which is not very large. (Although of course
one has to keep in mind that such a value is difficult to interpret
statistically, as clearly the $\chisq$ is not chi--square distributed
here!) However, when $\gmt$ is added (case \texttt{ALL}),
the best-fit value becomes about three times worse, giving
$\chisq/{\rm dof} > 6$, which is clearly unacceptable. This indicate
again a strong tension between $\gmt$ and the remaining observables,
which do not appear to be able to be fulfilled all at the same time
within the CMSSM.

Second, the best-fit $\chisq$ values and the posterior $\chisq$
average are almost invariably better (albeit often not dramatically
so) for the log prior scan. For the best-fit values, this is a
consequence of the finer detail with which the low mass region can be
explored with this prior, and therefore the scan is able to find
better fitting points that can be more easily missed by the flat prior scan. The
better average values reflect the fact that the log prior scan finds
in general better fitting points than the flat priors one.

Finally, the information gain with respect to both priors is dominated
by the CDM constraint, which alone accounts for about 80\% of the
combined constraining power of all the data in the log prior case and for about 95\% of the constraining power for the flat prior case. This follows from taking the ratio of the $\KL$ value for the case \texttt{+CDM}  with 
the \texttt{ALL} case. Taken on their
own, each of the $\brbsgamma$ and the $\gmt$ observables have less
than half the constraining power of the CDM abundance (compare the $\KL$ values of the \texttt{+CDM} case with either \texttt{+BSG} or \texttt{+GM2}). When added on
top of CDM, they only contribute about an extra 10\% information on
the parameters at most. This is also evident from the case \texttt{ALL but CDM}, where all the constraints have been applied except for the CDM abundance. In this case the information content is only very mildly increased from the \texttt{PHYS+NUIS} value. 

\begin{figure}[tbh!]
\begin{center}
\includegraphics[width=\ww]{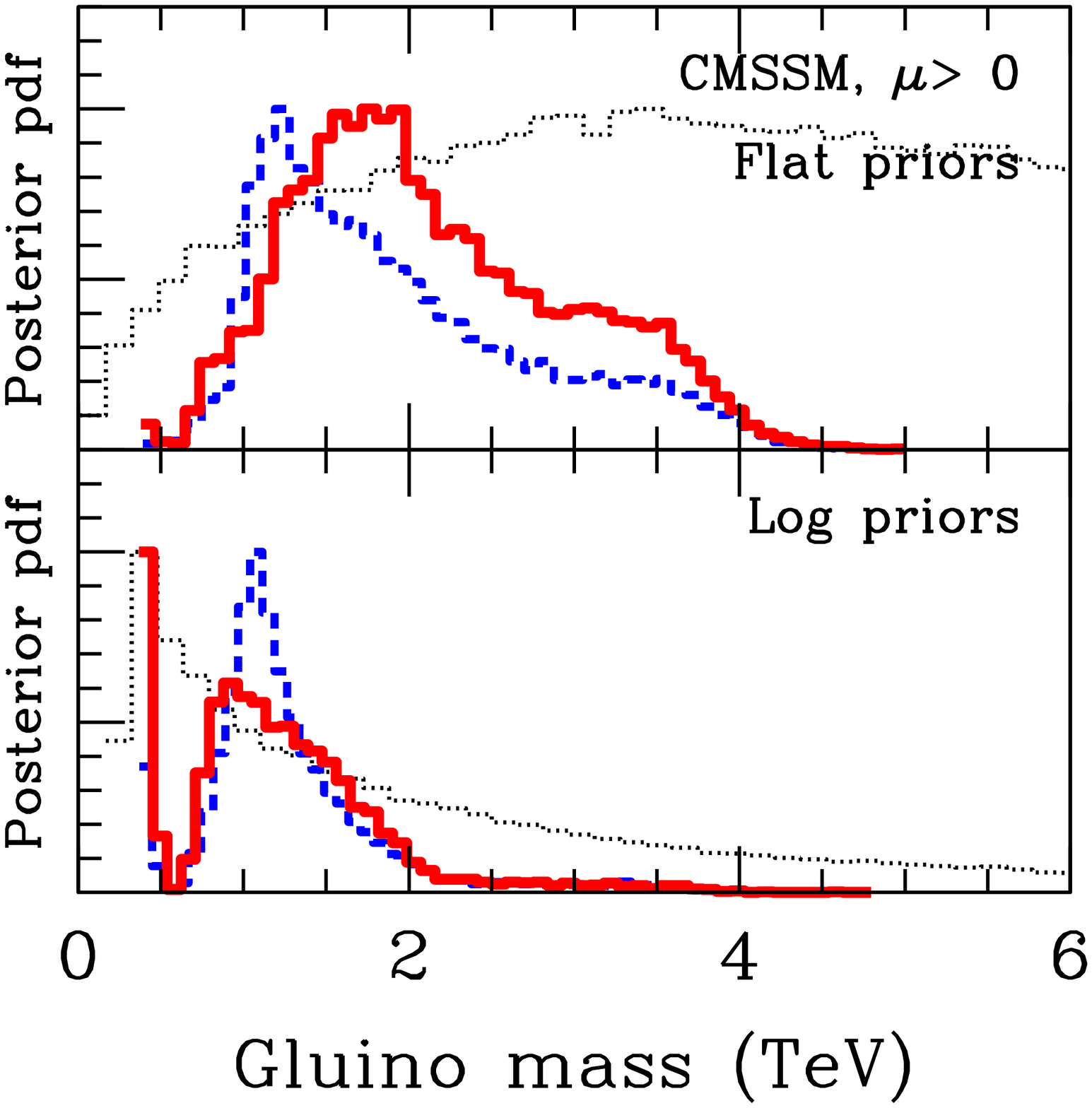}
\includegraphics[width=\ww]{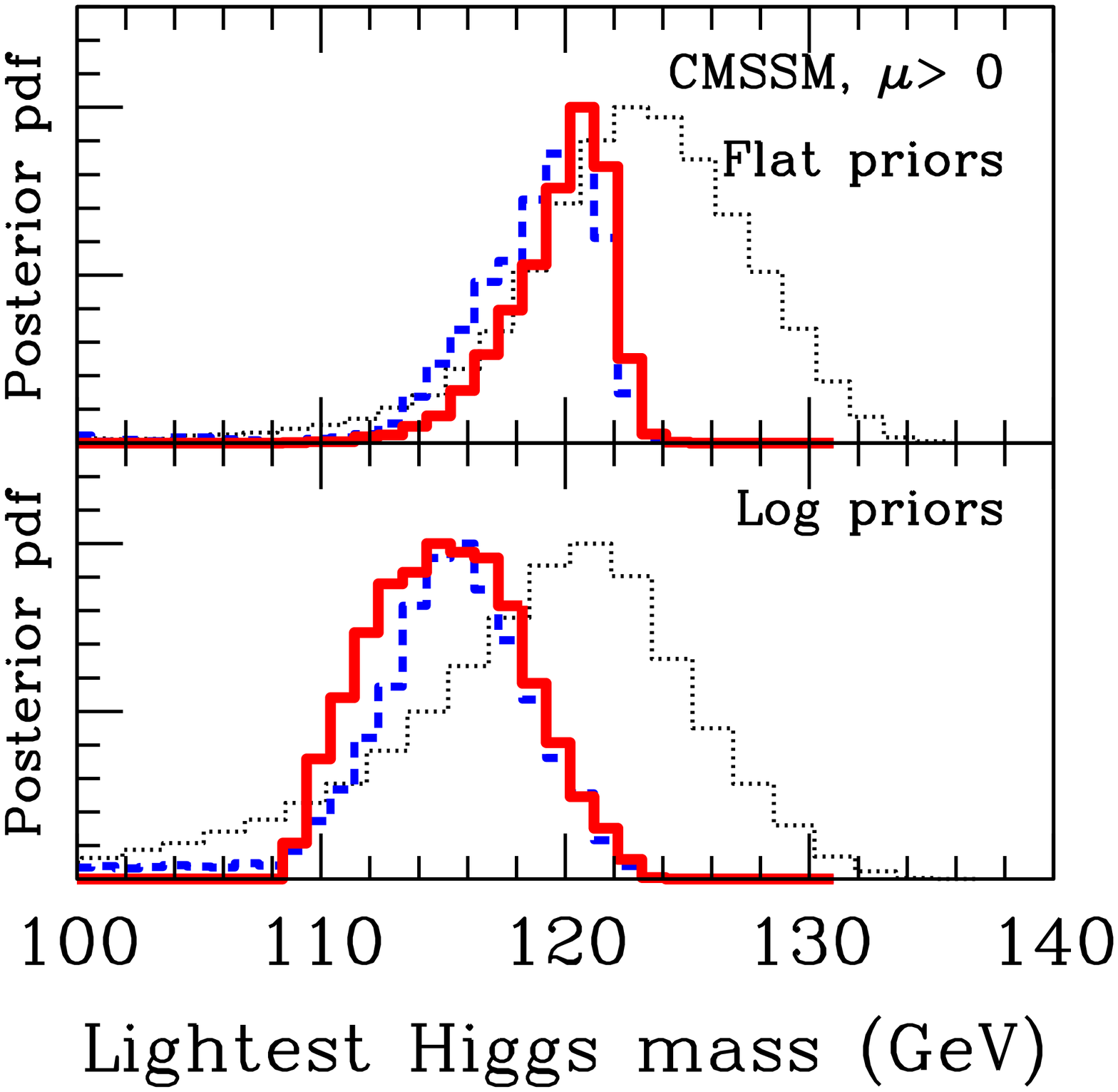}
\caption[aa]{Posterior pdf for the gluino mass and the lightest Higgs,
   for flat priors (top panels) and log priors (bottom panels) for
   different combinations of data. The constraints applied increase
   with increasing line thickness. Within each panel: the dotted black
   line has only physicality constraints (\texttt{PHYS}), the blue,
   dashed line has physicality constraints, SM parameters constraints,
   collider Higgs and superpartner masses limits and CDM abundance
   data imposed (\texttt{PHYS+NUIS+COLL+CDM}), the thickest, solid red
   line has all constraints applied (\texttt{ALL}). Though not plotted
   in the figure, the profile likelihood show a qualitatively similar
   behaviour.
\label{fig:1D_mass}}
\end{center}
\end{figure}

\section{Some implications for LHC and DM searches}\label{sec:det}

We now discuss some ensuing implications for prospects of experimental
CMSSM tests at the LHC and in DM searches.  We start by plotting in
fig.~\ref{fig:1D_mass} the posterior pdf for the gluino mass
$\mgluino$ and the lightest Higgs mass $\mhl$ for the flat and log
priors and for different combinations of data. (The profile likelihood
has a broadly similar behavior and is not shown in the figure.)  Since
$\mgluino\simeq 2.7\mhalf$, its posterior distributions (including
only physicality constraints \texttt{PHYS} marked with dotted black;
the case \texttt{PHYS+NUIS+COLL+CDM} with dashed blue; and all constraints,
\texttt{ALL}, with solid red) reflect the respective plots of $\mhalf$ in
figs.~\ref{fig:1Dpriors_base},~\ref{fig:1Dpriors+nuis+lep+cdm}
and~\ref{fig:1Dpriors+nuis+lep+cdm+bsg+gm2+ewo+bphys_base}.
(Although the plot only shows the range up to $6\tev$, the pdf for
\texttt{PHYS} remains approximately flat up to $\sim 8\tev$.) In
the case of the flat prior one can observe a significant narrowing of
the spread of $\mgluino$ due to the increasing number of constraints
applied (corresponding to increasing line thickness). The log prior
instead (bottom left panel of fig.~\ref{fig:1D_mass}) features a shift
of $\mgluino$ towards lower values ($\lsim2\tev$) almost independently
of the constraining power of the data applied -- a reflection of the
log prior giving more weight to lower values of $\mhalf$ and $\mzero$,
as mentioned earlier.  The dependence of $\mgluino$ on the prior
choice is still significant but, with the LHC reach expected to be
around $2.7-3\tev$, most of the gluino mass range will be explored
even in the less optimistic case of the flat prior~\cite{rtr1,rrt3}.

Turning next to the light Higgs, in the CMSSM in most cases its
couplings to $ZZ$ and $WW$ closely resemble those of the SM Higgs
boson with the same mass. (However, note some exceptions mentioned in
subsection~\ref{sec:colcdmbsggmt}.)  With both priors the posterior
pdf again peaks more strongly and shifts to the left with an
increasing number of constraints. After all the constraints have been
applied, the posterior features a rather sharp cutoff around
$122\gev$, similarly to the result of our detailed study~\cite{rrt2}.
(Note also that for the log prior much of the Higgs mass lies below
the LEP limit on the SM-like Higgs, a reflection of our more refined
treatment of the LEP limit.) This mass range is within reach of the
currently operating Tevatron but will actually be rather challenging
for the LHC where it may take several years to explore it.

\begin{figure}[tbh!]
\begin{center}
\includegraphics[width=\ww]{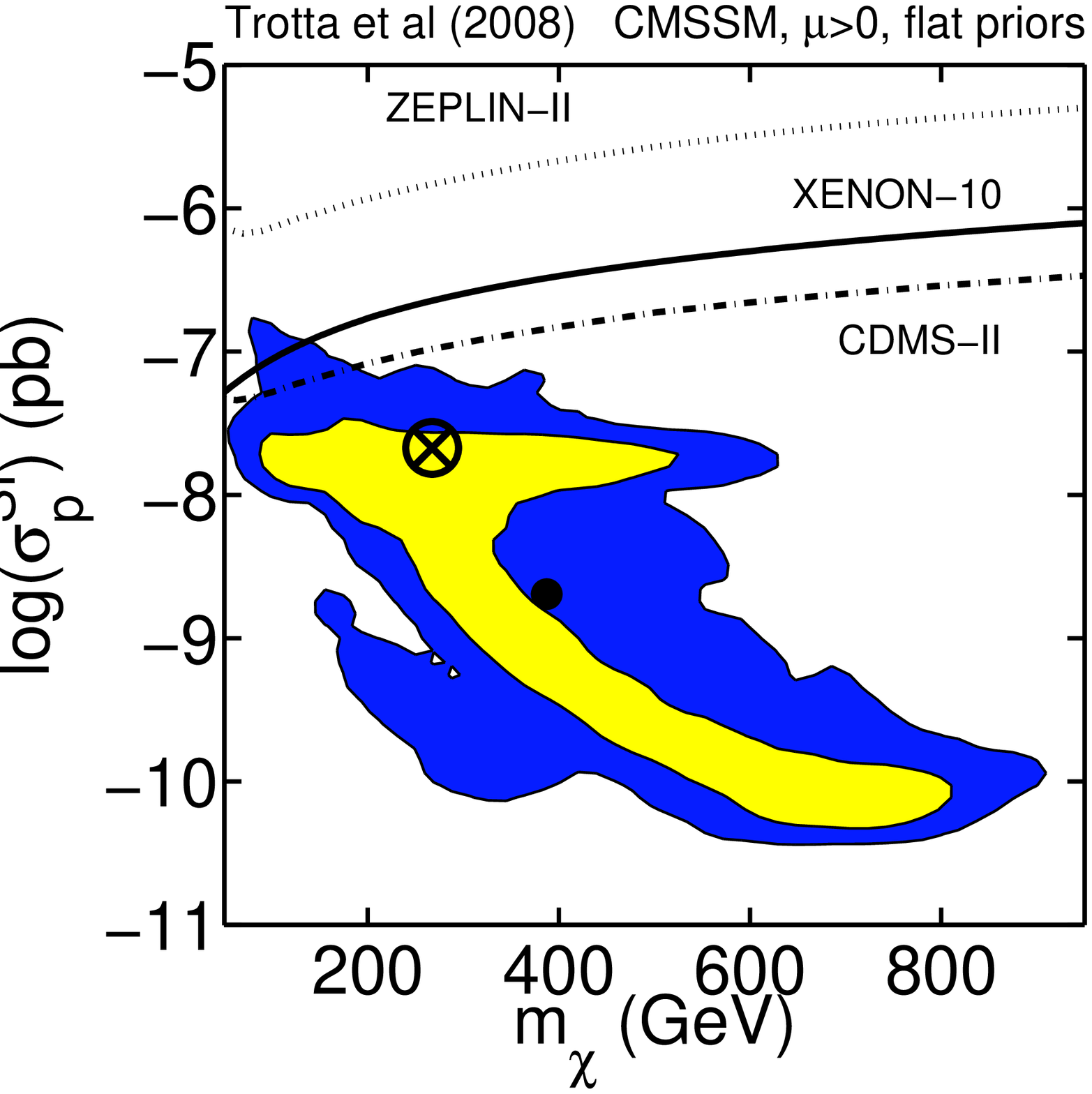}
\includegraphics[width=\ww]{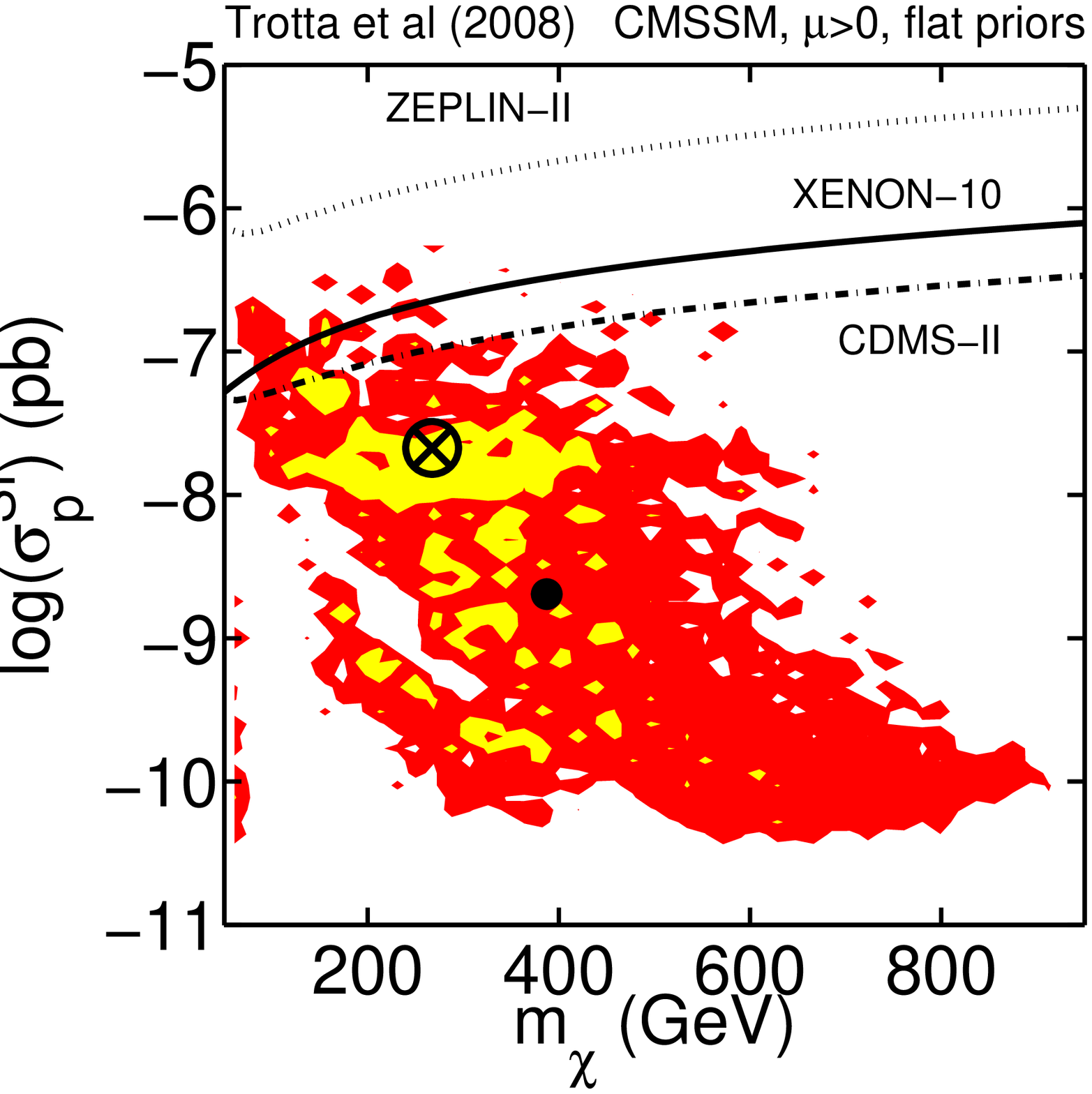} \\
\includegraphics[width=\ww]{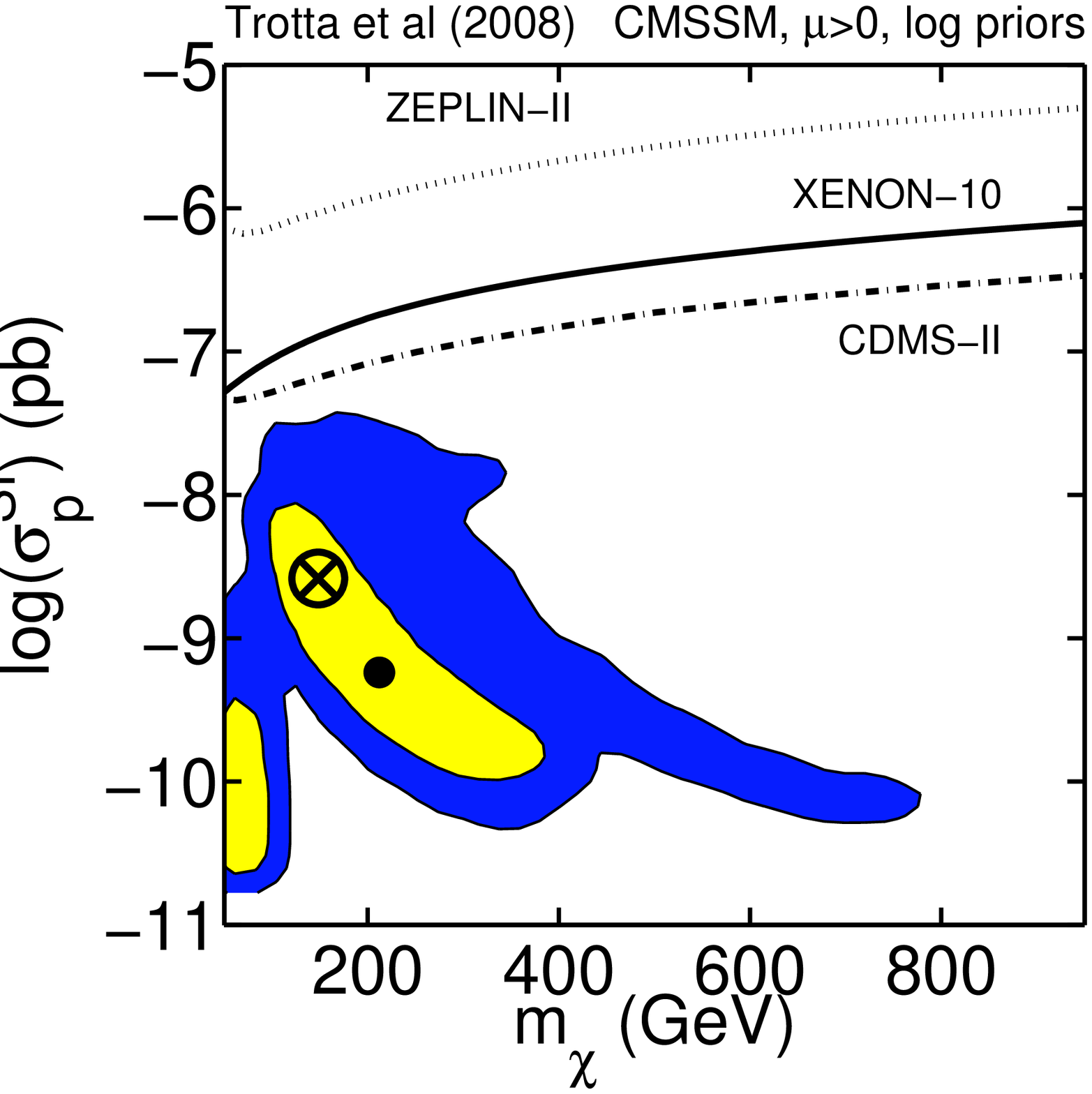}
\includegraphics[width=\ww]{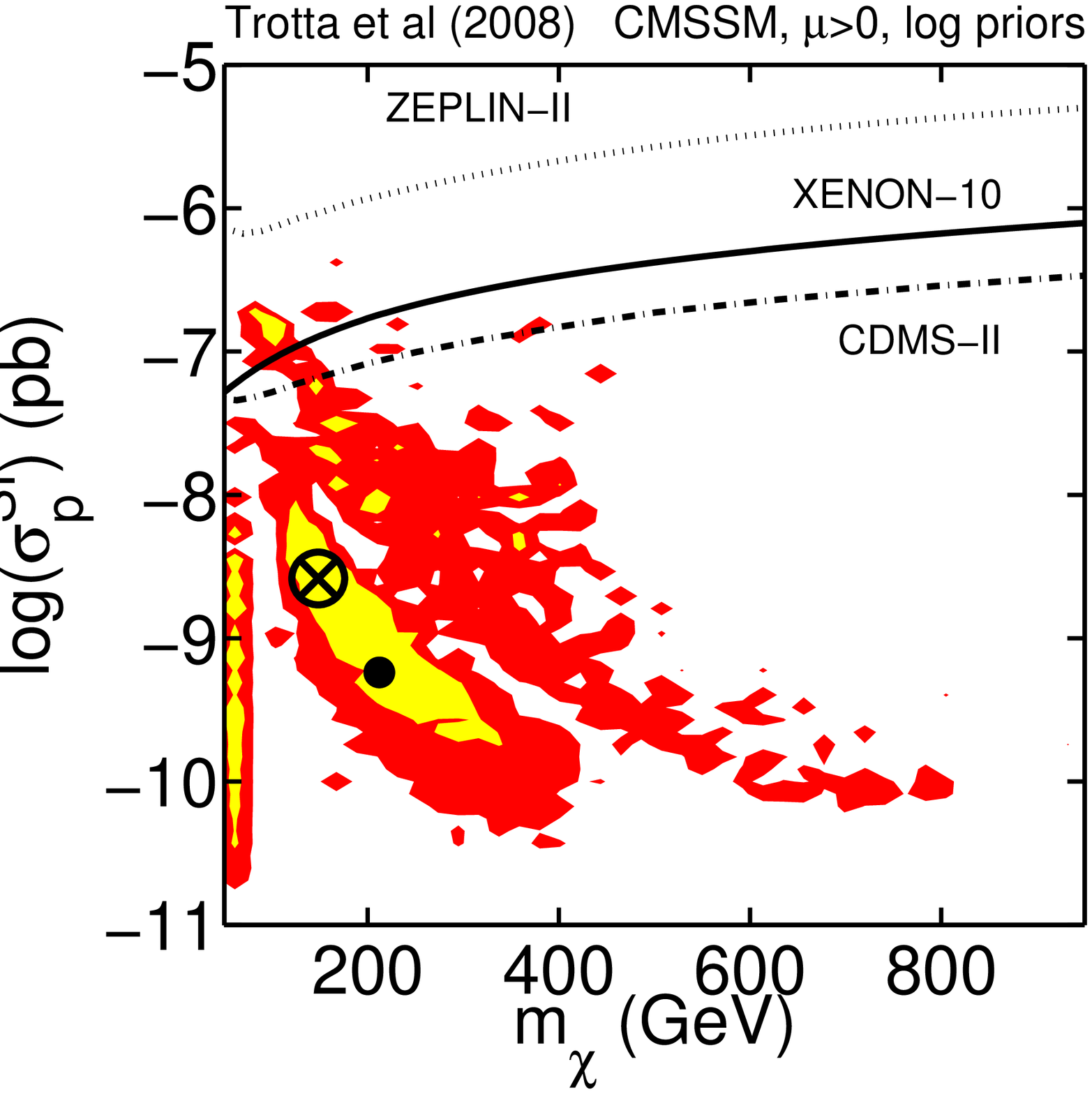}
\caption[aa]{
Posterior pdf (left column) and profile likelihood (right
column) for the spin-independent scattering cross section of the
neutralino WIMP off a proton versus the neutralino mass, for flat
priors (top row) and log priors (bottom row), for a scan including all
available constraints
(\texttt{ALL}). The inner and
outer contours enclose the respective 68\% and 95\% regions for both
statistics. The cross gives the best-fit point, the filled circle is
the posterior mean. We also plot some recent 90\% upper limits for
comparison (which, however, have not been included as constraints in
the scan).}
\label{fig:2D_DD}
\end{center}
\end{figure}
Finally, we investigate the implications for direct dark matter
detection experiments. In fig.~\ref{fig:2D_DD} in the plane spanned
by $\sigsip$ -- the spin-independent cross section for DM neutralino
scattering off a proton -- and the neutralino mass we plot the
posterior pdf (left panels) and the profile likelihood (right panels)
for the case of the flat (upper row) and log (lower row) priors. The
current strongest experimental 90\%~CL limits from CDMS~\cite{cdms08},
XENON-10~\cite{xenon-10} and ZEPLIN-II~\cite{zeplin-ii} have also been
marked for comparison (athough they have not been imposed as constraints in the analysis).

Our presentation here follows our earlier
studies~\cite{rtr1,Trotta:2006ew,rrt2,rrt3} where the direct detection
quantities were discussed, accounting fully for the first time for all
relevant particle physics sources of uncertainty and marginalising
over nuisance parameters. (There still remain hadronic uncertainties
which can change $\sigsip$ by up to a factor of ten~\cite{eos08}.) It
was shown that, with flat priors, the strong preference for the FP
region leads to a rather optimistic scenario for spin-independent
scattering off a nucleon, as most of the posterior probability was
found to be concentrated around $\sigsip \sim 10^{-8}\pb$.

Our updated results in fig.~\ref{fig:2D_DD} still show such relatively
high value (and a long $\mchi$--dependent tail) for the posterior pdf
for the flat prior. The profile likelihood follows a similar trend, but
shows a somewhat stronger preference for large values of $\sigsip$,
with the best-fit point around $\sigsip \sim 1.7 \times 10^{-8}\pb$.
Applying the log prior (which favors lower masses) reduces
significantly the contribution from the FP region. The best-fit point
shifts to a value which is about one order of magnitude below the
best-fit point found with the flat prior scan. (However notice from
table~\ref{tab:KL} that the quality of fit of both points is very
similar.) Finally, we have also investigated the case where all
constraints but the $\gmt$ observation are applied. Although this is
not shown here, this case yields very similar results to the case
\texttt{ALL} plotted in fig.~\ref{fig:2D_DD}.

The dependence on the choice of priors remains significant, which
calls for caution in drawing strong conclusions regarding prospects
for DM searches.\footnote{It was recently argued
in ref.~\cite{Allanach:2008iq} that, using a different parameterization of
the CMSSM leads to even more optimistic detection prospects. This
dependence on the choice of parameterization can be seen as another way of
phrasing the prior dependence and therefore the same caution applies
in this case.}  Despite this, with experiments aiming to reach down to
$10^{-10}\pb$ most of the high-probability range of $\sigsip$ will be
covered.

In conclusion, the current data are not yet constraining enough to
allow one to reliably predict values of some key observables discussed
here. However, even at present the predicted spread of their values
make prospects for LHC searches for gluino and light Higgs (the latter
also at the Tevatron) and DM searches in direct detection highly
encouraging.

\section{Summary and conclusions}
\label{sec:conclusions}

We have subjected current constraints for the CMSSM parameters to a
detailed scrutiny using a state-of-the art scanning technique
(MultiNest) which reduces the computational burden by over 2 orders of
magnitude with respect to previously employed MCMC techniques.  We
investigated the impact of prior choices and of applying different
combinations of constraints, both from the point of view of Bayesian
statistics and using the profile likelihood. We have updated and
applied all relevant constraints, from cosmology, collider limits, EW
observables, $\bsg$, $\gmt$ and $b$--physics.

We have found that current data are not yet constraining enough to allow drawing
statistically robust conclusions on allowed ranges for the CMSSM
parameters. Conclusions regarding the value of $\mzero$ and $\tanb$
are particularly sensitive to the choice of priors, statistics and
data included. We find that in general values of $\mhalf \lsim 2\tev$
are preferred, while for $\azero$ positive values are weakly favored. We have
highlighted the complex interplay between priors, observables and
statistics, which intrinsically limits the constraining power of the
observables on the value of the CMSSM parameters. 

For this reason we feel that it is difficult to argue that one choice
of parameters is in some sense or another superior to any other. In
particular, the standard choice of CMSSM parameters as given
by~(\ref{indeppars:eq}) is as good as the ``fundamental'' set in terms
of $\mu$ and $B$ advocated in~\cite{bclw07,Allanach:2008iq}. In fact,
if the choice of parameterization strongly impacts on the predictions
for the measurable quantities (e.g., $\sigsip$, as in
ref.~\cite{Allanach:2008iq}), this should be interpreted as a case in
which theoretical prejudice plays a stronger role than the constraints
from the data.  Clearly, better data are required in order to be able
to constrain univocally (i.e., independently of the choice of priors
and statistics) the parameters of the model. This conclusion is
expected to apply more generally to more complex phenomenological
models, with a larger number of free parameters than the CMSSM.

Among the observables, the most constraining role is played
by $\abundchi$, $\mhl$, $\brbsgamma$ and $\gmt$. The latter (still
somewhat controversial) constraint is singular in favoring smaller
$\mhalf$ and $\mzero$ but in a numerical analysis its impact becomes
outweighted by the other constraints, especially $\brbsgamma$ which
favors the FP region. The numerical measure of tension between the two
constraints is prior dependent but it is clear that both favor
different regions of the CMSSM parameter space.

In the light of our results, some comments are in order about the conclusions obtained in our previous works~\cite{rrt2, rrt3, Trotta:2006ew,
rrts}. Our previous findings regarding the posterior obtained with flat priors have been confirmed by the present analysis obtained using a different scanning algorithm. In particular, the preference for the FP region brought about by $\brbsgamma$~\cite{rrt3} has been exposed here more clearly, and the tension with the $\gmt$ measurement we had previously remarked has been further highlighted. As far as one is prepared to assume flat priors, these conclusions are therefore solid. This work has further investigated previous hints that current data are however not sufficiently strong to give conclusions that are fully independent on prior assumptions. This has allowed us to reinforce previous cautionary warnings on the interpretation of the posterior, which at present is still strongly influenced by the prior for some of the quantities. We also pointed out that the numerical evaluation of the profile likelihood is not immune from the influence of the chosen prior measure. Regarding direct and indirect detection prospects, we found that our previous predictions for direct detection experiments \cite{Trotta:2006ew} are robust with respect to changes in the prior and in the statistical measure. Although we have not addressed indirect detection prospects in this work (see \cite{rrts}, qualitatively we expect that the result will be dominated by residual astrophysical uncertainties (galactic halo profile, propagation parameters, boost factor) rather than by the statistical issues connected with the particle physics aspect. Therefore we can conclude that the results of~ \cite{rrts} qualitatively hold true.

We have quantified the information content of the different
combination of data using an information--theoretical measure and have
found that it is dominated (about 80\% for log priors and about 95\%
for flat priors) by the constraining power of the cosmological dark
matter abundance determination.

Finally, despite the above uncertainties, prospects for dark matter
direct detection and superpartner discovery at the LHC remain fairly
positive 

\medskip
{\bf Note added:} When this work was being finalized, a
paper~\cite{buchellislatest} appeared which employs an MCMC chi-square
analysis of the CMSSM and seems to be reaching rather different
conclusions. Ref.~\cite{buchellislatest} favor the region of much lower
$\mzero\lsim250\gev$ (at 68\%~CL) and it also claims that the determination of $\abundchi$ is
not very relevant in constraining the CMSSM parameters. We note that,
compared to~\cite{ehow06}, the chi-square expression employed
in~\cite{buchellislatest} no longer contains an extra term whose role
was to suppress (somewhat artificially) the weight of the FP
region. Also, contrary to refs~\cite{ehow06,buchellislatest},
$\abundchi$ cannot be used to unambigously determine $\mzero$ in terms
of the other CMSSM parameters if one also varies SM parameters, e.g.,
$\mtpole$ (compare fig.~4 in ref.~\cite{rrt3}).  Furthermore, there are
some indications that the code used in
refs~\cite{ehow06,buchellislatest} (\texttt{FeynHiggs}) to derive the
light Higgs mass value might disagree with the results obtained using
\texttt{SOFTSUSY} (employed here)~\cite{BenPC}.  However, without a
detailed comparison of the numerical outputs (which we have invited
the authors of~\cite{buchellislatest} to carry out), we are at present
unable to track down conclusively the reasons for the discrepancies between our
conclusions.

\medskip
{\bf Acknowledgements} \\ The authors wish to
thank Louis Lyons for many useful discussions and suggestions, as well
as Jim Berger, Merlise Clyde, Steffen Lauritzen, Tom Loredo and Nicolai Meinshausen,  
for comments and suggestions.   We are grateful to
Rachid Lemrani for setting up the online plotting tools and for developping the \texttt{SuperEGO} interactive routines (based on code by Sarah Bridle). R.T. is partially supported by the Lockyer Fellowship of the Royal
Astronomical Society, St Anne's College, Oxford, the Science and Technology Facilities Council (UK) and by the EU
FP6 Marie Curie Research \& Training Network ``UniverseNet"
(MRTN-CT-2006-035863). F.F. is supported by the Cambridge Commonwealth
Trust, Isaac Newton and the Pakistan Higher Education Commission
Fellowships. L.R. is partially supported
by the EC 6th Framework Programmes MRTN-CT-2004-503369 and
MRTN-CT-2006-035505. R.RdA is supported by the program ``Juan de la
Cierva'' of the Ministerio de Educaci\'{o}n y Ciencia of
Spain. The authors would like to thank the European Network of
Theoretical Astroparticle Physics ENTApP ILIAS/N6 under contract
number RII3-CT-2004-506222 for financial support.  The
computation was carried out largely on the the Cambridge High
Performance Computing Cluster Darwin and the authors would like to
thank Dr.~Stuart Rankin for computational assistance.

\begin{appendix}

\section{Nested Sampling and the MultiNest algorithm}\label{app:nest}

\FIGURE{{\includegraphics[width=0.3\columnwidth]{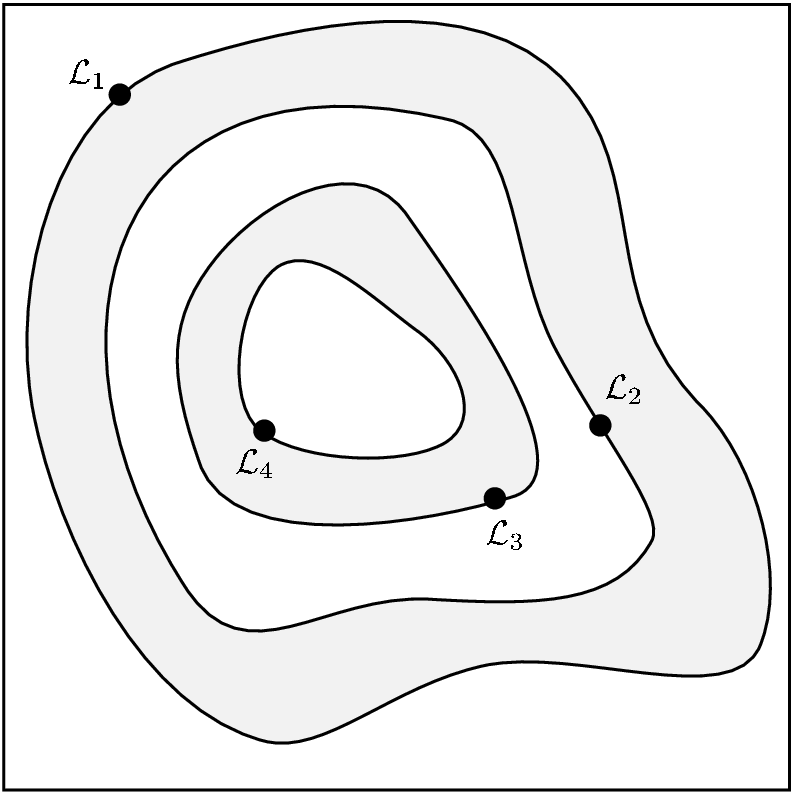}}
\hspace{0.3cm}
{\includegraphics[width=0.3\columnwidth]{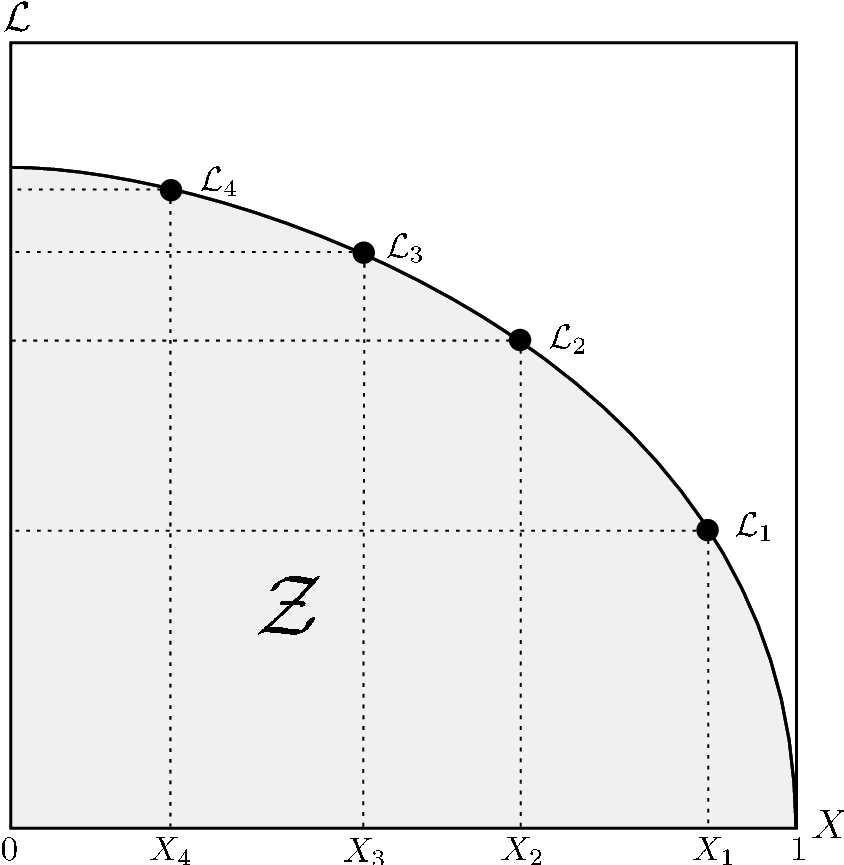}}
\caption{Cartoon illustrating (a) the posterior of a two dimensional problem; 
and (b) the transformed $L(X)$ function where the prior volumes $X_i$ are
associated with each likelihood $L_i$.}
\label{fig:nest}}

Nested sampling~\cite{SkillingNS} is a Monte Carlo technique aimed 
at efficient evaluation of the Bayesian evidence, but also produces posterior 
inferences as a by-product. It calculates the evidence by transforming the 
multi-dimensional evidence integral into a one--dimensional integral that is easy 
to evaluate numerically. This is accomplished by defining the prior volume $X$ as 
$dX = p(\basis)d^D \basis$, so that
\begin{equation}
X(\lambda) = \int_{L\left(\basis\right) > \lambda} p(\basis) d\basis,
\label{Xdef}
\end{equation}
where $\like(\basis) \equiv p(\data|\basis)$ is the likelihood function and the 
integral extends over the region(s) of parameter space contained within the 
iso-likelihood contour $\like(\basis) = \lambda$. Assuming that
$\like(X)$, i.e. the inverse of (\ref{Xdef}), is a monotonically decreasing 
function of $X$ (which is trivially satisfied for most posteriors), the evidence 
integral (\ref{evidence:eq}) can then be written as
\begin{equation} \label{eq:nested_integral}
\mathcal{Z} \equiv p(\data ) = \int_0^1 \like(X) \dr X,
\end{equation}
Thus, if one can evaluate the likelihoods $\like_{j}=\like(X_{j})$, where $X_{j}$
is a sequence of decreasing values,
\begin{equation}
0<X_{M}<\cdots <X_{2}<X_{1}< X_0=1,\label{eq:5}
\end{equation}
as shown schematically in fig.~\ref{fig:nest}, the evidence can be approximated 
numerically using standard quadrature methods as a weighted sum
\begin{equation}
\mathcal{Z}={\textstyle {\displaystyle
\sum_{i=1}^{M}}\like_{i}w_{i}}.\label{eq:6}
\end{equation} 
In the following we will use the simple trapezium rule, for which the weights 
are given by $w_i=\frac{1}{2}(X_{i-1}-X_{i+1})$. An example of a posterior in two 
dimensions and its associated function $\like(X)$ is shown in fig.~\ref{fig:nest}.

This technique allows to reduce the computational burden to about $10^5$ 
likelihood evaluations

\subsection{Evidence Evaluation}
\label{app:nested:evidence}

The nested sampling algorithm performs the summation (\ref{eq:6}) as follows. To 
begin, the iteration counter is set to $i=0$ and $N$ ``live'' (or ``active'') samples
are drawn from the full prior $p(\basis)$ (which is often simply the 
uniform distribution over the prior range), so the initial prior volume is $X_0=1$. 
The samples are then sorted in order of their likelihood and the smallest (with 
likelihood $\like_0$) is removed from the live set and replaced by a point drawn from 
the prior subject to the constraint that the point has a likelihood $\like>\like_0$. 
The corresponding prior volume contained within this iso-likelihood contour will be a 
random variable given by $X_1 = t_1 X_0$, where $t_1$ follows the distribution 
$\Pr(t) = Nt^{N-1}$ (i.e. the probability distribution for the largest of $N$ 
samples drawn uniformly from the interval $[0,1]$). At each subsequent iteration 
$i$, the discarding of the lowest likelihood point $\like_i$ in the live set, the 
drawing of a replacement with $L > L_i$ and the reduction of the corresponding 
prior volume $X_i=t_iX_{i-1}$ are repeated, until the entire prior volume has been
traversed. The algorithm thus travels through nested shells of likelihood as the 
prior volume is reduced.

The mean and standard deviation of $\log t$, which dominates the geometrical 
exploration, are:
\begin{equation}
E[\log t]=-\frac{1}{N},\qquad \sigma[\log t]=\frac{1}{N}.\label{eq:8}
\end{equation}
Since each value of $\log t$ is independent, after $i$ iterations the prior volume 
will shrink down such that $\log X_{i}\approx-(i\pm\sqrt{i})/N$. Thus, one takes
$X_i = \exp(-i/N)$.

\subsection{Stopping Criterion}
\label{nested:stopping}

The nested sampling algorithm should be terminated on determining the evidence to 
some specified precision. One way would be to proceed until the evidence estimated 
at each replacement changes by less than a specified tolerance. This could, however, 
underestimate the evidence in (for example) cases where the posterior contains any 
narrow peaks close to its maximum. \cite{SkillingNS} provides an adequate and robust
condition by determining an upper limit on the evidence that can be determined from 
the remaining set of current active points. By selecting the maximum-likelihood 
$\like_{\rm max}$ in the set of active points, one can safely assume that the largest 
evidence contribution that can be made by the remaining portion of the posterior is 
$\Delta{\mathcal{Z}}_{\rm i} = \like_{\rm max}X_{\rm i}$, i.e. the product of the 
remaining prior volume and maximum likelihood value. We choose to stop when this 
quantity would no longer change the final evidence estimate by some user-defined value 
(we use 0.5 in log-evidence).

\subsection{Posterior Inferences}
\label{nested:posterior}

Once the evidence $\mathcal{Z}$ is found, posterior inferences can be easily 
generated using the full sequence of discarded points from the nested sampling 
process, i.e. the points with the lowest likelihood value at each iteration $i$ of 
the algorithm. Each such point is simply assigned the probability weight 
\begin{equation}
p_{i}=\frac{\like_{i}w_{i}}{\mathcal{Z}}.\label{eq:12}
\end{equation}
These samples can then be used to calculate inferences of posterior parameters such as 
means, standard deviations, covariances and so on, or to construct marginalised 
posterior distributions.

\subsection{Ellipsoidal Nested Sampling}
\label{nested:ellipsoid}

The most challenging task in implementing the nested sampling algorithm is drawing samples 
from the prior within the hard constraint $\like > \like_i$ at each iteration $i$. 
Employing a naive approach that draws blindly from the prior would result in a steady 
decrease in the acceptance rate of new samples with decreasing prior volume (and increasing 
likelihood).

Ellipsoidal nested sampling \cite{Mukherjee:2005wg} tries to overcome the above problem by approximating 
the iso-likelihood contour of the point to be replaced by an $D$--dimensional ellipsoid 
determined from the covariance matrix of the current set of live points. New points are then 
selected from the prior within this (enlarged) ellipsoidal bound until one is obtained that 
has a likelihood exceeding that of the discarded lowest-likelihood point. In the limit that 
the ellipsoid coincides with the true iso-likelihood contour, the acceptance rate tends to
unity.

\subsection{MultiNest Algorithm}
\label{nested:MultiNest}

\FIGURE{
{\includegraphics[width=0.18\columnwidth]{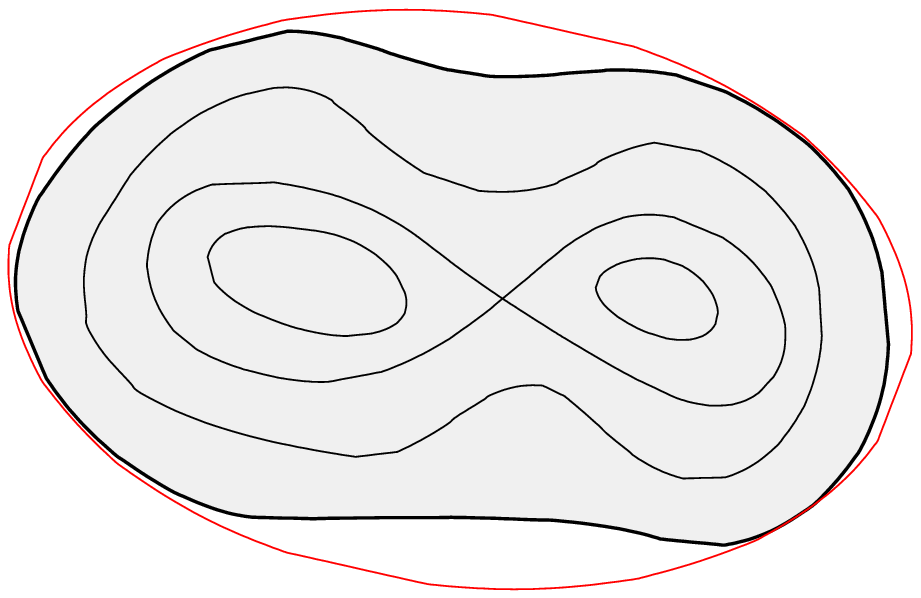}}
{\includegraphics[width=0.18\columnwidth]{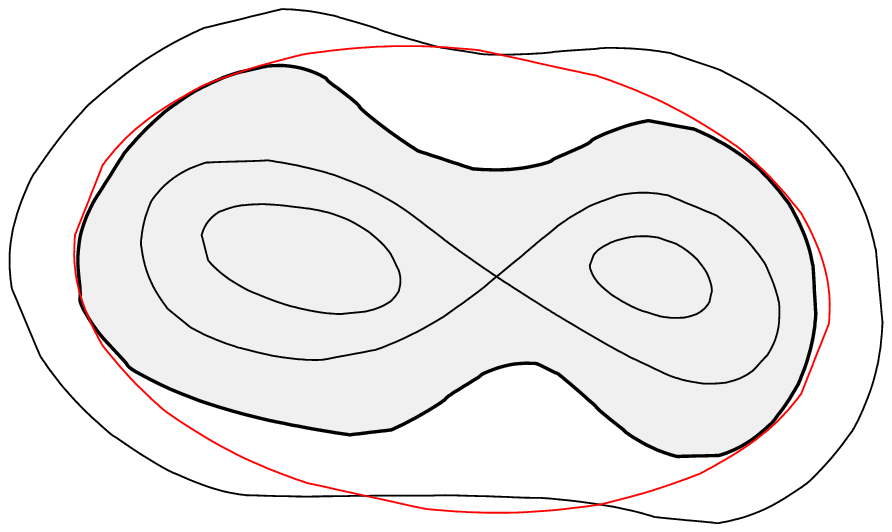}}
{\includegraphics[width=0.18\columnwidth]{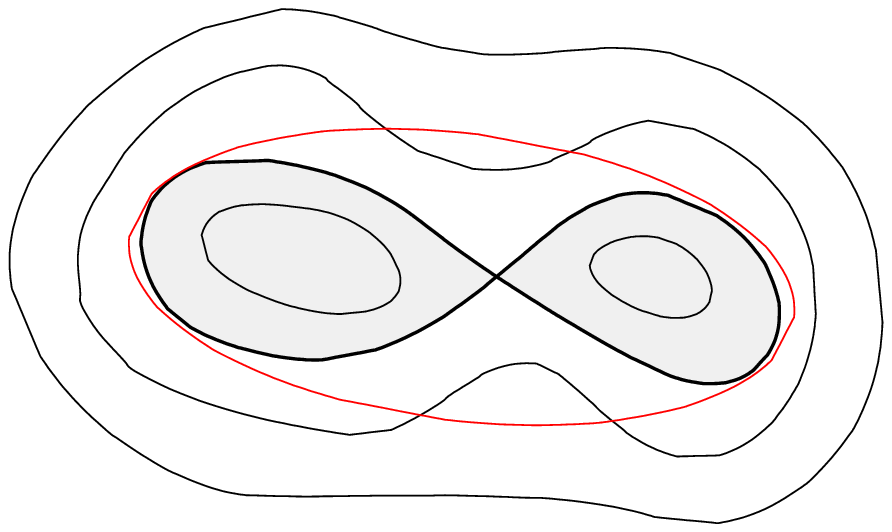}}
{\includegraphics[width=0.18\columnwidth]{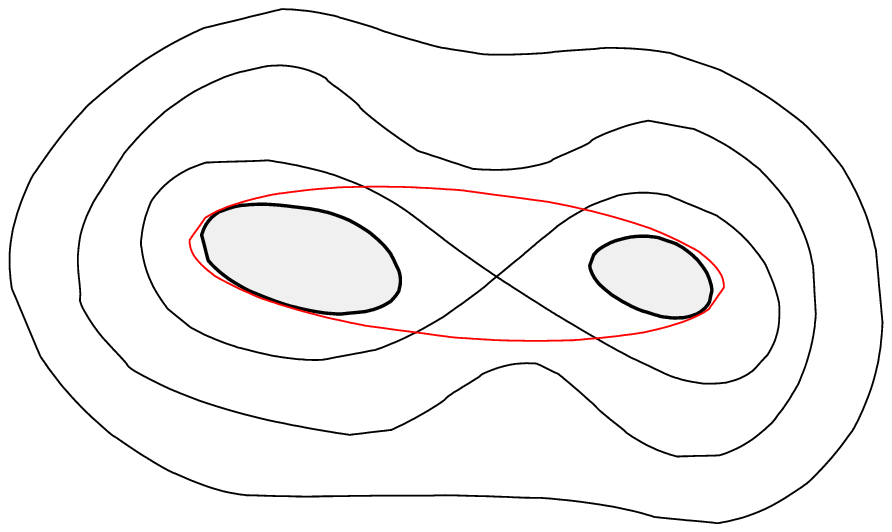}}
{\includegraphics[width=0.18\columnwidth]{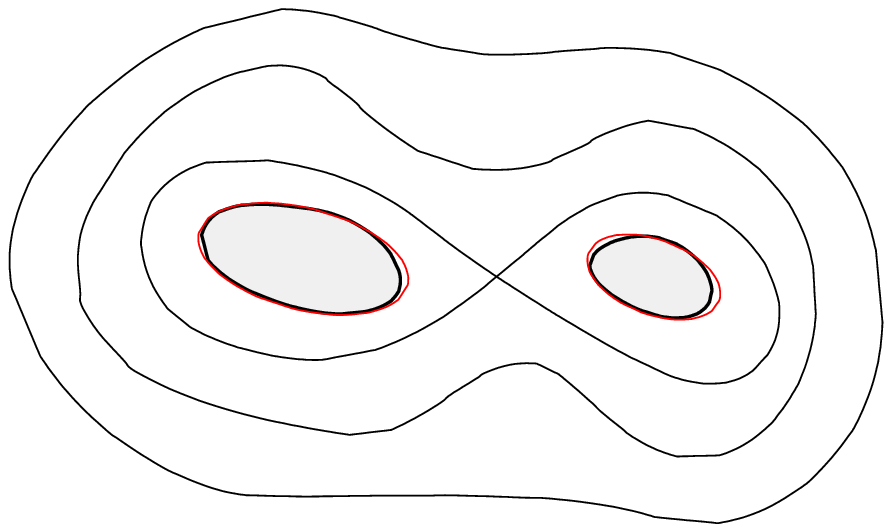}}
\caption{Cartoon of ellipsoidal nested sampling from a simple bimodal distribution. In 
the top left-hand panel, we see that the ellipsoid represents a good bound to the active 
region. Going towards the r.h.s., as we nest inward we can see that the acceptance 
rate will rapidly decrease as the bound steadily worsens. The final picture underneath 
illustrates the increase in efficiency obtained by sampling from each clustered region 
separately.}\label{fig:ellipsoid}}

\begin{figure}[tbh!]
\begin{center}
\includegraphics[width=5cm,angle=0]{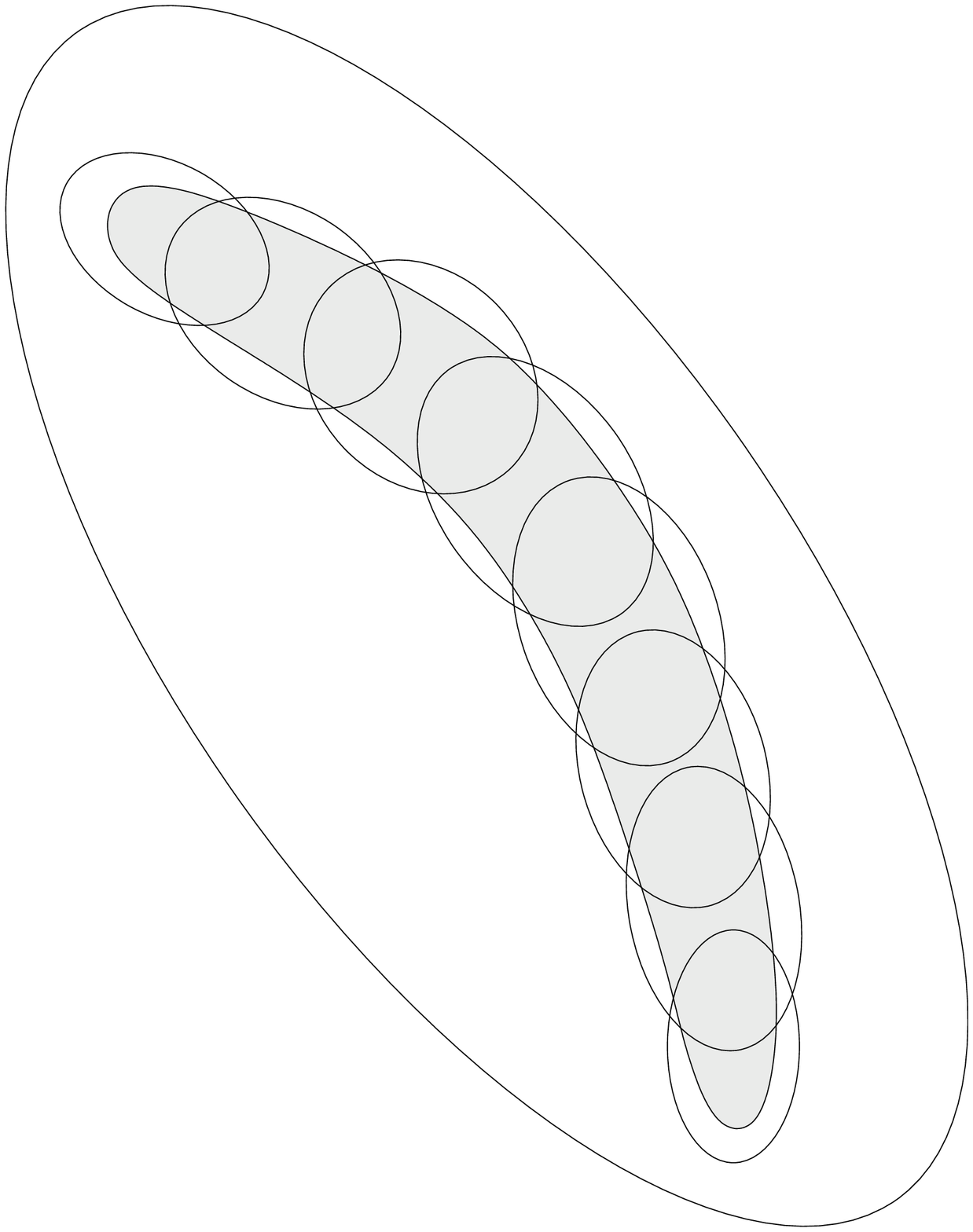}
\caption[short]{Cartoon of the sub-clustering approach used to deal with degeneracies. The true 
iso-likelihood contour contains the shaded region. The large enclosing ellipse is typical 
of that constructed using our basic method, whereas sub-clustering produces the set of small 
ellipses. \label{fig:degen}}
\end{center}
\end{figure}

Ellipsoidal nested sampling as described above is efficient for simple uni-modal posterior 
distributions without pronounced degeneracies, but is not well suited to multi-modal 
distributions. As advocated by \cite{Shaw:2007jj} and shown in fig.~\ref{fig:ellipsoid}, the 
sampling efficiency can be substantially improved by identifying distinct \emph{clusters} of 
live points that are well separated and constructing an individual ellipsoid for each cluster. 
In some problems, however, some modes of the posterior might possess a pronounced curving 
degeneracy so that it more closely resembles a (multi-dimensional) `banana'. Such features are 
problematic for all sampling methods, including the above mentioned clustered ellipsoidal 
sampling technique of \cite{Shaw:2007jj}. To sample with maximum efficiency from such distributions, 
MultiNest algorithm divides the live point set into sub-clusters which are then enclosed in 
ellipsoids and a new point is then drawn uniformly from the region enclosed by these `overlapping' 
ellipsoids. The no. of points in an individual sub-cluster and the total no. of sub-clusters is 
decided by a an `expectation-maximization' algorithm so that the total sampling volume, which is 
equal to the sum of volumes of the ellipsoids enclosing the sub-clusters, is minimized. This 
allows maximum flexibility and efficiency by breaking up a mode resembling a Gaussian into 
relatively fewer no. of sub-clusters, and if the posterior mode possesses a pronounced curving 
degeneracy so that it more closely resembles a (multi-dimensional) `banana' then it is broken into 
a relatively large no. of small `overlapping' ellipsoids. The essence of this modification is 
illustrated in fig.~\ref{fig:degen}.

The progress of the MultiNest algorithm is controlled by two main parameters: (i) the number of live 
points $N$; (ii) the maximum efficiency $f$. These values can be chosen quite easily as outlined 
below. First, $N$ should be large enough that, in the initial sampling from the full prior space, 
there is a high probability that at least one point lies in the `basin of attraction' of each mode 
of the posterior. In later iterations, live points will then tend to populate these modes. It 
should be remembered, of course, that $N$ must always exceed the dimensionality $D$ of the parameter 
space. Also, in order to calculate the evidence accurately, $N$ should be sufficiently higher so 
that all the regions of the parameter space are sampled adequately. The parameter $f$ controls the 
sampling volume $V_i$ at the $i^{th}$ iteration, which is equal to the sum of the volumes of the 
ellipoids enclosing the live point set, such that: 
\begin{equation}
V_i \geq X_i f
\end{equation} 
where $X_i$ is the prior volume at the $i^{th}$ iteration of MultiNest algorithm and $V_i > X_i f$ 
in the case when at the $i^{th}$ iteration, no set of ellipsoids enclosing the $N$ live points can 
be found such that the sum of their volumes, $V_i$, is smaller than the prior volume, $X_i$.

For all the models analysed in this paper, we used $4,000$ live points with maximum efficiency $f$ 
set to $1$. This corresponds to around $500,000$ likelihood evaluations taking approximately $48$ 
hours on $4$ $3.0$GHz Intel Woodcrest processors.

\end{appendix}


\end{document}